\documentclass[twocolumn,floatfix,eqsecnum,rmp]{revtex4}
\usepackage{graphicx}
\usepackage{amsmath}
\usepackage{bm}
\bibpunct{}{}{,}{s}{}{\textsuperscript{,}}
\newcommand{\be}{\begin{equation}}
\newcommand{\ee}{\end{equation}}
\begin{document}
\title{Quantum Transport in Semiconductor Nanostructures}
\author{C. W. J. Beenakker and H. van Houten}
\affiliation{Philips Research Laboratories, Eindhoven, The Netherlands}
\begin{abstract}
{\tt Published in Solid State Physics, {\bfseries 44}, 1-228 (1991)}
\end{abstract}
\maketitle

\tableofcontents

\section{\label{secI} Introduction}

\subsection{\label{sec1} Preface}

In recent years semiconductor nanostructures have become the model systems of choice for investigations of electrical conduction on short length scales. This development was made possible by the availability of semiconducting materials of unprecedented purity and crystalline perfection. Such materials can be structured to contain a thin layer of highly mobile electrons. Motion perpendicular to the layer is quantized, so that the electrons are constrained to move in a plane. As a model system, this {\em two-dimensional electron gas (2DEG)} combines a number of desirable properties, not shared by thin metal films. It has a low electron density, which may be readily varied by means of an electric field (because of the large screening length). The low density implies a large Fermi wavelength (typically 40 nm), comparable to the dimensions of the smallest structures (nanostructures) that can be fabricated today. The electron mean free path can be quite large (exceeding 10 $\mu{\rm m}$). Finally, the reduced dimensionality of the motion and the circular Fermi surface form simplifying factors.

Quantum transport is conveniently studied in a 2DEG because of the combination of a large Fermi wavelength and large mean free path. The quantum mechanical phase coherence characteristic of a microscopic object can be maintained at low temperatures (below 1 K) over distances of several microns, which one would otherwise have classified as macroscopic. The physics of these systems has been referred to as {\em mesoscopic},\cite{ref1} a word borrowed from statistical mechanics.\cite{ref2} Elastic impurity scattering does not destroy phase coherence, which is why the effects of quantum interference can modify the conductivity of a disordered conductor. This is the regime of diffusive transport, characteristic for disordered metals. Quantum interference becomes more important as the dimensionality of the conductor is reduced. Quasi-one dimensionality can readily be achieved in a 2DEG by lateral confinement.

Semiconductor nanostructures are unique in offering the possibility of studying quantum transport in an artificial potential landscape. This is the regime of {\em ballistic transport}, in which scattering with impurities can be neglected. The transport properties can then be tailored by varying the geometry of the conductor, in much the same way as one would tailor the transmission properties of a waveguide. The physics of this transport regime could be called {\em electron optics\/} in the solid state.\cite{ref3} The formal relation between conduction and transmission, known as the Landauer formula,\cite{ref1,ref4,ref5} has demonstrated its real power in this context. For example, the quantization of the conductance of a quantum point contact\cite{ref6,ref7} (a short and narrow constriction in the 2DEG) can be understood using the Landauer formula as resulting from the discreteness of the number of propagating modes in a waveguide.

Two-dimensional systems in a perpendicular magnetic field have the remarkable property of a quantized Hall resistance,\cite{ref8} which results from the quantization of the energy in a series of Landau levels. The magnetic length $(\hbar/eB)^{1/2}$ ($\approx$ 10 nm at $B = 5\,{\rm T}$) assumes the role of the wavelength in the quantum Hall effect. The potential landscape in a 2DEG can be adjusted to be smooth on the scale of the magnetic length, so that inter-Landau level scattering is suppressed. One then enters the regime of adiabatic transport. In this regime truly macroscopic behavior may not be found even in samples as large as 0.25 mm.

In this review we present a self-contained account of these three novel transport regimes in semiconductor nanostructures. The experimental and theoretical developments in this field have developed hand in hand, a fruitful balance that we have tried to maintain here as well. We have opted for the simplest possible theoretical explanations, avoiding the powerful --- but more formal --- Green's function techniques. If in some instances this choice has not enabled us to do full justice to a subject, then we hope that this disadvantage is compensated by a gain in accessibility. Lack of space and time has caused us to limit the scope of this review to metallic transport in the plane of a 2DEG at small currents and voltages. Transport in the regime of strong localization is
excluded, as well as that in the regime of a nonlinear current-voltage dependence. Overviews of these, and other, topics not covered here may be found in Refs. \onlinecite{ref9,ref10,ref11}, as well as in recent conference proceedings.\cite{ref12,ref13,ref14,ref15,ref16,ref17} 

We have attempted to give a comprehensive list of references to theoretical and experimental work on the subjects of this review. We apologize to those whose contributions we have overlooked. Certain experiments are discussed in some detail. In selecting these experiments, our aim has been to choose those that illustrate a particular phenomenon in the clearest fashion, not to establish priorities. We thank the authors and publishers for their kind permission to reproduce figures from the original publications. Much of the work reviewed here was a joint effort with colleagues at the Delft University of Technology and at the Philips Research Laboratories, and we are grateful for the stimulating collaboration.

The study of quantum transport in semiconductor nanostructures is motivated by more than scientific interest. The fabrication of nanostructures relies on sophisticated crystal growth and lithographic techniques that exist because of the industrial effort toward the miniaturization of transistors. Conventional transistors operate in the regime of classical diffusive transport, which breaks down on short length scales. The discovery of novel transport regimes in semiconductor nanostructures provides options for the development of innovative future devices. At this point, most of the proposals in the literature for a quantum interference device have been presented primarily as interesting possibilities, and they have not yet been critically analyzed. A quantitative comparison with conventional transistors will be needed, taking circuit design and technological considerations into account.\cite{ref18} Some proposals are very ambitious, in that they do not only consider a different principle of operation for a single transistor, but envision entire computer architectures in which arrays of quantum devices operate phase coherently.\cite{ref19}

We hope that the present review will convey some of the excitement that the workers in this rewarding field of research have experienced in its exploration. May the description of the variety of phenomena known at present, and of the simplest way in which they can be understood, form an inspiration for future investigations.

\subsection{\label{sec2} Nanostructures in Si inversion layers}

\begin{figure}
\centerline{\includegraphics[width=8cm]{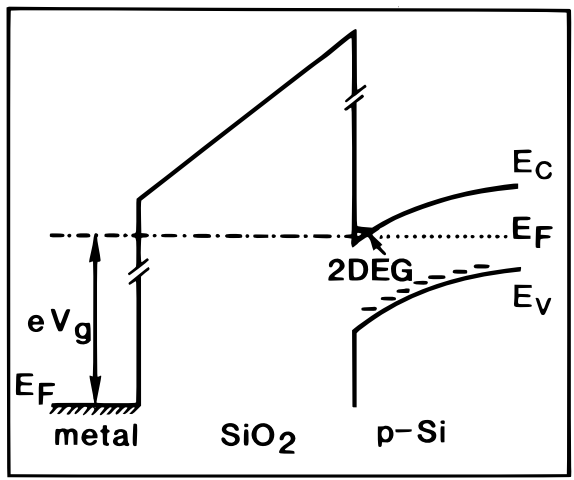}}
\caption{
Band-bending diagram (showing conduction band $E_{\rm c}$, valence band $E_{\rm v}$, and Fermi level $E_{\rm F}$) of a metal-oxide-semiconductor (MOS) structure. A 2DEG is formed at the interface between the oxide and the $p$-type silicon substrate, as a consequence of the positive voltage $V$ on the metal gate electrode.
\label{fig1}
}
\end{figure}

Electronic properties of the two-dimensional electron gas in Si MOSFET's (metal-oxide-semiconductor field-effect transistors) have been reviewed by Ando, Fowler, and Stern,\cite{ref20} while general technological and device aspects are covered in detail in the books by Sze\cite{ref21} and by Nicollian and Brew.\cite{ref22} In this section we only summarize those properties that are needed in the following. A typical device consists of a $p$-type Si substrate, covered by a ${\rm Si0}_{2}$ layer that serves as an insulator between the (100) Si surface and a metallic gate electrode. By application of a sufficiently strong positive voltage $V_{\rm g}$ on the gate, a 2DEG is induced electrostatically in the $p$-type Si under the gate. The band bending leading to the formation of this inversion layer is schematically indicated in Fig.\ \ref{fig1}. The areal electron concentration (or sheet density) $n_{\rm s}$ follows from $en_{\rm s}=C_{\rm ox}(V_{\rm g} - V_{\rm t})$, where $V_{\rm t}$ is the threshold voltage beyond which the inversion layer is created, and $C_{\rm ox}$ is the capacitance per unit area of the gate electrode with respect to the electron gas. Approximately, one has $C_{\rm ox} = \varepsilon_{\rm ox}/d_{\rm ox}$ (with $\varepsilon_{\rm ox} = 3.9\,\varepsilon_{0}$ the dielectric constant of the ${\rm Si0}_{2}$ layer),\cite{ref21} so
\begin{equation}
n_{\rm s}=\frac{\varepsilon_{\rm ox}}{ed_{\rm ox}}(V_{\rm g}-V_{\rm t}).\label{eq2.1}
\end{equation}
The linear dependence of the sheet density on the applied gate voltage is one of the most useful properties of Si inversion layers.

The electric field across the oxide layer resulting from the applied gate voltage can be quite strong. Typically, $V_{\rm g} - V_{\rm t} = 5\,{\rm V}$ and $d_{\rm ox} = 50\,{\rm nm}$, so the field strength is of order 1 MV/cm, at best a factor of 10 lower than typical fields for the dielectric breakdown of ${\rm Si0}_{2}$. It is possible to change the electric field at the interface, without altering $n_{\rm s}$, by applying an additional voltage across the $p$-$n$ junction that isolates the inversion layer from the $p$-type substrate (such a voltage is referred to as a substrate bias). At the Si-${\rm Si0}_{2}$ interface the electric field is continuous, but there is an electrostatic potential step of about 3 eV. An approximately triangular potential well is thus formed
at the interface (see Fig.\ \ref{fig1}). The actual shape of the potential deviates somewhat from the triangular one due to the electronic charge in the inversion layer, and has to be calculated self-consistently.\cite{ref20} Due to the confinement in one direction in this potential well, the three-dimensional conduction band splits into a series of two-dimensional subbands. Under typical conditions (for a sheet electron density $n_{\rm s} = 10^{11}-10^{12}\,{\rm cm}^{-2}$) only a single two-dimensional subband is occupied. Bulk Si has an indirect band gap, with six equivalent conduction band valleys in the (100) direction in reciprocal space. In inversion layers on the (100) Si surface, the degeneracy between these valleys is partially lifted. A twofold valley degeneracy remains. In the following, we treat these two valleys as completely independent, ignoring complications due to intervalley scattering. For each valley, the (one-dimensional) Fermi surface is simply a circle, corresponding to free motion in a plane with effective electron mass\cite{ref20} $m = 0.19\,m_{\rm e}$. For easy reference, this and other relevant numbers are listed in Table \ref{table1}.

\begin{figure}
\centerline{\includegraphics[width=8cm]{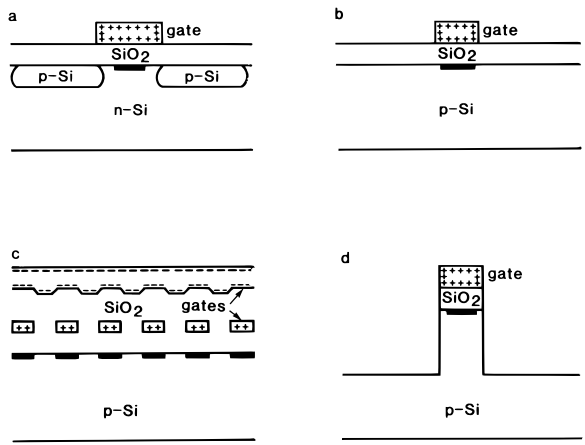}}
\caption{
Schematic cross-sectional views of the lateral pinch-off technique used to define a narrow electron accumulation layer (a), and of three different methods to define a narrow inversion layer in Si MOSFET's (b-d). Positive ($+$) and negative ($-$) charges on the gate electrodes are indicated. The location of the 2DEG is shown in black.
\label{fig2}
}
\end{figure}

The electronic properties of the Si inversion layer can be studied by capacitive or spectroscopic techniques (which are outside the scope of this review), as well as by transport measurements in the plane of the 2DEG. To determine the intrinsic transport properties of the 2DEG (e.g., the electron mobility), one defines a wide channel by fabricating a gate electrode with the appropriate shape. Ohmic contacts to the channel are then made by ion implantation, followed by a lateral diffusion and annealing process. The two current-carrying contacts are referred to as the source and the drain. One of these also serves as zero reference for the gate voltage. Additional side contacts to the channel are often fabricated as well (for example, in the Hall bar geometry), to serve as voltage probes for measurements of the longitudinal and Hall resistance. Insulation is automatically provided by the $p$-$n$ junctions surrounding the inversion layer. (Moreover, at the low temperatures of interest here, the substrate conduction vanishes anyway due to carrier freeze-out.) The electron mobility $\mu_{\rm e}$ is an important figure of merit for the quality of the device. At low temperatures the mobility in a given sample varies nonmonotonically\cite{ref20} with increasing electron density $n_{\rm s}$ (or increasing gate voltage), due to the opposite effects of enhanced screening (which reduces ionized impurity scattering) and enhanced confinement (which leads to an increase in surface roughness scattering at the Si-${\rm Si0}_{2}$ interface). The maximum low-temperature mobility of electrons in high-quality samples is around $10^4\,{\rm cm}^2 /{\rm Vs}$. This review deals with the modifications of the transport properties of the 2DEG in narrow geometries. Several lateral confinement schemes have been tried in order to achieve narrow inversion layer channels (see Fig. \ref{fig2}). Many more have been proposed, but here we discuss only those realized experimentally.

\begin{table*}
\caption{
Electronic properties of the 2DEG in GaAs-AlGaAs heterostructures and Si inversion layers.
\label{table1}
}
\begin{ruledtabular}
\begin{tabular}{lllll}
&&GaAs(100)&Si(100)&Units\\
\hline
Effective Mass&$m$&0.067&0.19&$m_{\rm e}=9.1\times10^{-28}\,{\rm g}$\\
Spin Degeneracy&$g_{\rm s}$&2&2&\\
Valley Degeneracy&$g_{\rm v}$&1&2&\\
Dielectric Constant&$\varepsilon$&13.1&11.9&$\varepsilon_{0}=8.9\times 10^{-12}\,{\rm Fm}^{-1}$\\
Density of States&$\rho(E)=g_{\rm s} g_{\rm v} (m/2\pi\hbar^{2})$&0.28&1.59&$10^{11}\,{\rm cm}^{-2}\,{\rm meV}^{-1}$\\
Electronic Sheet Density\footnotemark[1]
&$n_{\rm s}$&4&1--10&$10^{11}\,{\rm cm}^{-2}$\\
Fermi Wave Vector&$k_{\rm F} =(4\pi n_{\rm s} /g_{\rm s} g_{\rm v})^{1/2}$&1.58&0.56--1.77&$10^{6}\,{\rm cm}^{-1}$\\
Fermi Velocity&$v_{\rm F} =\hbar k_{\rm F} /m$&2.7&0.34--1.1&$10^7 \,{\rm cm/s}$\\
Fermi Energy&$E_{\rm F} =(\hbar k_{\rm F})^2/2m$&14&0.63--6.3&meV\\
Electron Mobility\footnotemark[1]&$\mu_{\rm e}$&$10^4-10^6$&$10^4$&${\rm cm}^2/{\rm Vs}$\\
Scattering Time&$\tau=m\mu_{\rm e}/e$&0.38--38&1.1&ps\\
Diffusion Constant&$D=v_{\rm F}^2\tau/2$&140--14000&6.4--64&${\rm cm}^2/{\rm s}$\\
Resistivity&$\rho=(n_{\rm s} e\mu_{\rm e})^{-1}$&1.6--0.016&6.3--0.63&${\rm k}\Omega$\\
Fermi Wavelength&$\lambda_{\rm F}=2\pi/k_{\rm F}$&40&112--35&nm\\
Mean Free Path&$l=v_{\rm F}\tau$&$10^2-10^4$&37--118&nm\\
Phase Coherence Length\footnotemark[2]&$l_{\phi}=(D\tau_{\phi})^{1/2}$&200--...&40--400&nm$(T/{\rm K})^{-1/2}$\\
Thermal Length&$l_{\rm T}=(\hbar D/k_{\rm B} T)^{1/2}$&330--3300&70--220&nm$(T/{\rm K})^{-1/2}$\\
Cyclotron Radius&$l_{\rm cycl}=\hbar k_{\rm F}/eB$&100&37--116&nm$(B/{\rm T})^{-1}$\\
Magnetic Length&$l_{\rm m}=(\hbar/eB)^{1/2}$&26&26&nm$(B/{\rm T})^{-1/2}$\\
&$k_{\rm F}l$&15.8--1580&2.1--21&\\
&$\omega_{\rm c}\tau$&1--100&1&$(B/{\rm T})$\\
&$E_{\rm F}/\hbar\omega_{\rm c}$&7.9&1--10&$(B/{\rm T})^{-1}$
\end{tabular}
\end{ruledtabular}
\footnotetext[1]{
A typical (fixed) density value is taken for GaAs-AlGaAs heterostructures, and a typical range of values in the metallic conduction regime for Si MOSFET's. For the mobility, a range of representative values is listed for GaAs-AlGaAs heterostructures, and a typical ``good'' value for Si MOSFET's. The variation in the other quantities reflects that in $n_{\rm s}$ and $\mu_{\rm e}$.}
\footnotetext[2]{
Rough estimate of the phase coherence length, based on weak localization experiments in laterally confined heterostructures\protect\cite{ref23,ref24,ref25,ref26,ref27} and Si MOSFET's.\protect\cite{ref28,ref29} The stated $T^{-1/2}$ temperature dependence should be regarded as an indication only, since a simple power law dependence is not always found (see, for example, Refs.\ \protect\onlinecite{ref30} and \protect\onlinecite{ref25}). For high-mobility GaAs-AlGaAs heterostructures the phase coherence length is not known, but is presumably\protect\cite{ref31} comparable to the (elastic) mean free path $l$.}
\end{table*}

Technically simplest, because it does not require electron beam lithography, is an approach first used by Fowler et al., following a suggestion by Pepper\cite{ref32,ref33,ref34} (Fig.\ \ref{fig2}a). By adjusting the negative voltage over $p$-$n$ junctions on either side of a relatively wide gate, they were able to vary the electron channel width as well as its electron density. This technique has been used to define narrow accumulation layers on $n$-type Si substrates, rather than inversion layers. Specifically, it has been used for the exploration of quantum transport in the strongly localized regime\cite{ref32,ref35,ref36,ref37} (which is not discussed in this review). Perhaps the technique is particularly suited to this highly resistive regime, since a tail of the diffusion profile inevitably extends into the channel, providing additional scattering centers.\cite{ref34} Some studies in the weak localization regime have also been reported.\cite{ref33}

The conceptually simplest approach (Fig.\ \ref{fig2}b) to define a narrow channel is to scale down the width of the gate by means of electron beam lithography\cite{ref38} or other advanced techniques.\cite{ref39,ref40,ref41} A difficulty for the characterization of the device is that fringing fields beyond the gate induce a considerable uncertainty in the channel width, as well as its density. Such a problem is shared to some degree by all approaches, however, and this technique has been quite successful (as we will discuss in Section \ref{secII}). For a theoretical study of the electrostatic confining potential induced by the narrow gate, we refer to the work by Laux and Stern.\cite{ref42} This is a complicated problem, which requires a self-consistent solution of the Poisson and Schr\"{o}dinger equations, and must be solved numerically.

The narrow gate technique has been modified by Warren et al.\cite{ref43,ref44} (Fig.\ \ref{fig2}c), who covered a multiple narrow-gate structure with a second dielectric followed by a second gate covering the entire device. (This structure was specifically intended to study one-dimensional superlattice effects, which is why multiple narrow gates were used.) By separately varying the voltages on the two gates, one achieves an increased control over channel width and density. The electrostatics of this particular structure has been studied in Ref.\ \onlinecite{ref43} in a semiclassical approximation.

Skocpol et al.\cite{ref29,ref45} have combined a narrow gate with a deep self-aligned mesa structure (Fig.\ \ref{fig2}d), fabricated using dry-etching techniques. One advantage of their method is that at least an upper bound on the channel width is known unequivocally. A disadvantage is that the deep etch exposes the sidewalls of the electron gas, so that it is likely that some mobility reduction occurs due to sidewall scattering. In addition, the deep etch may damage the 2DEG itself. This approach has been used successfully in the exploration of nonlocal quantum transport in multiprobe channels, which in addition to being narrow have a very small separation of the voltage probes.\cite{ref45,ref46} In another investigation these narrow channels have been used as instruments sensitive to the charging and discharging of a single electron trap, allowing a detailed study of the statistics of trap kinetics.\cite{ref46,ref47,ref48}

\subsection{\label{sec3} Nanostructures in GaAs-AlGaAs heterostructures}

\begin{figure}
\centerline{\includegraphics[width=8cm]{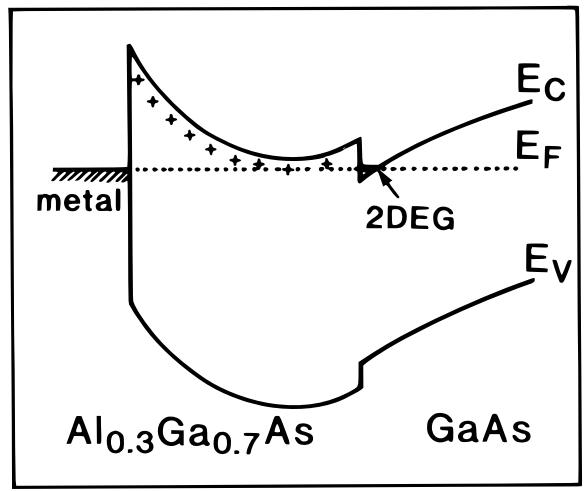}}
\caption{
Band-bending diagram of a modulation doped GaAs-${\rm Al}_x{\rm Ga}_{1-x}{\rm As}$ heterostructure. A 2DEG is formed in the undoped GaAs at the interface with the $p$-type doped AlGaAs. Note the Schottky barrier between the semiconductor and a metal electrode.
\label{fig3}
}
\end{figure}

In a modulation-doped\cite{ref49} GaAs-AlGaAs heterostructure, the 2DEG is present at the interface between GaAs and ${\rm Al}_x{\rm Ga}_{1-x}{\rm As}$ layers (for a recent review, see Ref.\ \onlinecite{ref50}). Typically, the Al mole fraction $x = 0.3$. As shown in the band-bending diagram of Fig.\ \ref{fig3}, the electrons are confined to the GaAs-AlGaAs interface by a potential well, formed by the repulsive barrier due to the conduction band offset of about 0.3 V between the two semiconductors, and by the attractive electrostatic potential due to the positively charged ionized donors in the $n$-doped AlGaAs layer. To reduce scattering from these donors, the doped layer is separated from the interface by an undoped AlGaAs spacer layer. Two-dimensional sub bands are formed as a result of confinement perpendicular to the interface and free motion along the interface. An important advantage over a MOSFET is that the present interface does not interrupt the crystalline periodicity. This is possible because GaAs and AlGaAs have almost the same lattice spacing. Because of the absence of boundary scattering at the interface, the electron mobility can be higher by many orders of magnitude (see Table \ref{table1}). The mobility is also high because of the low effective mass $m = 0.067\,m_{\rm e}$ in GaAs (for a review of GaAs material properties, see Ref.\ \onlinecite{ref51}). As in a Si inversion layer, only a single two-dimensional subband (associated with the lowest discrete confinement level in the well) is usually populated. Since GaAs has a direct band gap, with a single conduction band minimum, complications due to intervalley scattering (as in Si) are absent. The one-dimensional Fermi surface is a circle, for the commonly used (100) substrate orientation.

Since the 2DEG is present ``naturally'' due to the modulation doping (i.e., even in the absence of a gate), the creation of a narrow channel now requires the selective depletion of the electron gas in spatially separated regions. In principle, one could imagine using a combination of an undoped heterostructure and a narrow gate (similarly to a MOSFET), but in practice this does not work very well due to the lack of a natural oxide to serve as an insulator on top of the AlGaAs. The Schottky barrier between a metal and (Al)GaAs (see Fig.\ \ref{fig3}) is too low (only 0.9 V) to sustain a large positive voltage on the gate. For depletion-type devices, where a negative voltage is applied on the gate, the Schottky barrier is quite sufficient as a gate insulator (see, e.g., Ref.\ \onlinecite{ref52}).

\begin{figure}
\centerline{\includegraphics[width=8cm]{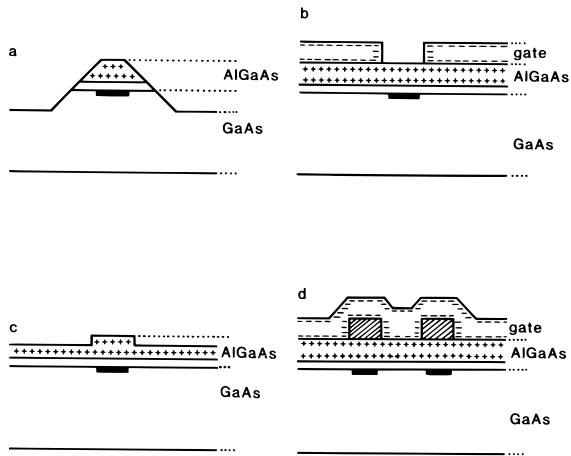}}
\caption{
Schematic cross-sectional views of four different ways to define narrow 2DEG
channels in a GaAs-AlGaAs heterostructure. Positive ionized donors and negative charges on a
Schottky gate electrode are indicated. The hatched squares in d represent unremoved resist used as a gate dielectric.
\label{fig4}
}
\end{figure}

The simplest lateral confinement technique is illustrated in Fig.\ \ref{fig4}a. The appropriate device geometry (such as a Hall bar) is realized by defining a deep mesa, by means of wet chemical etching. Wide Hall bars are usually fabricated in this way. This approach has also been used to fabricate the first micron-scale devices, such as the constrictions used in the study of the breakdown of the quantum Hall effect by Kirtley et al.\cite{ref53} and Bliek et al.,\cite{ref54} and the narrow channels used in the first study of quasi-one-dimensional quantum transport in heterostructures by Choi et al.\cite{ref55} The deep-mesa confinement technique using wet\cite{ref25,ref56} or dry\cite{ref57} etching is still of use for some experimental studies, but it is generally felt to be unreliable for channels less than 1 $\mu{\rm m}$ wide (in particular because of the exposed sidewalls of the structure).

The first working alternative confinement scheme was developed by Thornton et al.\cite{ref58} and Zheng et al.,\cite{ref24} who introduced the split-gate lateral confinement technique (Fig.\ \ref{fig4}b). On application of a negative voltage to a split Schottky gate, wide 2DEG regions under the gate are depleted, leaving a narrow channel undepleted. The most appealing feature of this confinement scheme is that the channel width and electron density can be varied continuously (but not independently) by increasing the negative gate voltage beyond the depletion threshold in the wide regions (typically about $-0.6$ V). The split-gate technique has become very popular, especially after it was used to fabricate the short and narrow constrictions known as quantum point contacts\cite{ref6,ref7,ref59} (see Section \ref{secIII}). The electrostatic confinement problem for the split-gate geometry has been studied numerically in Refs.\ \onlinecite{ref60} and \onlinecite{ref61}. A simple analytical treatment is given in in Ref. \protect\onlinecite{ref62}. A modification of the split-gate technique is the grating-gate technique, which may be used to define a 2DEG with a periodic density modulation.\cite{ref62}

The second widely used approach is the shallow-mesa depletion technique (Fig.\ \ref{fig4}c), introduced in Ref.\ \onlinecite{ref63}. This technique relies on the fact that a 2DEG can be depleted by removal of only a thin layer of the AlGaAs, the required thickness being a sensitive function of the parameters of the heterostructure material, and of details of the lithographic process (which usually involves electron beam lithography followed by dry etching). The shallow-mesa etch technique has been perfected by two groups,\cite{ref64,ref65,ref66} for the fabrication of multi probe electron waveguides and rings.\cite{ref67,ref68,ref69,ref70} Submicron trenches\cite{ref71} are still another way to define the channel. For simple analytical estimates of lateral depletion widths in the shallow-mesa geometry, see Ref.\ \onlinecite{ref72}.

A clever variant of the split-gate technique was introduced by Ford et al.\cite{ref73,ref74} A patterned layer of electron beam resist (an organic insulator) is used as a gate dielectric, in such a way that the separation between the gate and the 2DEG is largest in those regions where a narrow conducting channel has to remain after application of a negative gate voltage. As illustrated by the cross-sectional view in Fig.\ \ref{fig4}d, in this way one can define a ring structure, for example, for use in an Aharonov-Bohm experiment. A similar approach was developed by Smith et al.\cite{ref75} Instead of an organic resist they use a shallow-mesa pattern in the heterostructure as a gate dielectric of variable thickness. Initially, the latter technique was used for capacitive studies of one- and zero-dimensional confinement.\cite{ref75,ref76} More recently it was adopted for transport measurements as well.\cite{ref77} Still another variation of this approach was developed by Hansen et al.,\cite{ref78,ref79} primarily for the study of one-dimensional subband structure using infrared spectroscopy. Instead of electron beam lithography, they employ a photolithographic technique to define a pattern in the insulator. An array with a very large number of narrow lines is obtained by projecting the interference pattern of two laser beams onto light-sensitive resist. This technique is known as {\em holographic illumination\/} (see Section \ref{sec11b}).

\begin{figure}
\centerline{\includegraphics[width=8cm]{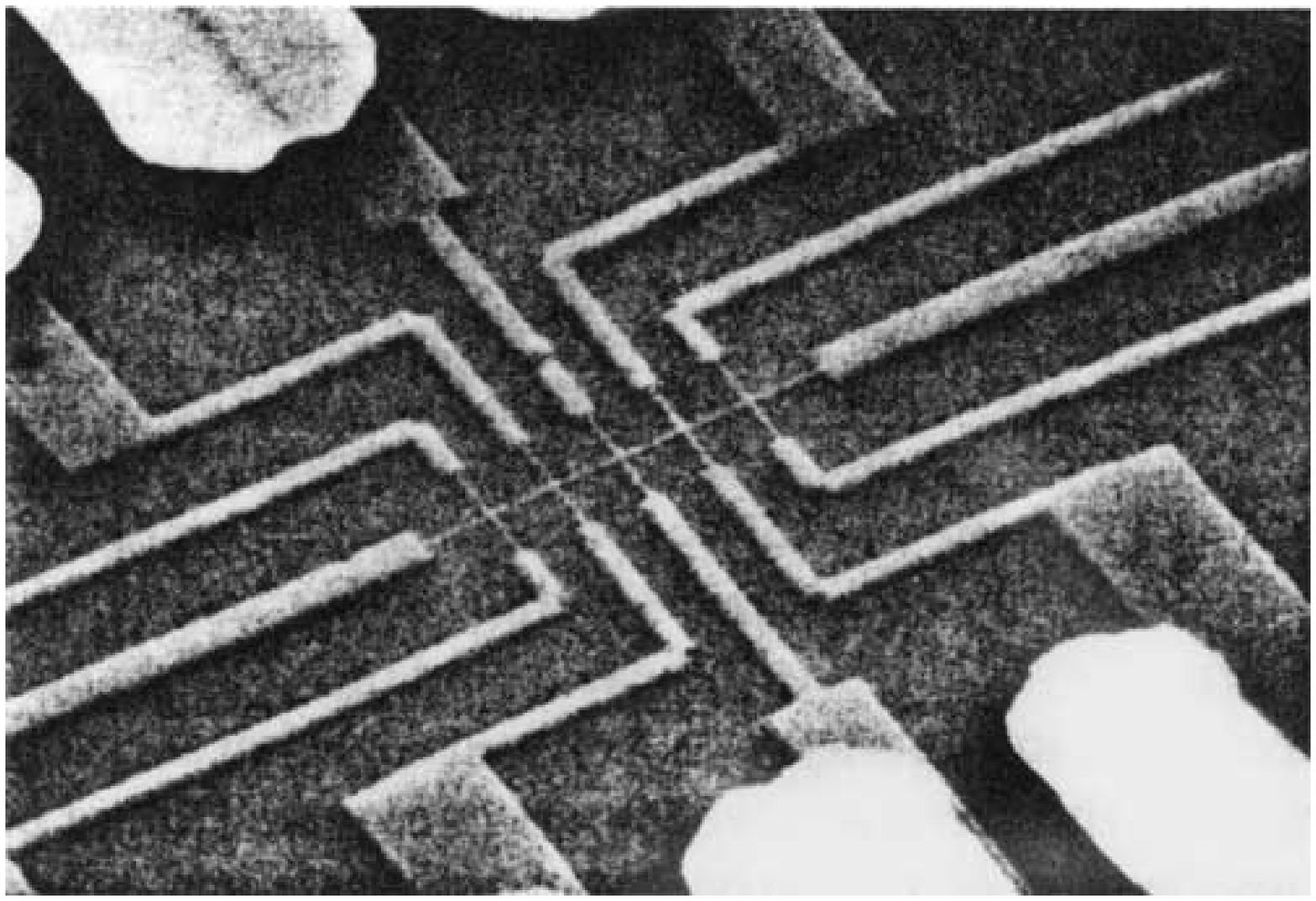}}\medskip

\centerline{\includegraphics[width=8cm]{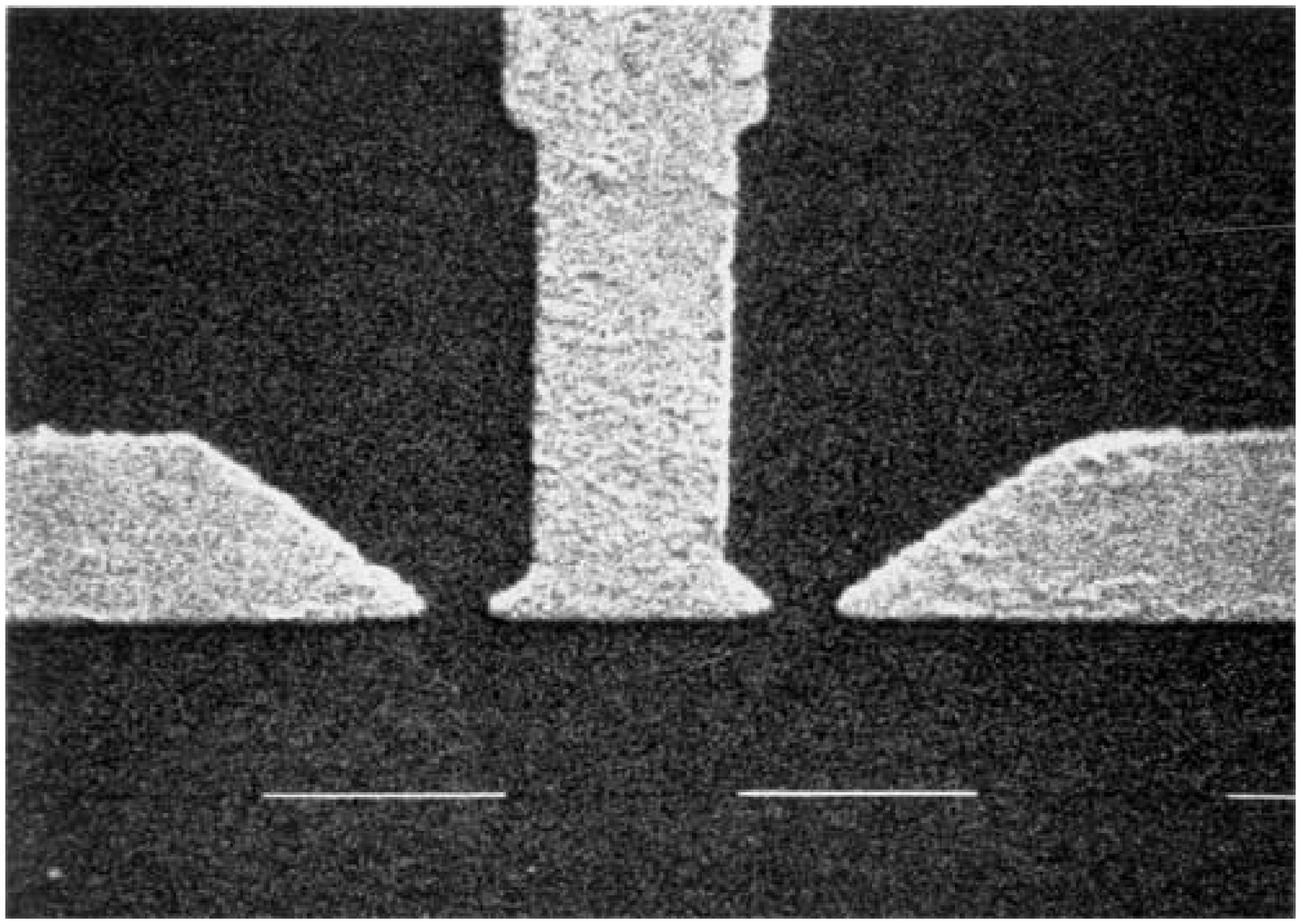}}
\caption{
Scanning electron micrographs of nanostructures in GaAs-AlGaAs heterostructures. (a, top) Narrow channel (width 75 nm), fabricated by means of the confinement scheme of Fig.\ \protect\ref{fig4}c. The channel has side branches (at a 2-$\mu{\rm m}$ separation) that serve as voltage probes. Taken from M. L. Roukes et al., Phys.\ Rev.\ Lett.\ {\bf 59}, 3011 (1987). (b, bottom) Double quantum point contact device, based on the confinement scheme of Fig. \protect\ref{fig4}b. The bar denotes a length of $1\,\mu{\rm m}$. Taken from H. van Houten et al., Phys.\ Rev.\ B {\bf 39}, 8556 (1989).
\label{fig5}
}
\end{figure}

As two representative examples of state-of-the-art nanostructures, we show in Fig.\ \ref{fig5}a a miniaturized Hall bar,\cite{ref67} fabricated by a shallow-mesa etch, and in Fig.\ \ref{fig5}b a double-quantum-point contact device,\cite{ref80} fabricated by means of the split-gate technique.

Other techniques have been used as well to fabricate narrow electron gas channels. We mention selective-area ion implantation using focused ion beams,\cite{ref81} masked ion beam exposure,\cite{ref82} strain-induced confinement,\cite{ref83} lateral $p$-$n$ junctions,\cite{ref84,ref85} gates in the plane of the 2DEG,\cite{ref86} and selective epitaxial growth.\cite{ref87,ref88,ref89,ref90,ref91,ref92} For more detailed and complete accounts of nanostructure fabrication techniques, we refer to Refs.\ \onlinecite{ref9} and \onlinecite{ref13,ref14,ref15}.

\subsection{\label{sec4} Basic properties}

\subsubsection{\label{sec4a} Density of states in two, one, and zero dimensions}

\begin{figure}
\centerline{\includegraphics[width=8cm]{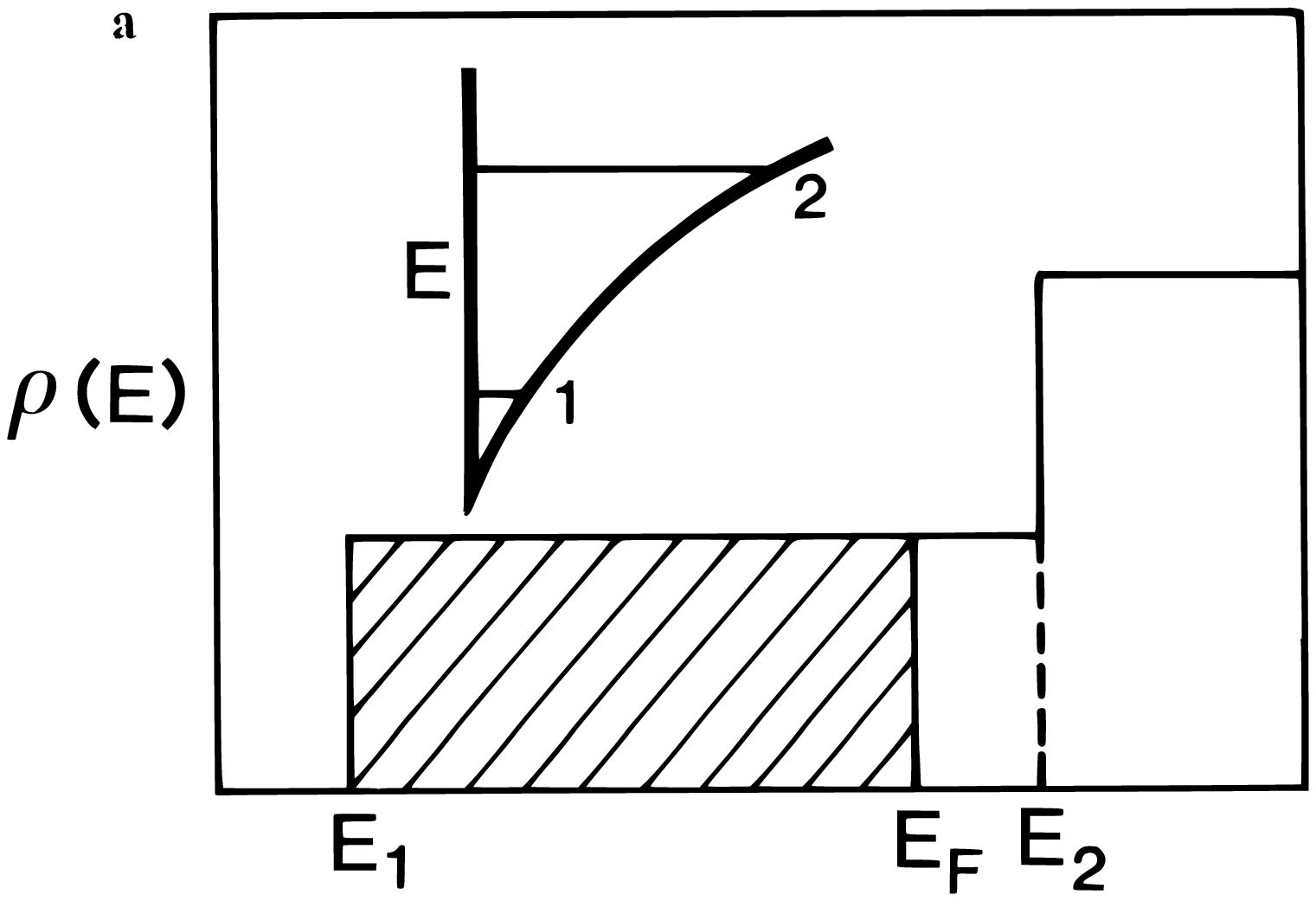}}\medskip

\centerline{\includegraphics[width=8cm]{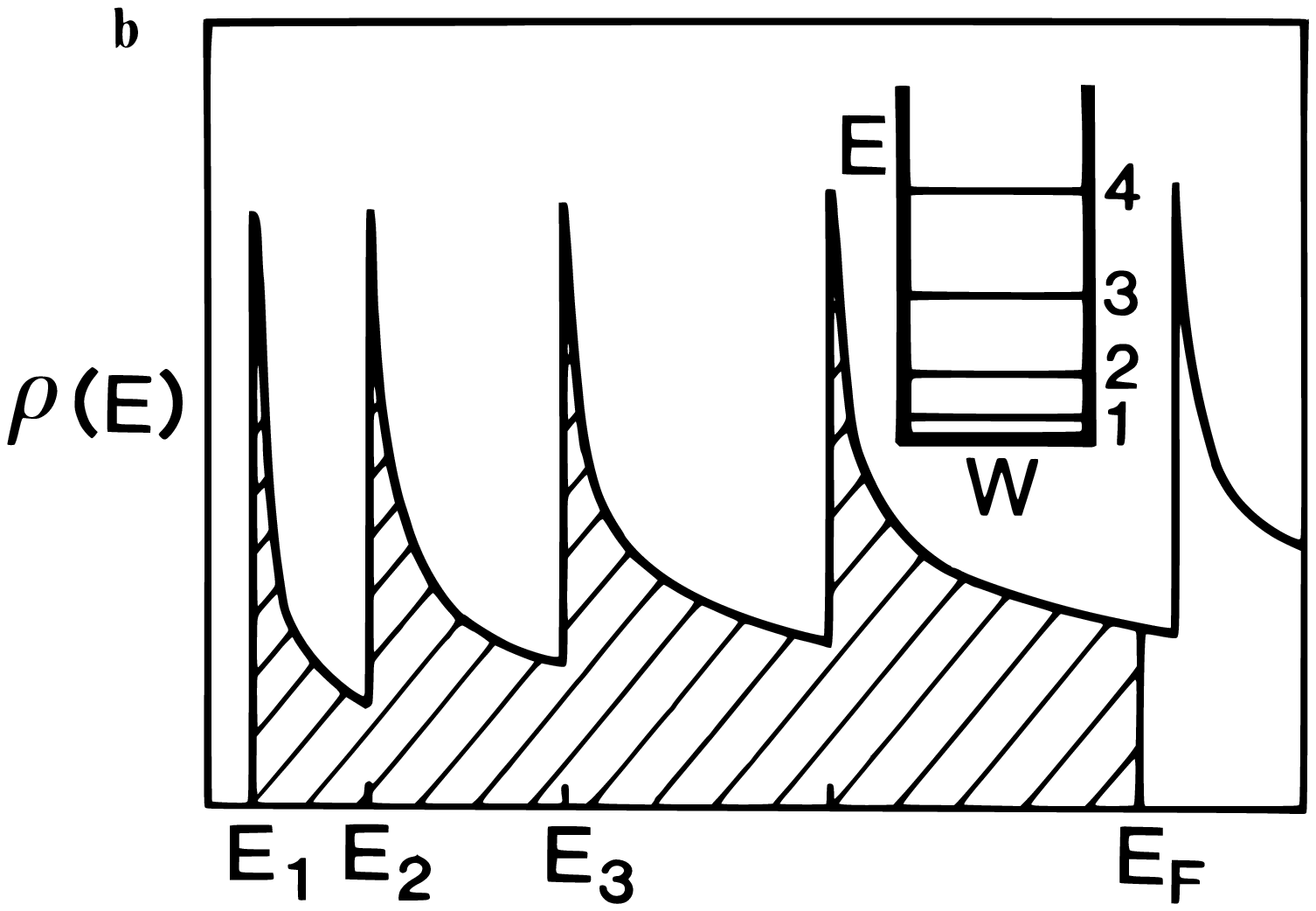}}\medskip

\centerline{\includegraphics[width=8cm]{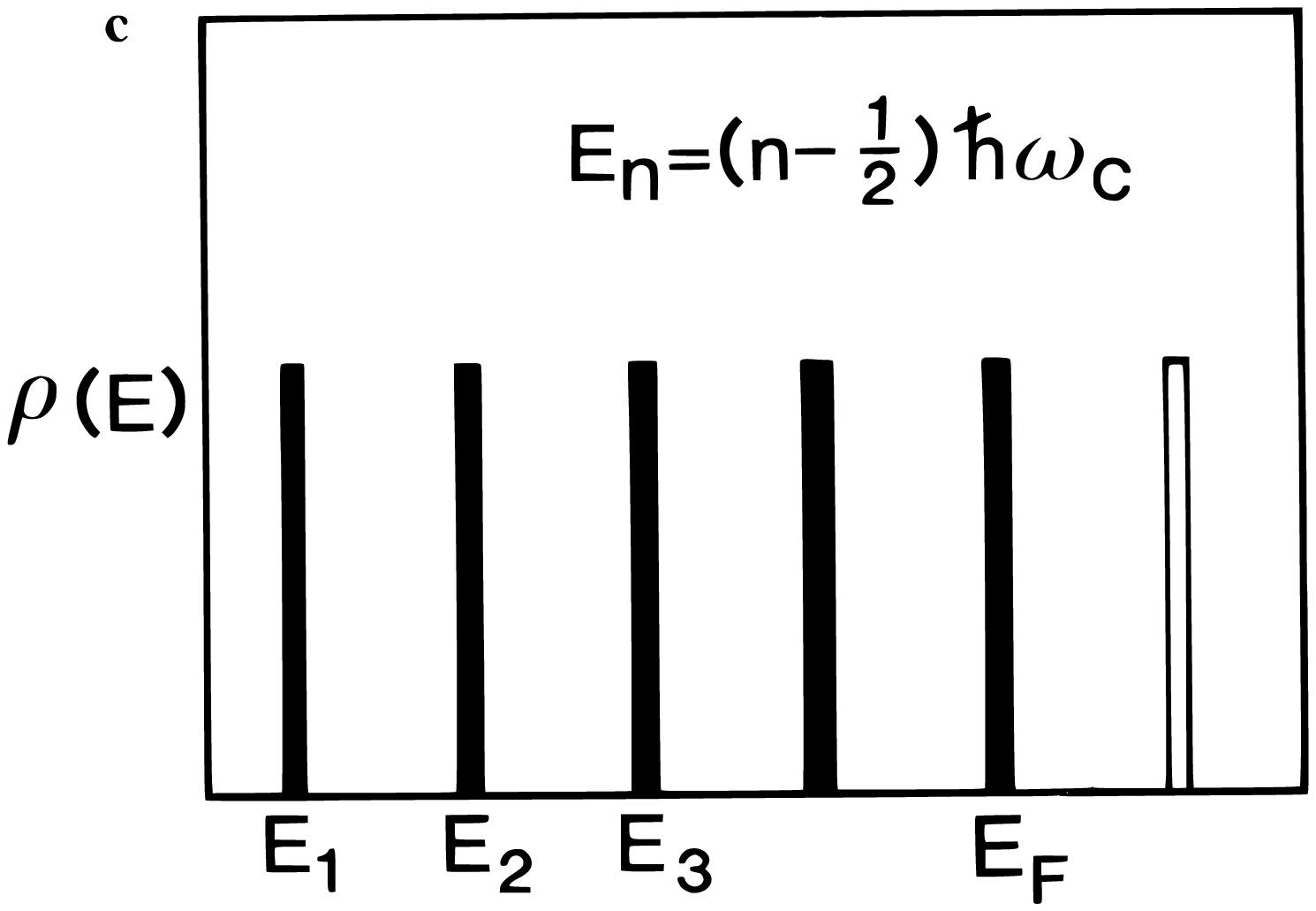}}
\caption{
Density of states $\rho(E)$ as a function of energy. (a) Quasi-2D density of states, with only the lowest subband occupied (hatched). Inset: Confinement potential perpendicular to the plane of the 2DEG. The discrete energy levels correspond to the bottoms of the first and second 2D subbands. (b) Quasi-1D density of states, with four 1D subbands occupied. Inset: Square-well lateral confinement potential with discrete energy levels indicating the 1D subband bottoms. (c) Density of states for a 2DEG in a perpendicular magnetic field. The occupied 0D subbands or Landau levels are shown in black. Impurity scattering may broaden the Landau levels, leading to a nonzero density of states between the peaks.
\label{fig6}
}
\end{figure}

The energy of conduction electrons in a single subband of an unbounded 2DEG, relative to the bottom of that subband, is given by
\begin{equation}
E(k)=\hbar^{2}k^{2}/2m,\label{eq4.1}
\end{equation}
as a function of momentum $\hbar k$. The effective mass $m$ is considerably smaller than the free electron mass $m_{e}$ (see Table \ref{table1}), as a result of interactions with the lattice potential. (The incorporation of this potential into an effective mass is an approximation\cite{ref20} that is completely justified for the present purposes.) The density of states $\rho(E)\equiv dn(E)/dE$ is the derivative of the number of electronic states $n(E)$ (per unit surface area) with energy smaller than $E$. In $k$-space, these states are contained within a circle of area $A=2\pi mE/\hbar^2$ [according to Eq.\ (\ref{eq4.1})], which contains a number $g_{\rm s} g_{\rm v} A/(2\pi)^2$ of distinct states. The factors $g_{\rm s}$ and $g_{\rm v}$ account for the spin degeneracy and valley degeneracy, respectively (Table \ref{table1}). One thus finds that $n(E) = g_{\rm s} g_{\rm v} mE/2\pi \hbar^2$, so the density of states corresponding to a single subband in a 2DEG,
\begin{equation}
\rho(E) = g_{\rm s} g_{\rm v} mE/2\pi \hbar^2,\label{eq4.2}
\end{equation}
is {\em independent\/} of the energy. As illustrated in Fig. \ref{fig6}a, a sequence of subbands is associated with the set of discrete levels in the potential well that confines the 2DEG to the interface. At zero temperature, all states are filled up to the Fermi energy $E_{\rm F}$ (this remains a good approximation at finite temperature if the thermal energy $k_{\rm B} T\ll E_{\rm F}$). Because of the constant density of states, the electron (sheet) density $n_{\rm s}$ is linearly related to $E_{\rm F}$ by $n_{\rm s} = E_{\rm F} g_{\rm s} g_{\rm v} m/2\pi \hbar^2$. The Fermi wave number $k_{\rm F} =(2mE_{\rm F} /\hbar^2)^{1/2}$ is thus related to the density by $k_{\rm F} = (4\pi n_{\rm s} /g_{\rm s} g_{\rm v})^{1/2}$. The second subband starts to be populated when $E_{\rm F}$ exceeds the energy of the second band bottom. The stepwise increasing density of states shown in Fig. \ref{fig6}a is referred to as {\em quasi}-two-dimensional. As the number of occupied subbands increases, the density of states eventually approaches the $\sqrt{E}$ dependence characteristic for a three-dimensional system. Note, however, that usually only a single subband is occupied.

If the 2DEG is confined laterally to a narrow channel, then Eq.\ (\ref{eq4.1}) only represents the kinetic energy from the free motion (with momentum $\hbar k$) {\em parallel\/} to the channel axis. Because of the lateral confinement, a single two-dimensional (2D) subband is split itself into a series of one-dimensional (1D) subbands, with band bottoms at $E_n$, $n = 1,2,\ldots$ The total energy $E_n (k)$ of an
electron in the $n$th 1D subband (relative to the bottom of the 2D subband) is given by
\begin{equation}
E_n (k)=E_n + \hbar^2 k^2 /2m.\label{eq4.3}
\end{equation}
Two frequently used potentials to model analytically the lateral confinement are the square-well potential (of width $W$, illustrated in Fig.\ \ref{fig6}b) and the parabolic potential well (described by $V(x)=\frac{1}{2}m\omega_{0}^2 x^2$). The confinement levels are then given either by $E_n = (n\pi \hbar)^2 /2mW^2$ for the square well or by $E_n = (n - \frac{1}{2})\hbar\omega_0$ for the parabolic well. When one considers electron transport through a narrow channel, it is useful to distinguish between states with positive and negative $k$, since these states move in opposite directions along the channel. We denote by $\rho_n^{+}(E)$ the density of states with $k > 0$ per unit channel length in the $n$th 1D subband. This quantity is given by
\begin{eqnarray}
\rho_{n}^{+}(E)&=&g_s g_v \left(2\pi\frac{dE_{n}(k)}{dk}\right)^{-1}\nonumber\\
&=&g_s g_v \frac{m}{2\pi\hbar^{2}}\left(\frac{\hbar^{2}}{2m(E-E_n)}\right)^{1/2}.\label{eq4.4}
\end{eqnarray}
The density of states $\rho_{n}^{-}$ with $k < 0$ is identical to $\rho_{n}^{+}$. (This identity holds because of time-reversal symmetry; In a magnetic field, $\rho_{n}^{+}\neq \rho_{n}^{-}$, in general.) The total density of states $\rho(E)$, drawn in Fig.\ \ref{fig6}b, is twice the result (\ref{eq4.4}) summed over all $n$ for which $E_n < E$. The density of states of a quasi-one-dimensional electron gas with many occupied 1D subbands may be approximated by the 2D result (\ref{eq4.2}).

If a magnetic field $B$ is applied perpendicular to an unbounded 2DEG, the energy spectrum of the electrons becomes fully discrete, since free translational motion in the plane of the 2DEG is impeded by the Lorentz force. Quantization of the circular cyclotron motion leads to energy levels at\cite{ref93}
\begin{equation}
E_n =(n-{\textstyle\frac{1}{2}})\hbar\omega_{\rm c},\label{eq4.5}
\end{equation}
with $\omega_{\rm c} = eB/m$ the cyclotron frequency. The quantum number $n = 1, 2,\ldots$ labels the Landau levels. The number of states is the same in each Landau level and equal to one state (for each spin and valley) per flux quantum $h/e$ through the sample. To the extent that broadening of the Landau levels by disorder can be neglected, the density of states (per unit area) can be approximated by
\begin{equation}
\rho(E)=g_s g_v \frac{eB}{h}\sum_{n=1}^{\infty}\delta(E-E_{n}),\label{eq4.6}
\end{equation}
as illustrated in Fig.\ \ref{fig6}c. The spin degeneracy contained in Eq.\ (\ref{eq4.6}) is resolved in strong magnetic fields as a result of the Zeeman splitting $g\mu_{\rm B}B$ of the Landau levels ($\mu_{\rm B}\equiv e\hbar/2m_{\rm e}$ denotes the Bohr magneton; the Land\'{e} $g$-factor is a complicated function of the magnetic field in these systems).\cite{ref20} Again, if a large number of Landau levels is occupied (i.e., at weak magnetic fields), one recovers approximately the 2D result (\ref{eq4.2}). The foregoing considerations are for an unbounded 2DEG. A magnetic field perpendicular to a narrow 2DEG channel causes the density of states to evolve gradually from the 1D form of Fig.\ \ref{fig6}b to the effectively 0D form of Fig.\ \ref{fig6}c. This transition is discussed in Section \ref{sec10}.

\subsubsection{\label{sec4b} Drude conductivity, Einstein relation, and Landauer formula}

In the presence of an electric field $\mathbf{E}$ in the plane of the 2DEG, an electron
acquires a drift velocity $\mathbf{v}=-e\mathbf{E}\Delta t/m$ in the time $\Delta t$ since the last impurity collision. The average of $\Delta t$ is the scattering time $\tau$, so the average drift
velocity $\mathbf{v}_{\mathrm{drift}}$ is given by
\begin{equation}
v_{\mathrm{drift}}=-\mu_{\mathrm{e}}\mathrm{E},\;\;\mu_{\mathrm{e}}=e\tau/m.\label{eq4.7}
\end{equation}
The electron mobility $\mu_{\mathrm{e}}$ together with the sheet density $n_{\mathrm{s}}$ determine the conductivity $\sigma$ in the relation $-en_{\mathrm{s}}\mathbf{v}_{\mathrm{drift}}=\sigma \mathbf{E}$. The result is the familiar Drude conductivity,\cite{ref94} which can be written in several equivalent forms:
\begin{equation} \sigma=en_{\mathrm{s}}\mu_{\mathrm{e}}=\frac{e^{2}n_{\mathrm{s}}\tau}{m}=g_{\mathrm{s}}
g_{\mathrm{v}}\frac{e^{2}}{h}\frac{k_{\mathrm{F}}l}{2}.\label{eq4.8}
\end{equation}
In the last equality we have used the identity $n_{\mathrm{s}}=g_{\mathrm{s}}g_{\mathrm{v}}k_{\mathrm{
F}}^{2}/4\pi$ (see Section \ref{sec4a}) and have defined the mean free path $l=v_{\mathrm{F}}\tau$. The dimensionless quantity $k_{\mathrm{F}}l$ is much greater than unity in metallic systems (see Table \ref{table1} for typical values in a 2DEG), so the conductivity is large compared with the quantum unit $e^{2}/h\approx(26\,\mathrm{k}\Omega)^{-1}$.

From the preceding discussion it is obvious that the current induced by the applied electric field is carried by {\it all\/} conduction electrons, since each electron acquires the same average drift velocity. Nonetheless, to determine the conductivity it is sufficient to consider the response of electrons {\it near the Fermi level\/} to the electric field. The reason is that the states that are more than a few times the thermal energy $k_{\mathrm{B}}T$ below $E_{\mathrm{F}}$ are all filled so that in response to a weak electric field only the distribution of electrons among
states at energies close to $E_{\mathrm{F}}$ is changed from the equilibrium Fermi-Dirac
distribution
\begin{equation}
f(E-E_{\mathrm{F}})=\left(1+ \exp\frac{E-E_{\mathrm{F}}}{k_{\mathrm{B}}T}\right)^{-1}.\label{eq4.9}
\end{equation}
The Einstein relation\cite{ref94}
\begin{equation}
\sigma=e^{2}\rho(E_{\mathrm{F}})D\label{eq4.10}
\end{equation}
is one relation between the conductivity and Fermi level properties (in this case the density of states $\rho(E)$ and the diffusion constant $D$, both evaluated at $E_{\mathrm{F}})$. The Landauer formula\cite{ref4} [Eq.\ (\ref{eq4.21})] is another such relation (in terms of the transmission probability at the Fermi level rather than in terms of the diffusion constant).

The Einstein relation (\ref{eq4.10}) for an electron gas at zero temperature follows
on requiring that the sum of the drift current density $-\sigma \mathbf{E}/e$ and the diffusion
current density $-D\nabla n_{\mathrm{s}}$ vanishes in thermodynamic equilibrium, characterized by a spatially constant electrochemical potential $\mu$:
\begin{equation}
-\sigma \mathbf{E}/e-D\nabla n_{\mathrm{s}}=0,\;\; \mathrm{when}\;\; \nabla\mu=0.\label{eq4.11}
\end{equation}
The electrochemical potential is the sum of the electrostatic potential energy $-eV$ (which determines the energy of the bottom of the conduction band) and the chemical potential $E_{\mathrm{F}}$ (being the Fermi energy relative to the conduction band bottom). Since (at zero temperature) $dE_{\mathrm{F}}/dn_{\mathrm{s}}=1/\rho(E_{\mathrm{F}})$, one has
\begin{equation}
\nabla\mu=e\mathbf{E}+\rho(E_{\mathrm{F}})^{-1}\nabla n_{\mathrm{s}}.\label{eq4.12}
\end{equation}
The combination of Eqs.\ (\ref{eq4.11}) and (\ref{eq4.12}) yields the Einstein relation (\ref{eq4.10}) between $\sigma$ and $D$. To verify that Eq.\ (\ref{eq4.10}) is consistent with the earlier expression (\ref{eq4.8}) for the Drude conductivity, one can use the result (see below)
for the 2D diffusion constant:
\begin{equation}
D={\textstyle\frac{1}{2}}v_{\mathrm{F}}^{2}\tau={\textstyle\frac{1}{2}}v_\mathrm{F}l,\label{eq4.13}
\end{equation}
in combination with Eq.\ (\ref{eq4.2}) for the 2D density of states.

At a finite temperature $T$, a chemical potential (or Fermi energy) gradient
$\nabla E_{\mathrm{F}}$ induces a diffusion current that is smeared out over an energy range of
order $k_{\mathrm{B}}T$ around $E_{\mathrm{F}}$. The energy interval between $E$ and $E+dE$ contributes to the diffusion current density $\mathbf{j}$ an amount $d\mathbf{j}$ given by
\begin{eqnarray}
d\mathbf{j}_{\mathrm{diff}}&=&-D\nabla\{ \rho(E)f(E-E_{\mathrm{F}})dE\}\nonumber\\
&=&-dED \rho(E)\frac{df}{dE_{\mathrm{F}}}\nabla E_{\mathrm{F}},\label{eq4.14}
\end{eqnarray}
where the diffusion constant $D$ is to be evaluated at energy $E$. The total
diffusion current density follows on integration over $E$:
\begin{equation} 
\mathbf{j}=-\nabla E_{\mathrm{F}}e^{-2}\int_{0}^{\infty} dE\, \sigma(E,0) \frac{df}{dE_{\mathrm{F}}},\label{eq4.15}
\end{equation}
with $\sigma(E,0)$ the conductivity (\ref{eq4.10}) at temperature zero for a Fermi energy equal to $E$. The requirement of vanishing current for a spatially constant
electrochemical potential implies that the conductivity $\sigma(E_{\mathrm{F}},T)$ at temperature $T$ and Fermi energy $E_{\mathrm{F}}$ satisfies
\[
\sigma(E_{\mathrm{F}},T)e^{-2}\nabla E_{\mathrm{F}}+\mathbf{j}=0. 
\]
Therefore, the finite-temperature conductivity is given simply by the energy average of the zero-temperature result
\begin{equation}
\sigma(E_{\mathrm{F}},T)=\int_{0}^{\infty} dE\, \sigma(E,0)\frac{df}{dE_{F}}.\label{eq4.16}
\end{equation}
As $T\rightarrow 0,$ $df/dE_{\mathrm{F}}\rightarrow\delta(E-E_{\mathrm{F}})$, so indeed only $E=E_{\mathrm{F}}$ contributes to the
energy average. Result (\ref{eq4.16}) contains exclusively the effects of a finite
temperature that are due to the thermal smearing of the Fermi-Dirac
distribution. A possible temperature dependence of the scattering processes is
not taken into account.

We now want to discuss one convenient way to calculate the diffusion
constant (and hence obtain the conductivity). Consider the diffusion current
density $j_{x}$ due to a small constant density gradient, $n(x)=n_{0}+cx$. We write
\begin{eqnarray}
j_{x}&=&\lim_{\Delta t\rightarrow\infty}\langle v_{x}(t=0)n(x(t=-\Delta t))\rangle\nonumber\\
&=&\lim_{\Delta t\rightarrow\infty}c\langle v_{x}(0)x(-\Delta t)\rangle\nonumber\\
&=& \lim_{\Delta t\rightarrow\infty}-c\int_{0}^{\Delta t}dt\langle v_{x}(0)v_{x}(-t)\rangle,\label{eq4.17}
\end{eqnarray}
where $t$ is time and the brackets $\langle\cdots\rangle$ denote an isotropic angular average over the Fermi surface. The time interval $\Delta t\rightarrow\infty$, so the velocity of the electron at time $0$ is uncorrelated with its velocity at the earlier time $-\Delta t$. This allows us to neglect at $x(-\Delta t)$ the small deviations from an isotropic velocity distribution induced by the density gradient [which could not have
been neglected at $x(0)]$. Since only the time difference matters in the velocity
correlation function, one has $\langle v_{x}(0)v_{x}(-t)\rangle=\langle v_{x}(t)v_{x}(0)\rangle$. We thus obtain for the diffusion constant $D=-j_{x}/c$ the familiar linear response formula\cite{ref95}
\begin{equation}
D= \int_{0}^{\infty}dt\langle v_{x}(t)v_{x}(0)\rangle.\label{eq4.18}
\end{equation}
Since, in the semiclassical relaxation time approximation, each scattering
event is assumed to destroy all correlations in the velocity, and since a
fraction $\exp(-t/\tau$) of the electrons has not been scattered in a time $t$, one has
(in 2D)
\begin{equation}
\langle v_{x}(t)v_{x}(0)\rangle=\langle v_{x}(0)^{2}\rangle \mathrm{e}^{-t/\tau}={\textstyle\frac{1}{2}}v^{2}_\mathrm{F}\mathrm{e}^{-t/\tau}. \label{eq4.19}
\end{equation}
Substituting this correlation function for the integrand in Eq.\ (\ref{eq4.18}), one
recovers on integration the diffusion constant (\ref{eq4.13}).

The Drude conductivity (4.8) is a {\it semiclassical\/} result, in the sense that
while the quantum mechanical Fermi-Dirac statistic is taken into account,
the dynamics of the electrons at the Fermi level is assumed to be classical. In
Section \ref{secII} we will discuss corrections to this result that follow from
correlations in the diffusion process due to quantum interference. Whereas
for classical diffusion correlations disappear on the time scale of the
scattering time $\tau$ [as expressed by the correlation function (\ref{eq4.19})], in quantum
diffusion correlations persist up to times of the order of the phase coherence
time. The latter time $\tau_{\phi}$ is associated with inelastic scattering and at low
temperatures can become much greater than the time $\tau$ associated with elastic
scattering.

\begin{figure}
\centerline{\includegraphics[width=8cm]{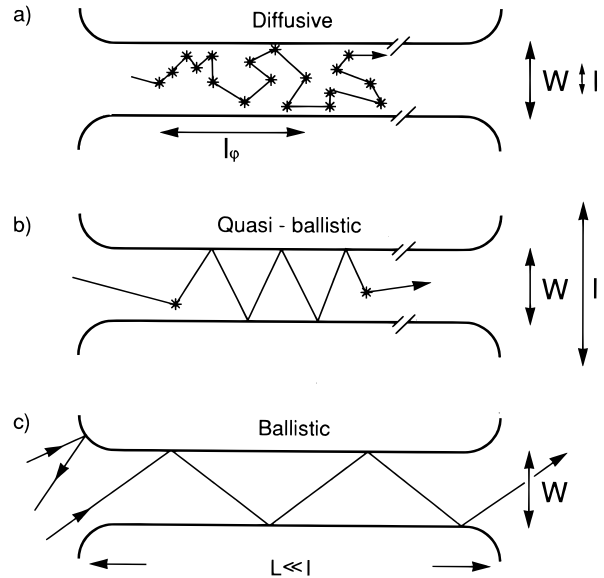}}
\caption{
Electron trajectories characteristic for the diffusive ($l < W, L$), quasi-ballistic ($W < l < L$), and ballistic ($W, L < l$) transport regimes, for the case of specular boundary scattering. Boundary scattering and internal impurity scattering (asterisks) are of equal importance in the quasi-ballistic regime. A nonzero resistance in the ballistic regime results from back scattering at the connection between the narrow channel and the wide 2DEG regions. Taken from H. van Houten et al., in {\em Physics and Technology of Submicron Structures\/} (H. Heinrich, G. Bauer, and F. Kuchar, eds.). Springer, Berlin, 1988.
\label{fig7}
}
\end{figure}

In an experiment one measures a {\it conductance\/} rather than a conductivity.
The conductivity $\sigma$ relates the local current density to the electric field,
$j=\sigma E$, while the conductance $G$ relates the total current to the voltage drop,
$I=GV$. For a large homogeneous conductor the difference between the two
is not essential, since Ohm's law tells us that
\begin{equation}
G=(W/L)\sigma\label{eq4.20}
\end{equation}
for a 2DEG of width $W$ and length $L$ in the current direction. (Note that $G$
and $\sigma$ have the same units in two dimensions.) If for the moment we disregard
the effects of phase coherence, then the simple scaling (\ref{eq4.20}) holds provided
both $W$ and $L$ are much larger than the mean free path $l$. This is the diffusive
transport regime, illustrated in Fig.\ \ref{fig7}a. When the dimensions of the sample
are reduced below the mean free path, one enters the {\it ballistic\/} transport
regime, shown in Fig.\ \ref{fig7}c. One can further distinguish an intermediate {\it quasi-ballistic\/} regime, characterized by $W<l<L$ (see Fig.\ \ref{fig7}b). In ballistic
transport only the conductance plays a role, not the conductivity. The
Landauer formula
\begin{equation}
G=(e^{2}/h)T\label{eq4.21}
\end{equation}
plays a central role in the study of ballistic transport because it expresses the
conductance in terms of a Fermi level property of the sample (the transmission probability $T$, see Section \ref{sec12}). Equation (\ref{eq4.21}) can therefore be
applied to situations where the conductivity does not exist as a local quantity,
as we will discuss in Sections \ref{secIII} and \ref{secIV}.

If phase coherence is taken into account, then the minimal length scale
required to characterize the conductivity becomes larger. Instead of the
(elastic) mean free path $l\equiv v_{\mathrm{F}}\tau$, the phase coherence length $l_{\phi}\equiv(D\tau_{\phi})^{1/2}$
becomes this characteristic length scale (up to a numerical coefficient
$l_{\phi}$ equals the average distance that an electron diffuses in the time $\tau_{\phi}$). Ohm's
law can now only be applied to add the conductances of parts of the sample
with dimensions greater than $l_{\phi}$. Since at low temperatures $l_{\phi}$ can become
quite large (cf.\ Table \ref{table1}), it becomes possible that (for a small conductor) phase coherence extends over a large part of the sample. Then only the conductance
(not the conductivity) plays a role, even if the transport is fully in the diffusive
regime. We will encounter such situations repeatedly in Section \ref{secII}.

\subsubsection{\label{sec4c} Magnetotransport}

In a magnetic field $B$ perpendicular to the 2DEG, the current is no longer
in the direction of the electric field due to the Lorentz force. Consequently,
the conductivity is no longer a scalar but a tensor $\bm{\sigma}$, related via the Einstein
relation $\bm{\sigma}=e^{2}\rho(E_{\mathrm{F}})\mathbf{D}$ to the diffusion tensor
\be
\mathbf{D}=\int_{0}^{\infty}dt\langle v(t)v(0)\rangle.\label{eq4.22}
\ee
Equation (\ref{eq4.22}) follows from a straightforward generalization of the argument leading to the scalar relation (\ref{eq4.18}) [but now the ordering of $v(t)$ and $v(0)$
matters]. Between scattering events the electrons at the Fermi level execute
circular orbits, with cyclotron frequency $\omega_{\mathrm{c}}=eB/m$ and cyclotron radius
$l_{\mathrm{cycl}}=mv_{\mathrm{F}}/eB$. Taking the 2DEG in the $x-y$ plane, and the magnetic field in
the positive $z$-direction, one can write in complex number notation
\be
\tilde{v}(t)\equiv v_{x}(t)+iv_{y}(t)=v_{\mathrm{F}}\exp(i\phi+i\omega_{\mathrm{c}}t).\label{eq4.23}
\ee
The diffusion tensor is obtained from
\begin{eqnarray}
D_{xx}+iD_{yx}&=& \int_{0}^{2\pi}\frac{d\phi}{2\pi}\int_{0}^{\infty}dt\,\tilde{v}(t)v_{\mathrm{F}}\cos\phi \mathrm{e}^{-t/\tau}\nonumber\\
&=&\frac{D}{1+(\omega_{\mathrm{c}}\tau)^{2}}(1+i\omega_{\mathrm{c}}\tau),\label{eq4.24}
\end{eqnarray}
where $D$ is the zero-field diffusion constant (\ref{eq4.13}). One easily verifies that
$D_{yy}=D_{xx}$ and $D_{xy}=-D_{yx}$. From the Einstein relation one then obtains the
conductivity tensor
\be 
{\bm\sigma}=\frac{\sigma}{1+(\omega_{\mathrm{c}}\tau)^{2}}\left(\begin{array}{ll}
1 & -\omega_{\mathrm{c}}\tau\\
\omega_{\mathrm{c}}\tau & 1
\end{array}\right),\label{eq4.25}
\ee
with $\sigma$ the zero-field conductivity (\ref{eq4.8}). The resistivity tensor ${\bm\rho}\equiv{\bm\sigma}^{-1}$ has the
form
\be
{\bm\rho}=\rho\left(\begin{array}{ll}
1 & \omega_{\mathrm{c}}\tau\\
-\omega_{\mathrm{c}}\tau & 1
\end{array}\right),   \label{eq4.26}
\ee
with $\rho=\sigma^{-1}=m/n_{\mathrm{s}}e^{2}\tau$ the zero-field resistivity.

The off-diagonal element $\rho_{xy}\equiv R_{\mathrm{H}}$ is the classical {\it Hall\/} resistance of a
2DEG:
\be
R_{\mathrm{H}}= \frac{B}{n_{\mathrm{s}}e}=\frac{1}{g_{\mathrm{s}}g_{\mathrm{v}}}\frac{h}{e^{2}}\frac{\hbar\omega_{\mathrm{c}}}{E_{\mathrm{F}}}.  \label{eq4.27}
\ee
Note that in a 2D channel geometry there is no distinction between the Hall
{\it resistivity\/} and the Hall {\it resistance}, since the ratio of the Hall voltage
$V_{\mathrm{H}}=WE_{x}$ across the channel to the current $I=Wj_{y}$ along the channel does
not depend on its length and width (provided transport remains in the
diffusive regime). The diagonal element $\rho_{xx}$ is referred to as the {\it longitudinal\/}
resistivity. Equation (\ref{eq4.26}) tells us that classically the magnetoresistivity is
zero (i.e., $\rho_{xx}(B)-\rho_{xx}(0)=0)$. This counterintuitive result can be understood
by considering that the force from the Hall voltage cancels the average
Lorentz force on the electrons. A general conclusion that one can draw from
Eqs.\ (\ref{eq4.25}) and (\ref{eq4.26}) is that the classical effects of a magnetic field are
important only if $\omega_{\mathrm{c}}\tau\gtrsim 1$. In such fields an electron can complete several
cyclotron orbits before being scattered out of orbit. In a high-mobility 2DEG
this criterion is met at rather weak magnetic fields (note that $\omega_{\mathrm{c}}\tau=\mu_{\mathrm{e}}B$, and
see Table \ref{table1}).

\begin{figure}
\centerline{\includegraphics[width=8cm]{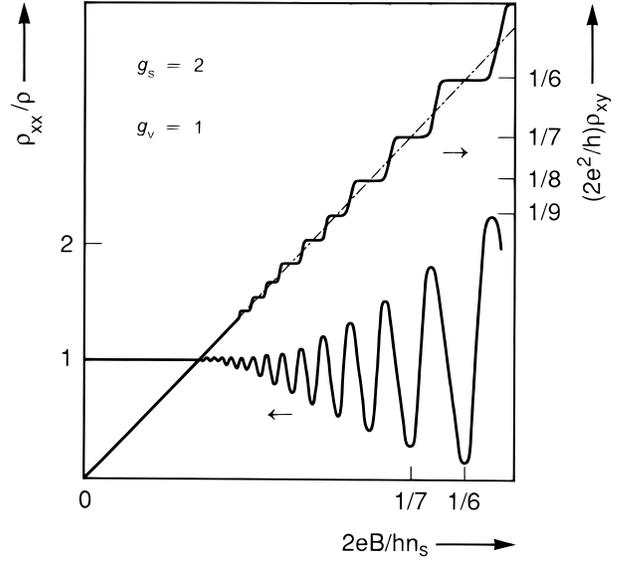}}
\caption{
Schematic dependence on the reciprocal filling factor $\nu^{-1}\equiv 2eB/hn_{\rm s}$ of the longitudinal resistivity $\rho_{xx}$ (normalized to the zero-field resistivity $\rho$) and of the Hall resistance $R_{\rm H}\equiv \rho_{xy}$ (normalized to $h/2e^{2}$). The plot is for the case of a single valley with twofold spin degeneracy. Deviations from the semiclassical result (\ref{eq4.26}) occur in strong magnetic fields, in the form of Shubnikov-De Haas oscillations in $\rho{xx}$ and quantized plateaus [Eq.\ (\ref{eq4.30})] in $\rho_{xy}$.
\label{fig8}
}
\end{figure}

In the foregoing application of the Einstein relation we have used the zero-field density of states. Moreover, we have assumed that the scattering time is
$B$-independent. Both assumptions are justified in weak magnetic fields, for
which $E_{\mathrm{F}}/h\omega_{\mathrm{c}}\gg  1$, but not in stronger fields (cf.\ Table \ref{table1}). As illustrated in Fig.\
\ref{fig8}, deviations from the semiclassical result (\ref{eq4.26}) appear as the magnetic field is
increased. These deviations take the form of an oscillatory magnetoresistivity
(the {\it Shubnikov-De Haas effect}) and plateaux in the Hall resistance (the
{\it quantum Hall effect}). The origin of these two phenomena is the formation of
Landau levels by a magnetic field, discussed in Section \ref{sec4a}, that leads to the $B$-dependent density of states (\ref{eq4.6}). The main effect is on the scattering rate $\tau^{-1}$,
which in a simple (Born) approximation\cite{ref96} is proportional to $\rho(E_{\mathrm{F}})$:
\be
\tau^{-1}=(\pi/\hbar)\rho(E_{\mathrm{F}})c_{\mathrm{i}}u^{2}. \label{eq4.28}
\ee
Here $c_{\mathrm{i}}$ is the areal density of impurities, and the impurity potential is
modeled by a 2D delta function of strength $u$. The diagonal element of the
resistivity tensor (\ref{eq4.26}) is $\rho_{xx}=(m/e^{2}n_{\mathrm{s}})\tau^{-1}\propto\rho(E_{\mathrm{F}})$. Oscillations in the
density of states at the Fermi level due to the Landau level quantization are
therefore observable as an oscillatory magnetoresistivity. One expects the
resistivity to be minimal when the Fermi level lies between two Landau levels,
where the density of states is smallest. In view of Eq.\ (\ref{eq4.6}), this occurs when
the Landau level {\it filling factor\/} $\nu\equiv(n_{\mathrm{s}}/g_{\mathrm{s}}g_{\mathrm{v}})(h/eB)$ equals an integer $N=1,2$,
$\ldots$ (assuming spin-degenerate Landau levels). The resulting Shubnikov-De
Haas oscillations are periodic in $1/B$, with spacing $\Delta(1/B)$ given by
\be 
\Delta(\frac{1}{B})=\frac{e}{h}\frac{g_{\mathrm{s}}g_{\mathrm{v}}}{n_{\mathrm{s}}},   \label{eq4.29}
\ee
providing a means to determine the electron density from a magnetoresistance measurement. This brief explanation of the Shubnikov-De Haas effect
needs refinement,\cite{ref20} but is basically correct. The quantum Hall effect,\cite{ref8} being
the occurrence of plateaux in $R_{\mathrm{H}}$ versus $B$ at precisely
\be
R_{\mathrm{H}}= \frac{1}{g_{\mathrm{s}}g_{\mathrm{v}}}\frac{h}{e^{2}}\frac{1}{N}, \;\;N=1,2,\ldots,   \label{eq4.30}
\ee
is a more subtle effect\cite{ref97} to which we cannot do justice in a few lines (see
Section \ref{sec18}). The quantization of the Hall resistance is related on a fundamental level to the quantization in zero magnetic field of the resistance of a
ballistic point contact.\cite{ref6,ref7} We will present a unified description of both these
effects in Sections \ref{sec12} and \ref{sec13}.

\section{\label{secII} Diffusive and quasi-ballistic transport}

\subsection{\label{sec5} Classical size effects}

In metals, the dependence of the resistivity on the size of the sample has been the subject of study for almost a century.\cite{ref98} Because of the small Fermi wave length in a metal, these are {\em classical\/} size effects. Comprehensive reviews of this field have been given by Chambers,\cite{ref99} Br\"{a}ndli and Olsen,\cite{ref100} Sondheimer,\cite{ref101} and, recently, Pippard.\cite{ref102} In semiconductor nanostructures both classical and quantum size effects appear, and an understanding of the former is necessary to distinguish them from the latter. Classical size effects in a 2DEG are of intrinsic interest as well. First of all, a 2DEG is an ideal model system to study known size effects without the complications of nonspherical Fermi surfaces and polycrystallinity, characteristic for metals. Furthermore, it is possible in a 2DEG to study the case of nearly complete specular boundary scattering, whereas in a metal diffuse scattering dominates. The much smaller cyclotron radius in a 2DEG, compared with a metal at the same magnetic field value, allows one to enter the regime where the cyclotron radius is comparable to the range of the scattering potential. The resulting modifications of known effects in the quasi-ballistic transport regime are the subject of this section. A variety of new classical size effects, not known from metals, appear in the ballistic regime, when the resistance is measured on a length scale below the mean free path. These are discussed in Section \ref{sec16}, and require a reconsideration of what is meant by a resistance on such a short length scale.

In the present section we assume that the channel length $L$ (or, more generally, the separation between the voltage probes) is much larger than the mean free path $l$ for impurity scattering so that the motion remains diffusive along the channel. Size effects in the resistivity occur when the motion across the channel becomes ballistic (i.e., when the channel width $W < l$). Diffuse boundary scattering leads to an increase in the resistivity in a zero magnetic field and to a nonmonotonic magnetoresistivity in a perpendicular magnetic field, as discussed in the following two subsections. The 2D channel geometry is essentially equivalent to the 3D geometry of a thin metal plate in a parallel magnetic field, with the current flowing perpendicular to the field. Size effects in this geometry were originally studied by Fuchs\cite{ref103} in a zero magnetic field and by MacDonald\cite{ref104} for a nonzero field. The alternative configuration in which the magnetic field is perpendicular to the thin plate, studied by Sondheimer\cite{ref105} does not have a 2D analog. We discuss in this section only the classical size effects, and thus the discreteness of the 1D subbands and of the Landau levels is ignored. Quantum size effects in the quasi-ballistic transport regime are treated in Section \ref{sec10}.

\subsubsection{\label{sec5a} Boundary scattering}

In a zero magnetic field, scattering at the channel boundaries increases the resistivity, unless the scattering is specular. {\em Specular scattering\/} occurs if the confining potential $V(x, y)$ does not depend on the coordinate $y$ along the channel axis. In that case the electron motion along the channel is not influenced at all by the lateral confinement, so the resistivity $\rho$ retains its 2D bulk value $\rho_0 =m/e^2 n_{\rm s}\tau$. More generally, specular scattering requires any roughness of the boundaries to be on a length scale smaller than the Fermi wavelength $\lambda_{\rm F}$. The confining potential created electrostatically by means of a gate electrode is known to cause predominantly specular scattering (as has been demonstrated by the electron focusing experiments\cite{ref59} discussed in Section \ref{sec14}). This is a unique situation, not previously encountered in metals, where as a result of the small $\lambda_{\rm F}$ (on the order of the interatomic separation) diffuse boundary scattering dominates.\cite{ref102}

{\em Diffuse\/} scattering means that the velocity distribution at the boundary is isotropic for velocity directions that point away from the boundary. Note that this implies that an incident electron is reflected with a (normalized) angular distribution $P(\alpha) = \frac{1}{2} \cos\alpha$, since the reflection probability is proportional to the flux normal to the boundary. Diffuse scattering increases the resistivity above $\rho_0$ by providing an upper bound $W$ to the effective mean free path. In order of magnitude, $\rho\sim(l/W)\rho_{0}$ if $l\gtrsim W$ (a more precise expression is derived later). In general, boundary scattering is neither fully specular nor fully diffuse and, moreover, depends on the angle of incidence (grazing incidence favors specular scattering since the momentum along the channel is large and not easily reversed). The angular dependence is often ignored for simplicity, and the boundary scattering is described, following Fuchs,\cite{ref103} by a single parameter $p$, such that an electron colliding with the boundary is reflected specularly with probability $p$ and diffusely with probability $1-p$.
This specularity parameter is then used as a fit parameter in comparison with experiments. Soffer\cite{ref106} has developed a more accurate, and more complicated, modeling in terms of an angle of incidence dependent specularity parameter.

In the extreme case of fully diffuse boundary scattering ($p = 0$), one is justified in neglecting the dependence of the scattering probability on the angle of incidence. We treat this case here in some detail to contrast it with fully specular scattering, and because diffuse scattering can be of importance in 2DEG channels defined by ion beam exposure rather than by gates.\cite{ref107,ref108} We calculate the resistivity from the diffusion constant by means of the Einstein relation. Fuchs takes the alternative (but equivalent) approach of calculating the resistivity from the linear response to an applied electric field.\cite{ref103} Impurity scattering is taken as isotropic and elastic and is described by a scattering time $\tau$ such that an electron is scattered in a time interval $dt$ with probability $dt/\tau$, regardless of its position and velocity, This is the commonly employed ``scattering time'' (or ``relaxation time'') approximation.

The channel geometry is defined by hard walls at $x=\pm W/2$ at which the
electrons are scattered diffusely. The stationary electron distribution function
at the Fermi energy $F(\mathbf{r}, \alpha)$ satisfies the Boltzmann equation
\be 
\mathbf{v}\cdot\frac{\partial}{\partial\mathbf{r}}F=-\frac{1}{\tau}F+\frac{1}{\tau}\int_{0}^{2\pi}\frac{d\alpha}{2\pi}F,  \label{eq5.1}
\ee
where $\mathbf{r}\equiv(x, y)$ is the position and $\alpha$ is the angle that the velocity $ \mathbf{v}\equiv v_{\mathrm{F}}(\cos \alpha, \sin \alpha)$ makes with the $x$-axis. The boundary condition corresponding to
diffuse scattering is that $F$ is independent of the velocity direction for
velocities pointing away from the boundary. In view of current conservation
this boundary condition can be written as
\begin{eqnarray}
F(\mathbf{r}, \alpha)&=& \frac{1}{2}\int_{-\pi/2}^{\pi/2}d\alpha^{\prime}\,F(\mathbf{r}, \alpha^{\prime})\cos \alpha^{\prime},\nonumber\\
&&{\rm for}\;\; x= \frac{W}{2},\; \frac{\pi}{2}<\alpha<\frac{3\pi}{2},\nonumber\\
&=& \frac{1}{2}\int_{\pi/2}^{3\pi/2}d\alpha^{\prime}\,F(\mathbf{r}, \alpha^{\prime})\cos \alpha^{\prime},\nonumber\\
&&{\rm for}\;\; x=- \frac{W}{2},\; - \frac{\pi}{2}<\alpha<\frac{\pi}{2}. \label{eq5.2}
\end{eqnarray}
To determine the diffusion constant, we look for a solution of  Eqs.\ (\ref{eq5.1}) and
(\ref{eq5.2}) corresponding to a constant density gradient along the channel,
$F(\mathbf{r}, \alpha)=-cy+f(x, \alpha)$. Since there is no magnetic field, we anticipate that the
density will be uniform across the channel width so that 
$\int_{0}^{2\pi}fd\alpha=0$. The
Boltzmann equation (\ref{eq5.1}) then simplifies to an ordinary differential equation
for $f$, which can be solved straightforwardly. The solution that satisfies the
boundary conditions (\ref{eq5.2}) is
\be
F(\mathbf{r}, \alpha)=-cy+cl\sin\alpha \left[1- \exp\left(-\frac{W}{2l|\cos \alpha|}-\frac{x}{l\cos \alpha}\right)\right],   \label{eq5.3}
\ee
where we have written $l\equiv v_{\mathrm{F}}\tau$. One easily verifies that $F$ has indeed a uniform
density along $x$. The diffusion current
\be
I_{y}=v_{\mathrm{F}} \int_{-W/2}^{W/2}dx\int_{0}^{2\pi}d\alpha\,F\sin\alpha \label{eq5.4}
\ee
along the channel in response to the density gradient $\partial n/\partial y=-2\pi c$
determines the diffusion constant $D=-(I_{y}/W)(\partial n/\partial y)^{-1}$. The resistivity
$\rho=E_{\mathrm{F}}/n_{\mathrm{s}}e^{2}D$ then follows from the Einstein relation (\ref{eq4.10}), with the 2D
density of states $n_{\mathrm{s}}/E_{\mathrm{F}}$. The resulting expression is
\be 
\rho=\rho_{0}\left[1-\frac{4l}{\pi W}\int_{0}^{1}d\xi\,\xi(1-\xi^{2})^{1/2}(1-\mathrm{e}^{-W/l\xi})\right]^{-1},\label{eq5.5}
\ee
which can be easily evaluated numerically. It is worth noting that the
above result\cite{ref109} for $\rho/\rho_{0}$ in a 2D channel geometry does not differ much
(less than 20\%) from the corresponding result\cite{ref103} in a 3D thin film.

For $l/W\ll 1$ one has
\be
\rho=\rho_0 \left( 1+\frac{4}{3\pi}\frac{l}{W}\right),\label{eq5.6}
\ee
which differs from Eq.\ (\ref{eq5.5}) by less than 10\% in the range $l/W\lesssim 10$. For $l/W\gg 1$ one has asymptotically
\begin{eqnarray}
\rho&=&\frac{\pi}{2}\rho_{0}\frac{l}{W}\frac{1}{\ln(l/W)}\nonumber\\
&=&\frac{\pi}{2}\frac{mv_{\rm F}}{n_{\rm s}e^2 W}\frac{1}{\ln(l/W)}.\label{eq5.7}
\end{eqnarray}
In the absence of impurity scattering (i.e., in the limit $l\rightarrow\infty$), Eq.\ (\ref{eq5.7}) predicts a vanishing resistivity. Diffuse boundary scattering is ineffective in establishing a finite resistivity in this limit, because electrons with velocities nearly parallel to the channel walls can propagate over large distances without collisions and thereby short out the current. As shown by Tesanovic et al.,\cite{ref110} a small but nonzero resistivity in the absence of impurity scattering is recovered if one goes beyond the semiclassical approximation and includes the effect of the quantum mechanical uncertainty in the transverse component of the electron velocity.

\subsubsection{\label{sec5b} Magneto size effects}

In an unbounded 2DEG, the longitudinal resistivity is magnetic-field independent in the semiclassical approximation (see Section \ref{sec4c}). We will discuss how a nonzero magnetoresistivity can arise classically as a result of boundary scattering. We consider the two extreme cases of specular and diffuse boundary scattering, and describe the impurity scattering in the scattering time approximation. Shortcomings of this approximation are discussed toward the end of this subsection.

We consider first the case of specular boundary scattering. In a zero magnetic field it is obvious that specular scattering cannot affect the resistivity, since the projection of the electron motion on the channel axis is not changed by the presence of the channel boundaries. If a magnetic field is applied perpendicular to the 2DEG, the electron trajectories in a channel cannot be mapped in this way on the trajectories in an unbounded system. In fact, in an unbounded 2DEG in equilibrium the electrons perform closed cyclotron orbits between scattering events, whereas a channel geometry supports open orbits that skip along the boundaries. One might suppose that the presence of these {\em skipping orbits\/} propagating along the channel would increase the diffusion constant and hence reduce the (longitudinal) resistivity below the value $\rho_0$ of a bulk 2DEG. That is not correct, at least in the scattering time approximation, as we now demonstrate.

The stationary Boltzmann equation in a magnetic field $\bf B$ in the $z$-direction (perpendicular to the 2DEG) is
\be
\mathbf{v}\cdot\frac{\partial}{\partial\mathbf{r}}F+\omega_{\rm c}\frac{\partial}{\partial\alpha}F=
-\frac{1}{\tau}F+\frac{1}{\tau}\int_{0}^{2\pi}\frac{d\alpha}{2\pi}F.\label{eq5.8}
\ee
Here, we have used the identity $-em^{-1}(\mathbf{v}\times\mathbf{B})\cdot\partial/\partial\mathbf{v}\equiv\omega_{\rm c}\partial/\partial\alpha$ (with $\omega_{\rm c}\equiv eB/m$ the cyclotron frequency) to rewrite the term that accounts for the Lorentz force. The distribution function $F(\mathbf{r},\alpha)$ must satisfy the boundary conditions for specular scattering,
\be
F(\mathbf{r},\alpha)=F(\mathbf{r},\pi-\alpha),\;\;{\rm for}\;\;x=\pm W/2.\label{eq5.9}
\ee
One readily verifies that
\be
F(\mathbf{r},\alpha)=-c(y+\omega_{\rm c}\tau x)+cl\sin\alpha\label{eq5.10}
\ee
is a solution of Eqs.\ (\ref{eq5.8}) and (\ref{eq5.9}). The corresponding diffusion current $I_y = \pi cWv_{\rm F}l$ and density gradient along the channel $\partial n/\partial y= -2\pi c$ are both the same as in a zero magnetic field. It follows that the diffusion constant $D = I_y /2\pi c W$ and, hence, the longitudinal resistivity $\rho = E_{\rm F} /n_{\rm s} e^2 D$ are $B$-independent; that is, $\rho=\rho_0 \equiv m/n_{\rm s} e^2 \tau$, as in an unbounded 2DEG. More generally, one can show that in the scattering time approximation the longitudinal resistivity is $B$-independent for {\em any\/} confining potential $V(x, y)$ that does not vary with the coordinate $y$ along the channel axis. (This statement is proven by applying the result of Ref.\ \onlinecite{ref111}, of a $B$-independent $\rho_{yy}$ for periodic $V(x)$, to a set of disjunct parallel channels (see Section \ref{sec11b}); the case of a single channel then follows from Ohm's law.)

\begin{figure}
\centerline{\includegraphics[width=8cm]{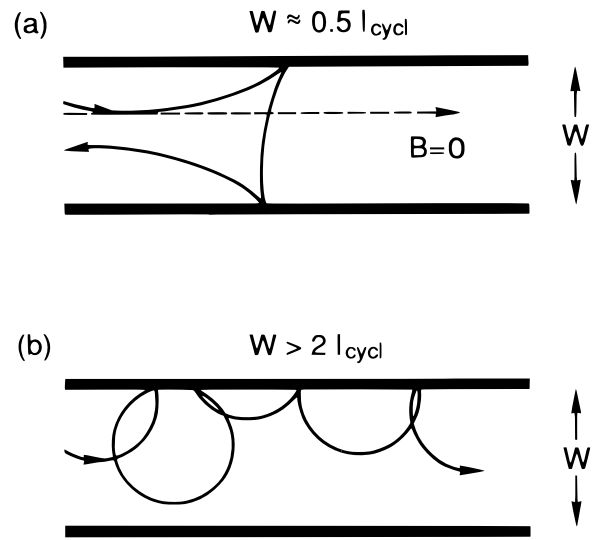}}
\caption{
Illustration of the effect of a magnetic field on motion through a channel with diffuse boundary scattering. (a) Electrons which in a zero field move nearly parallel to the boundary can reverse their motion in weak magnetic fields. This increases the resistivity. (b) Suppression of back scattering at the boundaries in strong magnetic fields reduces the resistivity.
\label{fig9}
}
\end{figure}

\begin{figure}
\centerline{\includegraphics[width=8cm]{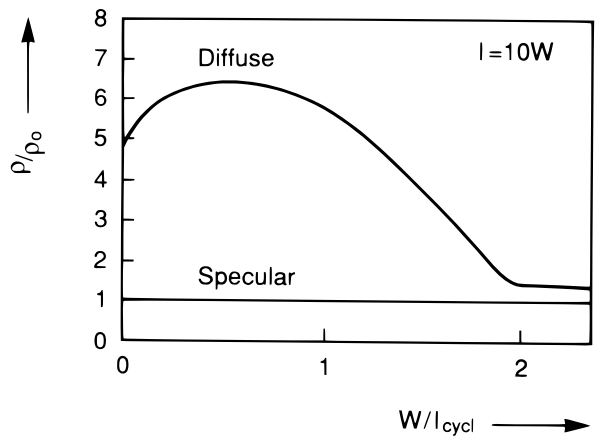}}
\caption{
Magnetic field dependence of the longitudinal resistivity of a channel for the two cases of diffuse and specular boundary scattering, obtained from the Boltzmann equation in the scattering time approximation. The plot for diffuse scattering is the result of Ref.\ \onlinecite{ref114} for a 3D thin film geometry with $l=10W$. (A 2D channel geometry is expected to give very similar results.\cite{ref102})
\label{fig10}
}
\end{figure}

In the case of diffuse boundary scattering, the zero-field resistivity is enhanced by approximately a factor $1 + l/2W$ [see Eq.\ (\ref{eq5.6})]. A sufficiently strong magnetic field suppresses this enhancement, and reduces the resistivity to its bulk value $\rho_0$. The mechanism for this negative magnetoresistance is illustrated in Fig.\ \ref{fig9}b. If the cyclotron diameter $2l_{\rm cycl}$ is smaller than the channel width $W$, diffuse boundary scattering cannot reverse the direction of motion along the channel, as it could for smaller magnetic fields. The diffusion current is therefore approximately the same as in the case of specular scattering, in which case we have seen that the diffusion constant and, hence, resistivity have their bulk values. Figure \ref{fig9} represents an example of {\em magnetic reduction of backscattering}. Recently, this phenomenon has been understood to occur in an extreme form in the quantum Hall effect\cite{ref112} and in ballistic transport through quantum point contacts.\cite{ref113} The effect was essentially known and understood by MacDonald\cite{ref104} in 1949 in the course of his magnetoresistivity experiments on sodium wires. The ultimate reduction of the resistivity is preceded by an initial increase in weak magnetic fields, due
to the deflection toward the boundary of electrons with a velocity nearly parallel to the channel axis (Fig.\ \ref{fig9}a). The resulting nonmonotonic $B$-dependence of the resistivity is shown in Fig.\ \ref{fig10}. The plot for diffuse scattering is based on a calculation by Ditlefsen and Lothe\cite{ref114} for a 3D thin­film geometry. The case of a 2D channel has been studied by Pippard\cite{ref102} in the limit $l/W\rightarrow\infty$, and he finds that the 2D and 3D geometries give very similar results.

\begin{figure}
\centerline{\includegraphics[width=8cm]{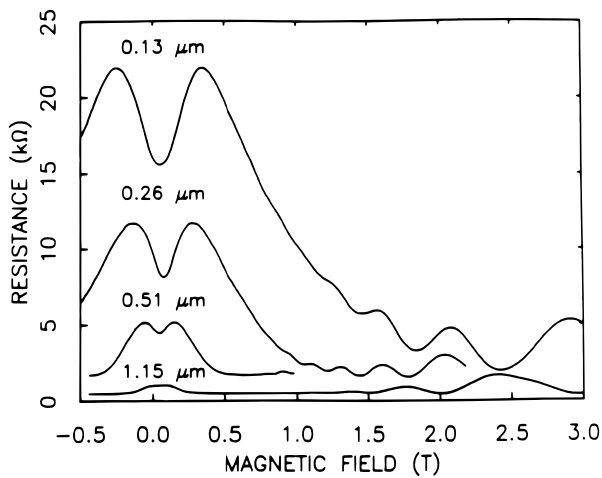}}
\caption{
Experimental magnetic field dependence of the resistance of channels of different widths, defined by ion beam exposure in the 2DEG of a GaAs-AlGaAs heterostructure ($L= 12\,\mu{\rm m}$, $T = 4.2\,{\rm K}$). The nonmonotonic magnetic field dependence below $1\,{\rm T}$ is a classical size effect due to diffuse boundary scattering, as illustrated in Fig.\ \ref{fig9}. The magnetoresistance oscillations at higher fields result from the quantum mechanical Shubnikov-De Haas effect. Taken from T. J. Thornton et al., Phys.\ Rev.\ Lett.\ {\bf 63}, 2128 (1989).
\label{fig11}
}
\end{figure}

\begin{figure}
\centerline{\includegraphics[width=8cm]{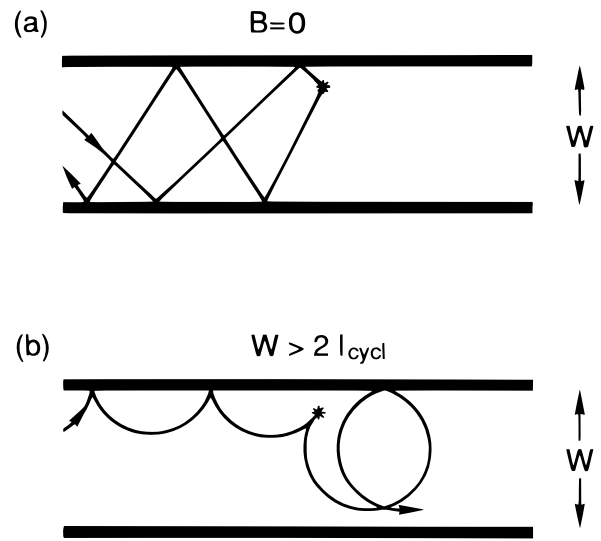}}
\caption{
Electron trajectories in a channel with specular boundary scattering, to illustrate how a magnetic field can suppress the back scattering by an isolated impurity close to a boundary. This effect would lead to a negative magneto resistivity if one would go beyond the scattering time approximation.
\label{fig12}
}
\end{figure}

An experimental study of this effect in a 2DEG has been performed by Thornton et al.\cite{ref107} In Fig.\ \ref{fig11} their magnetoresistance data are reproduced for channels of different widths $W$, defined by low-energy ion beam exposure. It was found that the resistance reaches a maximum when $W\approx 0.5\,l_{\rm cycl}$, in excellent agreement with the theoretical predictions.\cite{ref114,ref102} Thornton et al.\ also investigated channels defined electrostatically by a split gate, for which one expects predominantly specular boundary scattering.\cite{ref59} The foregoing analysis would then predict an approximately $B$-independent resistance (Fig.\ \ref{fig10}), and indeed only a small resistance maximum was observed in weak magnetic fields. At stronger fields, however, the resistance was found to decrease substantially. Such a monotonically decreasing resistance in channels with predominantly specular boundary scattering was first reported by Choi et al.,\cite{ref55} and studied for a narrower channel in Ref.\ \onlinecite{ref27} (see Section \ref{sec9b} for some of these experimental results). We surmise that a classical negative magnetoresistance in the case of specular boundary scattering can result if the cyclotron radius becomes smaller than some characteristic correlation length in the distribution of impurities (and in the resulting potential landscape). Correlations between the positions of impurities and the channel boundaries, which are neglected in the scattering time approximation, will then play a role. For an example, see Fig.\ \ref{fig12}, which shows how an isolated impurity near the boundary can reverse the direction of electron motion in a zero magnetic field but not in a sufficiently strong magnetic field. In metals, where the cyclotron radius is much larger than in a 2DEG, an electron will effectively experience a random impurity potential between subsequent boundary collisions, so the scattering can well be described in terms of an average relaxation time. The experiments in a 2DEG suggest that this approximation breaks down at relatively weak magnetic fields.

\subsection{\label{sec6} Weak localization}

\begin{figure}
\centerline{\includegraphics[width=8cm]{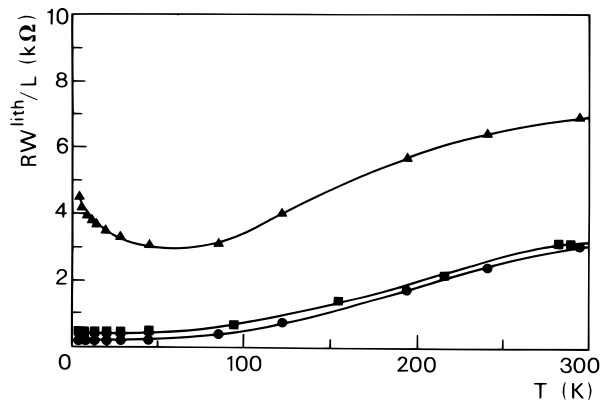}}
\caption{
Temperature dependence of the resistivity of a wide 2DEG in a GaAs-AlGaAs heterostructure (circles) and of two narrow channels of lithographic width $W_{\rm lith}= 1.5\,\mu{\rm m}$ (squares) and $W_{\rm lith}= 0.5\,\mu{\rm m}$ (triangles). The channel length $L = 10\,\mu{\rm m}$. The resistivity is estimated from the measured resistance $R$ by multiplying by $W_{\rm lith}/L$, disregarding the difference between the conducting and lithographic width in the narrow channels. Taken from H. van Houten et al., Appl.\ Phys.\ Lett.\ {\bf 49}, 1781 (1986).
\label{fig13}
}
\end{figure}

The temperature dependence of the Drude resistivity $\rho = m/n_{\rm s}e^2 \tau$ is contained in that of the scattering time $\tau$, since the electron density is constant in a degenerate electron gas. As one lowers the temperature, inelastic scattering processes (such as electron-phonon scattering) are suppressed, leading to a decrease in the resistivity. The residual resistivity is due entirely to elastic scattering (with stationary impurities or other crystalline defects) and is temperature-independent in the semiclassical theory. Experimentally, however, one finds that below a certain temperature the resistivity of the 2DEG starts to rise again. The increase is very small in broad samples, but becomes quite pronounced in narrow channels. This is illustrated in Fig.\ \ref{fig13}, where the temperature dependencies of the resistivities of wide and narrow GaAs-AlGaAs heterostructures are compared.\cite{ref63}

The anomalous resistivity increase is due to long-range correlations in the diffusive motion of an electron that are purely quantum mechanical. In the semiclassical theory it is assumed that a few scattering events randomize the electron velocity, so the velocity correlation function decays exponentially in time with decay time $\tau$ [see Eq.\ (\ref{eq4.19})]. As discussed in Section \ref{sec4c}, this assumption leads to the Drude formula for the resistivity. It is only in recent years that one has come to appreciate that purely elastic scattering is not effective in destroying correlations in the phase of the electron wave function. Such correlations lead to quantum interference corrections to the Drude result, which can explain the anomalous increase in the resistivity at low temperatures.

A striking effect of quantum interference is to enhance the probability for backscattering in a disordered system in the metallic regime. This effect has been interpreted as a precursor of localization in strongly disordered systems and has thus become known as {\em weak localization}.\cite{ref115,ref116,ref117} In Section \ref{sec6a} we describe the theory for weak localization in a zero magnetic field. The application of a magnetic field perpendicular to the 2DEG suppresses weak localization,\cite{ref118} as discussed in Section \ref{sec6b}. The resulting negative magnetoresistivity is the most convenient way to resolve experimentally the weak localization correction.\cite{ref119} The theory for a narrow channel in the quasi­ballistic transport regime\cite{ref109,ref120} differs in an interesting way from the theory for the diffusive regime,\cite{ref121} as a consequence of the flux cancellation effect.\cite{ref122} The diffusive and quasi-ballistic regimes are the subjects of Sections \ref{sec6b} and \ref{sec6c}, respectively.

\subsubsection{\label{sec6a} Coherent backscattering}

The theory of weak localization was developed by Anderson et al.\cite{ref116} and Gorkov et al.\cite{ref117} This is a diagrammatic perturbation theory that does not lend itself easily to a physical interpretation. The interpretation of weak localization as {\em coherent backscattering\/} was put forward by Bergmann\cite{ref123} and by Khmel'nitskii and Larkin,\cite{ref124,ref125} and formed the basis of the path integral theory of Chakravarty and Schmid.\cite{ref126} In this description, weak localization is understood by considering the interference of the probability amplitudes for the classical trajectories (or ``Feynman paths'') from one point to another, as discussed later. For reviews of the alternative diagrammatic approach, we refer to Refs.\ \onlinecite{ref127} and \onlinecite{ref128}.

\begin{figure}
\centerline{\includegraphics[width=8cm]{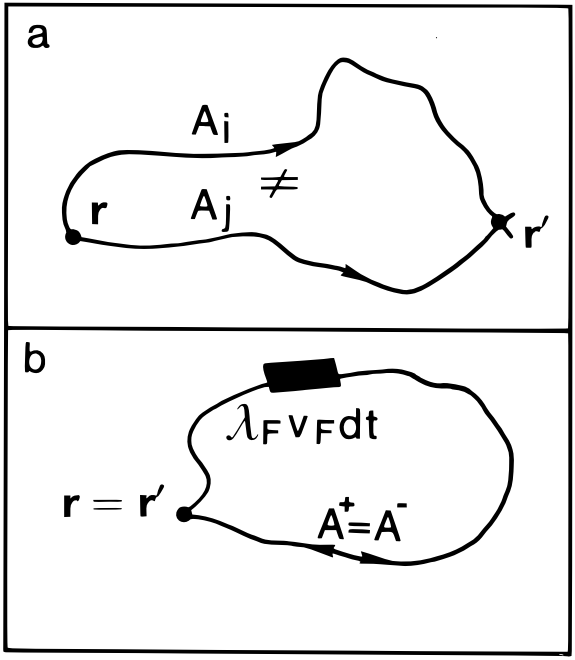}}
\caption{
Mechanism of coherent back scattering. The probability amplitudes $A_i$ and $A_j$ of two trajectories from $\mathbf{r}$ to $\mathbf{r}'$ have uncorrelated phases in general (a), but the amplitudes $A^+$ and $A^-$ of two time-reversed returning trajectories are equal (b). The constructive interference of $A^+$ and $A^-$ increases the probability for return to the point of departure, which is the origin of the weak localization effect. The volume indicated in black is the area $\lambda_{\rm F}v_{\rm F}dt$ covered by a flux tube in a time interval $dt$, which enters in Eq.\ (\ref{eq6.2}) for the conductivity correction.
\label{fig14}
}
\end{figure}

In a Feynman path description\cite{ref129} of diffusion, the probability $P(\mathbf{r},\mathbf{r}', t)$ for motion from point $\mathbf{r}$ to point $\mathbf{r}'$ in a time $t$ consists of the absolute value squared of the sum of probability amplitudes $A_i$, one for each trajectory from $\mathbf{r}$ to $\mathbf{r}'$ of duration $t$:
\be
P(\mathbf{r},\mathbf{r}',t)=\left| \sum_{i}A_i \right|^2 =\sum_i |A_i|^2 +\sum_{i\neq j}A_i A_j^{\ast}.\label{eq6.1}
\ee
The restriction to {\em classical\/} trajectories in the sum over Feynman paths is allowed if the separation between scattering events is much larger than the wavelength (i.e., if $k_{\rm F}l\gg 1$). The classical diffusion probability corresponds to the first term on the right-hand side of Eq.\ (\ref{eq6.1}), while the second term accounts for quantum interference. In the diffusive transport regime there is a very large number of different trajectories that contribute to the sum. One might suppose that for this reason the interference term averages out, because different trajectories have uncorrelated phases. This is correct if the beginning and end points $\mathbf{r}$ and $\mathbf{r}'$ are different (Fig.\ \ref{fig14}a), but not if the two coincide (Fig.\ \ref{fig14}b). In the latter case of ``backscattered'' trajectories, one can group the contributions to the sum (\ref{eq6.1}) in time-reversed pairs. Time-reversal invariance guarantees that the probability amplitudes $A^+$ and $A^-$ for clockwise and counterclockwise propagation around the closed loop are identical: $A^+ = A^- \equiv A$. The coherent backscattering probability $|A^+ + A^-|^2 =4|A|^2$ is then twice the classical result. The enhanced probability for return to the point of departure reduces the diffusion constant and, hence, the conductivity. This is the essence of weak localization. As phrased by Chakravarty and Schmid,\cite{ref126} ``it is one of those unique cases where the superposition principle of quantum mechanics leads to observable consequences at the macroscopic level.''

The magnitude of the weak localization correction $\delta\sigma_{\mathrm{loc}}$ to the Drude
conductivity $\sigma$ is proportional to the probability for return to the point of
departure.\cite{ref126} Since $\delta\sigma_{\mathrm{loc}}$ is assumed to be a small correction, one can estimate this probability from classical diffusion. Let $C(t)d\mathbf{r}$ denote the classical
probability that an electron returns after a time $t$ to within $d\mathbf{r}$ of its point of
departure. The weak localization correction is given by the time integral of
the return probability:
\be
\frac{\delta\sigma_{\mathrm{loc}}}{\sigma}=-\frac{2\hbar}{m}\int_{0}^{\infty}dt\,C(t)\mathrm{e}^{-t/\tau_{\phi}}. \label{eq6.2}
\ee 
The correction is negative because the conductivity is reduced by coherent
backscattering. The factor $\hbar/m\propto\lambda_{\mathrm{F}}v_{\mathrm{F}}$ follows in the path integral formalism
from the area covered by a flux tube of width $\lambda_{\mathrm{F}}$ and length $v_{\mathrm{F}}dt$ (see Fig.\ \ref{fig14}b). The factor $\exp(- t/\tau_{\phi})$ is inserted ``by hand'' to account for the loss of phase
coherence after a time $\tau_{\phi}$ (as a result of inelastic scattering). The return
probability $C(t)$ in a 2D channel of width $W$ is given for times $t\gg  \tau$ in the
diffusive regime by
\begin{subequations}
\label{eq6.3}
\begin{eqnarray}
C(t)&=&(4\pi Dt)^{-1},\;\;{\rm if}\;\;t\ll W^{2}/D, \label{eq6.3a}\\
C(t)&=&W^{-1}(4\pi Dt)^{-1/2},\;\;{\rm if}\;\;t\gg W^{2}/D. \label{eq6.3b}
\end{eqnarray}
\end{subequations}
The $1/t$ decay of the return probability (\ref{eq6.3a}) assumes unbounded diffusion in
two dimensions. A crossover to a lower $1/\sqrt{t}$ decay (\ref{eq6.3b}) occurs when the
root-mean-square displacement $(2Dt)^{1/2}$ exceeds the channel width, so
diffusion occurs effectively in one dimension only. Because the time integral of
$C(t)$ itself diverges, the weak localization correction (\ref{eq6.2}) is determined by the
behavior of the return probability on the phase coherence time $\tau_{\phi}$, which
provides a long-time cutoff. One speaks of 2D or 1D weak localization,
depending on whether the return probability $C(\tau_{\phi})$ on the time scale of $\tau_{\phi}$ is
determined by 2D diffusion (\ref{eq6.3a}) or by 1D diffusion (\ref{eq6.3b}). In terms of the
phase coherence length $l_{\phi}\equiv(D\tau_{\phi})^{1/2}$, the criterion for the dimensionality is
that 2D weak localization occurs for $l_{\phi}\ll W$ and 1D weak localization for
$l_{\phi}\gg  W$. On short time scales $t\lesssim\tau$, the motion is ballistic rather than diffusive,
and Eq.\ (\ref{eq6.3}) does not apply. One expects the return probability to go to zero
smoothly as one enters the ballistic regime. This short-time cutoff can be
accounted for heuristically by the factor $1- \exp(- t/\tau)$, to exclude those
electrons that at time $t$ have not been scattered.\cite{ref109}  The form of the short-time
cutoff becomes irrelevant for $\tau_{\phi}\gg  \tau$. (See Ref.\ \onlinecite{ref130} for a theoretical study of weak localization in the regime of comparable $\tau_{\phi}$ and $\tau$.)

The foregoing analysis gives the following expressions for the 2D and 1D
weak localization corrections:
\begin{widetext}
\begin{subequations}
\label{eq6.4}
\begin{eqnarray}
\delta\sigma_{\mathrm{loc}}&=&-\frac{2\hbar}{m}\sigma\int_{0}^{\infty}dt\,(4\pi Dt)^{-1}(1-e^{-t/\tau})e^{-t/\tau_{\phi}}=-g_{\mathrm{s}}g_{\mathrm{v}} \frac{e^{2}}{4\pi^{2}\hbar}\ln\left(1+\frac{\tau_{\phi}}{\tau}\right),\;\;{\rm if}\;\; l_{\phi}\ll W, \label{eq6.4a}\\
\delta\sigma_{\mathrm{loc}}&=&-\frac{2\hbar}{m}\sigma\int_{0}^{\infty}dt\,W^{-1}(4\pi Dt)^{-1/2}(1-e^{-t/\tau)}e^{-t/\tau_{\phi}}=-g_{\mathrm{s}}g_{ \mathrm{v}} \frac{e^{2}}{2\pi \hbar}\frac{l_{\phi}}{W}\left(1-\left(1+\frac{\tau_{\phi}}{\tau}\right)^{-1/2}\right),\;\;{\rm if}\;\; l_{\phi}\gg W,\nonumber\\&& \label{eq6.4b}
\end{eqnarray}
\end{subequations}
\end{widetext}
where we have used the expression for the Drude conductivity $\sigma=e^{2}\rho(E_{\mathrm{F}})D$
with the 2D density of states (\ref{eq4.2}). The ratio of the weak localization
correction to the Drude conductivity $\delta\sigma_{\mathrm{loc}}/\sigma$ is of order $1/k_{\mathrm{F}}l$ for 2D weak
localization and of order $(l_{\phi}/W)(1/k_{\mathrm{F}}l)$ for 1D weak localization. In the 2D
case, the correction is small (cf.\ the values of $k_{\mathrm{F}}l$ given in Table \ref{table1}), but still
much larger than in a typical metal. The correction is greatly enhanced in the
1D case $l_{\phi}\gg  W$. This is evident in the experimental curves in Fig.\ \ref{fig13}, in which
the resistivity increase at low temperatures is clearly visible only in the
narrowest channel.

The weak localization correction to the conductance $\delta G_{\mathrm{loc}}\equiv(W/L)\delta\sigma_{\mathrm{loc}}$ is
of order $(e^{2}/h)(W/L)$ in the 2D case and of order $(e^{2}/h)(l_{\phi}/L)$ in the 1D case. In
the latter case, the conductance correction does not scale with the channel
width $W$, contrary to what one would have classically. The conductance does
scale with the reciprocal of the channel length $L$, at least for $L\gg  l_{\phi}$. The factor
$l_{\phi}/L$ in $\delta G_{\mathrm{loc}}$ in the 1D case can be viewed as a consequence of the classical
series addition of $L/l_{\phi}$ channel sections. It will then be clear that the scaling
with $L$ has to break down when $L\lesssim l_{\phi}$, in which case the weak localization
correction saturates at its value for $L\approx l_{\phi}$. The maximum conductance
correction in a narrow channel is thus of order $e^{2}/h$, independent of the
properties of the sample. This ``universality'' is at the origin of the phenomenon of the universal conductance fluctuations discussed in Section \ref{sec7}.

\subsubsection{\label{sec6b} Suppression of weak localization by a magnetic field}

{\bf (a) Theory.} The resistance enhancement due to weak localization can be
suppressed by the application of a weak magnetic field oriented perpendicular to the 2DEG. The suppression results from the fact that a
magnetic field breaks time-reversal invariance. We recall that in a zero
magnetic field, time-reversal invariance guarantees that trajectories that form
a closed loop have equal probability amplitudes $A^{+}$ and $A^{-}$ for clockwise
and counterclockwise propagation around the loop. The resulting constructive interference enhances the backscattering probability, thereby leading to the weak localization effect. In a weak magnetic field, however, a phase
difference $\phi$ develops between $A^{+}$ and $A^{-}$, even if the curvature of the
trajectories by the Lorentz force can be totally negected. This Aharonov-Bohm phase results from the fact that the canonical momentum $\mathbf{p}=m \mathbf{v}-e\mathbf{A}$
of an electron in a magnetic field contains the vector potential $\mathbf{A}$. On
clockwise $(+)$ and counterclockwise $(-)$ propagation around a closed loop,
one thus acquires a phase difference
\begin{eqnarray}
\phi&=&\hbar^{-1}\oint_{+}\mathbf{p}^{+}\cdot d{\mathbf l}-\hbar^{-1}\oint_{-}\mathbf{p}^{-}\cdot d\mathbf{l}\nonumber\\
&=& \frac{2e}{\hbar}\int(\nabla\times \mathbf{A})\cdot d\mathbf{S}=\frac{2eBS}{\hbar}\equiv\frac{2S}{l_{\mathrm{m}}^{2}}\equiv 4\pi\frac{\Phi}{\Phi_{0}}.\nonumber\\
&& \label{eq6.5}
\end{eqnarray}
The phase difference is twice the enclosed area $S$ divided by the square of the
magnetic length $l_{\mathrm{m}}\equiv(\hbar/eB)^{1/2}$, or, alternatively, it is $4\pi$ times the enclosed
flux $\Phi$ in units of the elementary flux quantum $\Phi_{0}\equiv h/e$.

Many trajectories, with a wide distribution of loop areas, contribute to the
weak localization effect. In a magnetic field the loops with a large area $S\gtrsim l_{\mathrm{m}}^{2}$
no longer contribute, since on average the counterpropagating trajectories no
longer interfere constructively. Since trajectories enclosing a large area
necessarily take a long time to complete, the effect of a magnetic field is
essentially to introduce a long-time cutoff in the integrals of  Eqs.\ (\ref{eq6.2}) and
(\ref{eq6.4}), which is the magnetic relaxation time $\tau_{B}$. Recall that the long-time cutoff
in the absence of a magnetic field is the phase coherence time $\tau_{\phi}$. The
magnetic field thus begins to have a significant effect on weak localization if
$\tau_{B}$ and $\tau_{\phi}$ are comparable, which occurs at a characteristic field $B_{\mathrm{c}}$. The weak
localization effect can be studied experimentally by measuring the negative
magnetoresistance peak associated with its suppression by a magnetic field.
The significance of such experiments relies on the possibility of directly
determining the phase coherence time $\tau_{\phi}$. The experimental data are most
naturally analyzed in terms of the conductance. The magnitude of the zero-field conductance correction $\delta G_{\mathrm{loc}}(B=0)$ follows directly from the saturation
value of the magnetoconductance, according to
\be
G(B\gg  B_{\mathrm{c}})-G(B=0)=-\delta G_{\mathrm{loc}}(B=0). \label{eq6.6}
\ee 
Once $\delta G_{\mathrm{loc}}(B=0)$ is known, one can deduce the phase coherence length $l_{\phi}$
from Eq.\ (\ref{eq6.4}), since $D$ and $\tau$ are easily estimated from the classical part of the
conductance (which dominates at slightly elevated temperatures). The magnetoconductance contains, in addition, information on the channel width $W$,
which is a parameter difficult to determine otherwise, as will become clear in
the discussion of the experimental situation in subsection (b).

\begin{table*}
\caption{Magnetic relaxation time $\tau_{B}$ and characteristic field $B_{\rm c}$ for the suppression of 2D and 1D weak localization.
\label{table2}
}
\begin{ruledtabular}
\begin{tabular}{ccccc}
&\multicolumn{2}{c}{Dirty Metal\footnotemark[1]\footnotemark[2] $(l\ll W)$}&\multicolumn{2}{c}{Pure Metal\footnotemark[1]\footnotemark[3] $(W\ll l)$}\\
&2D ($l_{\phi}\ll W)$&1D ($W\ll l_{\phi}$)&1D weak field ($l_{\rm m}^{2}\gg Wl$)& 1D strong field ($Wl\gg l_{\rm m}^2 \gg W^2$)\\
\hline
&&&&\\
$\tau_{B}$&$\displaystyle\frac{l_{\rm m}^2}{2D}$&$\displaystyle\frac{3l_{\rm m}^4}{W^2 D}$&$\displaystyle\frac{C_1 l_{\rm m}^4}{W^3 v_{\rm F}}$&$\displaystyle\frac{C_2 l_{\rm m}^2 l}{W^2 v_{\rm F}}$\\
&&&&\\
$B_{\rm c}$&$\displaystyle\frac{\hbar}{e}\frac{1}{2 l_{\phi}^2}$&$\displaystyle\frac{\hbar}{e}\frac{3^{1/2}}{W l_{\phi}}$&$\displaystyle\frac{\hbar}{e}\frac{1}{W}\left(\frac{C_{1}}{Wv_{\rm F}\tau_{\phi}}\right)^{1/2}$&$\displaystyle\frac{\hbar}{e}\frac{C_{2}l}{W^2 v_{\rm F}\tau_{\phi}}$
\end{tabular}
\end{ruledtabular}
\footnotetext[1]{
All results assume a channel length $L\gg l_{\phi}$, a channel width $W\gg\lambda_{\rm F}$, as well as $\tau_{\phi}\gg\tau$.}
\footnotetext[2]{
From Refs.\ \onlinecite{ref118,ref131}, and \onlinecite{ref121}. The diffusion constant $D=\frac{1}{2}v_{\rm F}l$. If $W\ll l_{\phi}$, a transition to 2D weak localization occurs when $l_{\rm m}\lesssim W$.}
\footnotetext[3]{
From Ref.\ \onlinecite{ref109}. The constants are given by $C_1 =9.5$ and $C_2 = 24/5$ for specular boundary scattering ($C_1 =4\pi$ and $C_2 =3$ for a channel with diffuse boundary scattering). For pure metals, the case $l_{\rm m}<W$ is outside the diffusive transport regime for weak localization.}
\end{table*}

The effectiveness of a magnetic field in suppressing weak localization (as
contained in the functional dependence of $\tau_{B}$ on $B$, or in the expression for $B_{\mathrm{c}}$)
is determined by the average flux enclosed by backscattered trajectories of a
given duration. One can distinguish different regimes, depending on the
relative magnitude of the channel width $W$, the mean free path $l\equiv v_{\mathrm{F}}\tau$, the
magnetic length $l_{\mathrm{m}}$, and the phase coherence length $l_{\phi}\equiv(D\tau_{\phi})^{1/2}$. In Table \ref{table2}
the expressions for $\tau_{B}$ and $B_{\mathrm{c}}$ are summarized, as obtained by various
authors.\cite{ref109,ref118,ref121,ref131}  In the following, we present a simple physical interpretation that explains these results, except for the numerical prefactors. We will
not discuss the effects of spin-orbit scattering\cite{ref131} or of superconducting
fluctuations,\cite{ref132} since these may be neglected in the systems considered in this
review. In this subsection we only discuss the dirty metal regime $l\ll W$. The
pure metal regime $l\gg  W$, in which boundary scattering plays an important
role, will be discussed in Section \ref{sec6c}.

If $l_{\phi}\ll W$ the {\it two--dimensional\/} weak localization correction to the conductivity applies, given by Eq.\ (\ref{eq6.4a}) for a zero magnetic field. The typical area
$S$ enclosed by a backscattered trajectory on a time scale $\tau_{B}$ is then of the order
$S\sim D\tau_{B}$ (assuming diffusive motion on this time scale). The corresponding
phase shift is $\phi\sim D\tau_{B}/l_{\mathrm{m}}^{2}$, in view of Eq.\ (\ref{eq6.5}). The criteria $\phi\sim 1$ and $\tau_{B}\sim\tau_{\phi}$
thus imply
\be
\tau_{B}\sim l_{\mathrm{m}}^{2}/D;\;\;B_{\mathrm{c}}\sim h/eD\tau_{\phi}\equiv h/el_{\phi}^{2}. \label{eq6.7}
\ee
The full expression for the magnetoconductance due to weak localization
is\cite{ref118,ref131}
\begin{widetext}
\be
\delta G_{\mathrm{loc}}^{2\mathrm{D}}(B)-\delta G_{\mathrm{loc}}^{2\mathrm{D}}(0)=\frac{W}{L}g_{\mathrm{s}}g_{ \mathrm{v}}\frac{e^{2}}{4\pi^{2}\hbar}\left[\Psi\left(\frac{1}{2}+\frac{\tau_{B}}{2\tau_{\phi}}\right)-\Psi\left(\frac{1}{2}+\frac{\tau_{B}}{2\tau}\right)+\ln\left(\frac{\tau_{\phi}}{\tau}\right)\right],
\label{eq6.8}
\ee
\end{widetext}
where $\Psi(x)$ is the digamma function and $\tau_{B}=l_{\mathrm{m}}^{2}/2D$. The digamma function
has the asymptotic approximation $\Psi(x)\approx\ln(x)-1/x$ for large $x$; thus, in a
zero magnetic field result (\ref{eq6.4a}) is recovered (assuming also $\tau_{\phi}\gg  \tau$). In the
case of 2D weak localization the characteristic field $B_{\mathrm{c}}$ is usually very weak.
For example, if $l_{\phi}=1\,\mu \mathrm{m}$, then $B_{\mathrm{c}}\approx 1\,\mathrm{mT}$. The suppression of the weak
localization effect is complete when $\tau_{B}\lesssim\tau$, which occurs for
$B\gtrsim \hbar/eD\tau\sim \hbar/el^{2}$. These fields are still much weaker than classically strong
fields for which $\omega_{\mathrm{c}}\tau\gtrsim 1$ (as can be verified by noting that when $B=\hbar/el^{2}$, one
has $\omega_{\mathrm{c}}\tau=1/k_{\mathrm{F}}l\ll 1$). The neglect of the curvature of electron trajectories in
the theory of weak localization is thus entirely justified in the 2D case. The
safety margin is narrower in the 1D case, however, since the characteristic
fields can become significantly enhanced.

\begin{figure}
\centerline{\includegraphics[width=8cm]{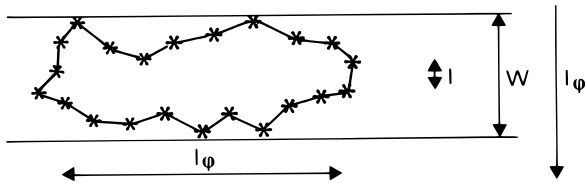}}
\caption{
Typical closed electron trajectory contributing to 1D weak localization ($l_{\phi}\gg W$) in the dirty metal regime ($l\ll W$). The asterisks denote elastic scattering events. Taken from H. van Houten et al., Acta Electronica {\bf 28}, 27 (1988).
\label{fig15}
}
\end{figure}

The {\it one-dimensional\/} case $W\ll l_{\phi}$ in a magnetic field has first been treated
by Al'tshuler and Aronov\cite{ref121} in the dirty metal regime. This refers to a narrow
channel with $l\ll W$ so that the wall-to-wall motion is diffusive. Since the
phase coherence length exceeds the channel width, the backscattered trajectories on a time scale $\tau_{B}$ have a typical enclosed area $S\sim W(D\tau_{B})^{1/2}$ (see Fig.
\ref{fig15}). Consequently, the condition $S\sim l_{\mathrm{m}}^{2}$ for a unit phase shift implies
\be
\tau_{B}\sim l_{\mathrm{m}}^{4}/DW^{2};\;\; B_{\mathrm{c}}\sim h/eWl_{\phi}. \label{eq6.9}
\ee
The difference with the 2D case is that the enclosed flux on a given time scale
is reduced, due to the lateral compression of the backscattered trajectories.
This leads to an enhancement by a factor $l_{\phi}/W$ of the characteristic field scale $B_{\mathrm{c}}$, compared with Eq.\ (\ref{eq6.7}). The full expression for the weak localization
correction if $l_{\phi}, l_{\mathrm{m}}\gg  W\gg  l$ is\cite{ref121}
\be
\delta G_{\mathrm{loc}}^{1\mathrm{D}}(B)=-g_{\mathrm{s}}g_{ \mathrm{v}}\frac{e^{2}}{h}\frac{1}{L}\left(\frac{1}{D\tau_{\phi}}+\frac{1}{D\tau_{B}}\right)^{-1/2},   \label{eq6.10}
\ee
with $\tau_{B}=3l_{\mathrm{m}}^{4}/W^{2}D$. For an elementary derivation of this result, see Ref.\ \onlinecite{ref109}.
At $l_{\mathrm{m}}\sim W$ a crossover from 1D to 2D weak localization occurs [i.e., from Eq.\
(\ref{eq6.10}) to Eq.\ (\ref{eq6.8})]. The reason for this crossover is that the lateral
confinement becomes irrelevant for the weak localization when $l_{\mathrm{m}}\lesssim W$,
because the trajectories of duration $\tau_{B}$ then have a typical extension
$(D\tau_{B})^{1/2}\lesssim W$, according to Eq.\ (\ref{eq6.9}). This crossover from 1D to 2D restricts
the available field range that can be used to study the magnetoconductance
associated with 1D weak localization.

The magnetic relaxation time $\tau_{B}$ in the dirty metal regime is found to be
inversely proportional to the diffusion constant $D$, in 2D as well as in 1D. The
reason for this dependence is clear: faster diffusion implies that less time is
needed to complete a loop of area $l_{\mathrm{m}}^{2}$. It is remarkable that in the pure metal
regime such a proportionality no longer holds. This is a consequence of the
flux cancellation effect discussed in Section \ref{sec6c}.

\begin{figure}
\centerline{\includegraphics[width=8cm]{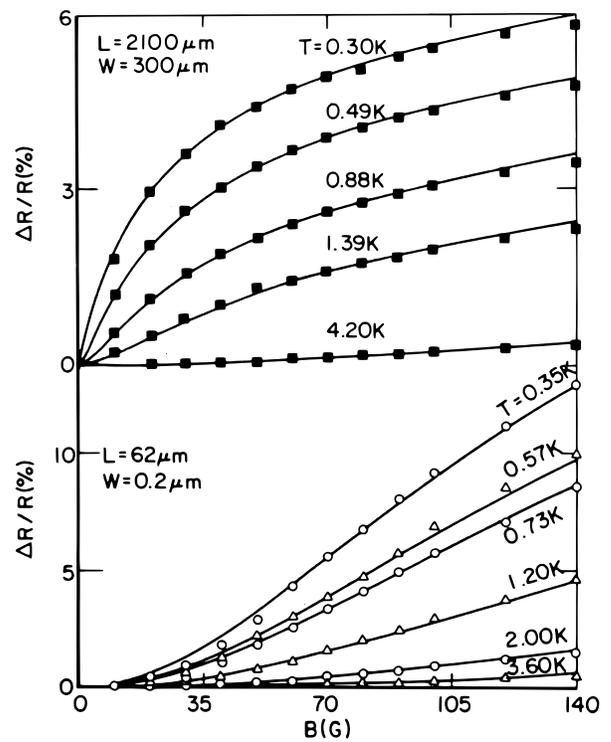}}
\caption{
A comparison between the magnetoresistance $\Delta R/R\equiv[R(0)-R(B)]/R(0)$ due to 2D weak localization in a wide channel (upper panel) and due to 1D weak localization in a narrow channel (lower panel), at various temperatures. The solid curves are fits based on Eqs.\ (\ref{eq6.8}) and (\ref{eq6.10}). Taken from K. K. Choi et al., Phys.\ Rev.\ B {\bf 36}, 7751 (1987).
\label{fig16}
}
\end{figure}

{\bf (b) Experiments in the dirty metal regime.} Magnetoresistance experiments
have been widely used to study the weak localization correction to the
conductivity of wide 2D electron gases in Si\cite{ref28,ref30,ref133,ref134,ref135} and GaAs.\cite{ref23,ref136,ref137}
Here we will discuss the experimental magnetoresistance studies of weak
localization in narrow channels in Si MOSFETs\cite{ref34,ref38,ref40,ref138} and GaAs-AlGaAs heterostructures.\cite{ref24,ref25,ref58} As an illustrative example, we reproduce in
Fig.\ \ref{fig16} a set of experimental results for $\delta R/R\equiv[R(0)-R(B)]/R(0)$ obtained
by Choi et al.\cite{ref25} in a wide and in a narrow GaAs-AlGaAs heterostructure.
The quantity $\delta R$ is positive, so the resistance decreases on applying a
magnetic field. The 2D results are similar to those obtained earlier by
Paalanen et al.\cite{ref137}  The qualitative difference in field scale for the suppression
of 2D (top) and 1D (bottom) weak localization is nicely illustrated by the data
in Fig.\ \ref{fig16}. The magnetoresistance peak is narrower in the 2D case, consistent
with the enhancement in 1D of the characteristic field $B_{\mathrm{c}}$ for the suppression
of weak localization, which we discussed in Section \ref{sec6b}(a). The solid curves in
Fig.\ \ref{fig16} were obtained from the 2D theoretical expression (\ref{eq6.8}) and the 1D
dirty metal result (\ref{eq6.10}), treating $W$ and $l_{\phi}$ as adjustable parameters. A
noteworthy finding of Choi et al.\cite{ref25} is that the effective channel width $W$ is
considerably reduced below the lithographic width $W_{\mathrm{lith}}$ in narrow channels
defined by a deep-etched mesa (as in Fig.\ \ref{fig4}a). Differences $W-W_{\mathrm{lith}}$ of about
$0.8\,\mu \mathrm{m}$ were found.\cite{ref25} Significantly smaller differences are obtained\cite{ref27,ref63} if a
shallow-etched mesa is used for the lateral confinement, as in Fig.\ \ref{fig4}c. A split-gate device (as in Fig.\ \ref{fig4}b) of variable width has been used by Zheng et al.\cite{ref24} to study weak localization in GaAs-AlGaAs heterostructure channels. Magnetoresistance experiments in a very narrow split-gate device (fabricated using
electron beam lithography) were reported by Thornton et al.\cite{ref58} and analyzed
in terms of the dirty metal theory. Unfortunately, in their experiment the
mean free path of $450\,\mathrm{nm}$ exceeded the width inferred from a fit to Eq.\ (\ref{eq6.10})
by an order of magnitude, so an analysis in terms of the pure metal theory
would have been required.

Early magnetoresistance experiments on narrow Si accumulation layers
were performed by Dean and Pepper,\cite{ref34} in which they observed evidence for a
crossover from the 2D to the 1D weak localization regime. A comparison of
weak localization in wide and narrow Si inversion layers was reported by
Wheeler et al.\cite{ref38} The conducting width of the narrow channel was taken to be
equal to the lithographic width of the gate (about $400\,\mathrm{nm}$), while the mean
free path was estimated to be about $100\,\mathrm{nm}$. This experiment on a low-mobility Si channel thus meets the requirement $l\ll W$ for the dirty metal
regime. The 1D weak localization condition $l_{\phi}\gg  W$ was only marginally
satisfied, however. Licini et al.\cite{ref40} reported a negative magnetoresistance peak
in 270-nm-wide Si inversion layers, which was well described by the 2D
theory at a temperature of $2.2\,\mathrm{K}$, where $l_{\phi}=120\,\mathrm{nm}$. Deviations from the 2D
form were found at lower temperatures, but the 1D regime was never fully
entered. A more recent study of 1D weak localization in a narrow Si
accumulation layer has been performed by Pooke et al.\cite{ref138} at low temperatures, and the margins are somewhat larger in their case.

We note a difficulty inherent to experiments on 1D weak localization in
semiconductor channels in the dirty metal regime. For 1D weak localization
it is required that the phase coherence length $l_{\phi}$ is much larger than the
channel width. If the mean free path is short, then the experiment is in the
dirty metal regime $l\ll W$, but the localization will be only marginally one-dimensional since the phase coherence length $l_{\phi}\equiv(D\tau_{\phi})^{1/2}=(v_{F}l\tau_{\phi}/2)^{1/2}$ will
be short as well (except for the lowest experimental temperatures). If the mean
free path is long, then the 1D criterion $l_{\phi}\gg  W$ is easily satisfied, but the
requirement $l\ll W$ will now be hard to meet so that the experiment will tend
to be in the pure metal regime. A quantitative comparison with the theory
(which would allow a reliable determination of $l_{\phi}$) is hampered because the
asymptotic regimes studied theoretically are not accessible experimentally
and because the channel width is not known a priori. Nanostructures are thus
not the best candidates for a quantitative study of the phase coherence length,
which is better studied in 2D systems. An altogether different complication is
that quantum corrections to the conductivity in semiconductor nanostructures can be remarkably large (up to 100\% at sufficiently low temperatures\cite{ref27,ref34}), which puts them beyond the range of validity of the perturbation theory.

\subsubsection{\label{sec6c} Boundary scattering and flux cancellation}

\begin{figure}
\centerline{\includegraphics[width=8cm]{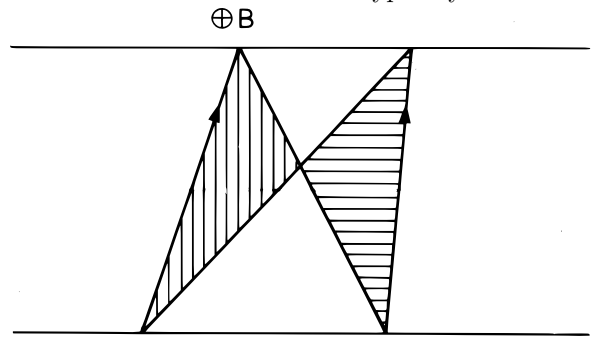}}
\caption{
Illustration of the flux cancellation effect for a closed trajectory of one electron in a narrow channel with diffuse boundary scattering. The trajectory is composed of two loops of equal area but opposite orientation, so it encloses zero flux. Taken from C. W. J. Beenakker and H. van Houten, Phys.\ Rev.\ B. {\bf 38}, 3232 (1988).
\label{fig17}
}
\end{figure}

\begin{figure}
\centerline{\includegraphics[width=8cm]{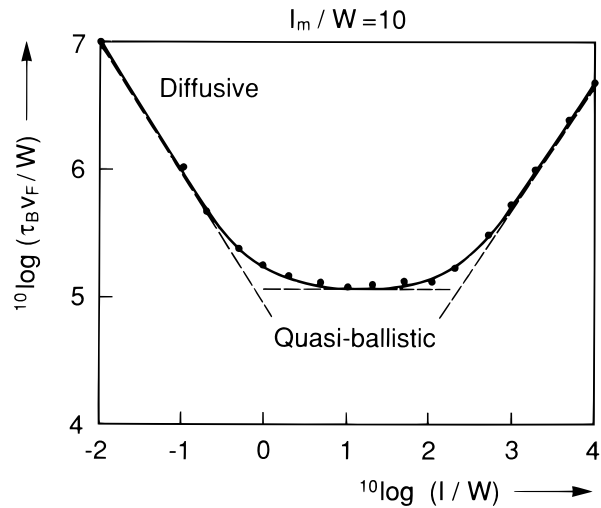}}
\caption{
Phase relaxation time $\tau_{B}$ in a channel with specular boundary scattering, as a function of the elastic mean free path $l$. The plot has been obtained by a numerical simulation of the phase relaxation process for a magnetic field such that $l_{\rm m}=10\,W$. The dashed lines are analytic formulas valid in the three asymptotic regimes (see Table \ref{table2}). Taken from C. W. J. Beenakker and H. van Houten, Phys.\ Rev.\ B {\bf 38}, 3232 (1988).
\label{fig18}
}
\end{figure}

{\bf (a) Theory.} In the previous subsection we noticed that the pure metal
regime, where $l\gg  W$, is characteristic for 1D weak localization in semiconductor nanostructures. This regime was first theoretically considered by
Dugaev and Khmel'nitskii,\cite{ref120} for the geometry of a thin metal film in a
parallel magnetic field and for diffuse boundary scattering. The geometry of a
narrow 2DEG channel in a perpendicular magnetic field, with either diffuse
or specular boundary scattering, was treated by the present authors.\cite{ref109}  Note
that the nature of the boundary scattering did not play a role in the dirty
metal regime of Section \ref{sec6b}, since there the channel walls only serve to impose
a geometrical restriction on the lateral diffusion.\cite{ref121} The {\it flux cancellation effect\/} is characteristic of the pure metal regime, where the electrons move
ballistically from one wall to the other. This effect (which also plays a role in
the superconductivity of thin films in a parallel magnetic field\cite{ref122}) leads to a
further enhancement of the characteristic field scale $B_{\mathrm{c}}$. Flux cancellation
results from the fact that typically backscattered trajectories for $l\gg  W$ self-intersect (cf.\ Fig.\ \ref{fig17}) and are thus composed of smaller loops that are
traversed in opposite directions. Zero net flux is enclosed by closed trajectories involving only wall collisions (as indicated by the shaded areas in Fig.\
\ref{fig17}, which are equal but of opposite orientation), so impurity collisions are
required for phase relaxation in a magnetic field. This is in contrast to the
dirty metal regime considered before, where impurity scattering hinders
phase relaxation by reducing the diffusion constant. The resulting nonmonotonous dependence of phase relaxation on impurity scattering in the dirty
and pure metal regimes is illustrated in Fig.\ \ref{fig18}, where the calculated\cite{ref109}
magnetic relaxation time $\tau_{B}$ is plotted as a function of $l/W$ for a fixed ratio
$l_{\mathrm{m}}/W$.

Before continuing our discussion of the flux cancellation effect, we give a
more precise definition of the phase relaxation time $\tau_{B}$. The effect of a
magnetic field on weak localization is accounted for formally by inserting the
term
\be
\langle {\mathrm e}^{\mathrm{i}\phi(t)}|\mathbf{r}(t)=\mathbf{r}(0)\rangle=e^{-t/\tau_{B}},\;\; W\ll l_{\mathrm{m}}, l_{\phi}, \label{eq6.11}
\ee 
in the integrand of Eq.\ (\ref{eq6.2}). The term (\ref{eq6.11}) is the conditional average over all
closed trajectories having duration $t$ of the phase factor $\mathrm{e}^{\mathrm{i}\phi(t)}$, with $\phi$ the phase
difference defined in Eq.\ (\ref{eq6.5}). It can be shown\cite{ref109} that in the case of 1D weak
localization (and for $l_{\mathrm{m}}\gg  W$), this term is given by an exponential decay
factor $\exp(- t/\tau_{B})$, which defines the magnetic relaxation time $\tau_{B}$. In this
regime the weak localization correction to the conductivity in the presence of
a magnetic field is then simply given by Eq.\ (\ref{eq6.4b}), after the substitution
\be
\tau_{\phi}^{-1}\rightarrow\tau_{\phi}^{-1}+\tau_{B}^{-1}. \label{eq6.12}
\ee 
Explicitly, one obtains
\begin{widetext}
\be
\delta G_{\mathrm{loc}}(B)=-g_{\mathrm{s}}g_{ \mathrm{v}}\frac{e^{2}}{h}\frac{1}{L}\left(\left[\frac{1}{D\tau_{\phi}}+\frac{1}{D\tau_{B}}\right]^{-1/2}-\left[\frac{1}{D\tau_{\phi}}+\frac{1}{D\tau_{B}}+\frac{1}{D\tau}\right]^{-1/2}\right).
\label{eq6.13}
\ee
\end{widetext}

One can see from Fig.\ \ref{fig18} and Table \ref{table2} that in the pure metal regime $l\gg  W$,
a weak and strong field regime can be distinguished, depending on the ratio
$Wl/l_{\mathrm{m}}^{2}$. This ratio corresponds to the maximum phase change on a closed
trajectory of linear extension $l$ (measured along the channel). In the {\it weak\/} field
regime $(Wl/l_{\mathrm{m}}^{2}\ll 1)$ many impurity collisions are required before a closed
electron loop encloses sufficient flux for complete phase relaxation. In this
regime a further increase of the mean free path does not decrease the phase
relaxation time (in contrast to the dirty metal regime), because as a
consequence of the flux cancellation effect, faster diffusion along the channel
does not lead to a larger enclosed flux. On comparing the result in Table \ref{table2}
for $B_{\mathrm{c}}$ in the weak field regime with that for the dirty metal regime, one sees an
enhancement of the characteristic field by a factor $(l/W)^{1/2}$. The {\it strong\/} field
regime is reached if $Wl/l_{\mathrm{m}}^{2}\gg  1$, while still $l_{\mathrm{m}}\gg  W$. Under these conditions, a
single impurity collision can lead to a closed trajectory that encloses sufficient
flux for phase relaxation. The phase relaxation rate $1/\tau_{B}$ is now proportional
to the impurity scattering rate $1/\tau$ and, thus, to $1/l$. The relaxation time $\tau_{B}$
accordingly {\it increases\/} linearly with $l$ in this regime (see Fig.\ \ref{fig18}). For
comparison with experiments in the pure metal regime, an analytic formula
that interpolates between the weak and strong field regimes is useful. The
following formula agrees well with numerical calculations:\cite{ref109}
\be
\tau_{B}=\tau_{B}^{\mathrm{weak}}+\tau_{B}^{\mathrm{strong}}. \label{eq6.14}
\ee 
Here $\tau_{B}^{\mathrm{weak}}$ and $\tau_{B}^{\mathrm{strong}}$ are the expressions for $\tau_{B}$ in the asymptotic weak and strong field regimes, as given in Table \ref{table2}.

So far, we have assumed that the transport is diffusive on time scales
corresponding to $\tau_{\phi}$. This will be a good approximation only if $\tau_{\phi}\gg  \tau$.
Coherent diffusion breaks down if $\tau_{\phi}$ and $\tau$ are of comparable magnitude (as
may be the case in high-mobility channels). The modification of weak
localization as one enters the ballistic transport regime has been investigated
by Wittmann and Schmid.\cite{ref130}  It would be of interest to see to what extent the
ad hoc short-time cutoff introduced in our Eq.\ (\ref{eq6.4}), which is responsible for
the second bracketed term in Eq.\ (\ref{eq6.13}), is satisfactory.

\begin{figure}
\centerline{\includegraphics[width=8cm]{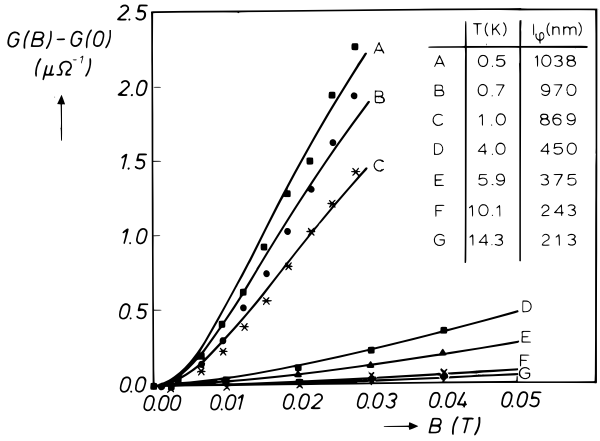}}
\caption{
Magnetoconductance due to 1D weak localization in the pure metal regime ($W = 120\,{\rm nm}$, $L = 350\,{\rm nm}$). The solid curves are one-parameter fits to Eq.\ (\ref{eq6.13}). Only the field range $l_{\rm m} > W$ is shown in accordance with the condition of coherent diffusion imposed by the theory. The phase coherence length $l_{\phi}$ obtained from the data at various temperatures is tabulated in the inset. Taken from H. van Houten et al., Surf.\ Sci.\ {\bf 196}, 144 (1988).
\label{fig19}
}
\end{figure}

{\bf (b) Experiments in the pure metal regime.} Because of the high mobility
required, the pure metal regime has been explored using GaAs-AlGaAs
heterostructures only. The first experiments on weak localization in the pure
metal regime were done by Thornton et al.,\cite{ref58} in a narrow split-gate device,
although the data were analyzed in terms of the theory for the dirty metal
regime. An experimental study specifically aimed at weak localization in the
pure metal regime was reported in Refs.\ \onlinecite{ref26} and \onlinecite{ref27}. In a narrow channel
defined by the shallow-mesa etch technique of Fig.\ \ref{fig4}c (with a conducting
width estimated at $0.12\,\mu \mathrm{m}$), a pronounced negative magnetoresistance effect
was found, similar to that observed by Thornton et al.\cite{ref58}  A good agreement of
the experimental results with the theory\cite{ref109} for weak localization in the pure
metal regime was obtained (see Fig.\ \ref{fig19}), assuming specular boundary
scattering (diffuse boundary scattering could not describe the data). The
width deduced from the analysis was consistent with independent estimates
from other magnetoresistance effects. Further measurements in this regime
were reported by Chang et al.\cite{ref70,ref139} and, more recently, by Hiramoto et al.\cite{ref81}
These experiments were also well described by the theory of Ref.\ \onlinecite{ref109}.

\subsection{\label{sec7} Conductance fluctuations}

Classically, sample-to-sample fluctuations in the conductance are negligible in the diffusive (or quasi-ballistic) transport regime. In a narrow-channel geometry, for example, the root-mean-square $\delta G_{\mathrm{class}}$ of the classical
fluctuations in the conductance is smaller than the average conductance $\langle G\rangle$
by a factor $(l/L)^{1/2}$, under the assumption that the channel can be subdivided
into $L/l\gg  1$ {\it independently\/} fluctuating segments. As we have discussed in the
previous section, however, quantum mechanical correlations persist over a
phase coherence length $l_{\phi}$ that can be much larger than the elastic mean free
path $l$. Quantum interference effects lead to significant sample-to-sample
fluctuations in the conductance if the size of the sample is not very much
larger than $l_{\phi}$. The Al'tshuler-Lee-Stone theory of {\it Universal Conductance Fluctuations}\cite{ref140,ref141} finds that $\delta G\approx e^{2}/h$ at $T=0$, when phase coherence is
maintained over the entire sample. Since $\langle G\rangle\propto L^{-1}$, it follows that
$\delta G/\langle G\rangle\propto L$ {\it increases\/} with increasing channel length; that is, there is a total
absence of self-averaging.

Experimentally, the large sample-to-sample conductance fluctuations
predicted theoretically are difficult to study in a direct way, because of
problems in the preparation of samples that differ in impurity configuration
only (to allow an ensemble average). The most convenient way to study the
effect is via the fluctuations in the conductance of a single sample as a
function of magnetic field, because a small change in field has a similar effect
on the interference pattern as a change in impurity configuration. Sections \ref{sec7c}
and \ref{sec7d} deal with theoretical and experimental studies of magnetoconductance fluctuations in narrow 2DEG channels, mainly in the quasi-ballistic
regime characteristic for semiconductor nanostructures. In Sections \ref{sec7a} and
\ref{sec7b} we discuss the surprising universality of the conductance fluctuations at
zero temperature and the finite-temperature modifications.

\subsubsection{\label{sec7a} Zero-temperature conductance fluctuations}

The most surprising feature of the conductance fluctuations is that their
magnitude at zero temperature is of order $e^{2}/h$, regardless of the size of the
sample and the degree of disorder,\cite{ref140,ref141}  provided at least that $L\gg  l$, so that
transport through the sample is diffusive (or possibly quasi-ballistic). Lee and
Stone\cite{ref141} coined the term {\it Universal Conductance Fluctuations\/} (UCF) for this
effect. In this subsection we give a simplified explanation of this universality
due to Lee.\cite{ref142}

\begin{figure}
\centerline{\includegraphics[width=8cm]{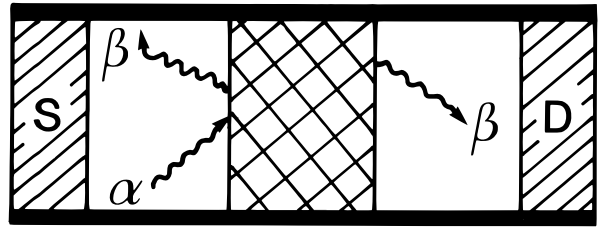}}
\caption{
Idealized conductor connecting source (S) and drain (D) reservoirs and containing a disordered region (crosshatched). The incoming quantum channels (or transverse waveguide modes) are labeled by $\alpha$, the transmitted and back scattered channels by $\beta$.
\label{fig20}
}
\end{figure}

Consider first the classical Drude conductance (\ref{eq4.8}) for a singe spin
direction (and a single valley):
\be
G=\frac{W}{L}\frac{e^{2}}{h}\frac{k_{\mathrm{F}}l}{2}=\frac{e^{2}}{h}\frac{\pi l}{2L} N,\;\;N \equiv\frac{k_{\mathrm{F}}W}{\pi}. \label{eq7.1}
\ee 
The number $N$ equals the number of transverse modes, or one-dimensional
subbands, that are occupied at the Fermi energy in a conductor of width $W$.
We have written the conductance in this way to make contact with the
Landauer approach\cite{ref4} to conduction, which relates the conductance to the
transmission probabilities of modes at the Fermi energy. (A detailed
discussion of this approach is given the context of quantum ballistic transport
in Section \ref{sec12b}). The picture to have in mind is shown in Fig.\ \ref{fig20}. Current is
passed from a source reservoir $\mathrm{S}$ to a drain reservoir $\mathrm{D}$, through a disordered
region (hatched) in which only elastic scattering takes place. The two
reservoirs are in thermal equilibrium and are assumed to be fully effective in
randomizing the phase via inelastic scattering, so there is no phase coherence
between the $N$ modes incident on the disordered region. The modes in this
context are called {\it quantum channels}. If $L\gg  l$, each channel has on average the
same transmission probability, given by $\pi l/2L$ according to  Eqs.\ (\ref{eq4.21}) and
(\ref{eq7.1}). We are interested in the fluctuations around this average. The resulting
fluctuations in $G$ then follow from the multichannel Landauer
formula\cite{ref1,ref143,ref144}
\be
G= \frac{e^{2}}{h}\sum_{\alpha,\beta=1}^{N}|t_{\alpha\beta}|^{2}, \label{eq7.2}
\ee 
where $t_{\beta\alpha}$ denotes the quantum mechanical transmission probability
amplitude from the incident channel $\alpha$ to the outgoing channel $\beta$ (cf.\ Fig.\ \ref{fig20}).
The ensemble averaged transmission probability $\langle|t_{\alpha\beta}|^{2}\rangle$ does not depend on
$\alpha$ or $\beta$, so the correspondence between  Eqs.\ (\ref{eq7.1}) and (\ref{eq7.2}) requires
\be
\langle|t_{\alpha\beta}|^{2}\rangle=\pi l/2NL. \label{eq7.3}
\ee 
The magnitude of the conductance fluctuations is characterized by its
variance $\mathrm{Var}\,(G)\equiv\langle(G-\langle G\rangle)^{2}\rangle$. As discussed by Lee, a difficulty arises in a
direct evaluation of $\mathrm{Var}\,(G)$ from Eq.\ (\ref{eq7.2}), because the correlation in the
transmission probabilities $|t_{\alpha\beta}|^{2}$ for different pairs of incident and outgoing
channels $\alpha, \beta$ may not be neglected.\cite{ref142} The reason is presumably that
transmission through the disordered region involves a large number of
impurity collisions, so a sequence of scattering events will in general be
shared by different channels. On the same grounds, it is reasonable to assume
that the reflection probabilities $|r_{\alpha\beta}|^{2}$ for different pairs $\alpha\beta$ of incident and
reflected channels are uncorrelated, since the reflection back into the source
reservoir would seem to be dominated by only a few scattering events.\cite{ref142}
(The formal diagrammatic analysis of Refs.\ \onlinecite{ref140} and \onlinecite{ref141} is required here for a
convincing argument.) The reflection and transmission probabilities are
related by current conservation
\be
\sum_{\alpha,\beta=1}^{N}|t_{\alpha\beta}|^{2}=N-\sum_{\alpha,\beta=1}^{N}|r_{\alpha\beta}|^{2}. \label{eq7.4}
\ee 
so the variance of the conductance equals
\begin{eqnarray}
\mathrm{Var}\,(G)&=&\left(\frac{e^{2}}{h}\right)^{2}\mathrm{Var}\,\left(\sum|r_{\alpha\beta}|^{2}\right)\nonumber\\
&=&\left(\frac{e^{2}}{h}\right)^{2}N^{2}\mathrm{Var}\,(|r_{\alpha\beta}|^{2}), \label{eq7.5}
\end{eqnarray}
assuming uncorrelated reflection probabilities. A large number $M$ of scattering sequences through the disordered region contributes with amplitude
$A(i)$ ($i=1,2, \ldots, M$) to the reflection probability amplitude $r_{\alpha\beta}$. (The different scattering sequences can be seen as independent Feynman paths
in a path integral formulation of the problem.\cite{ref142}) To calculate
$\mathrm{Var}\,(|r_{\alpha\beta}|^{2})=\langle|r_{\alpha\beta}|^{4}\rangle-\langle|r_{\alpha\beta}|^{2}\rangle^{2}$, one may then write (neglecting correlations
in $A(i)$ for different $i$)
\begin{eqnarray}
\langle|r_{\alpha\beta}|^{4}\rangle&=&\sum_{i,j,k,l=1}^{M}\langle A^{*}(i)A(j)A^{*}(k)A(l)\rangle\nonumber\\
&=& \sum_{i,j,k,l=1}^{M}\left\{\langle|A(i)|^{2}\rangle\langle|A(k)|^{2}\rangle\delta_{ij}\delta_{kl}\right.\nonumber\\
&&\left.\mbox{}+\langle|A(i)|^{2}\rangle\langle|A(j)|^{2}\rangle\delta_{il}\delta_{jk}\right\}\nonumber\\
&=&2\langle|r_{\alpha\beta}|^{2}\rangle^{2}, \label{eq7.6}
\end{eqnarray}
where we have neglected terms smaller by a factor $1/M$ (assuming $M\gg  1$).
One thus finds that the variance of the reflection probability is equal to the
square of its average:
\be
\mathrm{Var}\,(|r_{\alpha\beta}|^{2})=\langle|r_{\alpha\beta}|^{2}\rangle^{2}. \label{eq7.7}
\ee 
The average reflection probability $\langle|r_{\alpha\beta}|^{2}\rangle$ does not depend on $\alpha$ and $\beta$. Thus,
from  Eqs.\ (\ref{eq7.3}) and (\ref{eq7.4}) it follows that
\be
\langle|r_{\alpha\beta}|^{2}\rangle=N^{-1}(1- {\rm order}(l/L)). \label{eq7.8}
\ee
Combining  Eqs.\ (\ref{eq7.5}), (\ref{eq7.7}), and (\ref{eq7.8}), one obtains the result that the zero-temperature conductance has a variance $(e^{2}/h)^{2}$, independent of $l$ or $L$ (in the
diffusive limit $l\ll L$). We have discussed this argument of Lee in some detail,
because no other simple argument known to us gives physical insight in this
remarkable result.

The numerical prefactors follow from the diagrammatic analysis.\cite{ref140,ref141,ref145,ref146} The result of Lee and Stone\cite{ref141} for the root-mean-square
magnitude of the conductance fluctuations at $T=0$ can be written in the
form
\be
\delta G\equiv[\mathrm{Var}\,(G)]^{1/2}=\frac{g_{\mathrm{s}}g_{\mathrm{v}}}{2}\beta^{-1/2}C\frac{e^{2}}{h}. \label{eq7.9}
\ee 
Here $C$ is a constant that depends on the shape of the sample. Typically, $C$ is
of order unity; for example, $C\approx 0.73$ in a narrow channel with $L\gg  W$.
(However, in the opposite limit $W\gg  L$ of a wide and short channel, $C$ is of
order $(W/L)^{1/2}$.) The parameter $\beta=1$ in a zero magnetic field when time-reversal symmetry holds; $\beta=2$ when time-reversal symmetry is broken by a
magnetic field. The factor $g_{\mathrm{s}}g_{\mathrm{v}}$ assumes complete spin and valley degeneracy.
If the magnetic field is sufficiently strong that the two spin directions give
statistically independent contributions to the conductance, then the variances
add so that the factor $g_{\mathrm{s}}$ in $\delta G$ is to be replaced by a factor $g_{\mathrm{s}}^{1/2}$. We will return
to this point in Section \ref{sec7d}.

\subsubsection{\label{sec7b} Nonzero temperatures}

At nonzero temperatures, the magnitude of the conductance fluctuations is
reduced below $\delta G\approx e^{2}/h$. One reason is the effect of a finite phase coherence
length $l_{\phi}\equiv(D\tau_{\phi})^{1/2}$; another is the effect of thermal averaging, as expressed by
the thermal length $l_{\mathrm{T}}\equiv(hD/k_{B}T)^{1/2}$. The effect of a finite temperature,
contained in $l_{\phi}$ and $l_{\mathrm{T}}$, is to partially restore self-averaging, albeit that the
suppression of the fluctuation with sample size is much weaker than would be
the case classically. The theory has been presented clearly and in detail by
Lee, Stone, and Fukuyama.\cite{ref145} We limit the present discussion to the 1D
regime $W\ll l_{\phi}\ll L$, characteristic for narrow 2DEG channels.

The effects of thermal averaging may be neglected if $l_{\phi}\ll l_{\mathrm{T}}$ (see below).
The channel may then be thought to be subdivided in uncorrelated segments
of length $l_{\phi}$. The conductance fluctuation of each segment individually will be
of order $e^{2}/h$, as it is at zero temperature. The root-mean-square conductance
fluctuation of the entire channel is easily estimated. The segments are in
series, so their resistances add according to Ohm's law. We denote the
resistance of a channel segment of length $l_{\phi}$ by $R_{1}$. The variance of $R_{1}$ is
$\mathrm{Var}\,(R_{1})\approx\langle R_{1}\rangle^{4}\mathrm{Var}\,(R_{1}^{-1})\approx\langle R_{1}\rangle^{4}(e^{2}/h)^{2}$. The average resistance of the
whole channel $\langle R\rangle=(L/l_{\phi})\langle R_{1}\rangle$ increases linearly with the number $L/l_{\phi}$ of
uncorrelated channel segments, just as its variance
\[
\mathrm{Var}\,(R)=(L/l_{\phi})\mathrm{Var}\,(R_{1})\approx(L/l_{\phi})\langle R_{1}\rangle^{4}(e^{2}/h)^{2}.
\]
(The root-mean-square resistance fluctuation thus grows as $(L/l_{\phi})^{1/2}$, the square root of the number of channel
segments in series.) Expressed in terms of a conductance, one thus has
$\mathrm{Var}\,(G)\approx\langle R\rangle^{-4}\mathrm{Var}\,(R)\approx(l_{\phi}/L)^{3}(e^{2}/h)^{2}$, or
\be
\delta G= {\rm constant} 
\times\frac{e^{2}}{h}\left(\frac{l_{\phi}}{L}\right)^{3/2},\;\; {\rm if}\;\; l_{\phi}\ll l_{\mathrm{T}}. \label{eq7.10}
\ee
The constant prefactor is given in Table \ref{table3}.

We now turn to the second effect of the finite temperature, which is the
smearing of the fluctuations by the energy average within an interval of order
$k_{\mathrm{B}}T$ around the Fermi energy $E_{\mathrm{F}}$. Note that we did not have to consider this
thermal averaging in the context of the weak localization effect, since that is
a systematic, rather than a fluctuating, property of the sample. Two interfering Feynman paths, traversed with an energy difference $\delta E$, have to be considered as uncorrelated after a time $t_{1}$, if the acquired phase difference $t_{1}\delta E/\hbar$
is of order unity. In this time the electrons diffuse a distance
$L_{1}=(Dt_{1})^{1/2}\sim(\hbar D/\delta E)^{1/2}$. One can now define a correlation energy $E_{\mathrm{c}}(L_{1})$,
as the energy difference for which the phase difference following diffusion over
a distance $L_{1}$ is unity:
\be
E_{\mathrm{c}}(L_{1})\equiv \hbar D/L_{1}^{2}. \label{eq7.11}
\ee 
The thermal length $l_{\mathrm{T}}$ is defined such that $E_{\mathrm{c}}(l_{\mathrm{T}})\equiv k_{\mathrm{B}}T$, which implies
\be
l_{\mathrm{T}}\equiv(\hbar D/k_{B}T)^{1/2}. \label{eq7.12}
\ee

\begin{table*}
\caption{Asymptotic expressions for the root-mean-square conductance fluctuations in a narrow channel.
\label{table3}
}
\begin{ruledtabular}
\begin{tabular}{cccc}
&$T=0$\footnotemark[1]&\multicolumn{2}{c}{$T>0$\footnotemark[1]}\\
&$l_{\rm T},l_{\phi}\gg L$&$l_{\phi}\ll L,l_{\rm T}$&$l_{\rm T}\ll l_{\phi}\ll L$\\
\hline
&&&\\
$\displaystyle\delta G\times\frac{2}{g_{\rm s}g_{\rm v}}\beta^{1/2}$&$\displaystyle C\frac{e^{2}}{h}$&$\displaystyle C\frac{e^{2}}{h}\left(\frac{l_{\phi}}{L}\right)^{3/2}$&$\displaystyle C\frac{e^{2}}{h}\frac{l_{\rm T}l_{\phi}^{1/2}}{L^{3/2}}$\\
&&&\\
$C$&0.73&$\displaystyle\sqrt{12}$&$\displaystyle\left(\frac{8\pi}{3}\right)^{1/2}$
\end{tabular}
\end{ruledtabular}
\footnotetext[1]{
The results assume a narrow channel ($W\ll L$), with a 2D density of states ($W\gg\lambda_{\rm F}$), which is in the 1D limit for the conductance fluctuations ($W\ll l_{\phi}$). The expressions for $\delta G$ are from Refs.\ \onlinecite{ref140,ref141,ref145}, and \onlinecite{ref146}. The numerical prefactor $C$ for $T=0$ is from Ref.\ \onlinecite{ref141}, for $T>0$ from Ref.\ \onlinecite{ref147}. If time-reversal symmetry applies, then $\beta=1$, but in the presence of a magnetic field strong enough to suppress the cooperon contributions then $\beta=2$. If the spin degeneracy is lifted, $g_{\rm s}$ is to be replaced by $g_{\rm s}^{1/2}$.}
\end{table*}

(Note that this definition of $l_{\mathrm{T}}$ differs by a factor of $(2\pi)^{1/2}$ from that in Ref.\
\onlinecite{ref145}.) The thermal smearing of the conductance fluctuations is of importance
only if phase coherence extends beyond a length scale $l_{\mathrm{T}}$ (i.e., if $l_{\phi}\gg  l_{\mathrm{T}}$). In this
case the total energy interval $k_{\mathrm{B}}T$ around the Fermi level that is available for
transport is divided into subintervals of width $E_{\mathrm{c}}(l_{\phi})=\hbar/\tau_{\phi}$ in which phase
coherence is maintained. There is a number $N\approx k_{\mathrm{B}}T/E_{\mathrm{c}}(l_{\phi})$ of such subintervals, which we assume to be uncorrelated. The root-mean-square variation
$\delta G$ of the conductance is then reduced by a factor $N^{-1/2}\approx l_{\mathrm{T}}/l_{\phi}$ with respect
to the result (\ref{eq7.10}) in the absence of energy averaging. (A word of caution: as
discussed in Ref.\ \onlinecite{ref145}, the assumption of $N$ uncorrelated energy intervals is
valid in the 1D case $W\ll l_{\phi}$ considered here, but not in higher dimensions.)
From the foregoing argument it follows that
\be
\delta G=\mathrm{constant}\times\frac{e^{2}}{h}\frac{l_{\mathrm{T}}l_{\phi}^{1/2}}{L^{3/2}}\;\;{\rm if}\;\;l_{\phi}\gg  l_{\mathrm{T}}. \label{eq7.13}
\ee 

The asymptotic expressions (\ref{eq7.10}) and (\ref{eq7.13}) were derived by Lee, Stone,
and Fukuyama\cite{ref145} and by Al'tshuler and Khmel'nitskii\cite{ref146} up to unspecified
constant prefactors. These constants have been evaluated in Ref.\ \onlinecite{ref147}, and are
given in Table \ref{table3}. In that paper we also gave an interpolation formula
\begin{eqnarray}
\delta G&=&\frac{g_{\mathrm{s}}g_{\mathrm{v}}}{2}\beta^{-1/2}\sqrt{12}\frac{e^{2}}{h}\left(\frac{l_{\phi}}{L}\right)^{3/2}\nonumber\\
&&\mbox{}\times\left[1+\frac{9}{2\pi}\left(\frac{l_{\phi}}{l_{\mathrm{T}}}\right)^{2}\right]^{-1/2},   \label{eq7.14}
\end{eqnarray}
with $\beta$ defined in the previous subsection. This formula is valid (within 10\%
accuracy) also in the intermediate regime when $l_{\phi}\approx l_{\mathrm{T}}$, and is useful for
comparison with experiments, in which generally $l_{\phi}$ and $l_{\mathrm{T}}$ are not well
separated (cf.\ Table \ref{table1}).

\subsubsection{\label{sec7c} Magnetoconductance fluctuations}

Experimentally, one generally studies the conductance fluctuations resulting from a change in Fermi energy $E_{\mathrm{F}}$ or magnetic field $B$ rather than from a
change in impurity configuration. A comparison with the theoretical ensemble average becomes possible if one assumes that, insofar as the
conductance fluctuations are concerned, a sufficiently large change in $E_{\mathrm{F}}$ or $B$
is equivalent to a complete change in impurity configuration (this ``ergodic
hypothesis'' has been proven in Ref.\ \onlinecite{ref148}). The reason for this equivalence is
that, on one hand, the conductance at $E_{\mathrm{F}}+\Delta E_{\mathrm{F}}$ and $B+\Delta B$ is uncorrelated
with that at $E_{\mathrm{F}}$ and $B$, provided either $\Delta E_{\mathrm{F}}$ or $\Delta B$ is larger than a correlation
energy $\Delta E_{\mathrm{c}}$ or correlation field $\Delta B_{\mathrm{c}}$. On the other hand, the correlation
energies and fields are in general sufficiently small that the statistical
properties of the ensemble are not modified by the increment in $E_{\mathrm{F}}$ or $B$, so
one is essentially studying a new member of the same ensemble, without
changing the sample.

This subsection deals with the calculation of the correlation field $\Delta B_{\mathrm{c}}$. (The
correlation energy is discussed in Ref.\ \onlinecite{ref145} and will not be considered here.)
The magnetoconductance correlation function is defined as
\be
F(\Delta B)\equiv\langle[\delta G(B)-\langle G(B)\rangle][G(B+\Delta B)-\langle G(B+\Delta B)\rangle]\rangle,
\label{eq7.15}
\ee
where the angle brackets $\langle\cdots\rangle$ denote, as before, an ensemble average. The
root-mean-square variation $\delta G$ considered in the previous two subsections is
equal to $F(0)^{1/2}$. The correlation field $\Delta B_{\mathrm{c}}$ is defined as the half-width at half-height $F(\Delta B_{\mathrm{c}})\equiv F(0)/2$. The correlation function $F(\Delta B)$ is determined
theoretically\cite{ref141,ref145,ref146} by temporal and spatial integrals of two propagators:
the {\it diffuson\/} $P_{\mathrm{d}}(\mathbf{r}, \mathbf{r}^{\prime}, t)$ and the {\it cooperon\/} $P_{\mathrm{c}}(\mathbf{r}, \mathbf{r}^{\prime}, t)$. As discussed by Chakravarty and Schmid,\cite{ref126} these propagators consist of the product of three terms:
(1) the classical probability to diffuse from $\mathbf{r}$ to $\mathbf{r}^{\prime}$ in a time $t$ (independent of $B$
in the field range $\omega_{\mathrm{c}}\tau\ll 1$ of interest here); (2) the relaxation factor $\exp(- t/\tau_{\phi}$),
which describes the loss of phase coherence due to inelastic scattering events;
(3) the average phase factor $\langle\exp(i\Delta\phi)\rangle$, which describes the loss of phase
coherence due to the magnetic field. The average $\langle\cdots\rangle$ is taken over all
classical trajectories that diffuse from $\mathbf{r}$ to $\mathbf{r}^{\prime}$ in a time $t$. The phase difference
$\Delta\phi$ is different for a diffuson or cooperon:
\begin{subequations}
\label{eq7.16}
\begin{eqnarray}
\Delta\phi(\mathrm{diffuson})&=&\frac{e}{\hbar}\int_{\mathbf{r}}^{\mathbf{r}^{\prime}}\Delta \mathbf{A}\cdot d\mathbf{l}, \label{eq7.16a}\\
\Delta\phi(\mathrm{cooperon})&=&\frac{e}{\hbar}\int_{\mathbf{r}}^{\mathbf{r}^{\prime}}(2\mathbf{A}+\Delta \mathbf{A})\cdot d\mathbf{l}, \label{eq7.16b}
\end{eqnarray}
\end{subequations}
where the line integral is along a classical trajectory. The vector potential $\mathbf{A}$
corresponds to the magnetic field $\mathbf{B}=\nabla\times \mathbf{A}$, and the vector potential
increment $\Delta \mathbf{A}$ corresponds to the field increment $\Delta B$ in the correlation
function $F(\Delta B)$ (according to $\Delta \mathbf{B}=\nabla\times\Delta \mathbf{A}$). An explanation of the different
magnetic field dependencies of the diffuson and cooperon in terms of
Feynman paths is given shortly.

In Ref.\ \onlinecite{ref109} we have proven that in a narrow channel $(W\ll l_{\phi})$ the average
phase factor $\langle\exp(i\Delta\phi)\rangle$ does not depend on initial and final coordinates $\mathbf{r}$
and $\mathbf{r}^{\prime}$, provided that one works in the Landau gauge and that $t\gg  \tau$. This is a
very useful property, since it allows one to transpose the results for
$\langle \exp(i\Delta\phi)\rangle$ obtained for $\mathbf{r}=\mathbf{r}^{\prime}$ in the context of weak localization to the
present problem of the conductance fluctuations, where $\mathbf{r}$ can be different
from $\mathbf{r}^{\prime}$. We recall that for weak localization the phase difference $\Delta\phi$ is that of
the cooperon, with the vector potential increment $\Delta \mathbf{A}=0$ [cf.\ Eq.\ (\ref{eq6.5})]. The
average phase factor then decays exponentially as $\langle \exp(i\Delta\phi)\rangle=\exp(-t/\tau_{B}$)
[cf.\ Eq.\ (\ref{eq6.11})], with the relaxation time $\tau_{B}$ given as a function of magnetic
field $B$ in Table \ref{table2}. We conclude that the same exponential decay holds for the
average cooperon and diffuson phase factors after substitution of
$B\rightarrow B+\Delta B/2$ and $B\rightarrow\Delta B/2$, respectively, in the expressions for $\tau_{B}$:
\begin{subequations}
\label{eq7.17}
\begin{eqnarray}
\langle e^{i\Delta\phi}\rangle(\mathrm{diffuson})&=& \exp(-t/\tau_{\Delta B/2}), \label{eq7.17a}\\
\langle e^{i\Delta\phi}\rangle(\mathrm{cooperon})&=&\exp(-t/\tau_{B+\Delta B/2}). \label{eq7.17b}
\end{eqnarray}
\end{subequations}

\begin{figure}
\centerline{\includegraphics[width=8cm]{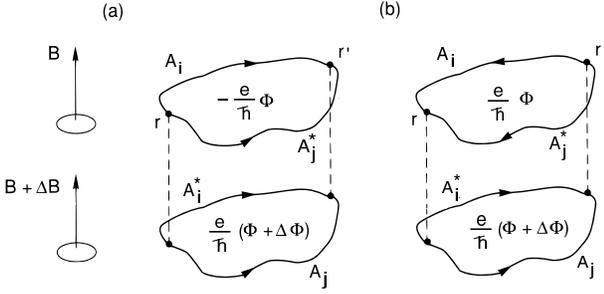}}
\caption{
Illustration of the different flux sensitivity of the interference terms of diffuson type (a) and of cooperon type (b). Both contribute to the conductance fluctuations in a zero magnetic field, but the cooperons are suppressed by a weak magnetic field, as discussed in the text.
\label{fig21}
}
\end{figure}

The cooperon is suppressed when $\tau_{B+\Delta B/2}\lesssim\tau_{\phi}$, which occurs on the same
field scale as the suppression of weak localization (determined by $\tau_{B}\lesssim\tau_{\phi}$).
The suppression of the cooperon can be seen as a consequence of the
breaking of the time-reversal invariance by the magnetic field, similar to the
suppression of weak localization. In a zero field the cooperons and the
diffusons contribute equally to the variance of the conductance; therefore,
when the cooperon is suppressed, $\mathrm{Var}\,(G)$ is reduced by a factor of 2. (The
parameter $\beta$ in Table \ref{table3} thus changes from 1 to 2 when $B$ increases beyond
$B_{\mathrm{c}}$.) In general, the magnetoconductance fluctuations are studied for $B>B_{\mathrm{c}}$
(i.e., for fields beyond the weak localization peak). Then only the diffuson
contributes to the conductance fluctuations, since the relaxation time of the
diffuson is determined by the field {\it increment\/} $\Delta B$ in the correlation function
$F(\Delta B)$, not by the magnetic field itself. This is the critical difference with weak
localization: The conductance fluctuations are {\it not suppressed\/} by a weak
magnetic field. The different behavior of cooperons and diffusons can
be understood in terms of Feynman paths. The correlation function $F(\Delta B)$
contains the product of four Feynman path amplitudes $A(i, B)$, $A^{*}(j, B)$,
$A(k, B+\Delta B)$, and $A^{*}(l, B+\Delta B)$ along various paths $i, j, k, l$ from $\mathbf{r}$ to $\mathbf{r}^{\prime}$.
Consider the diffuson term for which $i=l$ and $j=k$. The phase of this term
$A(i, B)A^{*}(j, B)A(j, B+\Delta B)A^{*}(i, B+\Delta B)$ is
\be
- \frac{e}{\hbar}\oint \mathbf{A}\cdot d\mathbf{l}+\frac{e}{\hbar}\oint(\mathbf{A}+\Delta \mathbf{A})\cdot d\mathbf{l}=\frac{e}{\hbar}\Delta\Phi. \label{eq7.18}
\ee 
where the line integral is taken along the closed loop formed by the two paths
$i$ and $j$ (cf.\ Fig.\ \ref{fig21}a). The phase is thus given by the flux {\it increment\/} $\Delta\Phi\equiv S\Delta B$
through this loop and does not contain the flux $\Phi\equiv SB$ itself. The fact that
the magnetic relaxation time of the diffuson depends only on $\Delta B$ and not on $B$
is a consequence of the cancellation contained in Eq.\ (\ref{eq7.18}). For the cooperon,
the relevant phase is that of the product of Feynman path amplitudes
$A_{-}(i, B)A_{-}^{*}(j, B)A_{+}(j, B+\Delta B)A_{+}^{*}(i, B+\Delta B)$, where the $-$ sign refers to a
trajectory from $\mathbf{r}^{\prime}$ to $\mathbf{r}$ and the $+$ sign to a trajectory from $\mathbf{r}$ to $\mathbf{r}^{\prime}$ (see Fig.\ \ref{fig21}b).
This phase is given by
\be
\frac{e}{\hbar}\oint \mathbf{A}\cdot d\mathbf{l}+\frac{e}{\hbar}\oint(\mathbf{A}+\Delta \mathbf{A})\cdot d\mathbf{l}=\frac{e}{\hbar}(2\Phi+\Delta\Phi). \label{eq7.19}
\ee
In contrast to the diffuson, the cooperon is sensitive to the flux $\Phi$ through the
loop and can therefore be suppressed by a weak magnetic field.

In the following, we assume that $B>B_{\mathrm{c}}$ so that only the diffuson
contributes to the magnetoconductance fluctuations. The combined effects of
magnetic field and inelastic scattering lead to a relaxation rate
\be
\tau_{\mathrm{eff}}^{-1}=\tau_{\phi}^{-1}+\tau_{\Delta B/2}^{-1}, \label{eq7.20}
\ee 
which describes the exponential decay of the average phase factor
$\langle e^{i\Delta\phi}\rangle=\exp(- t/\tau_{\mathrm{eff}})$. Equation (\ref{eq7.20}) contains the whole effect of the
magnetic field on the diffuson. Without having to do any diagrammatic
analysis, we therefore conclude\cite{ref147} that the correlation function $F(\Delta B)$ can be
obtained from the variance $F(0)\equiv \mathrm{Var}\,G=(\delta G)^{2}$ (given in Table \ref{table3}) by
simply replacing $\tau_{\phi}$ by the effective relaxation time $\tau_{\mathrm{eff}}$ defined in Eq.\ (\ref{eq7.20}).
The quantity $\tau_{\Delta B/2}$ corresponds to the magnetic relaxation time $\tau_{B}$ obtained
for weak localization (see Table \ref{table2}) after substitution of $B\rightarrow\Delta B/2$. For easy
reference, we give the results for the dirty and clean metal regimes
explicitly:\cite{ref109,ref147}
\begin{eqnarray}
\tau_{\Delta B/2}&=&12\left(\frac{\hbar}{e\Delta B}\right)^{2}\frac{1}{DW^{2}},\;\;{\rm if}\;\;l\ll W, \label{eq7.21}\\
\tau_{\Delta B/2}&=&4C_{1}\left(\frac{\hbar}{e\Delta B}\right)^{2}\frac{1}{v_{\mathrm{F}}W^{3}}+2C_{2}\left(\frac{\hbar}{e\Delta B}\right)\frac{l}{v_{\mathrm{F}}W^{2}},\nonumber\\
&&\;\;\;\;\;\;\;{\rm if}\;\;l\gg  W, \label{eq7.22}
\end{eqnarray}
where $C_{1}=9.5$ and $C_{2}=24/5$ for a channel with specular boundary
scattering ($C_{1}=4\pi$ and $C_{2}=3$ for a channel with diffuse boundary scattering). These results are valid under the condition $W^{2}\Delta B\ll \hbar/e$, which follows
from the requirement $\tau_{\mathrm{eff}}\gg  \tau$ that the electronic motion on the effective phase
coherence time scale $\tau_{\mathrm{eff}}$ be diffusive rather than ballistic, as well as from the
requirement $(D\tau_{\mathrm{eff}})^{1/2}\gg  W$ for one-dimensionality.

\begin{figure}
\centerline{\includegraphics[width=8cm]{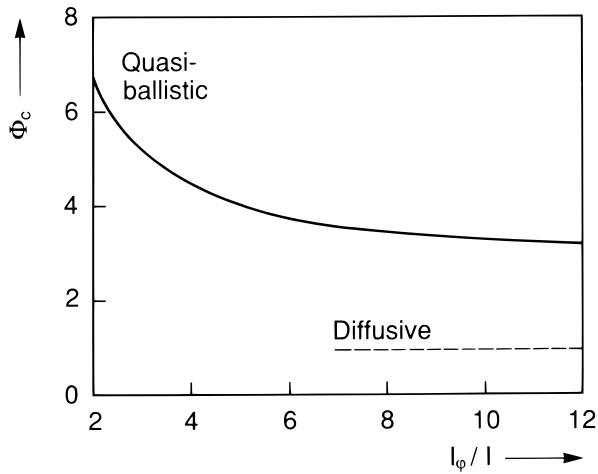}}
\caption{
Plot of the dimensionless correlation flux $\Phi_{\rm c}\equiv\Delta B_{\rm c}l_{\phi}We/h$ for the magnetoconductance fluctuations as a function of $l_{\phi}/l$ in the regime $l_{\rm T}\ll l_{\phi}$. The solid curve is for the case $l = 5\,W$; the dashed line is for $l\ll W$. Taken from C. W. J. Beenakker and H. van Houten, Phys.\ Rev.\ B {\bf 37}, 6544 (1988).
\label{fig22}
}
\end{figure}

With results (\ref{eq7.20})--(\ref{eq7.22}), the equation $F(\Delta B_{\mathrm{c}})=F(0)/2$, which defines the
correlation field $\Delta B_{\mathrm{c}}$, reduces to an algebraic equation that can be solved
straightforwardly. In the dirty metal regime one finds\cite{ref145}
\be
\Delta B_{\mathrm{c}}=2\pi C\frac{\hbar}{e}\frac{1}{Wl_{\phi}}, \label{eq7.23}
\ee 
where the prefactor $C$ decreases from\cite{ref147} 0.95 for $l_{\phi}\gg  l_{\mathrm{T}}$ to 0.42 for $l_{\phi}\ll l_{\mathrm{T}}$.
Note the similarity with the result (\ref{eq6.9}) for weak localization. Just as in weak
localization, one finds that the correlation field in the pure metal regime is
significantly enhanced above Eq.\ (\ref{eq7.23}) due to the flux cancellation effect
discussed in Section \ref{sec6c}. The enhancement factor increases from $(l/W)^{1/2}$ to
$l/W$ as $l_{\phi}$ decreases from above to below the length $l^{3/2}W^{-1/2}$. The relevant
expression is given in Ref.\ \onlinecite{ref147}. As an illustration, the dimensionless
correlation flux $\Delta B_{\mathrm{c}}Wl_{\phi}e/h$ in the pure and dirty metal regimes is plotted as a
function of $l_{\phi}/l$ in Fig.\ \ref{fig22} for $l_{\mathrm{T}}\ll l_{\phi}$.

In the following discussion of the experimental situation in semiconductor
nanostructures, it is important to keep in mind that the Al'tshuler-Lee-Stone theory of conductance fluctuations was formulated for an application to metals. This has justified the neglect of several possible complications,
which may be important in a 2DEG. One of these is the classical curvature of
the electron trajectories, which affects the conductance when $l_{\mathrm{cycl}} \lesssim\min(W, l)$.
A related complication is the Landau level quantization, which in a narrow
channel becomes important when $l_{\mathrm{m}}\lesssim W$. Furthermore, when $W\sim\lambda_{\mathrm{F}}$ the
lateral confinement will at low fields induce the formation of 1D subbands.
No quantization effects are taken into account in the theory of conductance
fluctuations discussed before. Finally, the present theory is valid only in the
regime of coherent diffusion $(\tau_{\phi}, \tau_{\mathrm{eff}}\gtrsim\tau)$. In high-mobility samples $\tau_{\phi}$ and $\tau$
may be comparable, however, as discussed in Section \ref{sec7d}. It would be of
interest to study the conductance fluctuations in this regime theoretically.

In the following discussion of experimental studies of conductance
fluctuations, we will have occasion to discuss briefly one further development.
This is the modification of the theory\cite{ref149,ref150,ref151,ref152,ref153,ref154} to account for the differences
between two- and four-terminal measurements of the conductance fluctuations, which becomes important when the voltage probes are separated by
less than the phase coherence length.\cite{ref155,ref156}

\subsubsection{\label{sec7d} Experiments}

The experimental observation of conductance fluctuations in semiconductors has preceded the theoretical understanding of this phenomenon.
Weak irregular conductance fluctuations in wide Si inversion layers were
reported in 1965 by Howard and Fang.\cite{ref157} More pronounced fluctuations
were found by Fowler et al.\ in narrow Si accumulation layers in the strongly
localized regime.\cite{ref32} Kwasnick et al.\ made similar observations in narrow Si
inversion layers in the metallic conduction regime.\cite{ref39}  These fluctuations in the
conductance as a function of gate voltage or magnetic field have been
tentatively explained by various mechanisms.\cite{ref158} One of the explanations
suggested is based on resonant tunneling,\cite{ref159} another on variable range
hopping. At the 1984 conference on ``Electronic Properties of Two-Dimensional Systems'' Wheeler et al.\cite{ref161} and Skocpol et al.\cite{ref162} reported
pronounced structure as a function of gate voltage in the low-temperature
conductance of narrow Si inversion layers, observed in the course of their
search for a quantum size effect.

\begin{figure}
\centerline{\includegraphics[width=8cm]{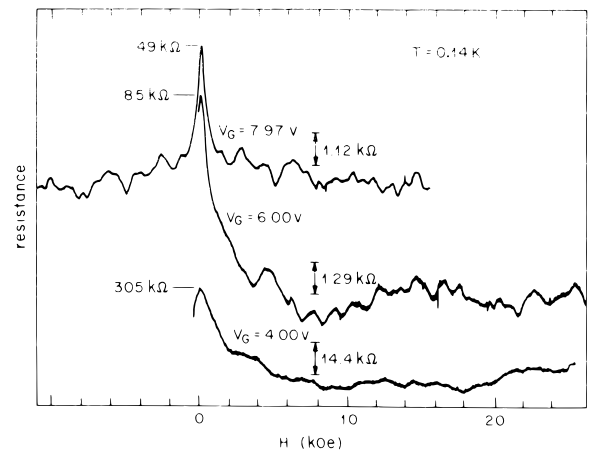}}
\caption{
Negative  magnetoresistance and aperiodic magnetoresistance fluctuations in a narrow Si inversion layer channel for several values of the gate voltage $V_{\rm G}$. Note that the vertical offset and scale is different for each $V_{\rm G}$. Taken from J. C. Licini et al., Phys.\ Rev.\ Lett.\ {\bf 55}, 2987 (1985).
\label{fig23}
}
\end{figure}

After the publication in 1985 of the Al'tshuler-Lee-Stone
 theory\cite{ref140,ref141,ref163} of universal conductance fluctuations, a consensus has
rapidly developed that this theory properly accounts for the conductance
fluctuations in the metallic regime, up to factor of two uncertainties in the
quantitative description.\cite{ref46,ref144,ref164} Following this theoretical work, Licini et
al.\cite{ref40} attributed the magnetoresistance oscillations that they observed in
narrow Si inversion layers to quantum interference in a disordered conductor. Their low-temperature measurements, which we reproduce in Fig.\ \ref{fig23},
show a large negative magnetoresistance peak due to weak localization at
low magnetic fields, in addition to aperiodic fluctuations that persist to high
fields. Such a clear weak localization peak is not found in shorter samples,
where the conductance fluctuations are larger. The reason is that the
magnitude of the conductance fluctuations $\Delta G$ is proportional to $(l_{\phi}/L)^{3/2}$
[for $l_{\phi}\ll l_{\mathrm{T}}$, cf.\ Eq.\ (\ref{eq7.10})], while the weak localization conductance correction scales with $l_{\phi}/L$ [as discussed below Eq.\ (\ref{eq6.4})]. Weak localization thus
predominates in long channels $(L\gg  l_{\phi})$ where the fluctuations are relatively
unimportant.

\begin{figure}
\centerline{\includegraphics[width=8cm]{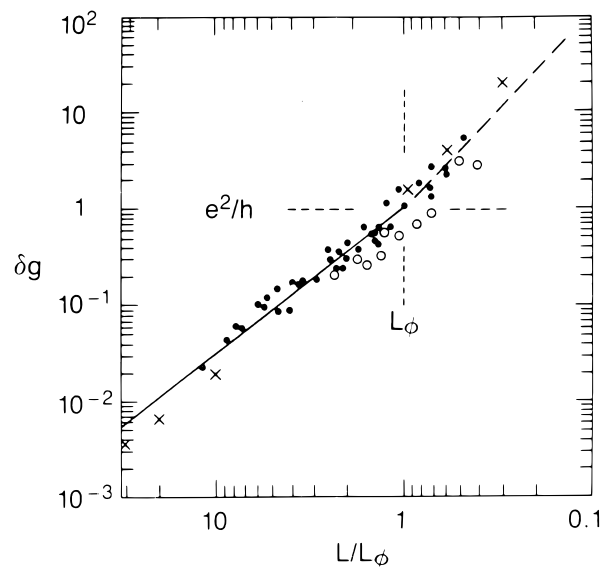}}
\caption{
Root-mean-square amplitude $\delta g$ of the conductance fluctuations (in units of $e^2 /h$) as a function of the ratio of the distance between the voltage probes $L$ to the estimated phase coherence length $l_{\phi}$ for a set of Si inversion layer channels under widely varying experimental conditions. The solid and dashed lines demonstrate the $(L/l_{\phi})^{-3/2}$ and $(L/l_{\phi})^{-2}$ scaling of $\delta g$ in the regimes $L > l_{\phi}$ and $L < l_{\phi}$, respectively. Taken from W. J. Skocpol, Physica Scripta {\bf T19}, 95 (1987).
\label{fig24}
}
\end{figure}

The most extensive quantitative study of the universality of the conductance fluctuations in narrow Si inversion layers (over a wide range of
channel widths, lengths, gate voltages, and temperatures) was made by
Skocpol et al.\cite{ref45,ref46,ref156} In the following, we review some of these experimental
results. We will not discuss the similarly extensive investigations by Webb et
al.\cite{ref155,ref164,ref165} on small metallic samples, which have played an equally
important role in the development of this subject. To analyze their experiments, Skocpol et al.\ estimated $l_{\phi}$ from weak localization experiments (with
an estimated uncertainty of about a factor of 2). They then plotted the root-mean-square variation $\delta G$ of the conductance as a function of $L/l_{\phi}$, with $L$ the
separation of the voltage probes in the channel. Their results are shown in
Fig.\ \ref{fig24}. The points for $L>l_{\phi}$ convincingy exhibit for a large variety of data
sets the $(L/l_{\phi})^{-3/2}$ scaling law predicted by the theory described in Section \ref{sec7c}
(for $l_{\phi}<l_{\mathrm{T}}$, which is usually the case in Si inversion layers).

For $L<l_{\phi}$ the experimental data of Fig.\ \ref{fig24} show a crossover to a $(L/l_{\phi})^{-2}$
scaling law (dashed line), accompanied by an increase of the magnitude of the
conductance fluctuations beyond the value $\delta G\approx e^{2}/h$ predicted by the
Al'tshuler-Lee-Stone theory for a conductor of length $L<l_{\phi}$. A similar
observation was made by Benoit et al.\cite{ref155} on metallic samples. The disagreement is explained\cite{ref155,ref156} by considering that the experimental geometry
differs from that assumed in the theory discussed in Section \ref{sec7c}. Use is made
of a long channel with voltage probes at different spacings. The experimental
$L$ is the spacing of two voltage probes, and not the length of a channel
connecting two phase-randomizing reservoirs, as envisaged theoretically. The
difference is irrelevant if $L>l_{\phi}$. If the probe separation $L$ is less than the
phase coherence length $l_{\phi}$, however, the measurement still probes a channel
segment of length $l_{\phi}$ rather than $L$. In this sense the measurement is
nonlocal.\cite{ref155,ref156}  The key to the $L^{-2}$ dependence of $\delta G$ found experimentally is
that the voltages on the probes fluctuate independently, implying that the
{\it resistance\/} fluctuations $\delta R$ are independent of $L$ in this regime so that
$\delta G\approx R^{-2}\delta R\propto L^{-2}$. This explanation is consistent with the anomalously
small correlation field $B_{\mathrm{c}}$ found for $L<l_{\phi}$.\cite{ref46,ref156} One might have expected
that the result $B_{\mathrm{c}}\approx h/eWl_{\phi}$ for $L>l_{\phi}$ should be replaced by the larger value
$B_{\mathrm{c}}\approx h/eWL$ if $L$ is reduced below $l_{\phi}$. The smaller value found experimentally
is due to the fact that the flux through parts of the channel adjacent to the
segment between the voltage probes, as well as the probes themselves, has to
be taken into account. These qualitative arguments\cite{ref155,ref156} are supported by
detailed theoretical investigations.\cite{ref149,ref150,ref151,ref152,ref153,ref154} The important message of these
theories and experiments is that the transport in a small conductor is phase
coherent over large length scales and that phase randomization (due to
inelastic collisions) occurs mainly as a result of the voltage probes. The
Landauer-B\"{u}ttiker formalism\cite{ref4,ref5} (which we will discuss in Section \ref{sec12}) is
naturally suited to study such problems theoretically. In that formalism,
current and voltage contacts are modeled by phase-randomizing reservoirs
attached to the conductor. We refer to a paper by B\"{u}ttiker\cite{ref149} for an
instructive discussion of conductance fluctuations in a multiprobe conductor
in terms of interfering Feynman paths.

Conductance fluctuations have also been observed in narrow-channel GaAs-AlGaAs
heterostructures.\cite{ref166,ref167} These systems are well in the pure
metal regime $(W<l)$, but unfortunately they are only marginally in the
regime of coherent diffusion (characterized by $\tau_{\phi}\gg  \tau$). This hampers a
quantitative comparison with the theoretical  results\cite{ref147} for the pure metal
regime discussed in Section \ref{sec7c}. (A phenomenological treatment of conductance fluctuations in the case that $\tau_{\phi}\sim\tau$ is given in Refs.\ \onlinecite{ref168} and \onlinecite{ref169}.)
The data of Ref.\ \onlinecite{ref167} are consistent with an enhancement of the correlation
field due to the flux cancellation effect, but are not conclusive.\cite{ref147} We note
that the flux cancellation effect can also explain the correlation field
enhancement noticed in a computer simulation by Stone.\cite{ref163}

In the analysis of the aforementioned experiments on magnetoconductance fluctuations, a twofold spin degeneracy has been assumed. The variance
$(\delta G)^{2}$ is reduced by a factor of 2 if the spin degeneracy is lifted by a strong
magnetic field $B>B_{\mathrm{c}2}$. The Zeeman energy $g\mu_{\mathrm{B}}B$ should be sufficiently large
than the spin-up and spin-down electrons give statistically independent
contributions to the conductance. The degeneracy factor $g_{\mathrm{s}}^{2}$ in $(\delta G)^{2}$ (introduced in Section \ref{sec7a}) should then be replaced by a factor $g_{\mathrm{s}}$, since the
variances of statistically independent quantities add. Since $g_{\mathrm{s}}=2$, one
obtains a factor-of-2 reduction in $(\delta G)^{2}$. Note that this reduction comes on
top of the factor-of-2 reduction in $(\delta G)^{2}$ due to the breaking of time-reversal
symmetry, which occurs at weak magnetic fields $B_{\mathrm{c}}$. Stone has  calculated\cite{ref170}
that the field $B_{\mathrm{c}2}$ in a narrow channel $(l_{\phi}\gg  W)$ is given by the criterion of unit
phase change $g\mu_{\mathrm{B}}B\tau_{\phi}/h$ in a coherence time, resulting in the estimate
$B_{\mathrm{c}2}\approx h/g\mu_{\mathrm{B}}\tau_{\phi}$. Surprisingy, the thermal energy $k_{\mathrm{B}}T$ is irrelevant for $B_{\mathrm{c}2}$ in
the 1D case $l_{\phi}\gg  W$ (but not in higher dimensions\cite{ref170}).

For the narrow-channel experiment of Ref.\ \onlinecite{ref167} just discussed, one finds
(using the estimates $\tau_{\phi}\approx 7\,\mathrm{ps}$ and $g\approx 0.4$) a crossover field $B_{\mathrm{c}2}$ of about $2\, \mathrm{T}$,
well above the field range used for the data analysis.\cite{ref147}  Most importantly, no
magnetoconductance fluctuations are observed if the magnetic field is applied
{\it parallel\/} to the 2DEG (see Section \ref{sec9}), demonstrating that the Zeeman splitting
has no effect on the conductance in this field regime. More recently, Debray et
al.\cite{ref171} performed an experimental study of the reduction by a perpendicular
magnetic field of the conductance fluctuations as a function of Fermi energy
(varied by means of a gate). The estimated value of $\tau_{\phi}$ is larger than that of
Ref.\ \onlinecite{ref167} by more than an order of magnitude. Consequently, a very small
$B_{\mathrm{c}2}\approx 0.07\,\mathrm{T}$ is estimated in this experiment. The channel is relatively wide
($2\,\mu \mathrm{m}$ lithographic width), so the field $B_{\mathrm{c}}$ for time-reversal symmetry breaking
is even smaller $(B_{\mathrm{c}}\approx 7\times 10^{-4}\,\mathrm{T})$. A total factor-of-4 reduction in $(\delta G)^{2}$ was
found, as expected. The values of the observed crossover fields $B_{\mathrm{c}}$ and $B_{\mathrm{c}2}$
also agree reasonably well with the theoretical prediction. Unfortunately, the
magnetoconductance in a parallel magnetic field was not investigated by
these authors, which would have provided a definitive test for the effect of
Zeeman splitting on the conductance above $B_{\mathrm{c}2}$. We note that related
experimental\cite{ref172,ref173} and theoretical\cite{ref174,ref175} work has been done on the
reduction of {\it temporal\/} conductance fluctuations by a magnetic field.

The Al'tshuler-Lee-Stone theory of conductance fluctuations ceases
to be applicable when the dimensions of the sample approach the mean
free path. In this ballistic regime observations of large aperiodic, as well
as quasi-periodic, magnetoconductance fluctuations have been reported.\cite{ref68,ref69,ref139,ref168,ref176,ref177,ref178,ref179} Quantum interference effects in this regime are
determined not by impurity scattering but by scattering off geometrical
features of the device, as will be discussed in Section \ref{sec3}.

\subsection{\label{sec8} Aharonov-Bohm effect}

Magnetoconductance fluctuations in a channel geometry in the diffusive
regime are {\it aperiodic}, since the interfering Feynman paths enclose a continuous range of magnetic flux values. A ring geometry, in contrast, encloses a
well-defined flux $\Phi$ and thus imposes a fundamental periodicity
\be
G(\Phi)=G(\Phi+n(h/e)),\;\; n=1,2,3, \ldots, \label{eq8.1}
\ee 
on the conductance as a function of perpendicular magnetic field $B$ (or flux
$\Phi=BS$ through a ring of area $S$). Equation (\ref{eq8.1}) expresses the fact that a flux
increment of an integer number of flux quanta changes by an integer multiple
of $2\pi$ the phase difference between Feynman paths along the two arms of the
ring. The periodicity (\ref{eq8.1}) would be an exact consequence of gauge invariance
if the magnetic field were nonzero only in the interior of the ring, as in the
original thought experiment of Aharonov and Bohm.\cite{ref180} In the present
experiments, however, the magnetic field penetrates the arms of the ring as
well as its interior so that deviations from Eq.\ (\ref{eq8.1}) can occur. Since in many
situations such deviations are small, at least in a limited field range, one still
refers to the magnetoconductance oscillations as an {\it Aharonov-Bohm effect}.

\begin{figure}
\centerline{\includegraphics[width=8cm]{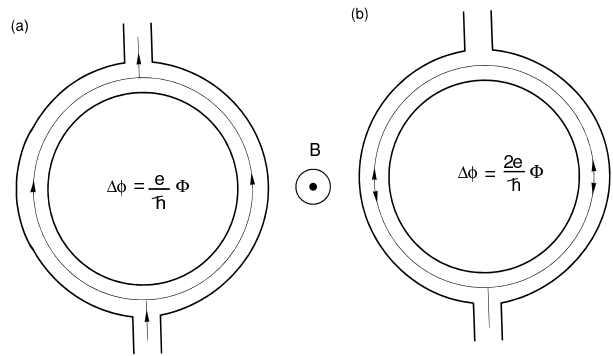}}
\caption{
Illustration of the Aharonov- Bohm effect in a ring geometry. Interfering trajectories responsible for the  magnetoresistance oscillations with $h/e$ periodicity in the enclosed flux $\Phi$ are shown (a). (b) The pair of time-reversed trajectories lead to oscillations with $h/2e$ periodicity.
\label{fig25}
}
\end{figure}

The fundamental periodicity
\be
\Delta B=\frac{h}{e}\frac{1}{S} \label{eq8.2}
\ee
is caused by interference between trajectories that make one half-revolution
around the ring, as in Fig.\ \ref{fig25}a. The first harmonic
\be
\Delta B=\frac{h}{2e}\frac{1}{S} \label{eq8.3}
\ee
results from interference after one revolution. A fundamental distinction
between these two periodicities is that the phase of the $h/e$ oscillations (\ref{eq8.2}) is
sample-specific, whereas the $h/2e$ oscillations (\ref{eq8.3}) contain a contribution
from time-reversed trajectories (as in Fig.\ \ref{fig25}b) that has a minimum conductance at $B=0$, and thus has a sample-independent phase. Consequently,
in a geometry with many rings in series (or in parallel) the $h/e$ oscillations
average out, but the $h/2e$ oscillations remain. The $h/2e$ oscillations can be
thought of as a periodic modulation of the weak localization effect due to
coherent backscattering.

The first observation of the Aharonov-Bohm effect in the solid state was
made by Sharvin and Sharvin\cite{ref181} in a long metal cylinder. Since this is
effectively a many-ring geometry, only the $h/2e$ oscillations were observed, in
agreement with a theoretical prediction by Al'tshuler, Aronov, and
Spivak,\cite{ref182} which motivated the experiment. (We refer to Ref.\ \onlinecite{ref125} for a simple
estimate of the order of magnitude of the $h/2e$ oscillations in the dirty metal
regime.) The effect was studied extensively by several groups.\cite{ref183,ref184,ref185}  The $h/e$
oscillations were first observed in single metal rings by Webb et al.\cite{ref186} and
studied theoretically by several authors.\cite{ref1,ref144,ref187,ref188}  The self-averaging of the
$h/e$ oscillations has been demonstrated explicitly in experiments with a
varying number of rings in series.\cite{ref189} Many more experiments have been
performed on one-and two-dimensional arrays and networks, as reviewed in
Refs.\ \onlinecite{ref190} and \onlinecite{ref191}.

In this connection, we mention that the development of the theory of
{\it aperiodic\/} conductance fluctuations (discussed in Section \ref{sec7}) has been much
stimulated by their observation in metal rings by Webb et al.,\cite{ref165}  in the course
of their search for the Aharonov-Bohm effect. The reason that aperiodic
fluctuations are observed in rings (in addition to periodic oscillations) is that
the magnetic field penetrates the width of the arms of the ring and is not
confined to its interior. By fabricating rings with a large ratio of radius $r$ to
width $W$, researchers have proven it is possible to separate\cite{ref190} the magnetic
field scales of the periodic and aperiodic oscillations (which are given by a
field interval of order $h/er^{2}$ and $h/eWl_{\phi}$, respectively). The penetration of the
magnetic field in the arms of the ring also leads to a broadening of the peak in
the Fourier transform at the $e/h$ and $2e/h$ periodicities, associated with a
distribution of enclosed flux. The width of the Fourier peak can be used as
a rough estimate for the width of the arms of the ring. In addition, the
nonzero field in the arms of the ring also leads to a damping of the amplitude
of the ensemble-averaged $h/2e$ oscillations when the flux through the arms is
sufficiently large to suppress weak localization.\cite{ref191}

Two excellent reviews of the Aharonov-Bohm effect in metal rings and
cylinders exist.\cite{ref190,ref191}  In the following we discuss the experiments in semiconductor nanostructures in the weak-field regime $\omega_{\mathrm{c}}\tau<1$, where the effect of
the Lorentz force on the trajectories can be neglected. The strong-field regime
$\omega_{\mathrm{c}}\tau>1$ (which is not easily accessible in the usual polycrystalline metal
rings) is only briefly mentioned; it is discussed more extensively in Section \ref{sec21}.
To our knowledge, no observation of Aharonov-Bohm magnetoresistance
oscillations in Si inversion layers has been reported. The first observation of
the Aharonov-Bohm effect in a 2DEG ring was published by Timp et al.,\cite{ref69}
who employed high-mobility GaAs-AlGaAs heterostructure material.
Similar results were obtained independently by Ford et al.\cite{ref73} and Ishibashi et
al.\cite{ref193}  More detailed studies soon followed.\cite{ref74,ref139,ref176,ref194,ref195} A characteristic
feature of these experiments is the large amplitude of the $h/e$ oscillations (up
to 10\% of the average resistance), much higher than in metal rings (where the
effect is at best\cite{ref192,ref196,ref197} of order 0.1\%). A similar difference in magnitude is
found for the aperiodic magnetoresistance fluctuations in metals and semiconductor nanostructures. The reason is simply that the amplitude $\delta G$ of the
periodic or aperiodic conductance oscillations has a maximum value of order
$e^{2}/h$, so the maximum relative resistance oscillation $\delta R/R\approx R\delta G\approx Re^{2}/h$ is
proportional to the average resistance $R$, which is typically much smaller in
metal rings.

In most studies only the $h/e$ fundamental periodicity is observed, although
Ford et al.\cite{ref73,ref74} found a weak $h/2e$ harmonic in the Fourier transform of the
magnetoresistance data of a very narrow ring. It is not quite clear whether
this harmonic is due to the Al'tshuler-Aronov-Spivak mechanism involving
the constructive interference of two time-reversed trajectories\cite{ref182} or to the
random interference of two non-time-reversed Feynman paths winding
around the entire ring.\cite{ref1,ref144,ref187} The relative weakness of the $h/2e$ effect in
single 2DEG rings is also typical for most experiments on single metal rings
(although the opposite was found to be true in the case of aluminum rings by
Chandrasekhar et al.,\cite{ref197} for reasons which are not understood). This is in
contrast to the case of arrays or cylinders, where, as we mentioned, the $h/2e$
oscillations are predominant the $h/e$ effect being ``ensemble-averaged'' to
zero because of its sample-specific phase. In view of the fact that the
experiments on 2DEG rings explore the borderline between diffusive and
ballistic transport, they are rather difficult to analyze quantitatively. A
theoretical study of the Aharonov-Bohm effect in the purely ballistic
transport regime was performed by Datta and Bandyopadhyay,\cite{ref198} in
relation to an experimental observation of the effect in a double-quantum-well device.\cite{ref199} A related study was published by Barker.\cite{ref200}

\begin{figure}
\centerline{\includegraphics[width=8cm]{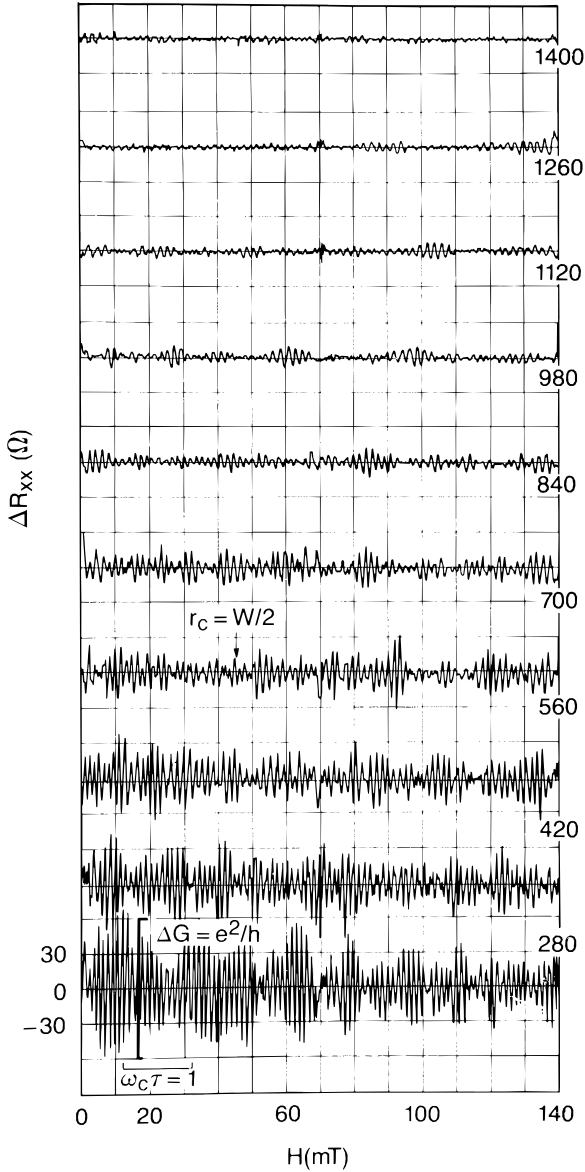}}
\caption{
Experimental  magnetoresistance of a ring of $2\,\mu{\rm m}$ diameter, defined in the 2DEG of a high-mobility GaAs-AlGaAs heterostructure ($T = 270\,{\rm mK}$). The different traces are consecutive parts of a magnetoresistance measurement from 0 to $1.4\,{\rm T}$, digitally filtered to suppress a slowly varying background. The oscillations are seen to persist for fields where $\omega_{\rm c}\tau>1$, but their amplitude is reduced substantially for magnetic fields where $2l_{\rm cycl}\ll W$. (The field value where $2l_{\rm cycl}\equiv 2r_{\rm c}=W$ is indicated). Taken from G. Timp et al., Surf.\ Sci.\ {\bf 196}, 68 (1988).
\label{fig26}
}
\end{figure}

The Aharonov-Bohm oscillations in the magnetoresistance of a small ring
in a high-mobility 2DEG are quite impressive. As an illustration, we
reproduce in Fig.\ \ref{fig26} the results obtained by Timp et al.\cite{ref201} Low-frequency
modulations were filtered out, so that the rapid oscillations are superimposed
on a constant background. The amplitude of the $h/e$ oscillations diminishes
with increasing magnetic field until eventually the Aharonov-Bohm effect is
completely suppressed. The reduction in amplitude is accompanied by a
reduction in frequency. A similar observation was made by Ford et al.\cite{ref74} In
metals, in contrast, the Aharonov-Bohm oscillations persist to the highest
experimental fields, with constant frequency. The different behavior in a
2DEG is a consequence of the effect of the Lorentz force on the electrons in
the ring, which is of importance when the cyclotron diameter $2l_{\mathrm{cycl}}$ becomes
smaller than the width $W$ of the arm of the ring, provided $W<l$ (note that
$l_{\mathrm{cycl}}=hk_{\mathrm{F}}/eB$ is much smaller in a 2DEG than in a metal, at the same
magnetic field value). We will return to these effects in Section \ref{sec21}.

An electrostatic potential $V$ affects the phase of the electron wave function
through the term $(e/\hbar) \int Vdt$ in much the same way as a vector potential
does. If the two arms of the ring have a potential difference $V$, and an
electron traverses an arm in a time $t$, then the acquired phase shift would lead
to oscillations in the resistance with periodicity $\Delta V=h/et$. The electrostatic
Aharonov-Bohm effect has a periodicity that depends on the transit time $t$,
and is not a geometrical property of the ring, as it is for the magnetic effect. A
distribution of transit times could easily average out the oscillations. Note
that the potential difference effectuates the phase difference by changing the
wavelength of the electrons (via a change in their kinetic energy), which also
distinguishes the electrostatic from the magnetic effect (where a phase shift is
induced by the vector potential without a change in wavelength). An
experimental search for the electrostatic Aharonov-Bohm effect in a small
metal ring was performed by Washburn et al.\cite{ref202} An electric field was applied
in the plane of the ring by small capacitive electrodes. They were able to shift
the phase of the magnetoresistance oscillations by varying the field, but the
effect was not sufficiently strong to allow the observation of purely electrostatic oscillations. Unfortunately, this experiment could not discriminate between the effect of the electric field penetrating in the arms of the ring (which
could induce a phase shift by changing the trajectories) and that of the electrostatic potential. Experiments have been reported by De Vegvar et al.\cite{ref203}
on the manipulation of the phase of the electrons by means of the voltage on a
gate electrode positioned across one of the arms of a heterostructure ring. In
this system a change in gate voltage has a large effect on the resistance of the
ring, primarily because it strongy affects the local density of the electron gas.
No clear periodic signal, indicative of an electrostatic Aharonov-Bohm
effect, could be resolved. As discussed in Ref.\ \onlinecite{ref203}, this is not too surprising, in
view of the fact that in that device 1D subband depopulation in the region
under the gate occurs on the same gate voltage scale as the expected
Aharonov-Bohm effect. The observation of an electrostatic Aharonov-Bohm effect thus remains an experimental challenge. A successful experiment
would appear to require a ring in which only a single 1D subband is occupied,
to ensure a unique transit time.\cite{ref198,ref200}

\subsection{\label{sec9} Electron-electron interactions}

\subsubsection{\label{sec9a} Theory}

In addition to the weak localization correction to the conductivity
discussed in Section \ref{sec6}, which arises from a single-electron quantum interference effect, the Coulomb interaction of the conduction electrons gives also
rise to a quantum correction.\cite{ref204,ref205} In two dimensions the latter correction
has a logarithmic temperature dependence, just as for weak localization [see
Eq.\ (\ref{eq6.4})]. A perpendicular magnetic field can be used to distinguish the two
quantum corrections, which have a different field dependence.\cite{ref118,ref204,ref205,ref206,ref207,ref208,ref209,ref210}
This field of research has been reviewed in detail by Al'tshuler and
Aronov,\cite{ref211} by Fukuyama,\cite{ref212} and by Lee and Ramakrishnan,\cite{ref127} with an
emphasis on the theory. A broader review of electronic correlation effects in
2D systems has been given by Isihara in this series.\cite{ref213} In the present
subsection we summarize the relevant theory, as a preparation for the
following subsection on experimental studies in semiconductor nanostructures. We do not discuss the diagrammatic perturbation theory, since it is
highly technical and does not lend itself to a discussion at the same level as for
the other subjects dealt with in this review.

An attempt at an intuitive interpretation of the Feynman diagrams was
made by Bergmann.\cite{ref214} It is argued that one important class of diagrams may
be interpreted as diffraction of one electron by the oscillations in the
electrostatic potential generated by the other electrons. The Coulomb
interaction between the electrons thus introduces a purely quantum mechanical correlation between their motion, which is observable in the conductivity.
The diffraction of one electron wave by the interference pattern generated
by another electron wave will only be of importance if their wavelength
difference, and thus their energy difference, is small. At a finite temperature $T$,
the characteristic energy difference is $k_{\mathrm{B}}T$. The time $\tau_{\mathrm{T}}\equiv \hbar/k_{\mathrm{B}}T$ enters as a
long-time cutoff in the theory of electron-electron interactions in a disordered conductor, in the usual case\cite{ref127,ref211} $\tau_{\mathrm{T}}\lesssim\tau_{\phi}$. (Fukuyama\cite{ref212} also
discusses the opposite limit $\tau_{\mathrm{T}}\gg  \tau_{\phi}$.) Accordingy, the magnitude of the
thermal length $l_{\mathrm{T}}\equiv(D\tau_{\mathrm{T}})^{1/2}$ compared with the width $W$ determines the
dimensional crossover from 2D to 1D [for $l_{\mathrm{T}}<l_{\phi}\equiv(D\tau_{\phi})^{1/2}]$. In the
expression for the conductivity correction associated with electron-electron
interactions, the long-time cutoff $\tau_{\mathrm{T}}$ enters logarithmically in 2D and as a
square root in 1D. These expressions thus have the same form as for weak
localization, but with the phase coherence time $\tau_{\phi}$ replaced by $\tau_{\mathrm{T}}$. The origin
of this difference is that a finite temperature does not introduce a long-time
cutoff for the single-electron quantum interference effect responsible for weak
localization, but merely induces an energy average of the corresponding
conductivity correction.

In terms of effective interaction parameters $g_{2\mathrm{D}}$ and $g_{1\mathrm{D}}$, the conductivity
corrections due to electron-electron interactions can be written as (assuming
$\tau\ll \tau_{\mathrm{T}}\ll \tau_{\phi})$
\begin{subequations}
\label{eq9.1}
\begin{eqnarray}
\delta\sigma_{\mathrm{ee}}&=&-\frac{e^{2}}{2\pi^{2}\hbar}g_{2\mathrm{D}}\ln\frac{\tau_{\mathrm{T}}}{\tau},\;\; {\rm for}\;\; l_{\mathrm{T}}\ll W, \label{eq9.1a}\\
\delta\sigma_{\mathrm{ee}}&=&-\frac{e^{2}}{2^{1/2}\pi \hbar}g_{1\mathrm{D}}\frac{l_{\mathrm{T}}}{W},\;\; {\rm for}\;\; W\ll l_{\mathrm{T}}\ll L. \label{eq9.1b}
\end{eqnarray}
\end{subequations}
Under typical experimental conditions,\cite{ref55} the constants $g_{2\mathrm{D}}$ and $g_{1\mathrm{D}}$ are
positive and of order unity. Theoretically, these effective interaction parameters depend in a complicated way on the ratio of screening length to Fermi
wavelength and can have either sign. We do not give the formulas here, but
refer to the reviews by Al'tshuler and Aronov\cite{ref211} and Fukuyama.\cite{ref212} In 2D
the interaction correction $\delta\sigma_{\mathrm{ee}}$ shares a logarithmic temperature dependence
with the weak localization correction $\delta\sigma_{\mathrm{loc}}$, and both corrections are of the
same order of magnitude. In 1D the temperature dependences of the two
effects are different (unless $\tau_{\phi}\propto T^{-1/2}$). Moreover, in the 1D case $\delta\sigma_{\mathrm{ee}}\ll \delta\sigma_{\mathrm{loc}}$
if $l_{\mathrm{T}}\ll l_{\phi}$.

A weak magnetic field fully suppresses weak localization, but has only a
small effect on the quantum correction from electron-electron interactions.
The conductance correction $\delta G_{\mathrm{ee}}$ contains contributions of diffuson type and
of cooperon type. The diffusons (which give the largest contributions to $\delta G_{\mathrm{ee}}$)
are affected by a magnetic field only via the Zeeman energy, which removes
the spin degeneracy when $g\mu_{\mathrm{B}}B\gtrsim k_{\mathrm{B}}T$. In the systems of interest here, spin
splitting can usually be ignored below $1\mathrm{T}$, so the diffusons are insensitive to a
weak magnetic field. Since the spin degeneracy is removed regardless of the
orientation of the magnetic field, the $B$-dependence of the diffuson is
isotropic. The smaller cooperon contributions exhibit a similar sensitivity as
weak localization to a weak perpendicular magnetic field, the characteristic
field being determined by $l_{\mathrm{m}}^{2}\approx l_{\mathrm{T}}^{2}$ in 2D and by $l_{\mathrm{m}}^{2}\approx Wl_{\mathrm{T}}$ in 1D (in the dirty
metal regime $W\gg  l$, so flux cancellation does not play a significant role). The
magnetic length $l_{\mathrm{m}}\equiv(\hbar/eB_{\perp})^{1/2}$ contains only the component $B_{\perp}$ of the field
perpendicular to the 2DEG, since the magnetic field affects the cooperon via
the phase shift induced by the enclosed flux. The anisotropy and the small
characteristic field are two ways to distinguish experimentally the cooperon
contribution from that of the diffuson. It is much more difficult to distinguish
the cooperon contribution to $\delta G_{\mathrm{ee}}$ from the weak localization correction,
since both effects have the same anisotropy, while their characteristic fields
are comparable ($l_{\mathrm{T}}$ and $l_{\phi}$ not being widely separated in the systems
considered here). This complication is made somewhat less problematic by
the fact that the cooperon contribution to $\delta G_{\mathrm{ee}}$ is often considerably smaller
than $\delta G_{\mathrm{loc}}$, in which case it can be ignored. In 1D the reduction factor\cite{ref55,ref211} is
of order $[1+\lambda\ln(E_{\mathrm{F}}/k_{\mathrm{B}}T)]^{-1}(l_{\mathrm{T}}/l_{\phi})$, with $\lambda$ a numerical coefficient of order unity.

There is one additional aspect to the magnetoresistance due to electron-electron interactions that is of little experimental relevance in metals but
becomes important in semiconductors in the classically strong-field regime
where $\omega_{\mathrm{c}}\tau>1$ (this regime is not easily accessible in metal nanostructures
because of the typically short scattering time). In such strong fields only the
diffuson contributions to the conductivity corrections survive. According to
Houghton et al.\cite{ref215} and Girvin et al.,\cite{ref216} the diffuson does not modify the off-diagonal elements of the conductivity tensor, but only the diagonal elements
\be
\delta\sigma_{xy}=\delta\sigma_{yx}=0,\;\;\delta\sigma_{xx}=\delta\sigma_{yy}\equiv\delta\sigma_{\mathrm{ee}}, \label{eq9.2}
\ee 
where $\delta\sigma_{\mathrm{ee}}$ is approximately field-independent (provided spin splitting does
not play a role). In a channel geometry one measures the longitudinal
resistivity $\rho_{xx}$, which is related to the conductivity tensor elements by
\begin{eqnarray}
\rho_{xx}&\equiv&\frac{\sigma_{yy}}{\sigma_{xx}\sigma_{yy}+\sigma_{xy}^{2}}\nonumber\\
&=&\rho_{xx}^{0}+\rho_{xx}^{0}\left(\frac{\delta\sigma_{\mathrm{ee}}}{\sigma_{xx}^{0}}-2\rho_{xx}^{0}\delta\sigma_{\mathrm{ee}}\right)+\mathrm{order}(\delta\sigma_{\mathrm{ee}})^{2}.\nonumber\\
&& \label{eq9.3}
\end{eqnarray}
Here $\rho_{xx}^{0}=\rho$ and $\sigma_{xx}^{0}=\sigma[1+(\omega_{\mathrm{c}}\tau)^{2}]^{-1}$ are the classical results (\ref{eq4.25}) and
(\ref{eq4.26}). In obtaining this result the effects of Landau level quantization on the
conductivity have been disregarded (see, however, Ref.\ \onlinecite{ref55}). The longitudinal
resistivity thus becomes magnetic-field-dependent:
\be
\rho_{xx}=\rho(1+[(\omega_{\mathrm{c}}\tau)^{2}-1]\delta\sigma_{\mathrm{ee}}/\sigma). \label{eq9.4}
\ee 
To the extent that the $B$-dependence of $\delta\sigma_{\mathrm{ee}}$ can be neglected, Eq.\ (\ref{eq9.4}) gives a
parabolic negative magnetoresistance, with a temperature dependence that is
that of the negative conductivity correction $\delta\sigma_{\mathrm{ee}}$. This effect can easily be
studied up to $\omega_{\mathrm{c}}\tau=10$, which would imply an enhancement by a factor of
100 of the resistivity correction in zero magnetic field. (The Hall resistivity $\rho_{xy}$
also contains corrections from $\delta\sigma_{\mathrm{ee}}$, but without the enhancement factor.) In
2D it is this enhancement that allows the small effect of electron-electron
interactions to be observable experimentally (in as far as the effect is due to
diffuson-type contributions).

Experimentally, the parabolic negative magnetoresistance associated with
electron-electron interactions was first identified by Paalanen et al.\cite{ref137} in
high-mobility GaAs-AlGaAs heterostructure channels. A more detailed
study was made by Choi et al.\cite{ref55} In that paper, as well as in Ref.\ \onlinecite{ref113}, it was
found that the parabolic magnetoresistance was less pronounced in narrow
channels than in wider ones. Choi et al.\ attributed this suppression to
specular boundary scattering. It should be noted, however, that specular
boundary scattering has no effect at all on the classical conductivity tensor $\sigma^{0}$
(in the scattering time approximation; cf.\ Section \ref{sec5b}). Since the parabolic
magnetoresistance results from the $(\omega_{\mathrm{c}}\tau)^{2}$ term in $1/\sigma_{xx}^{0}$ [see Eq.\ (\ref{eq9.4})], one
would expect that specular boundary scattering does not suppress the
parabolic magnetoresistance (assuming that the result $\delta\sigma_{xy}=\delta\sigma_{yx}=0$ still
holds in the pure metal regime $l>W$). Diffuse boundary scattering does
affect $\sigma^{0}$, but only for relatively weak fields such that $2l_{\mathrm{cycl}}>\sim W$ (see Section
\ref{sec5}); hence, diffuse boundary scattering seems equally inadequate in explaining
the observations. In the absence of a theory for electron-electron interaction
effects in the pure metal regime, this issue remains unsettled.

\subsubsection{\label{sec9b} Narrow-channel experiments}

Wheeler et al.\cite{ref38} were the first to use magnetoresistance experiments as a
tool to distinguish weak localization from electron-electron interaction
effects in narrow Si MOSFETs. As in most subsequent studies, the negative
magnetoresistance was entirely attributed to the suppression of weak
localization; the cooperon-type contributions from electron-electron interactions were ignored. After subtraction of the weak localization correction,
the remaining temperature dependence was found to differ from the simple
$T^{-1/2}$ dependence predicted by the theory for $W<l_{\mathrm{T}}<l_{\phi}$ [Eq. (\ref{eq9.1b})]. This
was attributed in Ref.\ \onlinecite{ref38} to temperature-dependent screening at the relatively
high temperatures of the experiment. Pooke et al.\cite{ref138} found a nice $T^{-1/2}$
dependence in similar experiments at lower temperatures in narrow Si
accumulation layers and in GaAs-AlGaAs heterostructures.

\begin{figure}
\centerline{\includegraphics[width=8cm]{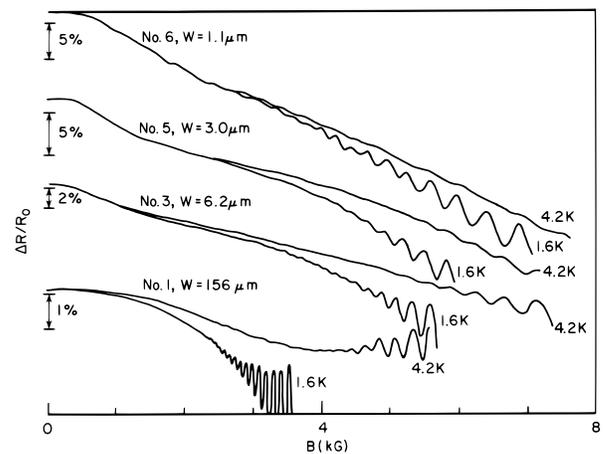}}
\caption{
Negative magnetoresistance in wide and narrow GaAs-AlGaAs channels at 4.2 and 1.6 K. The temperature-independent negative magnetoresistance at low fields is a classical size effect. The temperature-dependent parabolic  magnetoresistance at higher fields is a quantum interference effect associated with electron-electron interactions. Shubnikov-De Haas oscillations are visible for fields greater than about 0.3 T. Taken from K. K. Choi et al., Phys.\ Rev.\ B {\bf 33}, 8216 (1986).
\label{fig27}
}
\end{figure}

The most detailed study by far of the 2D to 1D crossover of the electron-electron interaction effect in narrow channels was made by Choi et al.\cite{ref55} in a
GaAs-AlGaAs heterostructure. In Fig.\ \ref{fig27} we reproduce some of their
experimental traces for channel widths from 156 to 1.1 $\mu \mathrm{m}$ and a channel
length of about 300 $\mu \mathrm{m}$. The weak localization peak in the magnetoresistance
is not resolved in this experiment, presumably because the channels are not in
the 1D regime for this effect (the 2D weak localization peak would be small
and would have a width of $10^{-4}\,\mathrm{T}$). The negative magnetoresistance that they
found below $0.1-0.2\, \mathrm{T}$ in the narrowest channels is temperature-independent
between 1 and $4\, \mathrm{K}$ and was therefore identified by Choi et al.\cite{ref55} as a classical
size effect. The classical negative magnetoresistance extends over a field range
for which $2l_{\mathrm{cycl}}\gtrsim W$. This effect has been discussed in Section \ref{sec5} in terms of
reduction of backscattering by a magnetic field. The electron-electron
interaction effect is observed as a (temperature-dependent) parabolic negative
magnetoresistance above $0.1\, \mathrm{T}$ for the widest channel and above $0.3\, \mathrm{T}$ for the
narrowest one. From the magnitude of the parabolic negative magnetoresistance, Choi et al.\cite{ref55} could find and analyze the crossover from 2D to 1D
interaction effects. In addition, they investigated the cross over to 0D by
performing experiments on short channels. As seen in Fig.\ \ref{fig27}, Shubnikov-De
Haas oscillations are superimposed on the parabolic negative magnetoresistance at low temperatures and strong magnetic fields. It is noteworthy that
stronger fields are required in narrower channels to observe the Shubnikov-De Haas oscillations, an effect discussed in terms of specular boundary
scattering by Choi et al. The Shubnikov-De Haas oscillations in narrow
channels are discussed further in Section \ref{sec10b}.

\begin{figure}
\centerline{\includegraphics[width=8cm]{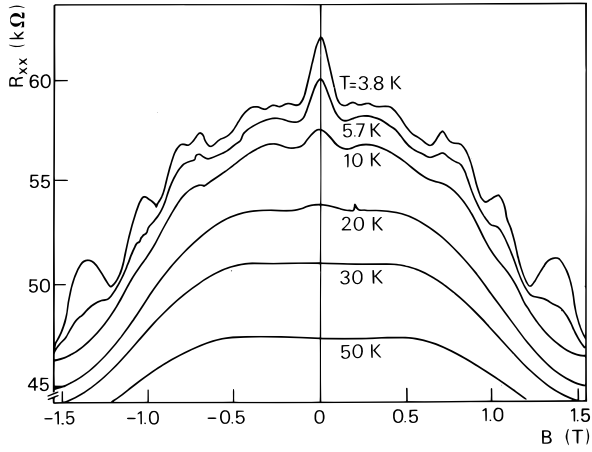}}
\caption{
Magnetoresistance at various temperatures of a GaAs-AlGaAs channel ($W = 0.12\,\mu{\rm m}$, $L = 10\,\mu{\rm m}$) defined by a shallow-mesa etch technique. The central negative magnetoresistance peak between $-0.1$ and $+0.1\,{\rm T}$ at low temperatures is due to 1D weak localization in the quasi-ballistic regime. Conductance fluctuations are seen at larger fields. The negative  magnetoresistance that persists to high temperatures is a classical size effect as in Fig.\ \ref{fig27}. The temperature dependence of the resistance at $B = 0$ is due to a combination of weak localization and electron-electron interaction effects (see Fig.\ \ref{fig30}). Taken from H. van Houten et al., Appl.\ Phys.\ Lett.\ {\bf 49}, 1781 (1986).
\label{fig28}
}
\end{figure}

In Refs.\ \onlinecite{ref63,ref167}, and \onlinecite{ref27} the work by Choi et al.\cite{ref55} was extended to even
narrower channels, well into the 1D pure metal regime. The results for a
conducting channel width of 0.12 $\mu \mathrm{m}$ are shown in Fig.\ \ref{fig28}. The 1D weak
localization peak in the magnetoresistance is quite large for this narrow
channel (even at the rather high temperatures shown) and clearly visible
below 0.1 T. The classical size effect due to reduction of backscattering now
leads to a negative magnetoresistance on a larger field scale of about $1\, \mathrm{T}$, in
agreement with the criterion $2l_{\mathrm{cycl}}\sim W$. This is best seen at temperatures
above $20\, \mathrm{K}$, where the quantum mechanical effects are absent. The
temperature-dependent parabolic negative magnetoresistance is no longer
clearly distinguishable in the narrow channel of Fig.\ \ref{fig28}, in contrast to wider
channels.\cite{ref27,ref55} The suppression of this effect in narrow channels is not yet
understood (see Section \ref{sec9a}). Superimposed on the smooth classical magnetoresistance, one sees large aperiodic fluctuations on a field scale of the
same magnitude as the width of the weak localization peak, in qualitative
agreement with the theory of universal conductance fluctuations in the pure
metal regime\cite{ref147} (see Section \ref{sec7d}). Finally, Shubnikov-De Haas oscillations
are beginning to be resolved around $1.2\, \mathrm{T}$, but they are periodic in $1/B$ at
stronger magnetic fields only (not shown). As discussed in Section \ref{sec10}, this
anomaly in the Shubnikov-De Haas effect is a manifestation of a quantum
size effect.\cite{ref167,ref217,ref218}  This one figure thus summarizes the wealth of classical
and quantum magnetoresistance phenomena in the quasi-ballistic transport
regime.

\begin{figure}
\centerline{\includegraphics[width=8cm]{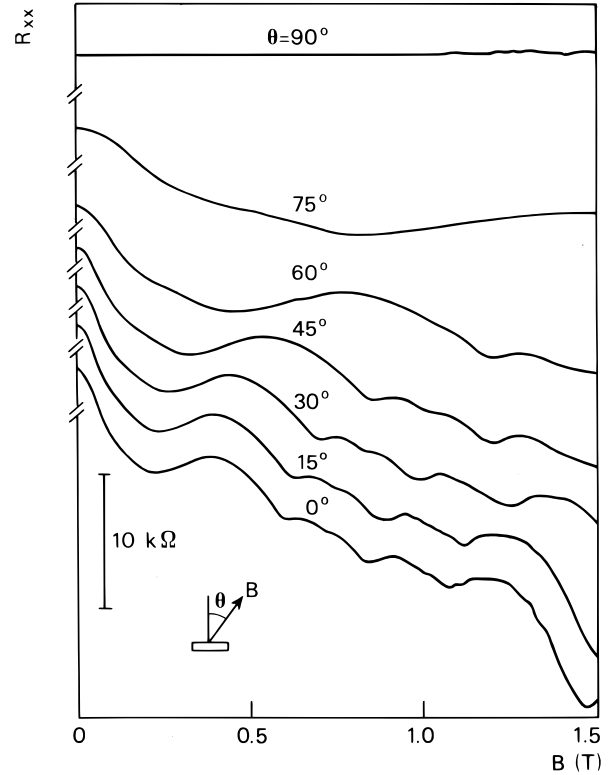}}
\caption{
Angular dependence of the magnetoresistance of Fig.\ \ref{fig28}, at 4 K, proving that it has a purely orbital origin. Taken from H. van Houten et al., Superlattices and Microstructures {\bf 3}, 497 (1987).
\label{fig29}
}
\end{figure}

Essentially similar results were obtained by Taylor et al.\cite{ref219} In the field
range of these experiments,\cite{ref27,ref55,ref63,ref167,ref219} the magnetoresistance is exclusively caused by the enclosed flux and the Lorentz force (so called {\it orbital\/} effects).
The Zeeman energy does not play a role. This is demonstrated in Fig.\ \ref{fig29},
where the magnetoresistance (obtained on the same sample as that used in
Fig.\ \ref{fig28}) is shown to vanish when $B$ is in the plane of the 2DEG (similar results
were obtained in Ref.\ \onlinecite{ref168}). In wide 2DEG channels a negative magnetoresistance has been found by Lin et al.\ in a parallel magnetic field.\cite{ref23}  This
effect has been studied in detail by Mensz and Wheeler,\cite{ref220} who attributed it
to a residual orbital effect associated with deviations of the 2DEG from a
perfectly flat plane. Fal'ko\cite{ref221} has calculated the effect of a magnetic field
parallel to the 2DEG on weak localization, and has found a negative
magnetoresistance, but only if the scattering potential does not have
reflection symmetry in the plane of the 2DEG.

\begin{figure}
\centerline{\includegraphics[width=8cm]{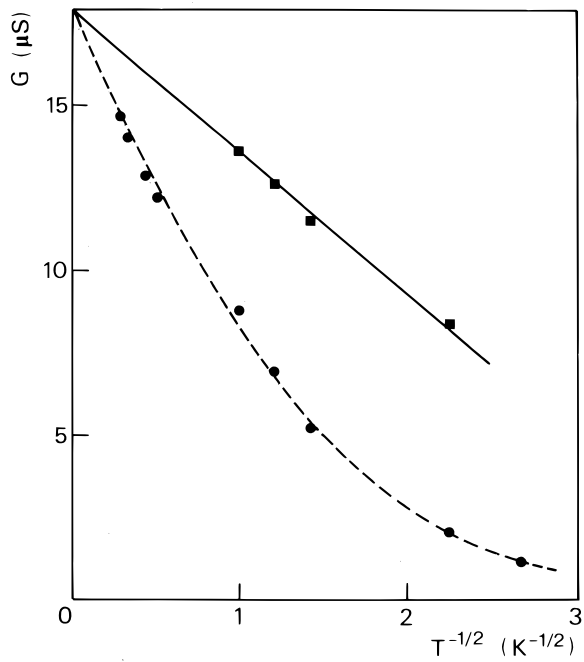}}
\caption{
Zero-field conductance (circles) and conductance corrected for the weak localization effect (squares) for the channel of Fig.\ \ref{fig28} as a function of $T^{-1/2}$, to demonstrate the $T^{-1/2}$ dependence on the temperature of the electron-electron interaction effect expected from Eq.\ (\ref{eq9.1b}). The solid and dashed lines are guides to the eye. The extrapolated value at high temperatures is the classical part of the conductance. Taken from H. van Houten et al., Acta Electronica {\bf 28}, 27 (1988).
\label{fig30}
}
\end{figure}

In Fig.\ \ref{fig30} the temperature dependence of the zero-field conductance\cite{ref27} is
plotted as a function of $T^{-1/2}$, together with the conductance after subtraction of the weak localization correction. The straight line through the
latter data points demonstrates that the remaining temperature dependence
may, indeed, be attributed to the electron-electron interactions. A similar
$T^{-1/2}$ dependence was found by Thornton et al.\cite{ref58} in a narrow GaAs-AlGaAs channel defined using the split-gate method. The slope of the straight
line in Fig.\ \ref{fig30} gives $g_{1\mathrm{D}}\approx 1.5$ in Eq.\ (\ref{eq9.1b}), which is close to the value found
by Choi et al.\cite{ref55}  It should be noted, however, that this experiment is already in
the regime where the quantum corrections are by no means small, so the
perturbation theory is of questionable validity. For this reason, and also in
view of other problems (such as the difficulty in determining the effective
channel width, the presence of channel width variations, and a frequently
observed saturation of the weak localization correction at low temperatures
due to loss of phase coherence associated with external noise or radio-frequency interference), a quantitative analysis of the parameters obtained
from the weak localization and electron-electron corrections in narrow
channels ($\tau_{\phi}$ and $g_{1\mathrm{D}}$) is not fully warranted. Indeed, most of the narrow-channel studies available today have not been optimized for the purpose of a
detailed quantitative analysis. Instead, they were primarily intended for a
phenomenological exploration, and as such we feel that they have been quite
successful.

\subsection{\label{sec10} Quantum size effects}

Quantum size effects on the resistivity result from modifications of the 2D
density of states in a 2DEG channel of width comparable to the Fermi
wavelength. The electrostatic lateral confinement in such a narrow channel
leads to the formation of 1D subbands in the conduction band of the 2DEG
(see Section \ref{sec4a}). The number $N\approx k_{\mathrm{F}}W/\pi$ of occupied 1D subbands is
reduced by decreasing the Fermi energy or the channel width. This depopulation of individual subbands can be detected via the resistivity. An
alternative method to depopulate the subbands is by means of a magnetic
field perpendicular to the 2DEG. The magnetic field $B$ has a negligible effect
on the density of states at the Fermi level if the cyclotron diameter $2l_{\mathrm{cycl}}\gg  W$
(i.e., for $B\ll B_{\mathrm{crit}}\equiv 2\hbar k_{\rm F}/eW$). If $B\gg  B_{\mathrm{crit}}$, the electrostatic confinement can
be neglected for the density of states, which is then described by Landau levels
[Eq.\ (\ref{eq4.6})]. The number of occupied Landau levels $N\approx B_{\mathrm{F}}/\hbar\omega_{\mathrm{c}}\approx k_{\mathrm{F}}l_{\mathrm{cycl}}/2$
decreases linearly with $B$ for $B\gg  B_{\mathrm{crit}}$. In the intermediate field range where $B$
and $B_{\mathrm{crit}}$ are comparable, the electrostatic confinement and the magnetic field
together determine the density of states. The corresponding {\it magnetoelectric\/}
subbands are depopulated more slowly by a magnetic field than are the
Landau levels, which results in an increased spacing of the Shubnikov-De
Haas oscillations in the magnetoresistivity (cf.\ Section \ref{sec4c}).

In the following subsection we give a more quantitative description of
magnetoelectric subbands. Experiments on the electric and magnetic depopulation of subbands in a narrow channel are reviewed in Section \ref{sec9b}. We
only consider here the case of a long channel $(L\gg  l)$ in the quasi-ballistic
regime. Quantum size effects in the fully ballistic regime $(L\lesssim l)$ are the
subject of Section \ref{secIII}.

\subsubsection{\label{sec10a} Magnetoelectric subbands}

Consider first the case of an unbounded 2DEG in a perpendicular
magnetic field $\mathbf{B}=\nabla\times \mathbf{A}$. The Hamiltonian for motion in the plane of the
2DEG is given by
\be
{\cal H}=\frac{(\mathbf{p}+e\mathbf{A})^{2}}{2m}, \label{eq10.1}
\ee 
for a single spin component. In the Landau gauge ${\bf A}=(0, Bx, 0)$, with $\mathbf{B}$ in the
$z$-direction, this may be written as
\be
{\cal H}=\frac{p_{x}^{2}}{2m}+\frac{m\omega_{\mathrm{c}}^{2}}{2}(x-x_{0})^{2}, \label{eq10.2}
\ee 
with $\omega_{\mathrm{c}}\equiv eB/m$ and $x_{0}\equiv-p_{y}/eB$. The $y$-momentum operator $p_{y}\equiv-i\hbar\partial/\partial y$
can be replaced by its eigenvalue $\hbar k_{y}$, since $p_{y}$ and ${\cal H}$ commute. The effect of
the magnetic field is then represented by a harmonic oscillator potential in
the $x$-direction, with center $x_{0}=-\hbar k_{y}/eB$ depending on the momentum in
the $y$-direction. The energy eigenvalues $E_{n}=(n- \frac{1}{2})\hbar\omega_{\mathrm{c}}$, $n=1,2,3, \ldots$, do
not depend on $k_{y}$ and are therefore highly degenerate. States with the same
quantum number $n$ are referred to collectively as {\it Landau levels}.\cite{ref93} The
number of Landau levels below energy $E$ is given by
\be
N=\mathrm{Int}[1/2+E/\hbar\omega_{\mathrm{c}}], \label{eq10.3}
\ee 
where $\mathrm{Int}$ denotes truncation to an integer.

\begin{figure}
\centerline{\includegraphics[width=8cm]{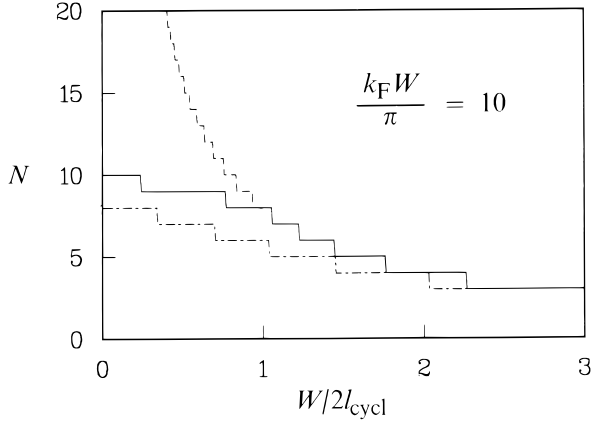}}
\caption{
Magnetic field dependence of the number $N$ of occupied subbands in a narrow channel for a parabolic confining potential according to Eq.\ (\ref{eq10.7}) (dot-dashed curve), and for a square-well confining potential according to Eq.\ (\ref{eq10.8}) (full curve). The dashed curve gives the magnetic depopulation of Landau levels in a wide 2DEG, which has a $1/B$ dependence. The calculations are done for a fixed Fermi energy and for channel width $W = W_{\rm par} = 10\pi/k_{\rm F}$.
\label{fig31}
}
\end{figure}

A narrow channel in the $y$-direction is defined by an electrostatic confining
potential $V(x)$. The case of a {\it parabolic\/} confinement is easily solved analytically.\cite{ref36,ref218,ref222,ref223} Adding a term $V(x)= \frac{1}{2}m\omega_{0}^{2}x^{2}$ to the hamiltonian
(\ref{eq10.2}), one finds, after a rearrangement of terms,
\be
{\cal H}=\frac{p_{x}^{2}}{2m}+\frac{m\omega^{2}}{2}(x-\overline{x}_{0})^{2}+\frac{\hbar^{2}k^{2}}{2M}, \label{eq10.4}
\ee 
with $\omega\equiv(\omega_{\mathrm{c}}^{2}+\omega_{0}^{2})^{1/2}$, $\overline{x}_{0}\equiv x_{0}\omega_{\mathrm{c}}/\omega$, and $M\equiv m\omega^{2}/\omega_{0}^{2}$. The first two terms
describe the motion in the $x$-direction in a harmonic potential with effective
frequency $\omega\geq\omega_{0}$, and the third term describes free motion in the $y$-direction
with an effective mass $M\geq m$. This last term removes the degeneracy of the
Landau levels, which become 1D subbands with energy
\be
E_{n}(k)=(n- {\textstyle\frac{1}{2}})\hbar\omega+\hbar^{2}k^{2}/2M. \label{eq10.5}
\ee 
The subband bottoms have energy $E_{n}=(n- \frac{1}{2})\hbar\omega$, and the number of
subbands occupied at energy $E$ is $N= \mathrm{Int}[\frac{1}{2}+E/\hbar\omega]$. The quasi-1D density
of states is obtained from Eq.\ (\ref{eq4.4}) on substituting $m$ for $M$. For the
comparison with experiments it is useful to define an effective width for the
parabolic potential. One can take the width $W_{\mathrm{par}}$ to be the separation between
the equipotentials at the Fermi energy,
\be
W_{\mathrm{par}}\equiv 2\hbar k_{\rm F}/m\omega_{0}. \label{eq10.6}
\ee 
(An alternative, which differs only in the numerical prefactor, is to take
$W_{\mathrm{par}}\equiv n_{1\mathrm{D}}/n_{\mathrm{s}}$, with $n_{\mathrm{s}}\equiv g_{\mathrm{s}}g_{\mathrm{v}}k_{\mathrm{F}}^{2}/4\pi$ the 2D sheet density and $n_{1\mathrm{D}}$ the number
of electrons per unit length in the narrow channel.\cite{ref218}) The number of
occupied magnetoelectric subbands at energy $E_{\mathrm{F}}$ in a {\it parabolic\/} confining
potential may then be written as
\be
N=\mathrm{Int}\left[
\frac{1}{2}+\frac{1}{4}k_{\rm F} W_{\rm par}[1+(W_{\mathrm{par}}/2l_{\mathrm{cycl}})^{2}]^{-1/2}\right], \label{eq10.7}
\ee
where $l_{\mathrm{cycl}}\equiv \hbar k_{\rm F}/eB$ is the cyclotron radius at the Fermi energy. For easy
reference, we also give the result for the number of occupied subbands at the
Fermi energy in a {\it square-well\/} confinement potential of width $W$:
\begin{widetext}
\begin{subequations}
\label{eq10.8}
\begin{eqnarray}
N&\approx& \mathrm{Int}\left[\frac{2}{\pi}\frac{E_{\mathrm{F}}}{\hbar\omega_{\mathrm{c}}}\left(\arcsin\frac{W}{2l_{\mathrm{cycl}}}+\frac{W}{2l_{\mathrm{cycl}}}\left[1-\left(\frac{W}{2l_{\mathrm{cycl}}}\right)^{2}\right]^{1/2}\right)\right],\;\;{\rm if}\;\; l_{\mathrm{cycl}}> \frac{W}{2},\label{eq10.8a}\\
N&\approx& \mathrm{Int}\left[\frac{1}{2}+\frac{E_{\mathrm{F}}}{\hbar\omega_{\mathrm{c}}}\right]\;\;{\rm if}\;\;l_{\mathrm{cycl}}< \frac{W}{2}. \label{eq10.8b}
\end{eqnarray}
\end{subequations}
\end{widetext}
(This result is derived in Section \ref{sec12a} in a semiclassical approximation. The
accuracy is $\pm 1$.) One easily verifies that, for $B\ll B_{\mathrm{crit}}\equiv 2\hbar k_{\rm F}/eW,$ Eq.\ (\ref{eq10.8})
yields $N\approx k_{\mathrm{F}}W/\pi$. The parabolic confining potential gives $N\approx k_{\mathrm{F}}W_{\mathrm{par}}/4$ in
the weak-field limit. In the strong-field limit $B\gg  B_{\mathrm{crit}}$, both potentials give
the result $N\approx E_{\mathrm{F}}/\hbar\omega_{\mathrm{c}}=k_{\mathrm{F}}l_{\mathrm{cycl}}/2$ expected for pure Landau levels. In Fig.\ \ref{fig31}
we compare the depopulation of Landau levels in an unbounded 2DEG with
its characteristic $1/B$ dependence of $N$ (dashed curve), with the slower
depopulation of magnetoelectric subbands in a narrow channel. The dash-dotted curve is for a parabolic confining potential, the solid curve for a
square-well potential. These results are calculated from  Eqs.\ (\ref{eq10.7}) and (\ref{eq10.8}),
with $k_{\mathrm{F}}W_{\mathrm{par}}/\pi=k_{\mathrm{F}}W/\pi=10$. A $B$-independent Fermi energy was assumed in
Fig.\ \ref{fig31} so that the density $n_{1\mathrm{D}}$ oscillates around its zero-field value. (For a
long channel, it is more appropriate to assume that $n_{1\mathrm{D}}$ is $B$-independent, to
preserve charge neutrality, in which case $E_{\mathrm{F}}$ oscillates. This case is studied in
Ref.\ \onlinecite{ref218}.) Qualitatively, the two confining potentials give similar results. The
numerical differences reflect the uncertainty in assigning an effective width to
the parabolic potential. Self-consistent solutions of the Poisson and Schr\"{o}dinger equations\cite{ref42,ref60,ref61,ref72,ref224} for channels defined by a split gate have shown
that a parabolic potential with a flat bottom section is a more realistic model.
The subband depopulation for this potential has been studied in a semiclassical approximation in Ref.\ \onlinecite{ref223}. A disadvantage of this more realistic model is
that an additional parameter is needed for its specification (the width of the
flat section). For this practical reason, the use of either a parabolic or a
square-well potential has been preferred in the analysis of most experiments.

\subsubsection{\label{sec10b} Experiments on electric and magnetic depopulation of subbands}

The observation of 1D subband effects unobscured by thermal smearing
requires low temperatures, such that $4k_{\mathbf{B}}T\ll \Delta E$, with $\Delta E$ the energy
difference between subband bottoms near the Fermi level ($4k_{\mathbf{B}}T$ being the
width of the energy averaging function $df/dE_{\mathrm{F}}$; see Section \ref{sec4b}; For a square
well $\Delta E\approx 2E_{\mathrm{F}}/N$, and for parabolic confinement $\Delta E\approx E_{\mathrm{F}}/N$). Moreover,
the formation of subbands requires the effective mean free path (limited by
impurity scattering and diffuse boundary scattering) to be much larger than
$W$ (cf.\ also Ref.\ \onlinecite{ref218}). The requirement on the temperature is not difficult to
meet, $\Delta E/4k_{\mathbf{B}}T$ being on the order of $50\, \mathrm{K}$ for a typical GaAs-AlGaAs
channel of width $W=100\,\mathrm{nm}$, and the regime $l>W$ is also well accessible.
These simple considerations seem to suggest that 1D subband effects should
be rather easily observed in semiconductor nanostructures. This conclusion is
misleading, however, and in reality manifestations of 1D subband structure
have been elusive, at least in the quasi-ballistic regime $W<l<L$. The main
reason for this is the appearance of large conductance fluctuations that mask
the subband structure if the channel is not short compared with the mean free
path.

Calculations\cite{ref225,ref226,ref227} of the {\it average\/} conductivity of an {\it ensemble\/} of narrow
channels do in fact show oscillations from the electric depopulation of
subbands [resulting from the modulation of the density of states at the Fermi
level, which determines the scattering time; see Eq.\ (\ref{eq4.28})]. The oscillations
are not as large as the Shubnikov-De Haas oscillations from the magnetic
depopulation of Landau levels or magnetoelectric subbands. One reason for
this difference is that the peaks in the density of states become narrower,
relative to their separation, on applying a magnetic field. (The quantum limit
of a single occupied 1D subband has been studied in Refs.\ \onlinecite{ref42} and \onlinecite{ref228,ref229,ref230}.)

\begin{figure}
\centerline{\includegraphics[width=6cm]{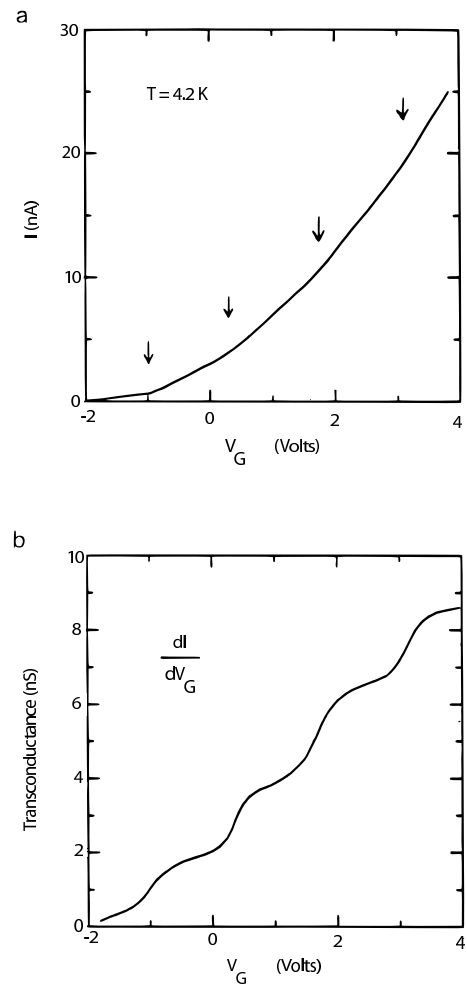}}
\caption{
(a) Dependence on the gate voltage of the current $I$ through 250 parallel narrow Si inversion layer channels at 1.2 K, showing the electric depopulation of subbands. (b) The effect is seen more clearly in the transconductance $dI/dV_{\rm G}$. Note the absence of universal conductance fluctuations, which have been averaged out by the large number of channels. Taken from A. C. Warren et al., IEEE Electron Device Lett.\ {\bf EDL-7}, 413 (1986).
\label{fig32}
}
\end{figure}

In an individual channel, aperiodic conductance fluctuations due to
quantum interference (see Section 7) are the dominant cause of structure in
the low-temperature conductance as a function of gate voltage (which
corresponds to a variation of the Fermi energy), as was found in experiments
on narrow Si inversion layers.\cite{ref46,ref161,ref162} Warren et al.\cite{ref44} were able to suppress
these fluctuations by performing measurements on an array of narrow
channels in a Si inversion layer. In Fig.\ \ref{fig32} we reproduce their results. The
structure due to the electric depopulation of 1D subbands is very weak in the
current-versus-gate-voltage plot, but a convincingly regular oscillation is
seen if the derivative of the current with respect to the gate voltage is taken
(this quantity is called the transconductance). Warren et al.\ pointed out that
the observation of a quantum size effect in an array of 250 channels indicates
a rather remarkable uniformity of the width and density of the individual
channels.

More recently a similar experimental study was performed by Ismail et
al.\cite{ref231} on 100 parallel channels defined in the 2DEG of a GaAs-AlGaAs
heterostructure. The effects were found to be much more pronounced than in
the earlier experiment on Si inversion layer channels, presumably because of
the much larger mean free path (estimated at 1 $\mu \mathrm{m}$), which was not much
shorter than the sample length (5 $\mu \mathrm{m}$). Quantum size effects in the quantum
ballistic transport regime (in particular, the conductance quantization of a
quantum point contact) are discussed extensively in Section \ref{sec13}.
In a wide 2DEG the minima of the Shubnikov-De Haas oscillations in the
magnetoresistance are periodic in $1/B$, with a periodicity $\Delta(1/B)$ determined
by the sheet density $n_{\mathrm{s}}$ according to Eq.\ (\ref{eq4.29}). In a narrow channel one
observes an increase in $\Delta(1/B)$ for weak magnetic fields because the
electrostatic confinement modifies the density of states, as discussed in
Section \ref{sec10a}. Such a deviation is of interest as a manifestation of magnetoelectric subbands, but also because it can be used to estimate the effective
channel width using the criterion $W\approx 2l_{\mathrm{cycl}}$ for the crossover field\cite{ref167} $B_{\mathrm{crit}}$
(the electron density in the channel, and hence $l_{\mathrm{cycl}}$, may be estimated from
the strong-field periodicity). The phenomenon has been studied in many
publications.\cite{ref36,ref56,ref57,ref74,ref79,ref167,ref217,ref218,ref223,ref232,ref233}

\begin{figure}
\centerline{\includegraphics[width=6cm]{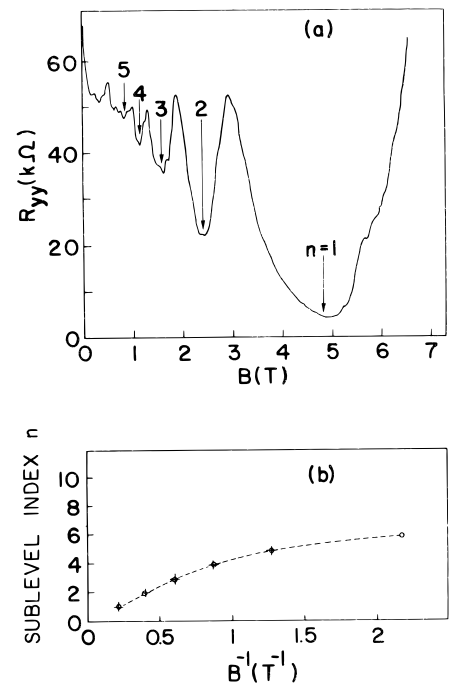}}
\caption{
(a) Magnetoresistance at 2.4 K of a narrow GaAs-AlGaAs channel (as in Fig.\ \ref{fig28}). The arrows indicate magnetic field values assigned to the depopulation of magnetoelectric subbands. (b) Subband index $n \equiv N - 1$ versus inverse magnetic field (crosses). The dashed line interpolates between theoretical points for a parabolic confining potential (circles). The electrostatic confinement causes deviations from a linear dependence of $n$ on $B^{-1}$. Taken from K.-F. Berggren et al., Phys.\ Rev.\ B {\bf 37}, 10118 (1988).
\label{fig33}
}
\end{figure}

As an illustration, we reproduce in Fig.\ \ref{fig33}a an experimental magnetoresistance trace\cite{ref167,ref218} obtained for a narrow ($W\approx 140\,\mathrm{nm}$) GaAs-AlGaAs channel, defined using a shallow-mesa etch.\cite{ref63} The arrows indicate
the magnetoresistance minima thought to be associated with magnetic
depopulation. The assignment becomes ambiguous in weak magnetic fields,
because of the presence of aperiodic conductance fluctuations. Nevertheless,
the deviation from a straight line in the $N$ versus $B^{-1}$ plot in Fig.\ \ref{fig33}b is
sufficiently large to be reasonably convincing. Also shown in Fig.\ \ref{fig33}b is the
result of a fit to a theoretical $N(B)$ function (assuming a parabolic confining
potential and a $B$-independent electron density). The parameter values found
from this fit for the width and electron density are reasonable and agree with
independent estimates.\cite{ref27}

We have limited ourselves to a discussion of transport studies, but wish to
point out that 1D subbands have been studied succesfully by capacitance\cite{ref75}
measurements and by infrared\cite{ref78} spectroscopy. As mentioned earlier, the
formation of 1D subbands also requires a reformulation of the theories of
weak localization and conductance fluctuations in the presence of boundary
scattering. Weak localization in the case of a small number of occupied
subbands has been studied by Tesanovic et al.\cite{ref110,ref234} (in a zero magnetic
field).

We will not discuss the subject of quantum size effects further in this part
of our review, since it has found more striking manifestations in the ballistic
transport regime (the subject of Section \ref{secIII}), where conductance fluctuations
do not play a role. The most prominent example is the conductance
quantization of a point contact.

\subsection{\label{sec11} Periodic potential}

\subsubsection{\label{sec11a} Lateral superlattices}

In a crystal, the periodic potential of the lattice opens energy gaps of zero
density of electronic states. An electron with energy in a gap is Bragg-reflected and hence cannot propagate through the crystal. Esaki and Tsu\cite{ref235}
proposed in 1970 that an artificial energy gap might be created by the
epitaxial growth of alternating layers of different semiconductors. In such a
{\it superlattice\/} a periodic potential of spacing $a$ is superimposed on the crystal
lattice potential. Typically, $a\approx 10\,\mathrm{nm}$ is chosen to be much larger than the
crystal lattice spacing ($0.5\, \mathrm{nm}$), leading to the formation of a large number of
narrow bands within the conduction band (minibands), separated by small
energy gaps (minigaps). Qualitatively new transport properties may then be
expected. For example, the presence of minigaps may be revealed under
strong applied voltages by a negative differential resistance phenomenon
predicted by Esaki and Tsu in their original proposal and observed
subsequently by Esaki and Chang.\cite{ref236,ref237} In contrast to a 3D crystal lattice, a
superlattice formed by alternating layers is 1D. As a consequence of the free
motion in the plane of the layers, the density of states is not zero in the
minigaps, and electrons may scatter between two overlapping minibands. Of
interest in the present context is the possibility of defining {\it lateral\/} superlattices\cite{ref238,ref239} by a periodic potential in the plane of a 2D electron gas. True
minigaps of zero density of states may form in such a system if the potential
varies periodically in two directions. Lateral superlattice effects may be
studied in the linear-response regime of small applied voltages (to which we
limit the discussion here) by varying $E_{\mathrm{F}}$ or the strength of the periodic
potential by means of a gate voltage. The conductivity is expected to vanish if
$E_{\mathrm{F}}$ is in a true minigap (so that electrons are Bragg-reflected). Calculations\cite{ref240,ref241} show pronounced minima also in the case of a 1D periodic
potential.

The conditions required to observe the minibands in a lateral superlattice
are similar to those discussed in Section \ref{sec10} for the observation of 1D
subbands in a narrow channel. The mean free path should be larger than the
lattice constant $a$, and $4k_{\mathbf{B}}T$ should be less than the width of a minigap near
the Fermi level. For a weak periodic potential,\cite{ref94} the $n$th minigap is
approximately $\Delta E_{n}\approx 2V_{n}$, with $V_{n}$ the amplitude of the Fourier component of
the potential at wave number $k_{n}=2\pi n/a$. The gap is centered at energy
$E_{n}\approx(\hbar k_{n}/2)^{2}/2m$. If we consider, for example, a 1D sinusoidal potential
$V(x, y)=V_{0}\sin(2\pi y/a)$, then the first energy gap $\Delta E_{1}\approx V_{0}$ occurs at
$E_{1}\approx(\hbar\pi/a)^{2}/2m$. (Higher-order minigaps are much smaller.) Bragg reflection
occurs when $E_{1}\approx E_{\mathrm{F}}$ (i.e., for a lattice periodicity $a\approx\lambda_{\mathrm{F}}/2$). Such a short-
period modulation is not easy to achieve lithographically, however (typically
$\lambda_{\mathrm{F}}=40\,\mathrm{nm})$, and the experiments on lateral superlattices discussed later are
not in this regime.

\begin{figure}
\centerline{\includegraphics[width=8cm]{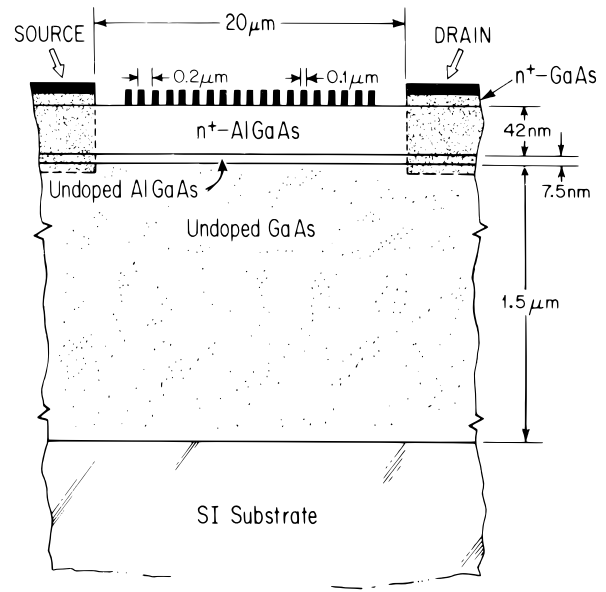}}
\caption{
Grating gate (in black) on top of a GaAs-AlGaAs heterostructure, used to define a 2DEG with a periodic density modulation. Taken from K. Ismail et al., Appl.\ Phys.\ Lett.\ {\bf 52}, 1071 (1988).\label{fig34}
}
\end{figure}

\begin{figure}
\centerline{\includegraphics[width=8cm]{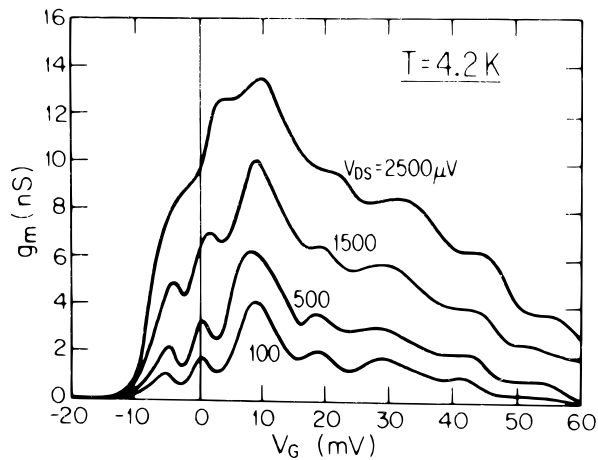}}
\caption{
Transconductance $g_{\rm m}\equiv \partial I/\partial V_{\rm SD}$ of the device of Fig.\ \ref{fig34} measured as a function of gate voltage for various values of the source-drain voltage. The oscillations, seen in particular at low source-drain voltages, are attributed to Bragg reflection in a periodic potential. Taken from K. Ismail et al., Appl.\ Phys.\ Lett.\ {\bf 52}, 1071 (1988).
\label{fig35}
}
\end{figure}

Warren et al.\cite{ref242} have observed a weak but regular structure in the
conductance of a 1D lateral superlattice with $a=0.2\,\mu \mathrm{m}$ defined in a Si
inversion layer (using the dual-gate arrangement of Fig.\ \ref{fig2}c). Ismail et al.\cite{ref62}
used a grating-shaped gate on top of a GaAs-AlGaAs heterostructure to
define a lateral superiattice. A schematic cross section of their device is shown in Fig.\ \ref{fig34}. The period of the grating is $0.2\,\mu{\rm m}$. One effect of the gate voltage is to change the overall carrier concentration, leading to a large but essentially smooth conductance variation (at 4.2 K). This variation proved to be essentially the same as that found for a continuous gate. As in the experiment by Warren et al., the transconductance as a function of the voltage on the grating revealed a regular oscillation. As an example, we reproduce the results of Ismail et al.\ (for various source-drain voltages) in Fig.\ \ref{fig35}. No such structure was found for control devices with a continuous, rather than a grating, gate. The observed structure is attributed to Bragg reflection in Ref.\ \onlinecite{ref62}. A 2D lateral superlattice was defined by Bernstein and Ferry,\cite{ref243} using a grid-shaped gate, but the transport properties in the linear response regime were not studied in detail. Smith et al.\cite{ref244} have used the split-gate technique to define a 2D array of 4000 dots in a high-mobility GaAs-AlGaAs heterostructure ($a = 0.5\,\mu{\rm m}$, $1= 10\,\mu{\rm m}$). When the 2DEG under the dots is depleted, a grid of conducting channels is formed. In this experiment the amplitude of
the periodic potential exceeds $E_{\mathrm{F}}$. Structure in the conductance is found
related to the depopulation of 1D subbands in the channels, as well as to
standing waves between the dots. The analysis is thus considerably more
complicated than it would be for a weak periodic potential. It becomes
difficult to distinguish between the effects due to quantum interference within
a single unit cell of the periodic potential and the effects due to the formation
of minibands requiring phase coherence over several unit cells. Devices with a
2D periodic potential with a period comparable to the Fermi wavelength and
much shorter than the mean free path will be required for the realization of
true miniband effects. It appears that the fabrication of such devices will have
to await further developments in the art of making nanostructures. Epitaxy
on tilted surfaces with a staircase surface structure is being investigated for
this purpose.\cite{ref87,ref88,ref169,ref179,ref245,ref246} Nonepitaxial growth on Si surfaces slightly
tilted from (100) is known to lead to miniband formation in the inversion
layer.\cite{ref20,ref247} A final interesting possibility is to use doping quantum wires, as
proposed in Ref.\ \onlinecite{ref248}.

As mentioned, it is rather difficult to discriminate experimentally between
true miniband effects and quantum interference effects occurring within one
unit cell. The reason is that both phenomena give rise to structure in the
conductance as a function of gate voltage with essentially the same periodicity. This difficulty may be circumvented by studying lateral superlattices with
a small number of unit cells. The miniband for a finite superlattice with $P$ unit
cells consists of a group of $P$ discrete states, which merge into a continuous
miniband in the limit $P\rightarrow\infty$. The discrete states give rise to closely spaced
resonances in the transmission probability through the superlattice as a
function of energy, and may thus be observed as a series of $P$ peaks in the
conductance as a function of gate voltage, separated by broad minima due to
the minigaps. Such an observation would demonstrate phase coherence over
the entire length $L=Pa$ of the finite superlattice and would constitute
conclusive evidence of a miniband. The conductance of a finite 1D superlattice in a narrow 2DEG channel in the ballistic transport regime has been
investigated theoretically by Ulloa et al.\cite{ref249} Similar physics may be studied in
the quantum Hall effect regime, where the experimental requirements are
considerably relaxed. A successful experiment of this type was recently
performed by Kouwenhoven et al.\cite{ref250} (see Section \ref{sec22}).

Weak-field magnetotransport in a 2D periodic potential (a grid) has been
studied by Ferry et al.\cite{ref251,ref252} and by Smith et al.\cite{ref244} Both groups reported
oscillatory structure in the magnetoconductance, suggestive of an
Aharonov-Bohm effect with periodicity $\Delta B=h/eS$, where $S$ is the area of a
unit cell of the ``lattice.'' In strong magnetic fields no such oscillations are
found. A similar suppression of the Aharonov-Bohm effect in strong fields is
found in single rings, as discussed in detail in Section \ref{sec21a}. Magnetotransport
in a 1D periodic potential is the subject of the next subsection.

\subsubsection{\label{sec11b} Guiding-center-drift resonance}

The influence of a magnetic field on transport through layered superlattices\cite{ref253} has been studied mainly in the regime where the (first) energy gap
$\Delta E\sim 100\,\mathrm{meV}$ exceeds the Landau level spacing $\hbar\omega_{\mathrm{c}}$ ($1.7\, \mathrm{meV}/\mathrm{T}$ in GaAs).
The magnetic field does not easily induce transitions between different
minibands in this regime. Magnetotransport through lateral superlattices is
often in the opposite regime $\hbar\omega_{\mathrm{c}}\gg  \Delta E$, because of the relatively large
periodicity $(a\sim 300\,\mathrm{nm})$ and small amplitude $(V_{0}\sim 1\,\mathrm{meV})$ of the periodic
potential. The magnetic field now changes qualitatively the structure of the
energy bands, which becomes richly complex in the case of a 2D periodic
potential.\cite{ref54} Much of this structure, however, is not easily observed, and the
experiments discussed in this subsection involve mostly the {\it classical\/} effect of
a weak periodic potential on motion in a magnetic field.

\begin{figure}
\centerline{\includegraphics[width=6cm]{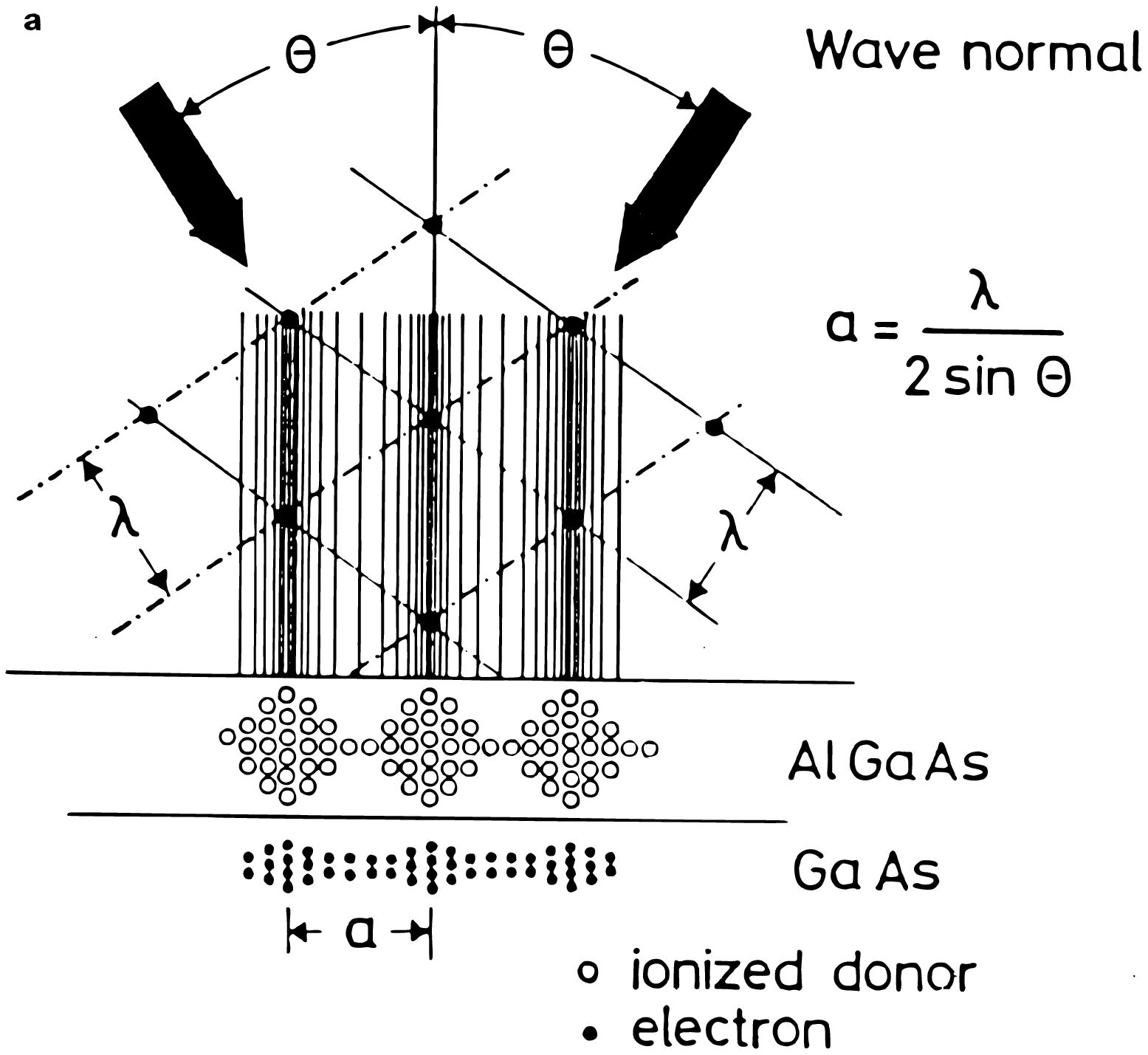}}

\centerline{\includegraphics[width=6cm]{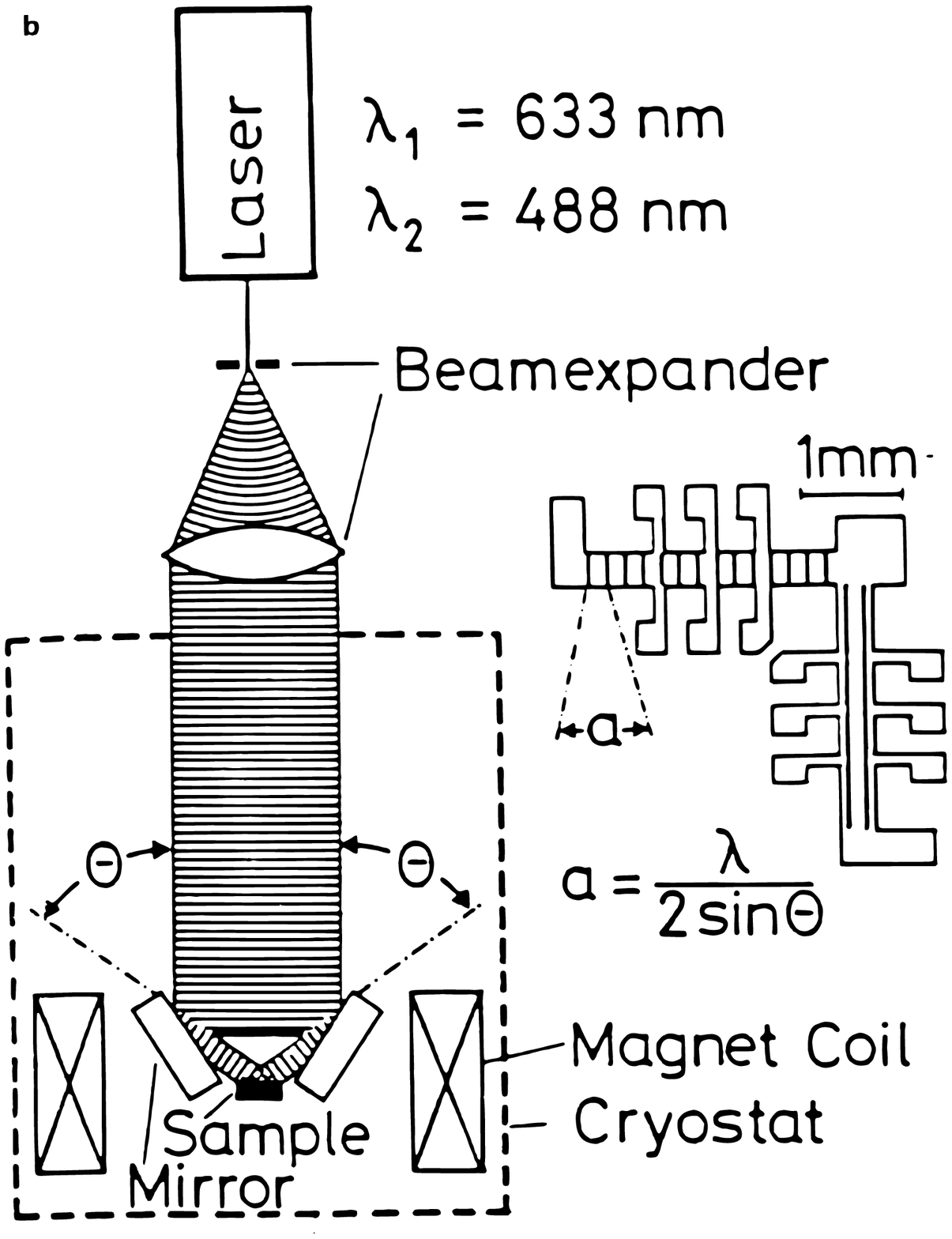}}
\caption{
A brief illumination of a GaAs-AlGaAs heterostructure with an interference pattern due to two laser beams (black arrows) leads to a persistent periodic variation in the concentration of ionized donors in the AlGaAs, thereby imposing a weak periodic potential on the 2DEG. The resulting spatial variation of the electron density in the 2DEG is indicated schematically. (b) Experimental arrangement used to produce a modulated 2DEG by means of the ``holographic illumination'' of (a). The sample layout shown allows measurements of the resistivity parallel and perpendicular to the equipotentials. Taken from D. Weiss et al., in ``High Magnetic Fields in Semiconductor Physics II'' (G. Landwehr, ed.). Springer, Berlin, 1989.
\label{fig36}
}
\end{figure}

\begin{figure}
\centerline{\includegraphics[width=8cm]{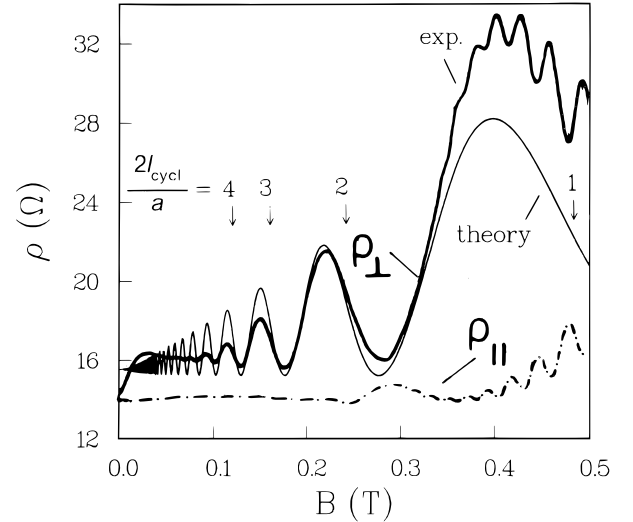}}
\caption{
Solid curves: Magnetic field dependence of the resistivity $\rho_{\perp}$ for current flowing perpendicular to a potential grating. The experimental curve is the measurement of Weiss et al.\cite{ref255} the theoretical curve follows from the guiding-center-drift resonance. Note the phase shift of the oscillations, indicated by the arrows at integer $2l_{\rm cycl}/a$. The potential grating has periodicity $a =382\,{\rm nm}$ and is modeled by a sinusoidal potential with root-mean-square amplitude of $\epsilon = 1.5$\% 
of the Fermi energy; The mean free path in the 2DEG is $12\,\mu{\rm m}$, much larger than $a$. The dash-dotted curve is the experimental resistivity $\rho_{||}$ for current flowing parallel to the potential grating, as measured by Weiss et al. Taken from C. W. J. Beenakker, Phys.\ Rev.\ Lett.\ {\bf 62}, 2020 (1989).
\label{fig37}
}
\end{figure}

Weiss et al.\cite{ref255,ref256} used an ingenious technique to impose a weak periodic
potential on a 2DEG in a GaAs-AlGaAs heterostructure. They exploit the
well-known persistent ionization of donors in AlGaAs after brief illumination
at low temperatures. For the illumination, two interfering laser beams are
used, which generate an interference pattern with a period depending on the
wavelength and on the angle of incidence of the two beams. This technique,
known as {\it holographic illumination}, is illustrated in Fig.\ \ref{fig36}. The interference
pattern selectively ionizes Si donors in the AlGaAs, leading to a weak
periodic modulation $V(y)$ of the bottom of the conduction band in the 2DEG,
which persists at low temperatures if the sample is kept in the dark. The
sample layout, also shown in Fig.\ \ref{fig36}, allows independent measurements of
the resistivity $\rho_{yy}(\equiv\rho_{\perp})$, perpendicular to, and $\rho_{xx}(\equiv\rho_{||})$ parallel to the
grating. In Fig.\ \ref{fig37} we show experimental results of Weiss et al.\cite{ref255} for the
magnetoresistivity of a 1D lateral superlattice $(a=382\,\mathrm{nm}$). In a zero
magnetic field, the resistivity tensor $\mathbf{\rho}$ is approximately isotropic: $\rho_{\perp}$ and $\rho_{||}$
are indistinguishable experimentally (see Fig.\ \ref{fig37}). This indicates that the
amplitude of $V(y)$ is much smaller than the Fermi energy $E_{\mathrm{F}}=11\,\mathrm{meV}$. On
application of a small magnetic field $B(\lesssim 0.4\,\mathrm{T})$ perpendicular to the 2DEG,
a large oscillation periodic in $1/B$ develops in the resistivity $\rho_{\perp}$ for current
flowing perpendicular to the potential grating. The resistivity is now strongly
anisotropic, showing only weak oscillations in $\rho_{||}$ (current parallel to the
potential grating). In appearance, the oscillations resemble the Shubnikov-De Haas oscillations at higher fields, but their different periodicity and much
weaker temperature dependence point to a different origin.

The effect was not anticipated theoretically, but now a fairly complete and
consistent theoretical picture has emerged from several analyses.\cite{ref111,ref227,ref257,ref258,ref259} The strong oscillations in $\rho_{\perp}$ result from a resonance\cite{ref111}
between the periodic cyclotron orbit motion and the oscillating $\mathbf{E}\times \mathbf{B}$ drift of
the orbit center induced by the electric field $\mathbf{E}\equiv-\nabla V$. Such {\it guiding-center-drift resonances\/} are known from plasma physics,\cite{ref260} and the experiment by
Weiss et al.\ appears to be the first observation of this phenomenon in the
solid state. Magnetic quantization is not essential for these strong oscillations, but plays a role in the transition to the Shubnikov-De Haas
oscillations at higher fields and in the weak oscillations in $\rho_{||}$.\cite{ref227,ref259} A
simplified physical picture of the guiding-center-drift resonance can be
obtained as follows.\cite{ref111}

\begin{figure}
\centerline{\includegraphics[width=8cm]{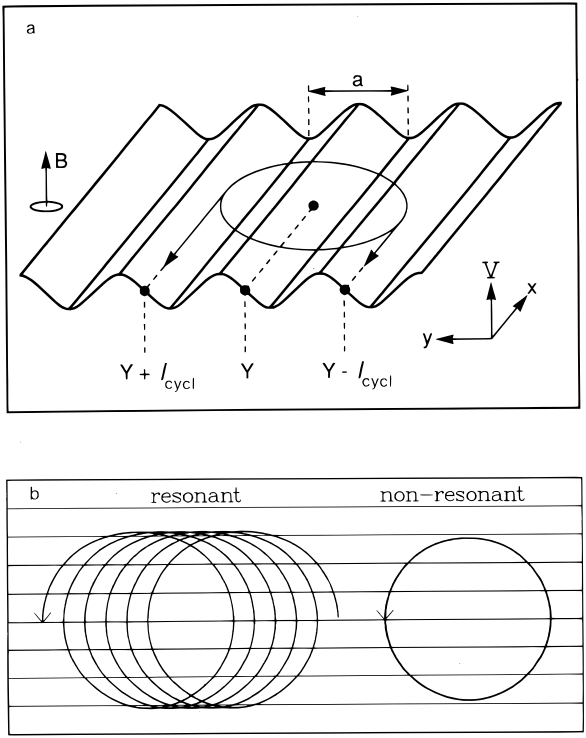}}
\caption{
(a) Potential grating with a cyclotron orbit superimposed. When the electron is close to the two extremal points $Y\pm l_{\rm cycl}$, the guiding center at $Y$ acquires an ${\bf E} \times {\bf B}$ drift in the direction of the arrows. (The drift along nonextremal parts of the orbit averages out, approximately.) A resonance occurs if the drift at one extremal point reinforces the drift at the other, as shown. (b) Numerically calculated trajectories for a sinusoidal potential ($\epsilon = 0.015$). The horizontal lines are equipotentials at integer $y/a$. On resonance ($2l_{\rm cycl}/a = 6.25$) the guiding center drift is maximal; off resonance ($2_{\rm lcycl}/a = 5.75$) the drift is negligible. Taken from C. W. J. Beenakker, Phys.\ Rev.\ Lett.\ {\bf 62}, 2020 (1989).
\label{fig38}
}
\end{figure}

The guiding center $(X, Y)$ of an electron at position $(x,y)$ having velocity
$(v_{x},v_{y})$ is given by $X=x-v_{y}/\omega_{\mathrm{c}}$, $Y=y+v_{x}/\omega_{\mathrm{c}}$. The time derivative of the
guiding center is $x=E(y)/B,\dot{Y}=0$, so its motion is parallel to the $x$-axis.
This is the ${\bf E} \times {\bf B}$ drift. In the case of a strong magnetic field and a slowly
varying potential $(l_{\mathrm{cycl}}\ll a)$, one may approximate $E(y)\approx E(Y)$ to close the
equations for $\dot{X}$ and $\dot{Y}$. This so-called adiabatic approximation cannot be
made in the weak-field regime $(l_{\mathrm{cycl}}\gtrsim a)$ of interest here. We consider the case
of a weak potential, such that $eV_{\mathrm{rms}}/E_{\mathrm{F}}\equiv\epsilon\ll 1$, with $V_{\mathrm{rms}}$ the root mean
square of $V(y)$. The guiding center drift in the $x$-direction is then simply
superimposed on the unperturbed cyclotron motion. Its time average $v_{\mathrm{drift}}$ is
obtained by integrating the electric field along the orbit
\be
v_{\mathrm{drift}}(Y)=(2 \pi B)^{-1}\int_{0}^{2\pi}d\phi\,E(Y+l_{\mathrm{cycl}}\sin\phi). \label{eq11.1}
\ee 
For $l_{\mathrm{cycl}}\gg  a$ the field oscillates rapidly, so only the drift acquired close to the
two extremal points $Y\pm l_{\mathrm{cycl}}$ does not average out. It follows that $v_{\mathrm{drift}}$ is
large or small depending on whether $E(Y+l_{\mathrm{cycl}})$ and $E(Y-l_{\mathrm{cycl}})$ have the
same sign or opposite sign (see Fig.\ \ref{fig38}). For a sinusoidal potential
$V(y)=\sqrt{2}V_{\mathrm{rms}}\sin(2\pi y/a)$, one easily calculates by averaging over $Y$ that, for
$l_{\mathrm{cycl}}\gg  a$, the mean square drift is
\be
\langle v_{\mathrm{drift}}^{2}\rangle=(v_{\mathrm{F}}\epsilon)^{2}\left(\frac{l_{\mathrm{cycl}}}{a}\right)\cos^{2}\left(\frac{2\pi l_{\mathrm{cycl}}}{a}-\frac{\pi}{4}\right). \label{eq11.2}
\ee 
The guiding center drift by itself leads, for $l_{\mathrm{cycl}}\ll l$, to 1D diffusion with
diffusion coefficient $\delta D$ given by
\be
\delta D=\int_{0}^{\infty}\langle v_{\mathrm{drift}}^{2}\rangle \mathrm{e}{}^{-t/\mathrm{t}}dt=\tau\langle v_{\mathrm{drift}}^{2}\rangle. \label{eq11.3}
\ee 
The term $\delta D$ is an additional contribution to the $xx$-element of the unperturbed diffusion tensor $\mathbf{D}^{0}$ given by $D_{xx}^{0}=D_{yy}^{0}=D_{0}$, $D_{xy}^{0}=-D_{xy}^{0}=-\omega_{\mathrm{c}}\tau D_{0}$, with $D_{0} \equiv\frac{1}{2}\tau v_{\mathrm{F}}^{2}[1+(\omega_{\mathrm{c}}\tau)^{2}]^{-1}$ (cf.\ Section \ref{sec4c}). At this point we
assume that for $l_{\mathrm{cycl}}\ll l$ the contribution $\delta D$ from the guiding center drift is
the dominant effect of the potential grating on the diffusion tensor ${\mathbf D}$. A
justification of this assumption requires a more systematic analysis of the
transport problem, which is given in Ref.\ \onlinecite{ref111}. Once $\mathbf{D}$ is known, the resistivity
tensor $\mathbf{\rho}$ follows from the Einstein relation $\mathbf{\rho}=\mathbf{D}^{-1}/e^{2}\rho(E_{\mathrm{F}})$, with $\rho(E_{\mathrm{F}})$ the
2D density of states (cf.\ Section \ref{sec4b}). Because of the large off-diagonal
components of $\mathbf{D}^{0}$, an additional contribution $\delta D$ to $D_{xx}^{0}$ modifies predominantly $\rho_{yy}\equiv\rho_{\perp}$. To leading order in $\epsilon$, one finds that
\be
\frac{\rho_{\perp}}{\rho_{0}}=1+2\epsilon^{2}\left(\frac{l^{2}}{al_{\mathrm{cycl}}}\right)\cos^{2}\left(2\pi\frac{l_{\mathrm{cycl}}}{a}-\frac{\pi}{4}\right),   \label{eq11.4}
\ee
with $\rho_{0}=m/n_{\mathrm{s}}e^{2}\tau$ the unperturbed resistivity. A rigorous solution\cite{ref111} of the
Boltzmann equation (for a $B$-independent scattering time) confirms this
simple result in the regime $a\ll l_{\mathrm{cycl}}\ll l$ and is shown in Fig.\ \ref{fig37} to be in quite
good agreement with the experimental data of Weiss et al.\cite{ref255} Similar
theoretical results have been obtained by Gerhardts et al.\cite{ref257} and by Winkler
et al.\cite{ref258} (using an equivalent quantum mechanical formulation; see below).

As illustrated by the arrows in Fig.\ \ref{fig37}, the maxima in $\rho_{\perp}$ are not at integer
$2l_{\mathrm{cycl}}/a$, but shifted somewhat toward lower magnetic fields. This phase shift is
a consequence of the finite extension of the segment of the orbit around the
extremal points $Y\pm l_{\mathrm{cycl}}$, which contributes to the guiding center drift
$v_{\mathrm{drift}}(Y)$. Equation (\ref{eq11.4}) implies that $\rho_{\perp}$ in a sinusoidal potential grating has
minima and maxima at approximately
\begin{eqnarray}
&&2l_{\mathrm{cycl}}/a\;\; ({\rm minima})=n - {\textstyle\frac{1}{4}},\nonumber\\
&&2l_{\mathrm{cycl}}/a\;\; ({\rm maxima})=n+{\textstyle\frac{1}{4}} - {\rm order}(1/n), \label{eq11.5}
\end{eqnarray}
with $n$ an integer. We emphasize that the phase shift is not universal, but
depends on the functional form of $V(y)$. The fact that the experimental phase
shift in Fig.\ \ref{fig37} agrees so well with the theory indicates that the actual
potential grating in the experiment of Weiss et al.\ is well modeled by a
sinusoidal potential. The maxima in $\rho_{\perp}/\rho_{0}$ have amplitude $\epsilon^{2}(l^{2}/al_{\mathrm{cycl}})$, which
for a large mean free path $l$ can be of order unity, even if $\epsilon\ll 1$. The guiding-center-drift resonance thus explains the surprising experimental finding that a
periodic modulation of the Fermi velocity of order $10^{-2}$ can double the
resistivity.

At low magnetic fields the experimental oscillations are damped more
rapidly than the theory would predict, and, moreover, an unexplained
positive magnetoresistance is observed around zero field in $\rho_{\perp}$ (but not in $\rho_{||}$).
Part of this disagreement may be due to nonuniformities in the potential
grating, which become especially important at low fields when the cyclotron
orbit overlaps many modulation periods. At high magnetic fields $B\gtrsim 0.4\,\mathrm{T}$
the experimental data show the onset of Shubnikov-De Haas oscillations,
which are a consequence of oscillations in the scattering time $\tau$ due to Landau
level quantization (cf.\ Section \ref{sec4c}). This effect is neglected in the semiclassical
analysis, which assumes a constant scattering time.

The quantum mechanical $B$-dependence of $\tau$ also leads to weak-field
oscillations in $\rho_{||}$, with the same periodicity as the oscillations in $\rho_{\perp}$ discussed
earlier, but of much smaller amplitude and shifted in phase (see Fig.\ \ref{fig37}, where
a maximum in the experimental $\sigma_{||}$ around $0.3\, \mathrm{T}$ lines up with a minimum in
$\rho_{\perp})$. These small antiphase oscillations in $\rho_{||}$ were explained by Vasilopoulos
and Peeters\cite{ref227} and by Gerhardts and Zhang\cite{ref259} as resulting from oscillations
in $\tau$ due to the oscillatory Landau bandwidth. The Landau levels
$E_{n}=(n-\frac{1}{2})\hbar\omega_{\mathrm{c}}$ broaden into a band of finite width in a periodic potential.\cite{ref261}
This Landau band is described by a dispersion law $E_{n}(k)$, where the wave
number $k$ is related to the classical orbit center $(X, Y)$ by $k=YeB/\hbar$ (cf.\ the
similar relation in Section \ref{sec12}). The classical guiding-center-drift resonance
can also be explained in these quantum mechanical terms, if one so desires, by
noticing that the bandwidth of the Landau levels is proportional to the root-mean-square average of $v_{\mathrm{drift}}=dE_{n}(k)/\hbar dk$. A maximal bandwidth thus
corresponds to a maximal guiding center drift and, hence, to a maximal $\rho_{\perp}$. A
maximum in the bandwidth also implies a minimum in the density of states at
the Fermi level and, hence, a maximum in $\tau$ [Eq.\ (\ref{eq4.28})]. A maximal
bandwidth thus corresponds to a minimal $\rho_{||}$, whereas the $B$-dependence of $\tau$
can safely by neglected for the oscillations in $\rho_{\perp}$ (which are dominated by the
classical guiding-center-drift resonance).

In a 2D periodic potential (a grid), the guiding center drift dominates the
magnetoresistivity in both diagonal components of the resistivity tensor.
Classically, the effect of a weak periodic potential $V(x, y)$ on $\rho_{xx}$ and $\rho_{yy}$
decouples if $V(x, y)$ is separable into $V(x, y)=f(x)+g(y)$. For the 2D
sinusoidal potential $V(x, y)\propto\sin(2\pi x/a)+\sin(2\pi y/b)$, one finds that the effect
of the grid is simply a superposition of the effects for two perpendicular
gratings of periods $a$ and $b$. (No such decoupling occurs quantum mechanically.\cite{ref254}) Experiments by Alves et al.\cite{ref262} and by Weiss et al.\cite{ref263} confirm this
expectation, except for a disagreement in the phase of the oscillations. As
noted, however, the phase is not universal but depends on the form of the
periodic potential, which need not be sinusoidal.

Because of the predominance of the classical guiding-center-drift resonance in a weak periodic potential, magnetotransport experiments are not
well suited to study miniband structure in the density of states. Magnetocapacitance measurements\cite{ref256,ref264,ref265} are a more direct means of investigation, but
somewhat outside the scope of this review.

\section{\label{secIII} Ballistic transport}

\subsection{\label{sec12} Conduction as a transmission problem}

In the ballistic transport regime, it is the scattering of electrons at the
sample boundaries which limits the current, rather than impurity scattering.
The canonical example of a ballistic conductor is the point contact illustrated
in Fig.\ \ref{fig7}c. The current $I$ through the narrow constriction in response to a
voltage difference $V$ between the wide regions to the left and right is {\it finite\/}
even in the absence of impurities, because electrons are scattered back at the
entrance of the constriction. The {\it contact conductance\/} $G=I/V$ is proportional
to the constriction width but independent of its length. One cannot therefore
describe the contact conductance in terms of a local conductivity, as one can
do in the diffusive transport regime. Consequently, the Einstein relation (\ref{eq4.10})
between the conductivity and the diffusion constant at the Fermi level, of
which we made use repeatedly in Section \ref{secII}, is not applicable in that form to
determine the contact conductance. The Landauer formula is an alternative
relation between the conductance and a Fermi level property of the sample,
without the restriction to diffusive transport. We discuss this formulation of
conduction in Section \ref{sec12b}. The Landauer formula expresses the conductance
in terms of transmission probabilities of propagating modes at the Fermi
level (also referred to as {\it quantum channels\/} in this context). Some elementary
properties of the modes are summarized in Section \ref{sec12a}.

\subsubsection{\label{sec12a} Electron waveguide}

We consider a conducting channel in a 2DEG (an ``electron waveguide''),
defined by a lateral confining potential $V(x)$, in the presence of a perpendicular magnetic field $B$ (in the $z$-direction). In the Landau gauge
$\mathbf{A}=(0, Bx, 0)$ the hamiltonian has the form
\be
{\cal H}=\frac{p_{x}^{2}}{2m}+\frac{(p_{y}+eBx)^{2}}{2m}+V(x) \label{eq12.1}
\ee
for a single spin component (cf.\ Section \ref{sec10a}). Because the canonical
momentum $p_{y}$ along the channel commutes with ${\cal H}$, one can diagonalize $p_{y}$
and ${\cal H}$ simultaneously. For each eigenvalue $\hbar k$ of $p_{y}$, the hamiltonian (\ref{eq12.1})
has a discrete spectrum of energy eigenvalues $E_{n}(k), n=1,2, \ldots$, corresponding to eigenfunctions of the form
\be
|n, k\rangle=\Psi_{n,k}(x)\mathrm{e}^{iky}. \label{eq12.2}
\ee 
In waveguide terminology, the index $n$ labels the modes, and the dependence
of the energy (or ``frequency'') $E_{n}(k)$ on the wave number $k$ is the dispersion
relation of the $n$th mode. A propagating mode at the Fermi level has cutoff
frequency $E_{n}(0)$ below $E_{\mathrm{F}}$. The wave function (\ref{eq12.2}) is the product of a
transverse amplitude profile $\Psi_{n,k}(x)$ and a longitudinal plane wave $e^{iky}$. The
average velocity $v_{n}(k)$ along the channel in state $|n, k\rangle$ is the expectation value
of the $y$-component of the velocity operator $\mathbf{p}+e\mathbf{A}$:
\begin{eqnarray}
v_{n}(k)&\equiv&\langle n, k| \frac{p_{y}+eA_{y}}{m}|n, k\rangle\nonumber\\
&=&\langle n, k| \frac{\partial{\cal H}}{\partial p_{y}}|n, k \rangle=\frac{dE_{n}(k)}{\hbar dk}. \label{eq12.3}
\end{eqnarray}
For a zero magnetic field, the dispersion relation $E_{n}(k)$ has the simple form
(\ref{eq4.3}). The {\it group velocity\/} $v_{n}(k)$ is then simply equal to the velocity $\hbar k/m$
obtained from the canonical momentum. This equality no longer holds in the
presence of a magnetic field, because the canonical momentum contains an
extra contribution from the vector potential. The dispersion relation in a
nonzero magnetic field was derived in Section \ref{sec10a} for a parabolic confining
potential $V(x)= \frac{1}{2}m\omega_{0}^{2}x^{2}$. From Eq.\ (\ref{eq10.5}) one calculates a group velocity
$\hbar k/M$ that is smaller than $\hbar k/m$ by a factor of $1+(\omega_{\mathrm{c}}/\omega_{0})^{2}$.

\begin{figure}
\centerline{\includegraphics[width=8cm]{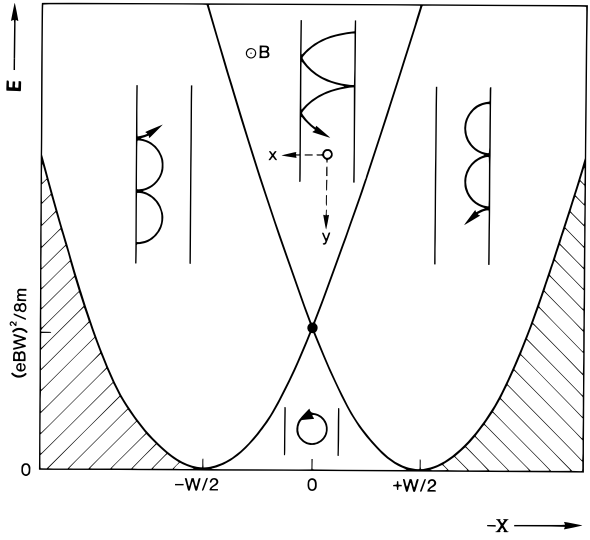}}
\caption{
Energy-orbit center phase space. The two parabolas divide the space into four regions, which correspond to different types of classical trajectories in a magnetic field (clockwise from left: skipping orbits on one edge, traversing trajectories, skipping orbits on the other edge, and cyclotron orbits). The shaded region is forbidden. The region at the upper center contains traversing trajectories moving in both directions, but only one region is shown for clarity. Taken from C. W. J. Beenakker et al., Superlattices and Microstructures {\bf 5}, 127 (1989).
\label{fig39}
}
\end{figure}

Insight into the nature of the wave functions in a magnetic field can be
obtained from the correspondence with classical trajectories. These are most
easily visualized in a square-well confining potential, as we now discuss
(following Ref.\ \onlinecite{ref266}). The position $(x, y)$ of an electron on the circle with center
coordinates $(X, Y)$ can be expressed in terms of its velocity $ \mathbf{v}$ by
\be
x=X+v_{y}/\omega_{\mathrm{c}},\;\; y=Y-v_{x}/\omega_{\mathrm{c}}, \label{eq12.4}
\ee 
with $\omega_{\mathrm{c}}\equiv eB/m$ the cyclotron frequency. The cyclotron radius is $(2mE)^{1/2}/eB$,
with $E \equiv\frac{1}{2}mv^{2}$ the energy of the electron. Both the energy $E$ and the
separation $X$ of the orbit center from the center of the channel are constants
of the motion. The coordinate $Y$ of the orbit center parallel to the channel
walls changes on each specular reflection. One can classify a trajectory as a
cyclotron orbit, skipping orbit, or traversing trajectory, depending on
whether the trajectory collides with zero, one, or both channel walls. In $(X, E)$
space these three types of trajectories are separated by the two parabolas
$(X\pm W/2)^{2}=2mE(eB)^{-2}$ (Fig.\ \ref{fig39}). The quantum mechanical dispersion
relation $E_{n}(k)$ can be drawn into this classical ``phase diagram'' because of the
correspondence $k=-XeB/h$.This correspondence exists because both $k$ and
$X$ are constants of the motion and it follows from the fact that the component
$\hbar k$ along the channel of the canonical momentum $\mathbf{p}=m \mathbf{v}-e\mathbf{A}$ equals
\be
\hbar k=mv_{y}-eA_{y}=mv_{y}-eBx=-eBx \label{eq12.5}
\ee
in the Landau gauge.

\begin{figure}
\centerline{\includegraphics[width=6cm]{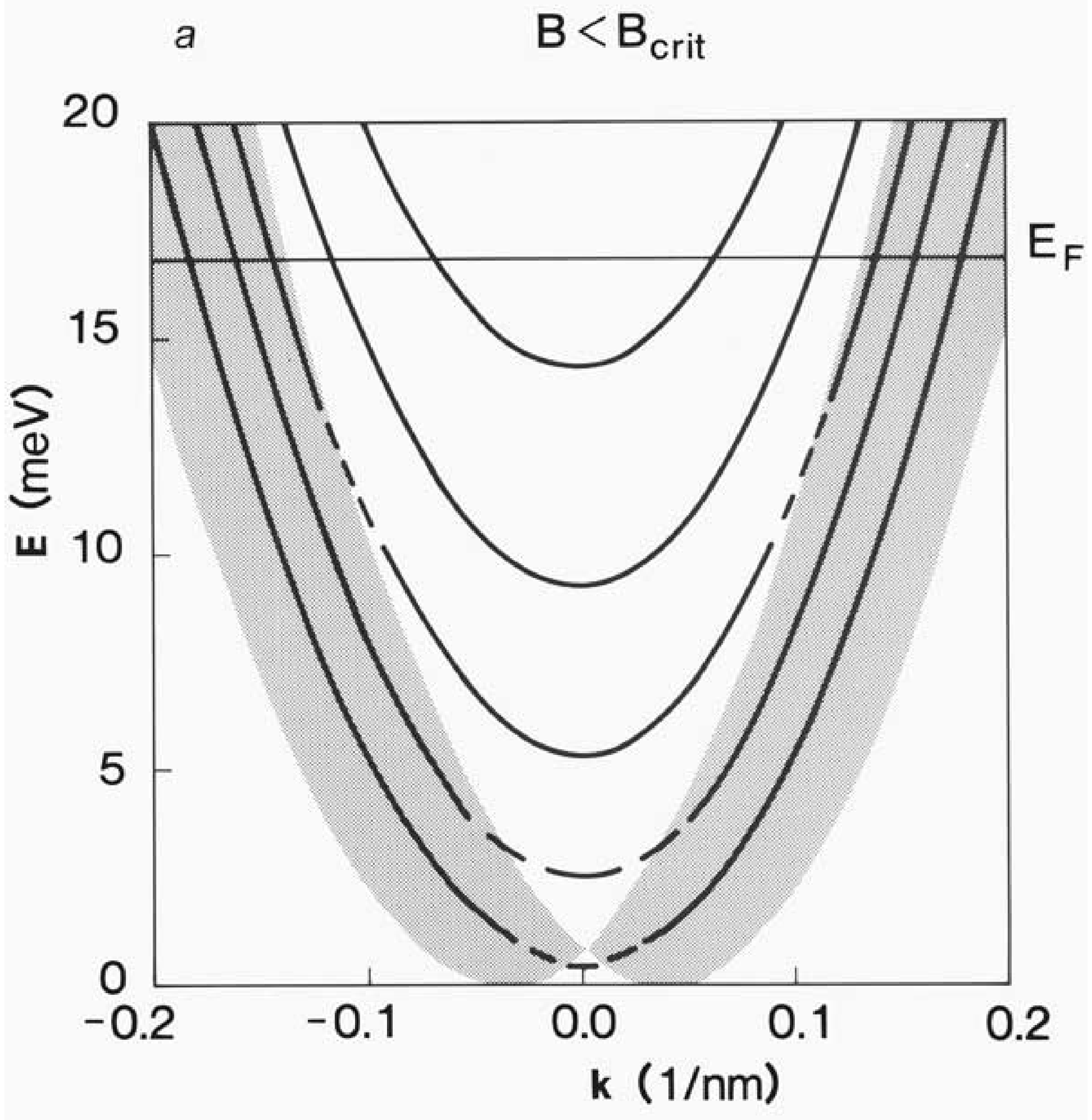}}

\centerline{\includegraphics[width=6cm]{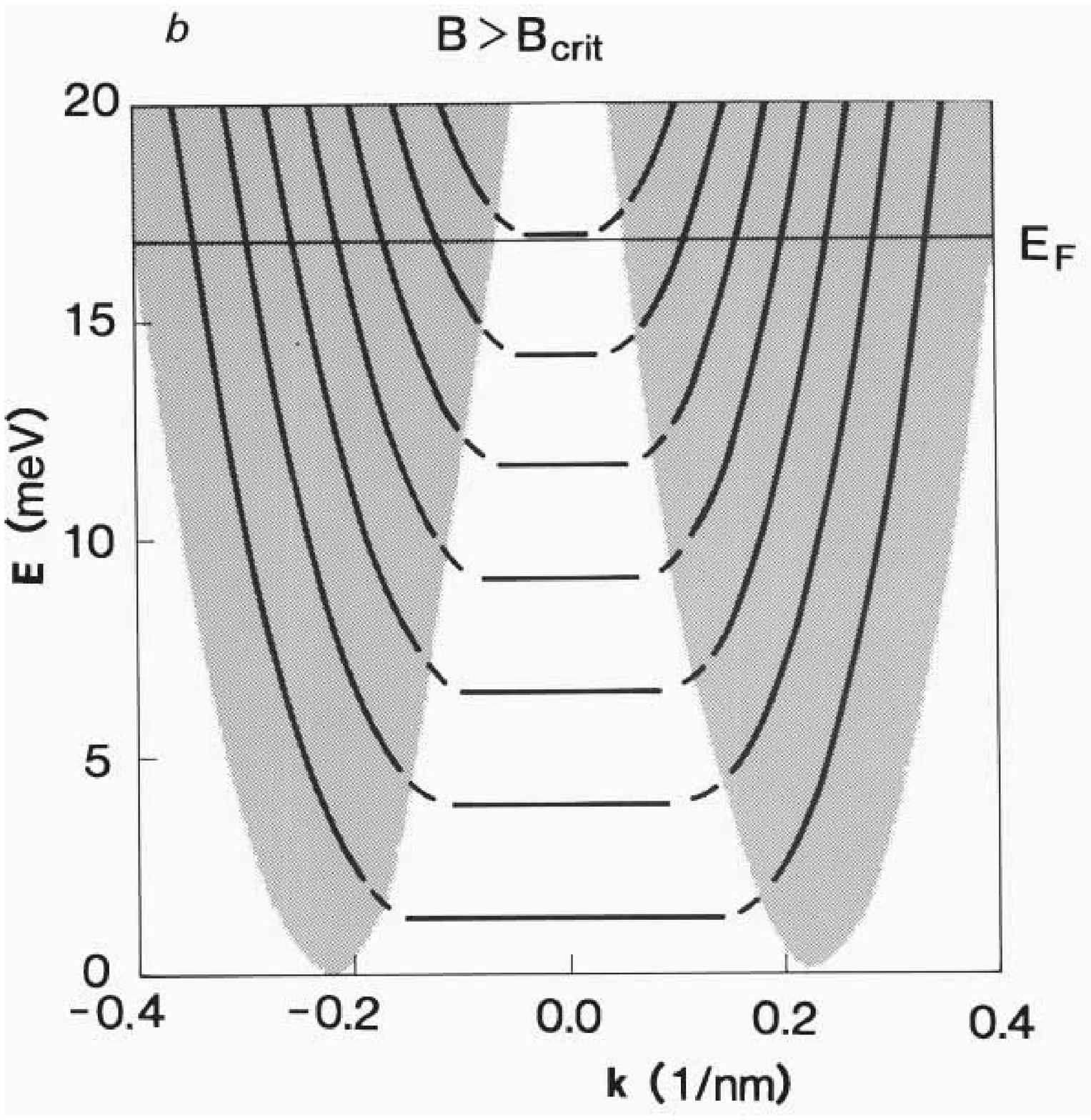}}
\caption{
Dispersion relation $E_{n}(k)$, calculated for parameters: (a) $W = 100\,{\rm nm}$, $B = 1\,{\rm T}$; (b) $W = 200\,{\rm nm}$, $B = 1.5\,{\rm T}$. The horizontal line at 17 meV indicates the Fermi energy. The shaded area is the region of classical skipping orbits and is bounded by the two parabolas shown in Fig.\ \ref{fig39} (with the correspondence $k = - X eB/\hbar$). Note that in (a) edge states coexist at the Fermi level with states interacting with both boundaries ($B < B_{\rm crit} \equiv 2\hbar k_{\rm F}/eB$), while in (b) all states at the Fermi level interact with one boundary only ($B > B_{\rm crit}$). Taken from C. W. J. Beenakker et al., Superlattices and Microstructures {\bf 5}, 127 (1989).
\label{fig40}
}
\end{figure}

In Fig.\ \ref{fig40} we show $E_{n}(k)$ both in weak and in strong magnetic fields,
calculated\cite{ref266} for typical parameter values from the Bohr-Sommerfeld
quantization rule discussed here. The regions in phase space occupied by
classical skipping orbits are shaded. The unshaded regions contain cyclotron
orbits (at small $E$) and traversing trajectories (at larger $E$) (cf.\ Fig.\ \ref{fig39}). The
{\it cyclotron orbits\/} correspond quantum mechanically to states in {\it Landau levels}.
These are the flat portions of the dispersion relation at energies
$E_{n}=(n- \frac{1}{2})\hbar\omega_{\mathrm{c}}$. The group velocity (\ref{eq12.3}) is zero in a Landau level, as one
would expect from the correspondence with a circular orbit. The {\it traversing trajectories\/} correspond to states in {\it magnetoelectric subbands}, which interact
with both the opposite channel boundaries and have a nonzero group
velocity. The {\it skipping orbits\/} correspond to {\it edge states}, which interact with a
single boundary only. The two sets of edge states (one for each boundary) are
disjunct in $(k, E)$ space. Edge states at opposite boundaries move in opposite
directions, as is evident from the correspondence with skipping orbits or by
inspection of the slope of $E_{n}(k)$ in the two shaded regions in Fig.\ \ref{fig40}.

If the Fermi level lies between two Landau levels, the states at the Fermi
level consist only of edge states if $B>B_{\mathrm{crit}}$, as in Fig.\ \ref{fig40}b. The ``critical'' field
$B_{\mathrm{crit}}=2\hbar k_{\rm F}/eW$ is obtained from the classical correspondence by requiring
that the channel width $W$ should be larger than the cyclotron diameter
$2\hbar k_{\rm F}/eB$ at the Fermi level. This is the same characteristic field that played a
role in the discussion of magneto size effects in Sections \ref{sec5} and \ref{sec10}. At fields
$B<B_{\mathrm{crit}}$, as in Fig.\ \ref{fig40}a, edge states coexist at the Fermi level with
magnetoelectric subbands. In still lower fields $B<B_{\mathrm{thres}}$ {\it all\/} states at the
Fermi level interact with both edges. The criterion for this is that $W$ should be
smaller than the transverse wavelength\cite{ref267} $\lambda_{\mathrm{t}}=(\hbar/2k_{\mathrm{F}}eB)^{1/3}$ of the edge
states, so the threshold field $B_{\mathrm{thres}}\sim \hbar/ek_{\mathrm{F}}W^{3}$. Contrary to initial expectations,\cite{ref268} this lower characteristic field does not appear to play a decisive role
in magneto size effects.

\begin{figure}
\centerline{\includegraphics[width=6cm]{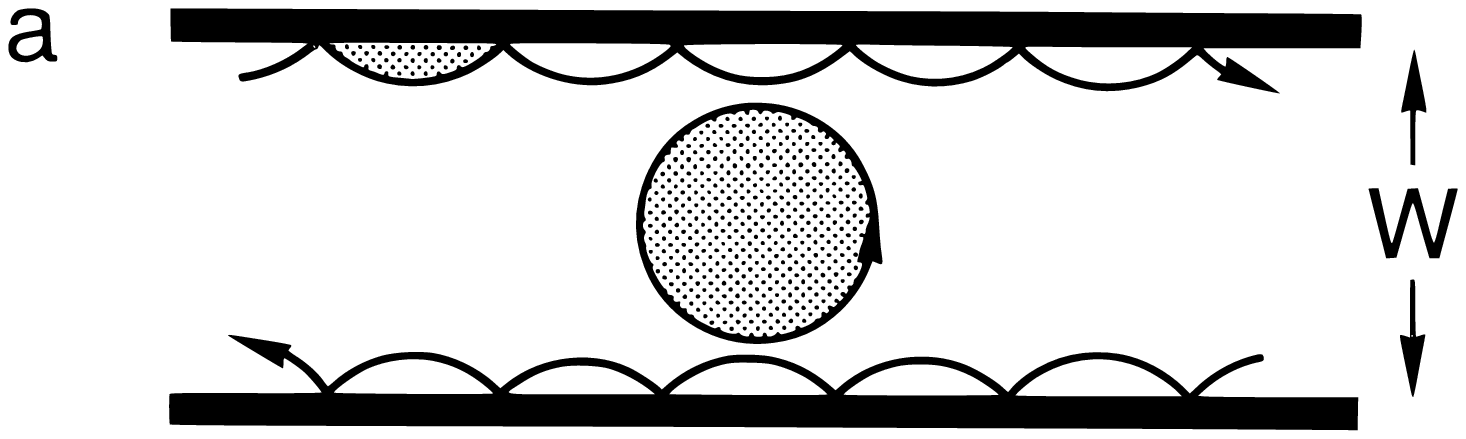}}

\centerline{\includegraphics[width=6cm]{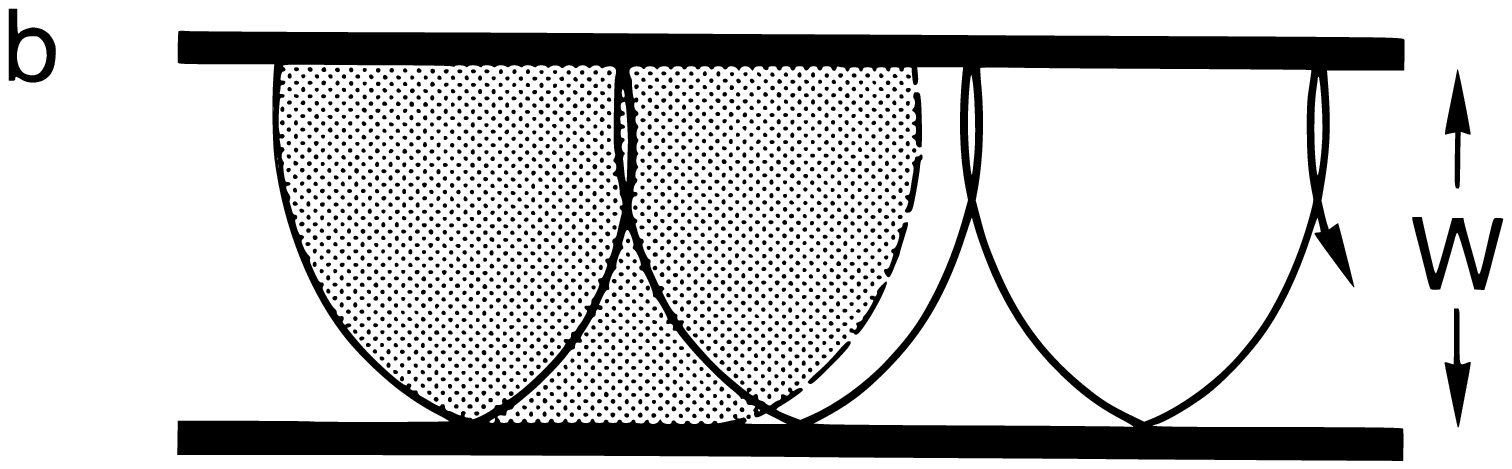}}
\caption{
Classical trajectories in a magnetic field. The flux through the shaded area is quantized according to the Bohr-Sommerfeld quantization rule (\ref{eq12.7}). The shaded area in (b) is bounded by the channel walls and the circle formed by the continuation (dashed) of one circular arc of the traversing trajectory.
\label{fig41}
}
\end{figure}

A quick way to arrive at the dispersion relation $E_{n}(k)$, which is sufficiently
accurate for our purposes, is to apply the Bohr-Sommerfeld quantization
rule\cite{ref80,ref269} to the classical motion in the $x$-direction:
\be
\frac{1}{\hbar}\oint p_{x}dx+\gamma=2\pi n,\;\;n=1,2, \ldots. \label{eq12.6}
\ee 
The integral is over one period of the motion. The phase shift $\gamma$ is the sum of
the phase shifts acquired at the two turning points of the projection of the
motion on the $x$-axis. The phase shift upon reflection at the boundary is $\pi$, to
ensure that incident and reflected waves cancel (we consider an infinite
barrier potential at which the wave function vanishes). The other turning
points (at which $v_{x}$ varies smoothly) have a phase shift of $-\pi/2$.\cite{ref93} Consequently, for a traversing trajectory $\gamma=\pi+\pi=0$ $(\mathrm{mod}\,2\pi)$, for a skipping
orbit $\gamma=\pi-\pi/2=\pi/2$, and for a cyclotron orbit $\gamma=-\pi/2-\pi/2=\pi$
$(\mathrm{mod}\,2\pi)$. In the Landau gauge one has $p_{x}=mv_{x}=eB(Y-y)$, so Eq.\ (\ref{eq12.6})
takes the form
\be
B \oint(Y-y)dx=\frac{h}{e}(n-\frac{\gamma}{2\pi}). \label{eq12.7}
\ee 
This quantization condition has the appealing geometrical interpretation
that $n-\gamma/2\pi$ flux quanta $h/e$ are contained in the area bounded by the
channel walls and a circle of cyclotron radius $(2mE)^{1/2}/eB$ centered at $X$ (cf.\
Fig.\ \ref{fig41}). It is now straightforward to find for each integer $n$ and coordinate $X$
the energy $E$ that satisfies this condition. The dispersion relation $E_{n}(k)$ then
follows on identifying $k=-XeB/h$, as shown in Fig.\ \ref{fig40}.

The total number $N$ of propagating modes at energy $E$ is determined by
the maximum flux $\Phi_{\rm max}$ contained in an area bounded by the channel walls
and a circle of radius $(2mE)^{1/2}/eB$, according to $N=\mathrm{Int}[e\Phi_{\rm max}/h+\gamma/2\pi]$.
Note that a maximal enclosed flux is obtained by centering the circle on the
channel axis. Some simple geometry then leads to the result\cite{ref80} (\ref{eq10.8}), which is
plotted together with that for a parabolic confinement in Fig.\ \ref{fig31}. Equation
(\ref{eq10.8}) has a discontinuity at magnetic fields for which the cyclotron diameter
equals the channel width, due to the jump in the phase shift $\gamma$ as one goes
from a cyclotron orbit to a traversing trajectory. This jump is an artifact of
the present semiclassical approximation in which the extension of the wave
function beyond the classical orbit is ignored. Since the discontinuity in $N$ is
at most $\pm 1$, it is unimportant in many applications. More accurate formulas
for the phase shift $\gamma$, which smooth out the discontinuity, have been derived in
Ref.\ \onlinecite{ref270}. If necessary, one can also use more complicated but exact solutions
of the Schr\"{o}dinger equation, which are known.\cite{ref267}

\subsubsection{\label{sec12b} Landauer formula}

\begin{figure}
\centerline{\includegraphics[width=6cm]{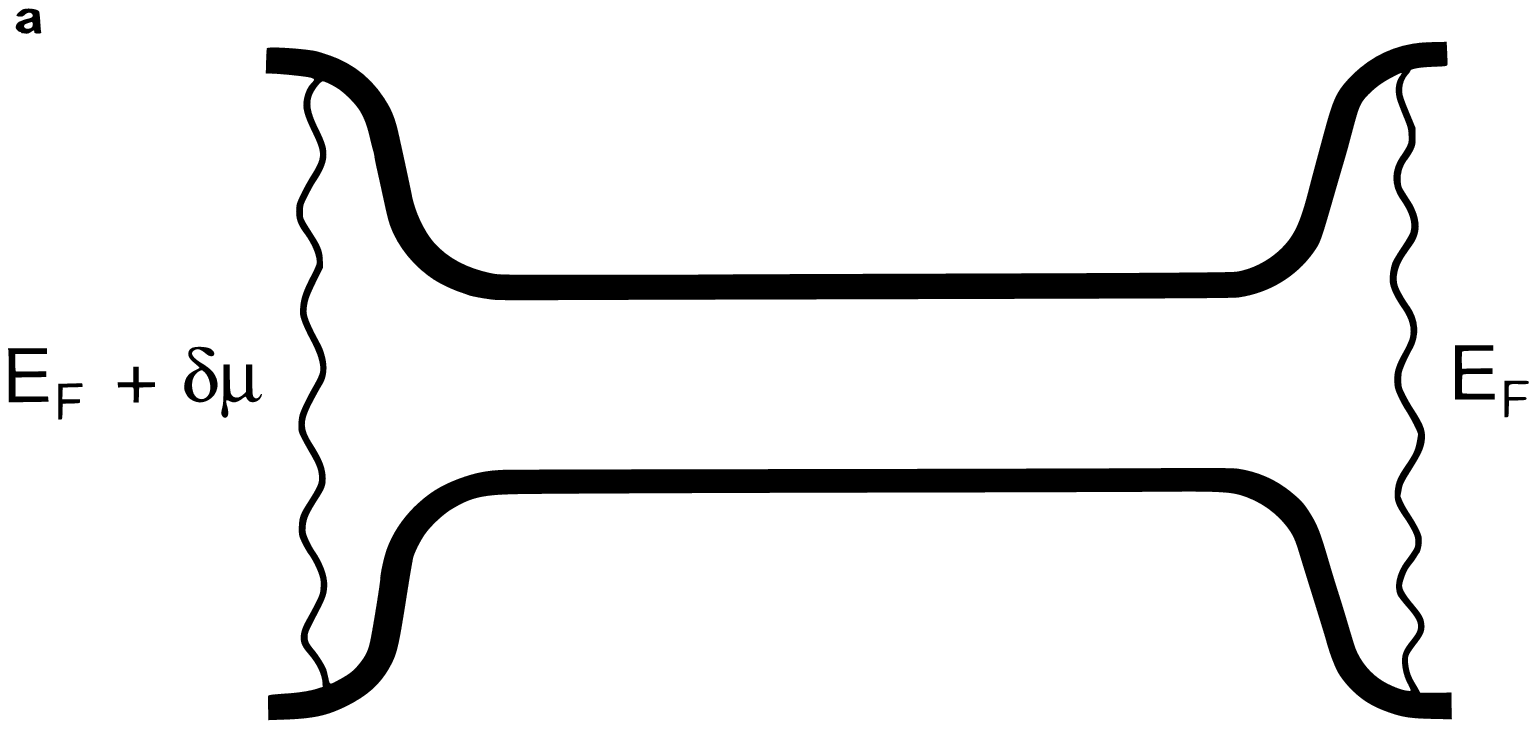}}

\centerline{\includegraphics[width=6cm]{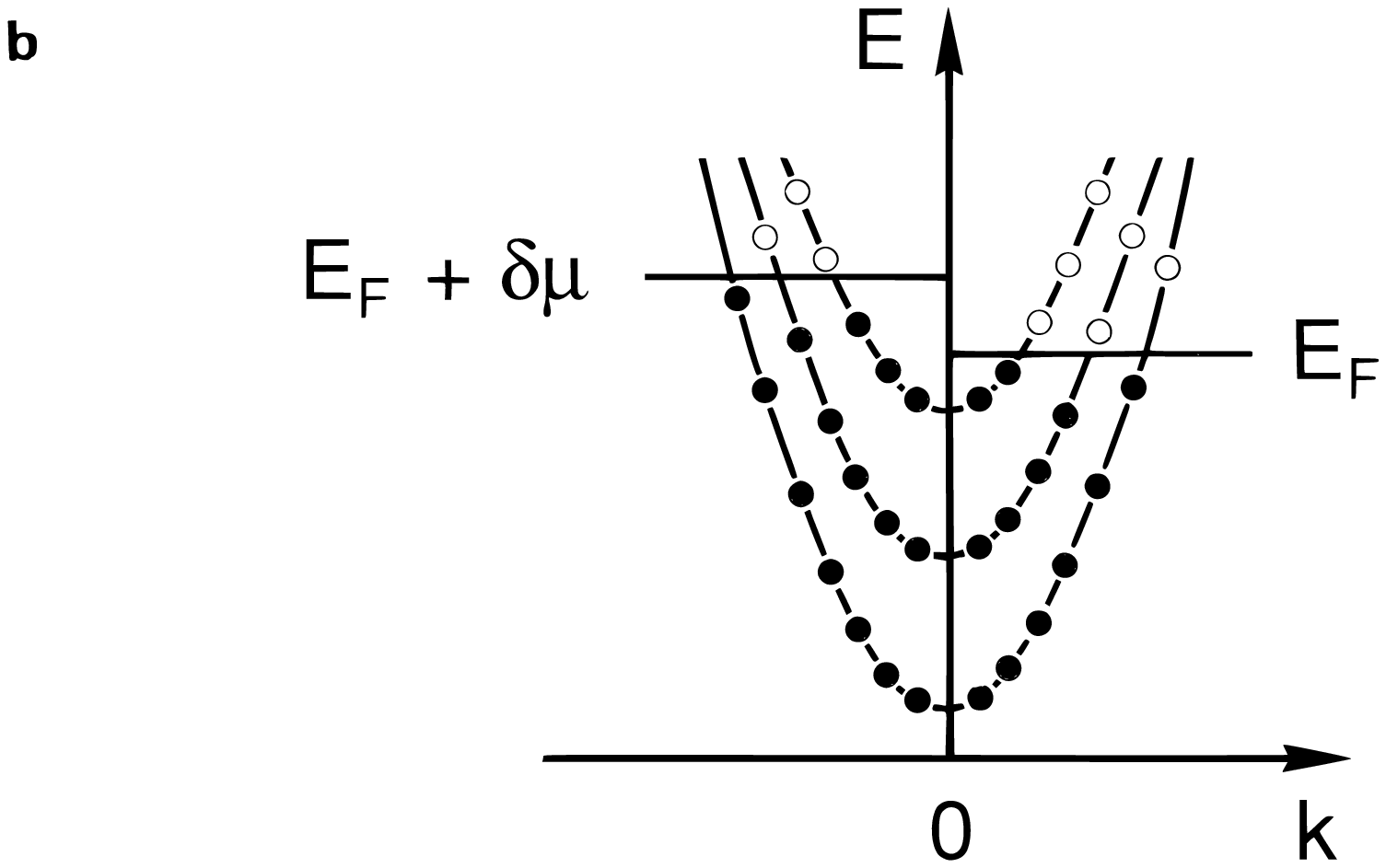}}
\caption{
(a) Narrow channel connecting two wide electron gas regions, having a chemical potential difference $\delta\mu$. (b) Schematic dispersion relation in the narrow channel. Left-moving states ($k > 0$) are filled up to chemical potential $E_{\rm F}$, right-moving states up to $E_{\rm F} + \delta\mu$ (solid dots). Higher-lying states are unoccupied (open dots).
\label{fig42}
}
\end{figure}

Imagine two wide electron gas reservoirs having a slight difference $\delta n$ in
electron density, which are brought into contact by means of a narrow
channel, as in Fig.\ \ref{fig42}a. A diffusion current $J$ will flow in the channel, carried
by electrons with energies between the Fermi energies $E_{\mathrm{F}}$ and $E_{\mathrm{F}}+\delta\mu$ in the
low- and high-density regions. For small $\delta n$, one can write for the Fermi
energy difference (or chemical potential difference) $\delta\mu=\delta n/\rho(E_{\mathrm{F}})$, with $\rho(E_{\mathrm{F}})$
the density of states at $E_{\mathrm{F}}$ in the reservoir (cf.\ Section \ref{sec4a}). The diffusion
constant (or ``diffusance'')\cite{ref4} $\tilde{D}$ is defined by $J=\tilde{D}\delta n$ and is related to the
conductance $G$ by
\be
G=e^{2}\rho(E_{\mathrm{F}})\tilde{D}. \label{eq12.8}
\ee 
Equation (\ref{eq12.8}) generalizes the Einstein relation (\ref{eq4.10}) and is derived in a
completely analogous way [by requiring that the sum of drift current $GV/e$
and diffusion current $\tilde{D}\delta n$ be zero when the sum of the electrostatic potential
difference $eV$ and chemical potential difference $\delta n/\rho(E_{\mathrm{F}})$ vanishes].

Since the diffusion current (at low temperatures) is carried by electrons
within a narrow range $\delta\mu$ above $E_{\mathrm{F}}$, the diffusance can be expressed in terms
of Fermi level properties of the channel (see below). The Einstein relation
(\ref{eq12.8}) then yields the required Fermi level expression of the conductance. This
by no means implies that the drift current induced by an electrostatic
potential difference is carried entirely by electrons at the Fermi energy. To the
contrary, all electrons (regardless of their energy) acquire a nonzero drift
velocity in an electric field. This point has been the cause of some confusion in
the literature on the quantum Hall effect, so we will return to it in Section
\ref{sec18c}. In the following we will refer to electrons at the Fermi energy as
``current-carrying electrons'' and show that ``the current in the channel is
shared equally among the modes at the Fermi level.'' These and similar
statements should be interpreted as referring to the diffusion problem, where
the current is induced by density differences without an electric field. We
make no attempt here to evaluate the distribution of current in response to an
electric field in a system of uniform density. That is a difficult problem, for
which one has to determine the electric field distribution self-consistently
from Poisson's and Boltzmann's equations. Such a calculation for a quantum
point contact has been performed in Refs.\ \onlinecite{ref271} and \onlinecite{ref272}. Fortunately, there is
no need to know the actual current distribution to determine the conductance, in view of the Einstein relation (\ref{eq12.8}). The distribution of current
(and electric field) is of importance only beyond the regime of a linear relation
between current and voltage. We will not venture beyond this linear response
regime.

To calculate the diffusance, we first consider the case of an ideal electron
waveguide between the two reservoirs. By ``ideal'' it is meant that within the
waveguide the states with group velocity pointing to the right are occupied
up to $E_{\mathrm{F}}+\delta\mu$, while states with group velocity to the left are occupied up to
$E_{\mathrm{F}}$ and empty above that energy (cf.\ Fig.\ \ref{fig42}b). This requires that an electron
near the Fermi energy that is injected into the waveguide by the reservoir at
$E_{\mathrm{F}}+\delta\mu$ propagates into the other reservoir without being reflected. (The
physical requirements for this to happen will be discussed in Section \ref{sec13}.) The
amount of diffusion current per energy interval carried by the right-moving
states (with $k<0$) in a mode $n$ is the product of density of states $\rho_{n}^{-}$ and group
velocity $v_{n}$. Using  Eqs.\ (\ref{eq4.4}) and (\ref{eq12.3}), we find the total current $J_{n}$ carried by
that mode to be
\begin{eqnarray}
J_{n}&=& \int_{E_{\rm F}}^{E_{\rm F}+\delta\mu}g_{\mathrm{s}}g_{\mathrm{v}}\left(2\pi\frac{dE_{n}(k)}{dk}\right)^{-1}\frac{dE_{n}(k)}{\hbar dk}\nonumber\\
&=&\frac{g_{\mathrm{s}}g_{\mathrm{v}}}{h}\delta\mu, \label{eq12.9}
\end{eqnarray}
independent of mode index and Fermi energy. The current in the channel is
shared equally among the $N$ modes at the Fermi level because of the
cancellation of group velocity and density states. We will return to this
{\it equipartition rule\/} in Section \ref{sec13}, because it is at the origin of the quantization\cite{ref6,ref7} of the conductance of a point contact.

Scattering within the narrow channel may reflect part of the injected
current back into the left reservoir. If a fraction $T_{n}$ of $J_{n}$ is transmitted to the
reservoir at the right, then the total diffusion current in the channel becomes
$J=(2/h) \delta\mu\sum_{n=1}^{N}T_{n}$. (Unless stated otherwise, the formulas in the remainder
of this review refer to the case $g_{\mathrm{s}}=2, g_{\mathrm{v}}=1$ of twofold spin degeneracy and a
single valley, appropriate for most of the experiments.) Using $\delta\mu=\delta n/\rho(E_{F})$,
$J=\tilde{D}\delta n$, and the Einstein relation (\ref{eq12.8}), one obtains the result
\begin{subequations}
\label{eq12.10}
\be
G= \frac{2e^{2}}{h}\sum_{n=1}^{N}T_{n}, \label{eq12.10a}
\ee
which can also be written in the form
\be
G= \frac{2e^{2}}{h}\sum_{n,m=1}^{N}|t_{mn}|^{2}\equiv\frac{2e^{2}}{h}\mathrm{Tr}\,{\bf tt}^{\dagger}, \label{eq12.10b}
\ee
\end{subequations}
where $T_{n}= \sum_{m=1}^{N}|t_{mn}|^{2}$ is expressed in terms of the matrix $\mathbf{t}$ of transmission
probability amplitudes from mode $n$ to mode $m$. This relation between
conductance and transmission probabilities at the Fermi energy is referred to
as the {\it Landauer formula\/} because of Landauer's pioneering 1957 paper.\cite{ref4}
Derivations of Eq.\ (\ref{eq12.10}) based on the Kubo formula of linear response
theory have been given by several authors, both for zero\cite{ref143,ref273,ref274} and
nonzero\cite{ref275,ref276} magnetic fields. The identification of $G$ as a {\it contact\/} conductance is due to Imry.\cite{ref1} In earlier work Eq.\ (\ref{eq12.10}) was considered
suspect\cite{ref277,ref228,ref279} because it gives a {\it finite\/} conductance for an ideal (ballistic)
conductor, and alternative expressions were proposed\cite{ref188,ref280,ref281,ref282} that were
considered to be more realistic. (In one dimension these alternative formulas
reduce to the original Landauer formula\cite{ref4} $G=(e^{2}/h)T(1-T)^{-1}$, which gives
infinite conductance for unit transmission since the contact conductance $e^{2}/h$
is ignored.\cite{ref1}) For historical accounts of this controversy, from two different
points of view, we refer the interested reader to papers by Landauer\cite{ref283} and
by Stone and Szafer.\cite{ref274} We have briefly mentioned this now-settled controversy, because it sheds some light onto why the quantization of the contact
conductance had not been predicted theoretically prior to its experimental
discovery in 1987.

\begin{figure}
\centerline{\includegraphics[width=8cm]{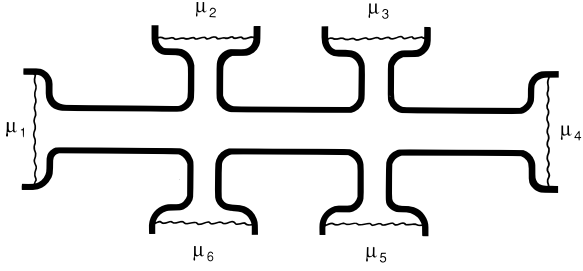}}
\caption{
Generalization of the geometry of Fig.\ \ref{fig42}a to a multireservoir conductor.
\label{fig43}
}
\end{figure}

Equation (\ref{eq12.10}) refers to a {\it two-terminal\/} resistance measurement, in which
the same two contacts (modeled by reservoirs in Fig.\ \ref{fig42}a) are used to drive a
current through the system and to measure the voltage drop. More generally,
one can consider a multireservoir conductor as in Fig.\ \ref{fig43} to model, for
example, {\it four-terminal\/} resistance measurements in which the current source
and drain are distinct from the voltage probes. The generalization of the
Landauer formula (\ref{eq12.10}) to multiterminal resistances is due to B\"{u}ttiker.\cite{ref5}  Let
$T_{\alpha\rightarrow\beta}$ be the total transmission probability from reservoir $\alpha$ to $\beta$:
\be
T_{\alpha\rightarrow\beta}= \sum_{n=1}^{N_{\alpha}}\sum_{m=1}^{N_{\beta}}|t_{\beta\alpha,mn}|^{2}. \label{eq12.11}
\ee 
Here $N_{\alpha}$ is the number of propagating modes in the channel (or ``lead'')
connected to reservoir $\alpha$ (which in general may be different from the number
$N_{\beta}$ in lead $\beta$), and $t_{\beta\alpha,mn}$ is the transmission probability amplitude from mode
$n$ in lead $\alpha$ to mode $m$ in lead $\beta$. The leads are modeled by ideal electron
waveguides, in the sense discussed before, so that the reservoir $\alpha$ at chemical
potential $\mu_{\alpha}$ above $E_{\mathrm{F}}$ injects into lead $\alpha$ a (charge) current $(2e/h)N_{\alpha}\mu_{\alpha}$. A
fraction $T_{\alpha\rightarrow\beta}/N_{\alpha}$ of that current is transmitted to reservoir $\beta$, and a fraction
$T_{\alpha\rightarrow a}/N_{\alpha}\equiv R_{\alpha}/N_{\alpha}$ is reflected back into reservoir $\alpha$, before reaching one of the
other reservoirs. The net current $I_{\alpha}$ in lead $\alpha$ is thus given by\cite{ref5}
\be
\frac{h}{2e}I_{\alpha}=(N_{\alpha}-R_{\alpha})\mu_{\alpha}-\sum_{\beta(\beta\neq \alpha)}T_{\rho\rightarrow \alpha}\mu_{\beta}. \label{eq12.12}
\ee 
The chemical potentials of the reservoirs are related to the currents in the
leads via a matrix of transmission and reflection coefficients. The sums of
columns or rows of this matrix vanish:
\begin{eqnarray}
&&N_{\alpha}-R_{\alpha}- \sum_{\beta(\beta\neq \alpha)}T_{\alpha\rightarrow\beta}=0, \label{eq12.13}\\ 
&&N_{\alpha}-R_{\alpha}- \sum_{\beta(\beta\neq \alpha)}T_{\beta\rightarrow \alpha}=0. \label{eq12.14}
\end{eqnarray}
Equation (\ref{eq12.13}) follows from current conservation, and Eq.\ (\ref{eq12.14}) follows
from the requirement that an increase of all the chemical potentials by the
same amount should have no effect on the net currents in the leads.
Equation (\ref{eq12.12}) can be applied to a measurement of the four-terminal
resistance $R_{\alpha\beta,\gamma\delta}=V/I$, in which a current $I$ flows from contact $\alpha$ to $\beta$ and a
voltage difference $V$ is measured between contacts $\gamma$ and $\delta$. Setting
$I_{\alpha}=I=-I_{\beta}$, and $I_{\alpha^{\prime}}=0$ for $\alpha^{\prime}\neq \alpha, \beta$, one can solve the set of linear
equations (\ref{eq12.12}) to determine the chemical potential difference $\mu_{\gamma}-\mu_{\delta}$.
(Only the {\it differences\/} in chemical potentials can be obtained from the $n$
equations (\ref{eq12.12}), which are not independent in view of Eq.\ (\ref{eq12.14}). By fixing
one chemical potential at zero, one reduces the number of equations to $n-1$
independent ones.) The four-terminal resistance $R_{\alpha\beta,\gamma\delta}=(\mu_{\gamma}-\mu_{\delta})/eI$ is then
obtained as a rational function of the transmission and reflection probabilities. We will refer to this procedure as the {\it Landauer-B\"{u}ttiker formalism}. It
provides a unified description of the whole variety of electrical transport
experiments discussed in this review.

The transmission probabilities have the symmetry
\be
t_{\beta\alpha,nm}(B)=t_{\alpha\beta,mn}(-B)\Rightarrow T_{\alpha\rightarrow\beta}(B)=T_{\rho\rightarrow \alpha}(-B). \label{eq12.15}
\ee 
Equation (\ref{eq12.15}) follows by combining the unitarity of the scattering matrix
$\mathbf{t}^{\dagger}=\mathbf{t}^{-1}$, required by current conservation, with the symmetry
$\mathbf{t}^{*}(-B)=\mathbf{t}^{-1}(B)$, required by time-reversal invariance ($*$ and $\dagger$ denote
complex and Hermitian conjugation, respectively). As shown by B\"{u}ttiker,\cite{ref5,ref284} the symmetry (\ref{eq12.15}) of the coefficients in Eq.\ (\ref{eq12.12}) implies a
{\it reciprocity relation\/} for the four-terminal resistance:
\be
R_{\alpha\beta,\gamma\delta}(B)=R_{\gamma\delta,\alpha\beta}(-B). \label{eq12.16}
\ee 
The resistance is unchanged if current and voltage leads are interchanged
with simultaneous reversal of the magnetic field direction. A special case of
Eq.\ (\ref{eq12.16}) is that the two-terminal resistance $R_{\alpha\beta,\alpha\beta}$ is {\it even\/} in $B$. In the
diffusive transport regime, the reciprocity relation for the resistance follows
from the Onsager-Casimir relation\cite{ref285} $\mathbf{\rho}(B)=\mathbf{\rho}^{\mathrm{T}}(-\beta)$ for the resistivity
tensor ($\mathrm{T}$ denotes the transpose). Equation (\ref{eq12.16}) holds also in cases that the
concept of a local resistivity breaks down. One experimental demonstration\cite{ref80} of the validity of the reciprocity relation in the quantum ballistic
transport regime will be discussed in Section \ref{sec14}. Other demonstrations have
been given in Refs.\ \onlinecite{ref286,ref287,ref288,ref289}. We emphasize that strict reciprocity holds only
in the linear response limit of infinitesimally small currents and voltages.
Deviations from Eq.\ (\ref{eq12.16}) can occur experimentally\cite{ref290} due to nonlinearities
from quantum interference,\cite{ref146,ref291} which in the case of a long phase
coherence time $\tau_{\phi}$ persist down to very small voltages $V\gtrsim \hbar/e\tau_{\phi}$. Magnetic
impurities can be another source of deviations from reciprocity if the applied
magnetic field is not sufficiently strong to reverse the magnetic moments on
field reversal. The large $\pm B$ asymmetry of the two-terminal resistance of a
point contact reported in Ref.\ \onlinecite{ref292} has remained unexplained (see Section \ref{sec21}).

The scattering matrix $\mathbf{t}$ in Eq.\ (\ref{eq12.15}) describes {\it elastic\/} scattering only.
Inelastic scattering is assumed to take place exclusively in the reservoirs. That
is a reasonable approximation at temperatures that are sufficiently low that
the size of the conductor is smaller than the inelastic scattering length (or the
phase coherence length if quantum interference effects play a role). Reservoirs
thus play a dual role in the Landauer-B\"{u}ttiker formalism: On the one hand,
a reservoir is a model for a current or voltage contact; on the other hand, a
reservoir brings inelastic scattering into the system. The reciprocity relation
(\ref{eq12.16}) is unaffected by adding reservoirs to the system and is not restricted to
elastic scattering.\cite{ref284} More realistic methods to include inelastic scattering in
a distributed way throughout the system have been proposed, but are not yet
implemented in an actual calculation.\cite{ref293,ref294}

\subsection{\label{sec13} Quantum point contacts}

Many of the principal phenomena in ballistic transport are exhibited in
the cleanest and most extreme way by quantum point contacts. These are
short and narrow constrictions in a 2DEG, with a width of the order of the
Fermi wavelength.\cite{ref6,ref7,ref59} The conductance of quantum point contacts is
quantized in units of $2e^{2}/h$. This quantization is reminiscent of the quantization of the Hall conductance, but is measured in the absence of a magnetic
field. The zero-field conductance quantization and the smooth transition to
the quantum Hall effect on applying a magnetic field are essentially
consequences of the equipartition of current among an integer number of
propagating modes in the constriction, each mode carrying a current of $2e^{2}/h$
times the applied voltage $V$. Deviations from precise quantization result from
nonunit transmission probability of propagating modes and from nonzero
transmission probability of evanescent (nonpropagating) modes. Experiment
and theory in a zero magnetic field are reviewed in Section \ref{sec13a}. The effect of a
magnetic field is the subject of Section \ref{sec13b}, which deals with depopulation of
subbands and suppression of backscattering by a magnetic field, two
phenomena that form the basis for an understanding of magnetotransport in
semiconductor nanostructures.

\subsubsection{\label{sec13a} Conductance quantization}

\begin{figure}
\centerline{\includegraphics[width=8cm]{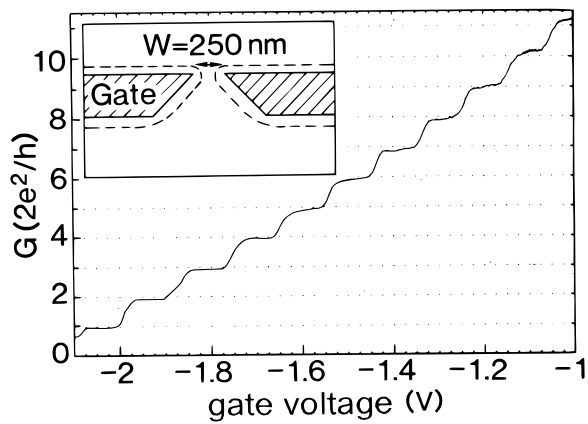}}
\caption{
Point contact conductance as a function of gate voltage at 0.6 K, demonstrating the conductance quantization in units of $2e^{2}/h$. The data are obtained from the two-terminal resistance after subtraction of a background resistance. The constriction width increases with increasing voltage on the gate (see inset). Taken from B. J. van Wees et al., Phys.\ Rev.\ Lett.\ {\bf 60}, 848 (1988).
\label{fig44}
}
\end{figure}

{\bf (a) Experiments.} The study of electron transport through point contacts in
metals has a long history, which goes back to Maxwell's investigations\cite{ref295} of
the resistance of an orifice in the diffusive transport regime. Ballistic transport
was first studied by Sharvin,\cite{ref296} who proposed and subsequently realized\cite{ref297}
the injection and detection of a beam of electrons in a metal by means of point
contacts much smaller than the mean free path. With the possible exception
of the scanning tunneling microscope, which can be seen as a point contact on
an atomic scale,\cite{ref298,ref299,ref300,ref301,ref302,ref303} these studies in metals are essentially restricted to the
{\it classical\/} ballistic transport regime because of the extremely small Fermi
wavelength $(\lambda_{\mathrm{F}}\approx 0.5\,\mathrm{nm})$. Point contacts in a 2DEG cannot be fabricated by
simply pressing two wedge- or needle-shaped pieces of material together (as
in bulk semiconductors\cite{ref304} or metals\cite{ref305}), since the electron gas is confined at
the GaAs-AlGaAs interface in the interior of the heterostructure. Instead,
they are defined electrostatically\cite{ref24,ref58} by means of a split gate on top of the
heterostructure (a schematical cross-sectional view was given in Fig.\ \ref{fig4}b, while
the micrograph Fig.\ \ref{fig5}b shows a top view of the split gate of a double-point
contact device; see also the inset in Fig.\ \ref{fig44}). In this way one can define short
and narrow constrictions in the 2DEG, of variable width $0\lesssim W\lesssim 250\,\mathrm{nm}$
comparable to the Fermi wavelength $\lambda_{\mathrm{F}}\approx 40\,\mathrm{nm}$ and much shorter than the
mean free path $l\approx 10\,\mu \mathrm{m}$.

Van Wees et al.\cite{ref6} and Wharam et al.\cite{ref7} independently discovered a sequence
of steps in the conductance of such a point contact as its width was varied by
means of the voltage on the split gate (see Fig.\ \ref{fig44}). The steps are near integer
multiples of $2e^{2}/h\approx(13\,k\Omega)^{-1}$, after correction for a gate-voltage-independent series resistance from the wide 2DEG regions. An elementary
explanation of this effect relies on the fact that each 1D subband in the
constriction contributes $2e^{2}/h$ to the conductance because of the cancellation
of the group velocity and the 1D density of states discussed in Section \ref{sec12}.
Since the number $N$ of occupied subbands is necessarily an integer, it follows
from this simple argument that the conductance $G$ is quantized,
\be
G=(2e^{2}/h)N, \label{eq13.1}
\ee 
as observed experimentally. A more complete explanation requires an explicit
treatment of the mode coupling at the entrance and exit of the constriction, as
discussed later.

The zero-field conductance quantization of a quantum point contact is not
as accurate as the Hall conductance quantization in strong magnetic fields.
The deviations from exact quantization are typically\cite{ref6,ref7,ref306} 1\%, while in the
quantum Hall effect one obtains routinely\cite{ref97} an accuracy of 1 part in $10^{7}$. It is
unlikely that a similar accuracy will be achieved in the case of the zero-field
quantization, one reason being the additive contribution to the point contact
resistance of a background resistance whose magnitude cannot be determined precisely. The largest part of this background resistance originates in
the ohmic contacts\cite{ref307} and can thus be eliminated in a four-terminal
measurement of the contact resistance. The position of the additional voltage
probes on the wide 2DEG regions has to be more than an inelastic scattering
length away from the point contact so that a local equilibrium is established.
A residual background resistance\cite{ref307} of the order of the resistance $\rho$ of a
square is therefore unavoidable. In the experiments of Refs.\ \onlinecite{ref6} and \onlinecite{ref7} one has
$\rho\approx 20\,\Omega$, but lower values are possible for higher-mobility material. It would
be of interest to investigate experimentally whether resistance plateaux
quantized to such an accuracy are achievable. It should be noted, however,
that the degree of flatness of the plateaux and the sharpness of the steps in
the present experiments vary among devices of identical design, indicating
that the detailed shape of the electrostatic potential defining the constriction
is important. There are many uncontrolled factors affecting this shape, such
as small changes in the gate geometry, variations in the pinning of the Fermi
level at the free GaAs surface or at the interface with the gate metal, doping
inhomogeneities in the heterostructure material, and trapping of charge in
deep levels in AlGaAs.

On increasing the temperature, one finds experimentally that the plateaux
acquire a finite slope until they are no longer resolved.\cite{ref308} This is a
consequence of the thermal smearing of the Fermi-Dirac distribution (\ref{eq4.9}). If
at $T=0$ the conductance $G(E_{\mathrm{F}}, T)$ has a step function dependence on the
Fermi energy $E_{\mathrm{F}}$, at finite temperatures it has the form\cite{ref309}
\begin{eqnarray}
G(E_{\mathrm{F}}, T)&=& \int_{0}^{\infty}G(E, 0) \frac{df}{dE_{\mathrm{F}}} dE\nonumber\\
&=& \frac{2e^{2}}{h}\sum_{n=1}^{\infty}f(E_{n}-E_{\mathrm{F}}). \label{eq13.2}
\end{eqnarray}
Here $E_{n}$ denotes the energy of the bottom of the $n$th subband [cf.\ Eq.\ (\ref{eq4.3})].
The width of the thermal smearing function $df/dE_{\mathrm{F}}$ is about $4k_{\mathbf{B}}T$, so the
conductance steps should disappear for $T\gtrsim\Delta E/4k_{B}\sim 4\,\mathrm{K}$ (here $\Delta E$ is the
subband splitting at the Fermi level). This is confirmed both by experiment\cite{ref308} and by numerical calculations (see below).

Interestingly, it was found experimentally\cite{ref6,ref7} that in general a finite
temperature yielded the most pronounced and flat plateaux as a function of
gate voltage in the zero-field conductance. If the temperature is increased
beyond this optimum (which is about $0.5\, \mathrm{K}$), the plateaux disappear because
of the thermal averaging discussed earlier. Below this temperature, an
oscillatory structure may be superimposed on the conductance plateaux. This
phenomenon depends on the precise shape of the constriction, as discussed
later. A small but finite voltage drop across the constriction has an effect that
is qualitatively similar to that of a finite temperature.\cite{ref309} This is indeed borne
out by experiment.\cite{ref308} (Experiments on conduction through quantum point
contacts at larger applied voltages in the nonlinear transport regime have
been reviewed in Ref.\ \onlinecite{ref307}).

Theoretically, one would expect the conductance quantization to be
preserved in longer channels than those used in the original experiment\cite{ref6,ref7}
(in which typically $L\sim W\sim 100\,\mathrm{nm}$). Experiments on channels longer than
about 1 $\mu \mathrm{m}$ did not show the quantization,\cite{ref306,ref307,ref310} however, although their
length was well below the transport mean free path in the bulk (about 10 $\mu \mathrm{m}$).
The lack of clear plateaux in long constrictions is presumably due to
enhanced backscattering inside the constriction, either because of impurity
scattering (which may be enhanced\cite{ref306,ref310} due to the reduced screening in a
quasi-one-dimensional electron gas\cite{ref72}) or because of boundary scattering at
channel wall irregularities. As mentioned in Section \ref{sec5}, Thornton et al.\cite{ref107}
have found evidence for a small (5\%) fraction of diffuse, rather than specular,
reflections at boundaries defined electrostatically by a gate. In a 200-nm-wide
constriction this leads to an effective mean free path of about
$200\, \mathrm{nm}/0.05\approx 4\,\mu \mathrm{m}$, comparable to the constriction length of devices that do
not exhibit the conductance quantization.\cite{ref113,ref307}

\begin{figure}
\centerline{\includegraphics[width=6cm]{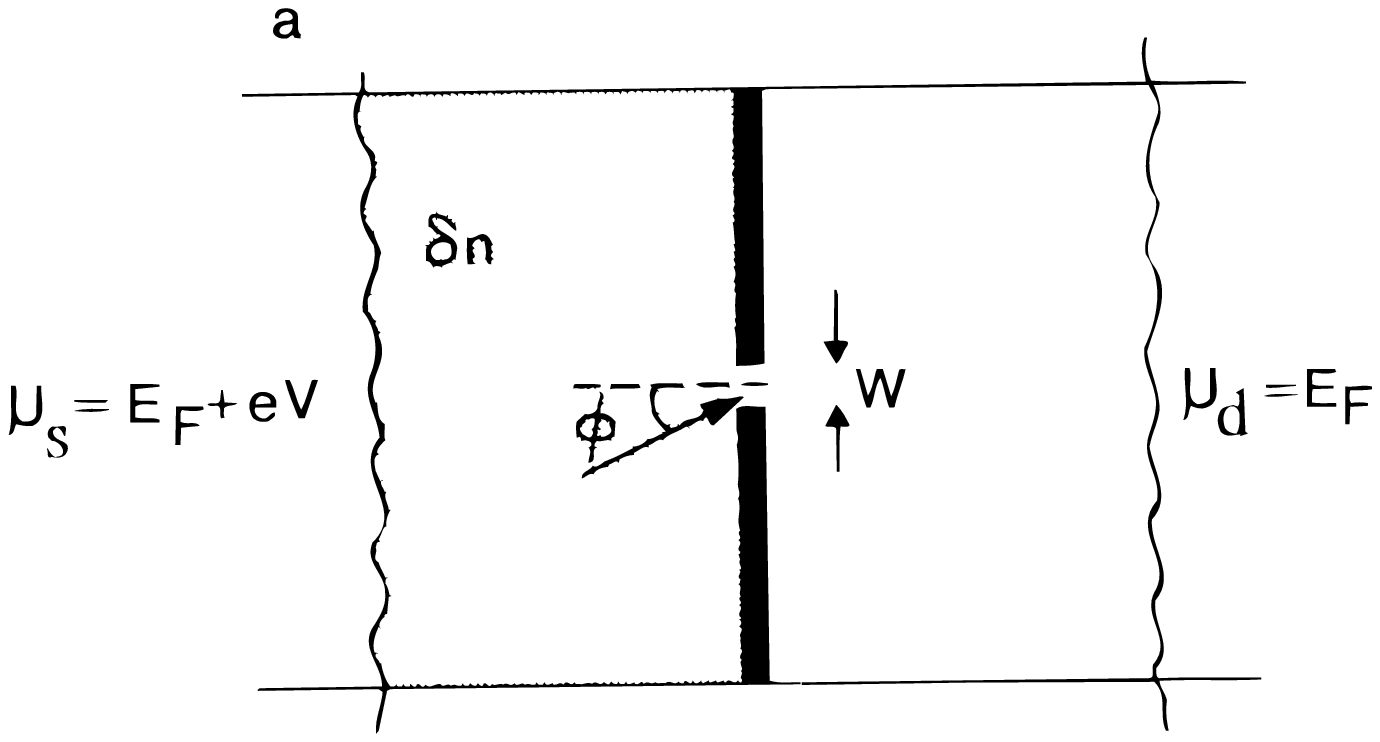}}

\centerline{\includegraphics[width=6cm]{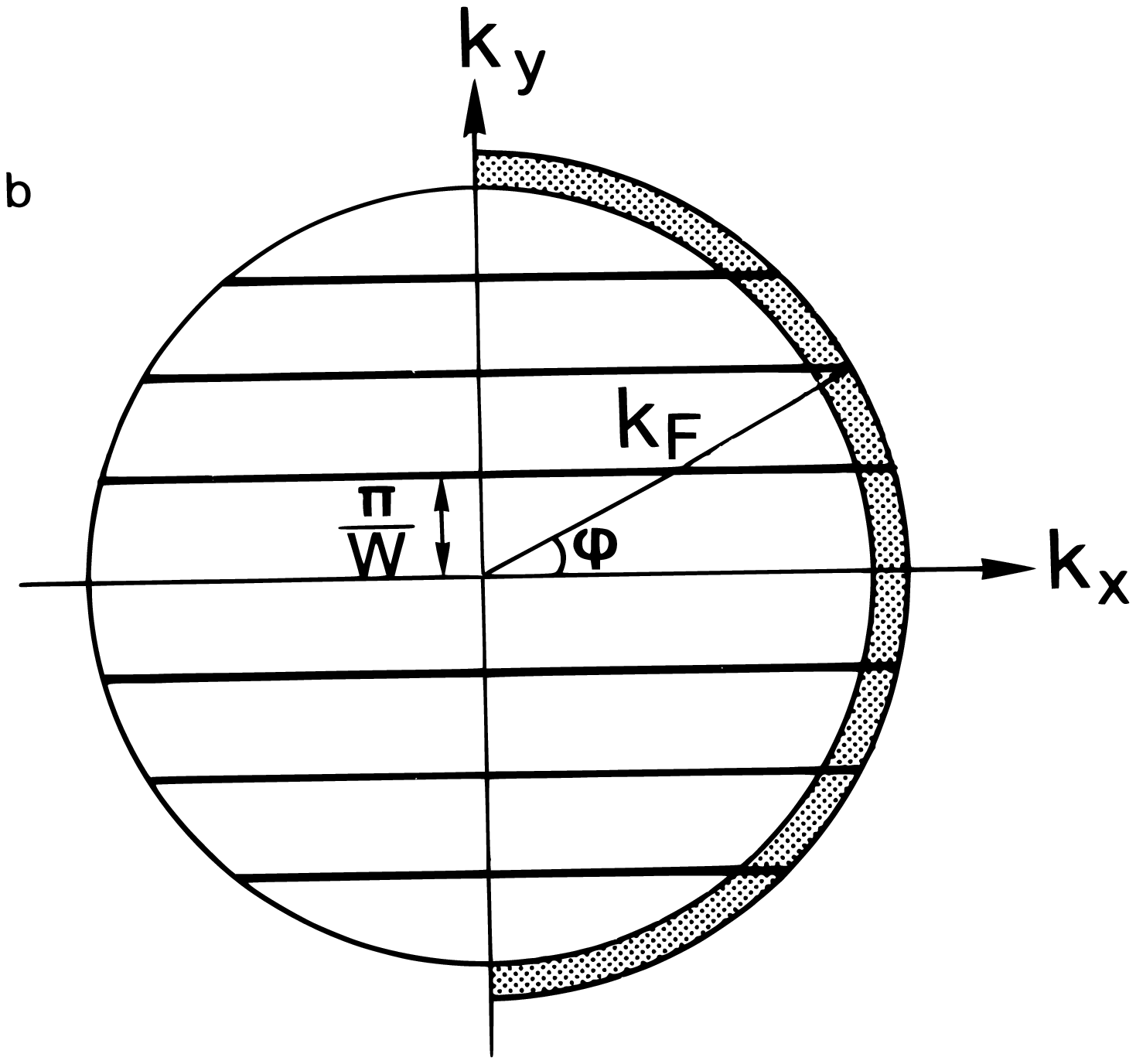}}
\caption{
(a) Classical ballistic transport through a point contact induced by a concentration difference $\delta n$, or electrochemical potential difference $e V$, between source (s) and drain (d). (b) The net current through a quantum point contact is carried by the shaded region in $k$-space. In a narrow channel the allowed states lie on the horizontal lines, which correspond to quantized values for $k_{y} = \pm n\pi/W$, and continuous values for $k_{x}$. The formation of these 1D subbands gives rise to a quantized conductance. Taken from H. van Houten et al., in ``Physics and Technology of Submicron Structures'' (H. Heinrich, G. Bauer, and F. Kuchar, eds.). Springer, Berlin, 1988; and in ``Nanostructure Physics and Fabrication'' (M. Reed and W. P. Kirk, eds.). Academic, New York, 1989.
\label{fig45}
}
\end{figure}

{\bf (b) Theory.} It is instructive to first consider {\it classical\/} 2D point contacts in
some detail.\cite{ref31,ref311} The ballistic electron flow through a point contact is
illustrated in Fig.\ \ref{fig45}a in real space, and in Fig.\ \ref{fig45}b in $k$-space, for a small
excess electron density $\delta n$ at one side of the point contact. At low temperatures this excess charge moves with the Fermi velocity $v_{\mathrm{F}}$. The flux normally
incident on the point contact is $\delta nv_{\mathrm{F}}\langle\cos\phi\,\theta(\cos\phi)\rangle$, where $\theta(x)$ is the unit
step function and the symbol $\langle\; \rangle$ denotes an isotropic angular average (the
angle $\phi$ is defined in Fig.\ \ref{fig45}a). In the ballistic limit $l\gg  W$ the incident flux is
fully transmitted, so the total diffusion current $J$ through the point contact is
given by
\be
J=W \delta nv_{\mathrm{F}}\int_{-\pi/2}^{\pi/2}\cos\phi\frac{d\phi}{2\pi}=\frac{1}{\pi}Wv_{\mathrm{F}}\delta n. \label{eq13.3}
\ee 
The diffusance $\tilde{D}\equiv J/\delta n=(1/\pi)Wv_{\mathrm{F}}$; therefore, the conductance $G=
e^{2}\rho(E_{\mathrm{F}})\tilde{D}$ becomes (using the 2D density of states (\ref{eq4.2}) with the appropriate
degeneracy factors $g_{\mathrm{s}}=2$, $g_{\mathrm{v}}=1$)
\be
G= \frac{2e^{2}}{h}\frac{k_{\mathrm{F}}W}{\pi},\;\;\mbox{in 2D}. \label{eq13.4}
\ee
Eq.\ (\ref{eq13.4}) is the 2D analogue\cite{ref6} of Sharvin's well-known expression\cite{ref296} for the
point contact conductance in three dimensions,
\be
G= \frac{2e^{2}}{h}\frac{k_{\mathrm{F}}^{2}S}{4\pi},\;\;\mbox{in 3D}, \label{eq13.5}
\ee
where now $S$ is the area of the point contact. The number of propagating
modes for a square-well lateral confining potential is $N=\mathrm{Int}[k_{\mathrm{F}}W/\pi]$ in 2D,
so Eq.\ (\ref{eq13.4}) is indeed the classical limit of the quantized conductance (\ref{eq13.1}).

Quantum mechanically, the current through the point contact is equipartitioned among the 1D subbands, or transverse modes, in the constriction. The
equipartitioning of current, which is the basic mechanism for the conductance
quantization, is illustrated in Fig.\ \ref{fig45}b for a square-well lateral confining
potential of width $W$. The 1D subbands then correspond to the pairs of
horizontal lines at $k_{y}=\pm n\pi/W$, with $n=1,2, \ldots, N$ and $N=\mathrm{Int}[k_{\mathrm{F}}W/\pi]$.
The group velocity $v_{n}=\hbar k_{x}/m$ is proportional to $\cos\phi$ and thus decreases
with increasing $n$. However, the decrease in $v_{n}$ is compensated by an increase
in the 1D density of states. Since $\rho_{n}$ is proportional to the length of the
horizontal lines within the dashed area in Fig.\ \ref{fig45}b, $\rho_{n}$ is proportional to
$1/\cos\phi$ so that the product $v_{n}\rho_{n}$ does not depend on the subband index. We
emphasize that, although the classical formula (\ref{eq13.4}) holds only for a square-well lateral confining potential, the quantization (\ref{eq13.1}) is a general result for
any shape of the confining potential. The reason is simply that the
fundamental cancellation of the group velocity $v_{n}=dE_{n}(k)/\hbar dk$ and the 1D
density of states $\rho_{n}^{+}=(\pi dE_{n}(k)/dk)^{-1}$ holds {\it regardless\/} of the form of the
dispersion relation $E_{n}(k)$. For the same reason, Eq.\ (\ref{eq13.1}) is equally applicable
in the presence of a magnetic field, when magnetic edge channels at the Fermi
level take over the role of 1D subbands. Equation (\ref{eq13.1}) thus implies a
continuous transition from the zero-field quantization to the quantum Hall
effect, as we will discuss in Section \ref{sec13b}.

To analyze deviations from Eq.\ (\ref{eq13.1}) it is necessary to solve the
Schr\"{o}dinger equation for the wave functions in the narrow point contact and
the adjacent wide regions and to match the wave functions and their
derivatives at the entrance and exit of the constriction. The resulting
transmission coefficients determine the conductance via the Landauer formula (\ref{eq12.10}). This mode coupling problem has been solved numerically for
point contacts of a variety of shapes\cite{ref312,ref313,ref314,ref315,ref316,ref317,ref318,ref319,ref320,ref321} and analytically in special
geometries.\cite{ref322,ref323,ref324} When considering the mode coupling at the entrance and
exit of the constriction, one must distinguish gradual ({\it adiabatic}) from {\it abrupt\/}
transitions from wide to narrow regions.

The case of an {\it adiabatic\/} constriction has been studied by Glazman et
al.,\cite{ref325} by Yacoby and Imry\cite{ref326} and by Payne.\cite{ref272} If the constriction width
$W(x)$ changes sufficiently gradually, the transport through the constriction is
adiabatic (i.e., without intersubband scattering). The transmission coefficients
then vanish, $|t_{nm}|^{2}=0$, unless $n=m\leq N_{\min}$, with $N_{\min}$ the smallest number of
occupied subbands in the constriction. The conductance quantization (\ref{eq13.1})
now follows immediately from the Landauer formula (\ref{eq12.10}). The criterion
for adiabatic transport is\cite{ref326} $dW/dx\lesssim 1/N(x)$, with $N(x)\approx k_{\mathrm{F}}W(x)/\pi$ the local
number of subbands. As the constriction widens, $N(x)$ increases and adiabaticity is preserved only if $W(x)$ increases more and more slowly. In practice,
adiabaticity breaks down at a width $W_{\max}$, which is at most a factor of 2 larger
than the minimum width $W_{\min}$ (cf.\ the collimated beam experiment of Ref.\
\onlinecite{ref327}, discussed in Section \ref{sec15}). This does not affect the conductance of the
constriction, however, if the breakdown of adiabaticity results in a mixing of
the subbands without causing reflection back through the constriction. If
such is the case, the total transmission probability through the constriction
remains the same as in the hypothetical case of fully adiabatic transport. As
pointed out by Yacoby and Imry,\cite{ref326} a relatively small adiabatic increase in
width from $W_{\min}$ to $W_{\max}$ is sufficient to ensure a drastic suppression of
reflections at $W_{\max}$. The reason is that the subbands with the largest reflection
probability are close to cutoff, that is, they have subband index close to $N_{\max}$,
the number of subbands occupied at $W_{\max}$. Because the transport is adiabatic
from $W_{\min}$ to $W_{\max}$, only the $N_{\min}$ subbands with the smallest $n$ arrive at $W_{\max}$,
and these subbands have a small reflection probability. In the language of
waveguide transmission, one has impedance-matched the constriction to the
wide 2DEG regions.\cite{ref328} The filtering of subbands by a gradually widening
constriction has an interesting effect on the angular distribution of the
electrons injected into the wide 2DEG. This {\it horn collimation\/} effect\cite{ref329} is
discussed in Section \ref{sec15}.

\begin{figure}
\centerline{\includegraphics[width=8cm]{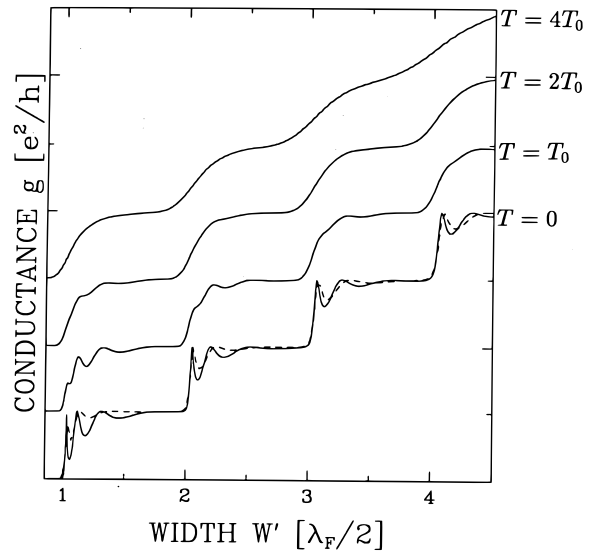}}
\caption{
Transmission resonances exhibited by theoretical results for the conductance of a quantum point contact of abrupt (rectangular) shape. A smearing of the resonances occurs at nonzero temperatures ($T_{0}=0.02\,E_{\rm F}/k_{\rm B}\approx 2.8\,{\rm K}$). The dashed curve is an exact numerical result; the full curves are approximate. Taken from A. Szafer and A. D. Stone, Phys.\ Rev.\ Lett.\ {\bf 62}, 300 (1989).
\label{fig46}
}
\end{figure}

An adiabatic constriction improves the accuracy of the conductance
quantization, but is not required to observe the effect. Calculations\cite{ref312,ref313,ref314,ref315,ref316,ref317,ref318,ref319,ref320,ref321,ref322,ref323,ref324}
show that well-defined conductance plateaux persist for {\it abrupt\/} constrictions,
especially if they are neither very short nor very long. The optimum length for
the observation of the plateaux is given by\cite{ref313} $L_{\mathrm{opt}}\approx 0.4(W\lambda_{\mathrm{F}})^{1/2}$. In shorter
constrictions the plateaux acquire a finite slope, although they do not
disappear completely even at zero length. For $L>L_{\mathrm{opt}}$ the calculations
exhibit regular oscillations that depress the conductance periodically below
its quantized value. The oscillations are damped and have usually vanished
before the next plateau is reached. As a representative illustration, we
reproduce in Fig.\ \ref{fig46} a set of numerical results for the conductance as a
function of width (at fixed Fermi wave vector), obtained by Szafer and
Stone.\cite{ref315}  Note that a finite temperature improves the flatness of the plateaux,
as observed experimentally. The existence of an optimum length can be
understood as follows.

Because of the abrupt widening of the constriction, there is a significant
probability for reflection at the exit of the constriction, in contrast to the
adiabatic case considered earlier. The conductance as a function of width, or
Fermi energy, is therefore not a simple step function. On the $n$th conductance
plateau backscattering occurs predominantly for the $n$th subband, since it
is closest to cutoff. Resonant transmission of this subband occurs if
the constriction length $L$ is approximately an integer multiple of half
the longitudinal wavelength $\lambda_{n}=h[2m(E_{\mathrm{F}}-E_{n})]^{-1/2}$, leading to oscillations
on the conductance plateaux. These transmission resonances are
damped, because the reflection probability decreases with decreasing $\lambda_{n}$. The
shortest value of $\lambda_{n}$ on the $N\mathrm{th}$ conductance plateau is
$h[2m(E_{N+1}-E_{N})]^{-1/2}\approx(W\lambda_{\mathrm{F}})^{1/2}$ (for a square-well lateral confining potential). The transmission resonances are thus suppressed if $L\lesssim(W\lambda_{\mathrm{F}})^{1/2}$.
Transmission through evanescent modes (i.e., subbands above $E_{\mathrm{F}}$) is predominant for the $(N+1)\mathrm{th}$ subband, since it has the largest decay length
$\Lambda_{N+1}=h[2m(E_{N+1}-E_{\mathrm{F}})]^{-1/2}$. The observation of that plateau requires
that the constriction length exceeds this decay length at the population
threshold of the $N$ th mode, or $L\gtrsim h[2m(E_{N+1}-E_{N})]^{-1/2}\approx(W\lambda_{\mathrm{F}})^{1/2}$. The
optimum length\cite{ref313} $L_{\mathrm{opt}}\approx 0.4(W\lambda_{\mathrm{F}})^{1/2}$ thus separates a short constriction
regime, in which transmission via evanescent modes cannot be ignored, from
a long constriction regime, in which transmission resonances obscure the
plateaux.

\begin{figure}
\centerline{\includegraphics[width=8cm]{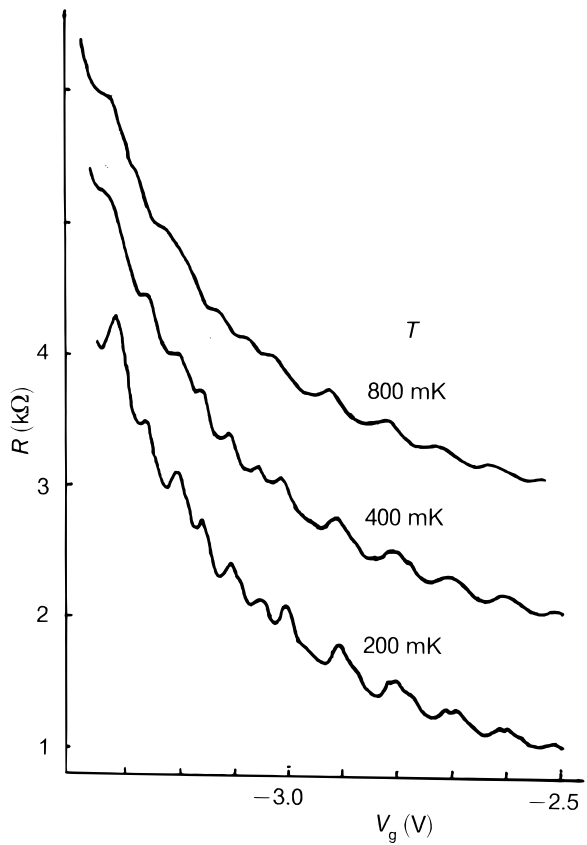}}
\caption{
Resistance as a function of gate voltage for an elongated quantum point contact ($L = 0.8\,\mu{\rm m}$) at temperatures of 0.2, 0.4, and 0.8 K, showing transmission resonances. Subsequent curves from the bottom are offset by 1 ${\rm k}\Omega$. Taken from R. J. Brown et al., Solid State Electron.\ {\bf 32}, 1179 (1989).
\label{fig47}
}
\end{figure}

Oscillatory structure was resolved in low-temperature experiments on the
conductance quantization of one quantum point contact by van Wees et
al.,\cite{ref308} but was not clearly seen in other devices. A difficulty in the
interpretation of these and other experiments is that oscillations can also be
caused by quantum interference processes involving impurity scattering near
the constriction. Another experimental observation of oscillatory structure
was reported by Hirayama et al.\cite{ref330} for short (100-nm) quantum point
contacts of fixed width (defined by means of focused ion beam lithography).
To observe the plateaux, they slowly varied the electron density by weakly
illuminating the sample. The oscillations were quite reproducible, also after
thermal cycling of the sample, but again they were found in some of the
devices only (this was attributed to variations in the abruptness of the
constrictions\cite{ref330,ref331}). Brown et al.\cite{ref332} have studied the conductance of split-gate constrictions of lengths $L\approx 0.3,0.8$, and 1 $\mu \mathrm{m}$, and they observed
pronounced oscillations instead of the flat conductance plateaux found for
shorter quantum point contacts. The observed oscillatory structure (reproduced in Fig.\ \ref{fig47}) is quite regular, and it correlates with the sequence of
plateaux that is recovered at higher temperatures (around $0.8\, \mathrm{K}$). The effect
was seen in all of the devices studied in Ref.\ \onlinecite{ref332}. Measurements by Timp et
al.\cite{ref306} on rather similar $0.9$-$\mu \mathrm{m}$-long constrictions did not show periodic
oscillations, however. Brown et al.\ conclude that their oscillations are due to
transmission resonances associated with reflections at entrance and exit of
the constriction. Detailed comparison with theory is difficult because the
transmission resonances depend sensitively on the shape of the lateral
confining potential and on the presence of a potential barrier in the
constriction (see Section \ref{sec13b}). A calculation that comes close to the
observation of Brown et al.\ has been published by Martin-Moreno and
Smith.\cite{ref333}

\begin{figure}
\centerline{\includegraphics[width=8cm]{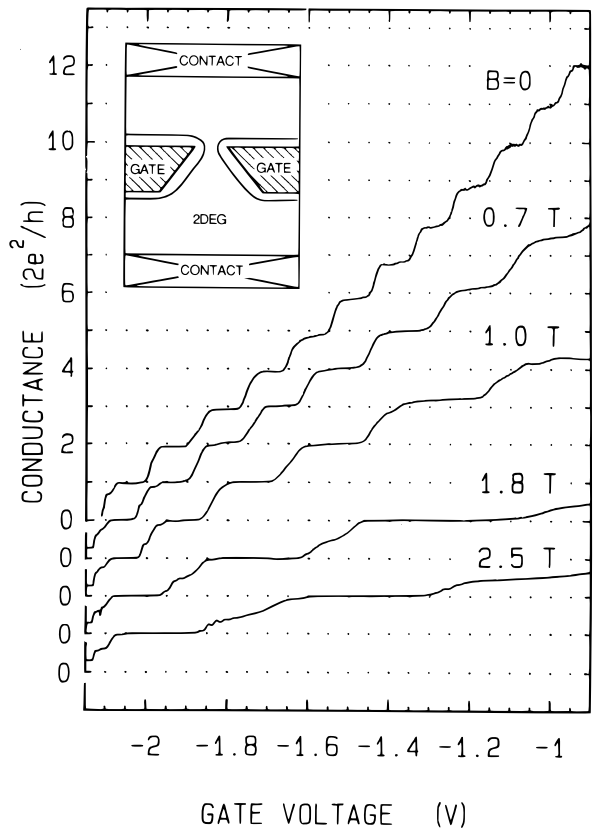}}
\caption{
Point contact conductance (corrected for a background resistance) as a function of gate voltage for several magnetic field values, illustrating the transition from zero-field quantization to quantum Hall effect. The curves have been offset for clarity. The inset shows the device geometry. Taken from B. J. van Wees et al., Phys.\ Rev.\ B. {\bf 38}, 3625 (1988).
\label{fig48}
}
\end{figure}

\subsubsection{\label{sec13b} Depopulation of subbands and suppression of backscattering by a magnetic field}

The effect of a magnetic field (perpendicular to the 2DEG) on the
quantized conductance of a point contact is shown in Fig.\ \ref{fig48}, as measured by
van Wees et al.\cite{ref334} First of all, Fig.\ \ref{fig48} demonstrates that the conductance
quantization is conserved in the presence of a magnetic field and shows a
smooth transition from zero-field quantization to quantum Hall effect. The
most noticeable effect of the magnetic field is to reduce the number of
plateaux in a given gate voltage interval. This provides a demonstration of
depopulation of magnetoelectric subbands, which is more direct than that
provided by the experiments discussed in Section \ref{sec10}. In addition, one
observes that the flatness of the plateaux improves in the presence of the field.
This is due to the reduction of the reflection probability at the point contact,
which is revealed most clearly in a somewhat different (four-terminal)
measurement configuration. These two effects of a magnetic field will be
discussed separately. We will return to the magnetic suppression of back-scattering in Section \ref{sec18} in connection with the edge channel theory\cite{ref112} of the
quantum Hall effect.

{\bf (a) Depopulation of subbands.} Because the equipartitioning of current
among the 1D subbands holds regardless of the nature of the subbands
involved, one can conclude that in the presence of a magnetic field $B$ the
conductance remains quantized according to $G=(2e^{2}/h)N$ (ignoring spin
splitting of the subbands, for simplicity). Explicit calculations\cite{ref335} confirm this
expectation. The number of occupied subbands $N$ as a function of $B$ has been
studied in Sections \ref{sec10} and \ref{sec12} and is given by  Eqs.\ (\ref{eq10.7}) and (\ref{eq10.8}) for a
parabolic and a square-well potential, respectively. In the high-magnetic-field
regime $W\gtrsim 2l_{\mathrm{cycl}}$, the number $N\approx E_{\mathrm{F}}/\hbar\omega_{\mathrm{c}}$ is just the number of occupied
Landau levels. The conductance quantization is then a manifestation of the
quantum Hall effect.\cite{ref8} (The fact that $G$ is not a Hall conductance but a two-terminal conductance is not an essential distinction for this effect; see Section
\ref{sec18}.) At lower magnetic fields, the conductance quantization provides a direct
and extremely straightforward method to measure via $N=G(2e^{2}/h)^{-1}$ the
depopulation of magnetoelectric subbands in the constriction.

\begin{figure}
\centerline{\includegraphics[width=8cm]{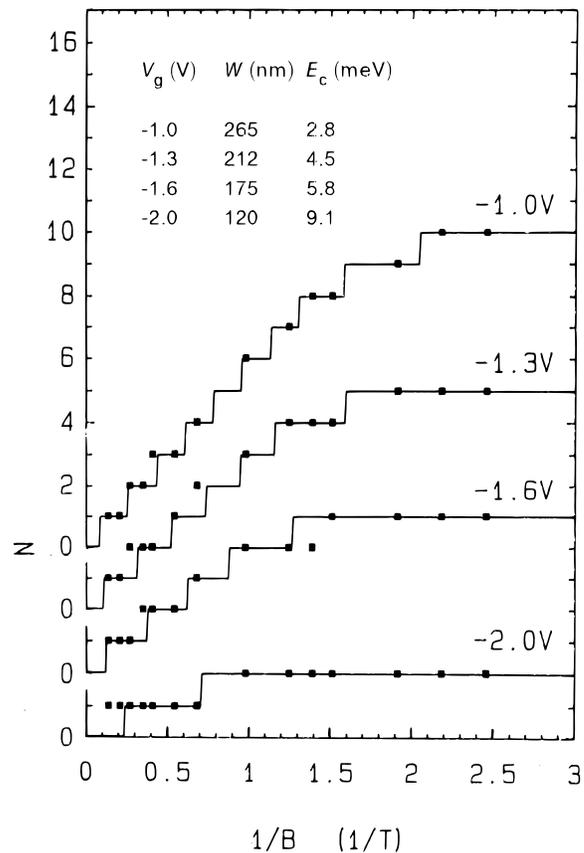}}
\caption{
Number of occupied subbands as a function of reciprocal magnetic field for several values of the gate voltage. Data points have been obtained directly from the quantized
conductance (Fig.\ \ref{fig48}); solid curves are calculated for a square-well confining potential of width $W$ and well bottom $E_{\rm c}$ as tabulated in the inset. Taken from B. J. van Wees et al., Phys.\ Rev.\ B {\bf 38}, 3625 (1988).
\label{fig49}
}
\end{figure}

Figure \ref{fig49} shows $N$ versus $B^{-1}$ for various gate voltages, as it follows from
the experiment of Fig.\ \ref{fig48}. Also shown are the theoretical curves for a square-well confining potential, with the potential barrier in the constriction taken
into account by replacing $E_{\mathrm{F}}$ by $E_{\mathrm{F}}-E_{\mathrm{c}}$ in Eq.\ (\ref{eq10.8}). The $B$-dependence of
$E_{\mathrm{F}}$ has been ignored in the calculation. The barrier height $E_{\mathrm{c}}$ is obtained from
the high-field conductance plateaux [where $N\approx(E_{\mathrm{F}}-E_{\mathrm{c}})/\hbar\omega_{\mathrm{c}}]$, and the constriction width $W$ then follows from the zero-field conductance (where
$N\approx[2m(E_{\mathrm{F}}-E_{\mathrm{c}})/h^{2}]^{1/2}W/\pi)$. The good agreement found over the entire
field range confirms the expectation that the quantized conductance is
exclusively determined by the number of occupied subbands, irrespective of
their electric or magnetic origin. The analysis in Fig.\ \ref{fig49} is for a square-well
confining potential.\cite{ref334} For the narrowest constrictions a parabolic potential
should be more appropriate,\cite{ref61} which has been used to analyze the data of
Fig.\ \ref{fig48} in Refs.\ \onlinecite{ref336} and \onlinecite{ref308}. Wharam et al.\cite{ref337} have analyzed their
depopulation data using the intermediate model of a parabolic potential
with a flattened bottom (cf.\ also Ref.\ \onlinecite{ref336}). Because of the uncertainties in the
actual shape of the potential, the parameter values tabulated in Fig.\ \ref{fig49} should
be considered as rough estimates only.

In strong magnetic fields the spin degeneracy of the energy levels is
removed, and additional plateaux appear\cite{ref7,ref334} at {\it odd\/} multiples of $e^{2}/h$.
Wharam et al.\cite{ref7} have demonstrated this effect in a particularly clear fashion,
using a magnetic field parallel (rather than perpendicular) to the 2DEG.
Rather strong magnetic fields turned out to be required to fully lift the spin
degeneracy in this experiment (about $10\, \mathrm{T}$).

{\bf (b) Suppression of backscattering.} Only a small fraction of the electrons
injected by the current source into the 2DEG is transmitted through the
point contact. The remaining electrons are scattered back into the source
contact. This is the origin of the nonzero resistance of a ballistic point
contact. In this subsection we shall discuss how a relatively weak magnetic
field leads to a suppression of the {\it geometrical backscattering\/} caused by the
finite width of the point contact, while the amount of backscattering caused
by the potential barrier in the point contact remains essentially unaffected.

\begin{figure}
\centerline{\includegraphics[width=8cm]{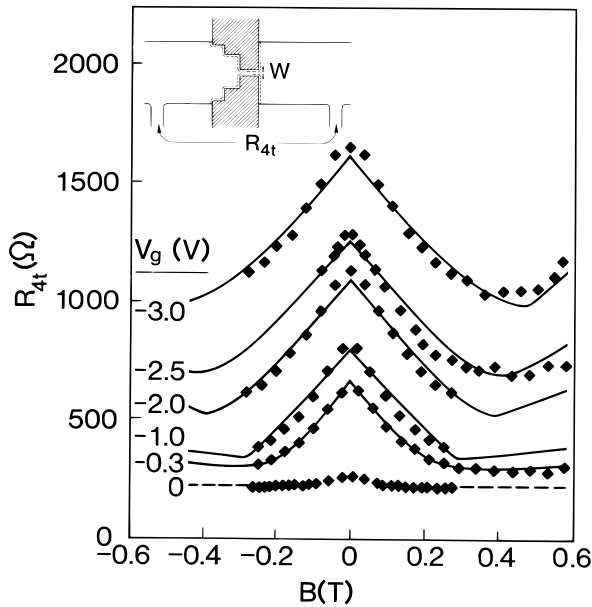}}
\caption{
Four-terminal longitudinal  magnetoresistance $R_{4{\rm t}}\equiv R_{\rm L}$ of a constriction for a series of gate voltages. The negative  magnetoresistance is temperature independent between 50 mK and 4 K. Solid lines are according to Eqs.\ (\ref{eq13.7}) and (\ref{eq10.8}), with the constriction width as adjustable parameter. The inset shows schematically the device geometry, with the two voltage probes used to measure $R_{\rm L}$. Taken from H. van Houten et al., Phys.\ Rev.\ B {\bf 37}, 8534 (1988).
\label{fig50}
}
\end{figure}

The reduction of backscattering by a magnetic field is observed as a
{\it negative\/} magnetoresistance [i.e., $R(B)-R(0)<0]$ in a {\it four-terminal\/} measurement of the longitudinal point contact resistance $R_{\mathrm{L}}$. The voltage probes in
this experiment\cite{ref113} are positioned on wide 2DEG regions, well away from the
constriction (see the inset in Fig.\ \ref{fig50}). This allows the establishment of local
equilibrium near the voltage probes, at least in weak magnetic fields (cf.\
Sections \ref{sec18} and \ref{sec19}), so that the measured four-terminal resistance does not
depend on the properties of the probes. The experimental results for $R_{\mathrm{L}}$ in this
geometry are plotted in Fig.\ \ref{fig50}. The negative magnetoresistance is
temperature-independent (between $50\, \mathrm{mK}$ and 4 K) and is observed in weak
magnetic fields once the narrow constriction is defined (for $V_{\mathrm{g}}\lesssim-0.3\,\mathrm{V}$). At
stronger magnetic fields $(B>0.4\,\mathrm{T})$, a crossover is observed to a positive
magnetoresistance. The zero-field resistance, the magnitude of the negative
magnetoresistance, the slope of the positive magnetoresistance, as well as the
crossover field, all increase with increasing negative gate voltage.

\begin{figure}
\centerline{\includegraphics[width=8cm]{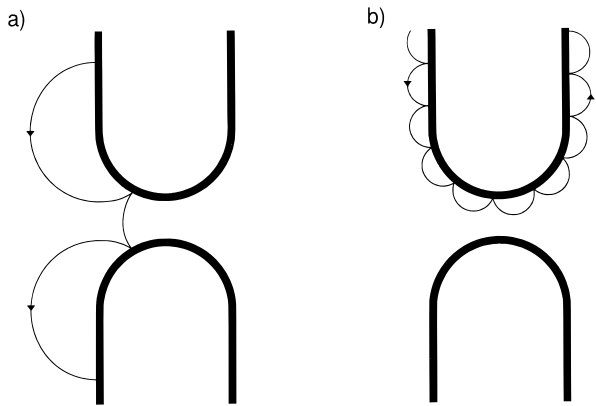}}
\caption{
Illustration of the reduction of backscattering by a magnetic field, which is responsible for the negative  magnetoresistance of Fig.\ \ref{fig50}. Shown are trajectories approaching a constriction without a potential barrier, in a weak (a) and strong (b) magnetic field. Taken from H. van Houten et al., in Ref.\ \onlinecite{ref9}.
\label{fig51}
}
\end{figure}

The magnetic field dependence of the four-terminal resistance shown in
Fig.\ \ref{fig50} is qualitatively different from that of the two-terminal resistance
$R_{2\mathrm{t}}\equiv G^{-1}$ considered in the previous subsection. In fact, $R_{2\mathrm{t}}$ is approximately $B$-independent in weak magnetic fields (below the crossover fields of Fig.\
\ref{fig50}). The reason is that $R_{2\mathrm{t}}$ is given by [cf.\ Eq.\ (\ref{eq13.1})]
\be
R_{2\mathrm{t}}= \frac{h}{2e^{2}}\frac{1}{N_{\min}}, \label{eq13.6}
\ee 
with $N_{\min}$ the number of occupied subbands in the constriction (at the point
where it has its minimum width and electron gas density). In weak magnetic
fields such that $2l_{\mathrm{cycl}}>W$, the number of occupied subbands remains
approximately constant [cf.\ Fig.\ \ref{fig31} or Eq.\ (\ref{eq10.8})], so $R_{2\mathrm{t}}$ is only weakly
dependent on $B$ in this field regime. For stronger fields Eq.\ (\ref{eq13.6}) describes a
{\it positive\/} magnetoresistance, because $N_{\min}$ decreases due to the magnetic
depopulation of subbands discussed earlier. (A similar positive magnetoresistance is found in a Hall bar with a cross gate; see Ref.\ \onlinecite{ref338}.) Why then does one
find a {\it negative\/} magnetoresistance in the four-terminal measurements of Fig.\
\ref{fig50}? Qualitatively, the answer is shown in Fig.\ \ref{fig51}, for a constriction without a
potential barrier. In a magnetic field the left-and right-moving electrons are
spatially separated by the Lorentz force at opposite sides of the constriction.
Quantum mechanically the skipping orbits in Fig.\ \ref{fig51} correspond to magnetic
edge states (cf.\ Fig.\ \ref{fig41}). Backscattering thus requires scattering across the
width of the constriction, which becomes increasingly improbable as $l_{\mathrm{cycl}}$
becomes smaller and smaller compared with the width (compare Figs.\ \ref{fig51}a,b).
For this reason a magnetic field suppresses the {\it geometrical\/} constriction
resistance in the ballistic regime, but not the resistance associated with the
constriction in energy space, which is due to the potential barrier.

These effects were analyzed theoretically in Ref.\ \onlinecite{ref113}, with the simple result
\be
R_{\mathrm{L}}= \frac{h}{2e^{2}}\left(\frac{1}{N_{\min}}-\frac{1}{N_{\mathrm{wide}}}\right). \label{eq13.7}
\ee 
Here $N_{\mathrm{wide}}$ is the number of occupied Landau levels in the wide 2DEG
regions. The simplest (but incomplete) argument leading to Eq.\ (\ref{eq13.7}) is that
the additivity of voltages on reservoirs (ohmic contacts) implies that the two-terminal resistance $R_{2\mathrm{t}}=(h/2e^{2})N_{\min}^{-1}$ should equal the sum of the Hall
resistance $R_{\mathrm{H}}=(h/2e^{2})N_{\mathrm{wide}}^{-1}$ and the longitudinal resistance $R_{\mathrm{L}}$. This argument is incomplete because it assumes that the Hall resistance in the wide
regions is not affected by the presence of the constriction. This is correct in
general only if inelastic scattering has equilibrated the edge states transmitted
through the constriction before they reach a voltage probe. Deviations from
Eq.\ (\ref{eq13.7}) can occur in the absence of local equilibrium near the voltage
probes, depending on the properties of the probes themselves. We discuss this
in Section \ref{sec19}, following a derivation of Eq.\ (\ref{eq13.7}) from the Landauer-B\"{u}ttiker formalism.

At small magnetic fields $N_{\min}$ is approximately constant, while
$N_{\mathrm{wide}}\approx E_{\mathrm{F}}/\hbar\omega_{\mathrm{c}}$ decreases linearly with $B$. Equation (\ref{eq13.7}) thus predicts a
{\it negative\/} magnetoresistance. If the electron density in the wide and narrow
regions is equal (i.e., the barrier height $E_{\mathrm{c}}=0$), then the resistance $R_{\mathrm{L}}$ {\it vanishes}
for fields $B>B_{\mathrm{crit}}\equiv 2\hbar k_{\rm F}/eW$. This follows from Eq.\ (\ref{eq13.7}), because in this
case $N_{\min}$ and $N_{\mathrm{wide}}$ are identical. If the electron density in the constriction is
less than its value in the wide region, then Eq.\ (\ref{eq13.7}) predicts a crossover at
$B_{\mathrm{crit}}$ to a strong-field regime of {\it positive\/} magnetoresistance described by
\be
R_{\mathrm{L}} \approx\frac{h}{2e^{2}}\left(\frac{\hbar\omega_{\mathrm{c}}}{E_{\mathrm{F}}-E_{\mathrm{c}}}-\frac{\hbar\omega_{\mathrm{c}}}{E_{\mathrm{F}}}\right)\;\;{\rm if}\;\;B>B_{\mathrm{crit}}. \label{eq13.8}
\ee  

The experimental results are well described by the solid curves following
from Eq.\ (\ref{eq13.7}) (with $N_{\min}$ given by the square-well result (\ref{eq10.8}), and with an
added constant background resistance). The constriction in the present
experiment is relatively long $(L\approx 3.4\,\mu \mathrm{m})$, and wide ($W$ ranging from 0.2 to
1.0 $\mu \mathrm{m}$) so that it does not exhibit quantized two-terminal conductance
plateaux in the absence of a magnetic field. For this reason the discreteness of
$N_{\min}$ was ignored in the theoretical curves in Fig.\ \ref{fig50}. We emphasize, however,
that Eq.\ (\ref{eq13.7}) is equally applicable to the quantized case, as observed by
several groups\cite{ref307,ref339,ref340,ref341,ref342} (see Section \ref{sec19}).

The negative magnetoresistance (\ref{eq13.7}) due to the suppression of the
contact resistance is an additive contribution to the magnetoresistance of a
long and narrow channel in the quasi-ballistic regime (if the voltage probes
are positioned on two wide 2DEG regions, connected by the channel). For a
channel of length $L$ and a mean free path $l$ the zero-field contact resistance is a
fraction $\sim l/L$ of the Drude resistance and may thus be ignored for $L\gg  l$. The
strong-field positive magnetoresistance (\ref{eq13.8}) resulting from a different
electron density in the channel may still be important, however. The effect of
the contact resistance may be suppressed to a large extent by using narrow
voltage probes attached to the channel itself rather than to wide 2DEG
regions. As we will see in Section \ref{sec16}, such a solution no longer works in the
ballistic transport regime, because of the additional scattering induced\cite{ref289} by
the voltage probes.

\subsection{\label{sec14} Coherent electron focusing}

A magnetic field may be used to focus the electrons injected by a point
contact onto a second point contact. Electron focusing in metals was
originally conceived by Sharvin\cite{ref296} as a method to investigate the shape of the
Fermi surface. It has become a powerful tool in the study of surface
scattering\cite{ref343} and the electron-phonon interaction,\cite{ref344} as reviewed in Refs.\
\onlinecite{ref305,ref345}, and \onlinecite{ref346}. The experiment is the analogue in the solid state of
magnetic focusing of electrons in vacuum. Required is a large mean free path
for the carriers at the Fermi surface, to ensure ballistic motion as in vacuum.
The mean free path should be much larger than the separation $L$ of the two
point contacts. Moreover, $L$ should be much larger than the point contact
width $W$, to achieve optimal resolution. In metals, electron focusing is
essentially a {\it classical\/} phenomenon because the Fermi wavelength
$\lambda_{\mathrm{F}}\sim 0.5\,\mathrm{nm}$ is much smaller than both $W\sim 1\,\mu \mathrm{m}$ and $L\sim 100\,\mu \mathrm{m}$. The
ratios $\lambda_{\mathrm{F}}/L$ and $\lambda_{\mathrm{F}}/W$ are much larger in a 2DEG than in a metal, typically by
factors of $10^{4}$ and $10^{2}$, respectively. {\it Coherent\/} electron focusing\cite{ref59,ref80,ref347} is
possible in a 2DEG because of this relatively large value of the Fermi
wavelength, and turns out to be strikingy different from classical electron
focusing in metals.

Electron focusing can be seen as a transmission experiment in electron
optics (cf.\ Ref.\ \onlinecite{ref3} for a discussion from this point of view). An alternative point
of view (emphasized in Refs.\ \onlinecite{ref80} and \onlinecite{ref348}) is that coherent electron focusing is a
prototype of a nonlocal resistance measurement in the quantum ballistic
transport regime, such as studied extensively in narrow-channel geometries.\cite{ref310} Longitudinal resistances that are negative (not $\pm B$ symmetric) and
dependent on the properties of the current and voltage contacts as well as on
their separation, periodic and aperiodic magnetoresistance oscillations,
absence of local equilibrium are all characteristic features of this transport
regime that appear in a most extreme and bare form in the electron focusing
geometry. One reason for the simplification offered by this geometry is that
the current and voltage contacts, being point contacts, are not nearly as
invasive as the wide leads in a Hall bar geometry. Another reason is that the
electrons interact with only one boundary (instead of two in a narrow
channel).

The outline of this section is as follows.In Section \ref{sec14a} the experimental
results on coherent electron focusing\cite{ref59,ref80} are presented. A theoretical
description\cite{ref80,ref347} is given in Section \ref{sec14b}, in terms of mode interference in the
waveguide formed by the magnetic field at the 2DEG boundary. Apart from
the intrinsic interest of electron focusing in a 2DEG, the experiment can also
be seen as a method to study electron scattering, as in metals. Two such
applications\cite{ref108,ref349} are discussed in Section \ref{sec14c}. We restrict ourselves in this
section to focusing by a magnetic field. Electrostatic focusing\cite{ref350} is discussed
in Section \ref{sec15b}.

\subsubsection{\label{sec14a} Experiments}

\begin{figure}
\centerline{\includegraphics[width=8cm]{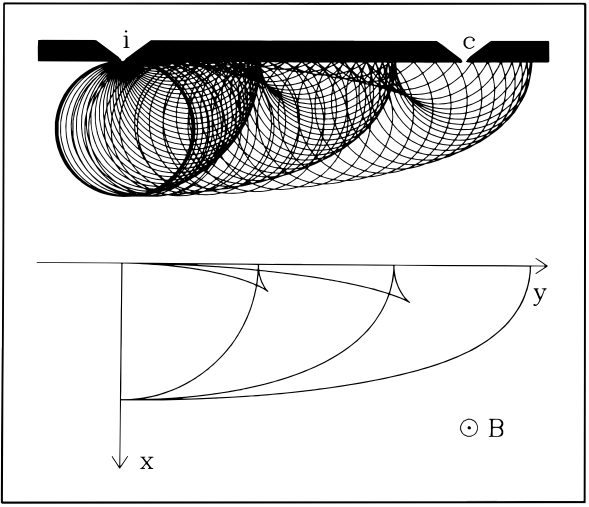}}
\caption{
Illustration of classical electron focusing by a magnetic field. Top: Skipping orbits along the 2DEG boundary. The trajectories are drawn up to the third specular reflection. Bottom: Plot of the caustics, which are the collection of focal points of the trajectories. Taken from H. van Houten et al., Phys.\ Rev.\ B {\bf 39}, 8556 (1989).
\label{fig52}
}
\end{figure}

The geometry of the experiment\cite{ref59} in a 2DEG is the transverse focusing
geometry of Tsoi\cite{ref343} and consists of two point contacts on the same boundary
in a perpendicular magnetic field. (In metals one can also use the geometry of
Sharvin\cite{ref296} with opposite point contacts in a longitudinal field. This is not
possible in two dimensions.) Two point contacts and the intermediate 2DEG
boundary are created electrostatically by means of the two split gates shown
in Fig.\ \ref{fig5}b. Figure \ref{fig52} illustrates electron focusing in two dimensions as it
follows from the classical mechanics of electrons at the Fermi level. The
injector (i) injects a divergent beam of electrons ballistically into the 2DEG.
Electrons are detected if they reach the adjacent collector (c), after one or
more specular reflections at the boundary connecting $\mathrm{i}$ and $\mathrm{c}$. (These are the
{\it skipping orbits\/} discussed in Section \ref{sec12a}.) The focusing action of the magnetic
field is evident in Fig.\ \ref{fig52} (top) from the black lines of high density of
trajectories. These lines are known in optics as {\it caustics\/} and they are plotted
separately in Fig.\ \ref{fig52} (bottom). The caustics intersect the 2DEG boundary at
multiples of the cyclotron diameter from the injector. As the magnetic field is
increased, a series of these focal points shifts past the collector. The electron
flux incident on the collector thus reaches a maximum whenever its
separation $L$ from the injector is an integer multiple of $2l_{\mathrm{cycl}}=2\hbar k_{\rm F}/eB$. This
occurs when $B=pB_{\mathrm{focus}}$, $p=1,2, \ldots$, with
\be
B_{\mathrm{focus}}=2\hbar k_{\rm F}/eL. \label{eq14.1}
\ee 
For a given injected current $I_{\mathrm{i}}$ the voltage $V_{\mathrm{c}}$ on the collector is proportional
to the incident flux. The classical picture thus predicts a series of equidistant
peaks in the collector voltage as a function of magnetic field.

\begin{figure}
\centerline{\includegraphics[width=8cm]{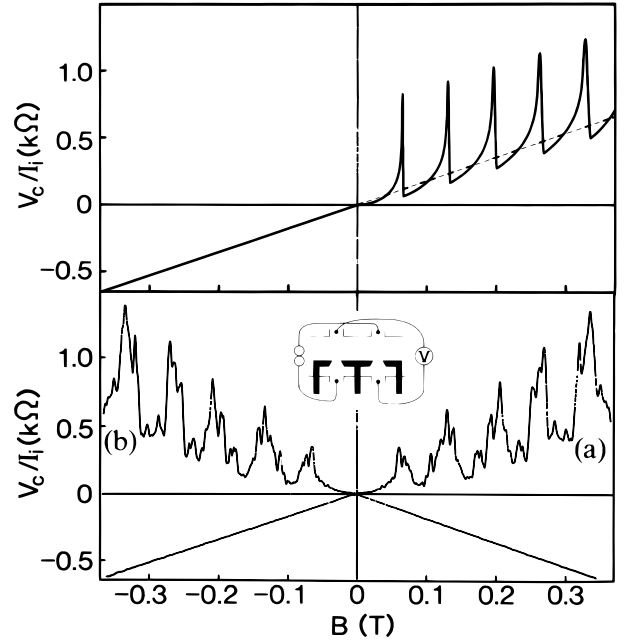}}
\caption{
Bottom: Experimental electron focusing spectrum ($T= 50\,{\rm mK}$, $L= 3.0\,\mu{\rm m}$) in the generalized Hall resistance configuration depicted in the inset. The two traces $a$ and $b$ are measured with interchanged current and voltage leads, and demonstrate the injector-collector reciprocity as well as the reproducibility of the fine structure. Top: Calculated classical focusing spectrum corresponding to the experimental trace $a$ (50-nm-wide point contacts were assumed). The dashed line is the extrapolation of the classical Hall resistance seen in reverse fields. Taken from H. van Houten et al., Phys.\ Rev.\ B {\bf 39}, 8556 (1989).
\label{fig53}
}
\end{figure}

In Fig.\ \ref{fig53} (top) we show such a classical focusing spectrum, calculated for
parameters corresponding to the experiment discussed later $(L=3.0\,\mu \mathrm{m}$,
$k_{\mathrm{F}}=1.5\times 10^{8}\,\mathrm{m}^{-1})$. The spectrum consists of equidistant focusing peaks of
approximately equal magnitude superimposed on the Hall resistance (dashed
line). The $p\mathrm{th}$ peak is due to electrons injected perpendicularly to the
boundary that have made $p-1$ specular reflections between injector and
collector. Such a classical focusing spectrum is commonly observed in
metals,\cite{ref351,ref352} albeit with a decreasing height of subsequent peaks because of
partially diffuse scattering at the metal surface. Note that the peaks occur in
one field direction only. In reverse fields the focal points are at the wrong side
of the injector for detection, and the normal Hall resistance is obtained. The
experimental result for a 2DEG is shown in the bottom half of Fig.\ \ref{fig53} (trace a;
trace b is discussed later). A series of five focusing peaks is evident at the
expected positions. The observation of multiple focusing peaks immediately
implies that the electrostatically defined 2DEG boundary scatters predominantly specularly. (This finding\cite{ref59} is supported by the magnetoresistance experiments of Thornton et a.\cite{ref107} in a narrow split-gate channel; cf.\ Section \ref{sec5}.) Figure \ref{fig53} is obtained in a measuring configuration (inset) in which an imaginary line connecting the voltage probes crosses that between the current source and drain. This is the configuration for a generalized Hall resistance measurement. If the crossing is avoided, one measures a longitudinal resistance, which shows the focusing peaks without a superimposed Hall slope. This longitudinal resistance periodically becomes negative. This is a classical result\cite{ref80} of magnetic defocusing, which causes the probability density near the point contact voltage probe to be reduced with respect to the spatially averaged probability density that determines the voltage on the wide voltage probe (cf.\ the regions of reduced density between lines of focus in Fig.\ \ref{fig52}).

On the experimental focusing peaks a fine structure is resolved at low temperatures (below 1 K). The fine structure is well reproducible but sample-dependent. A nice demonstration of the reproducibility of the fine structure is obtained upon interchanging current and voltage leads, so that the injector becomes the collector, and vice versa. The resulting focusing spectrum shown in Fig.\ \ref{fig53} (trace b) is almost the precise mirror image of the original one (trace a), although this particular device had a strong asymmetry in the widths of injector and collector. The symmetry in the focusing spectra is an example of the general reciprocity relation (\ref{eq12.16}). If one applies the B\"{u}ttiker equations (\ref{eq12.12}) to the electron focusing geometry (as is done in Section \ref{sec19}), one finds that the ratio of collector voltage $V_{\rm c}$ to injector current $I_{\rm i}$ is given by
\be
\frac{V_{\rm c}}{I_{\rm i}}=\frac{2e^{2}}{h}\frac{T_{{\rm i}\rightarrow{\rm c}}}{G_{\rm i}G_{\rm c}},\label{eq14.2}
\ee
where $T_{{\rm i}\rightarrow{\rm c}}$ is the transmission probability from injector to collector, and $G_{\rm i}$ and $G_{\rm c}$ are the conductances of the injector and collector point contact. Since $T_{{\rm i}\rightarrow{\rm c}}(B)=T_{{\rm c}\rightarrow{\rm i}}(-B)$ and $G(B) = G( - B)$, this expression for the focusing spectrum is manifestly symmetric under interchange of injector and collector with reversal of the magnetic field.

\begin{figure}
\centerline{\includegraphics[width=8cm]{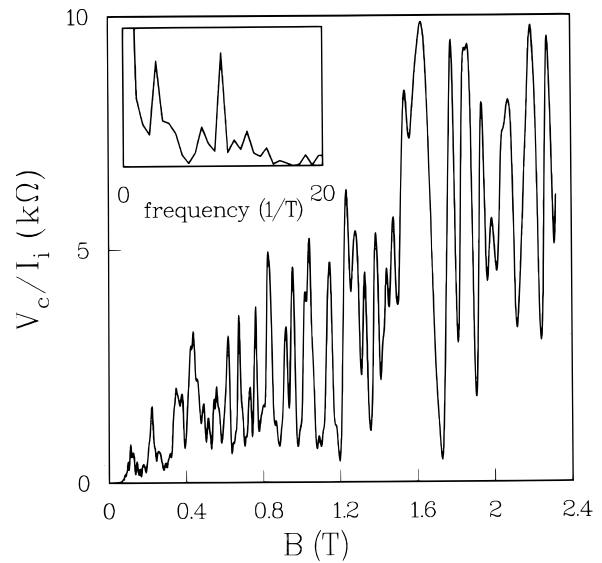}}
\caption{
Experimental electron focusing spectrum over a larger field range and for very narrow point contacts (estimated width 20--40 nm; $T= 50\,{\rm mK}$, $L= 1.5\,\mu{\rm m}$). The inset gives the Fourier transform for $B\geq 0.8\,{\rm T}$. The high-field oscillations have the same dominant periodicity as the low-field focusing peaks, but with a much larger amplitude. Taken from H. van Houten et al., Phys.\ Rev.\ B {\bf 39}, 8556 (1989).
\label{fig54}
}
\end{figure}

The fine structure on the focusing peaks in Fig.\ \ref{fig53} is the first indication that electron focusing in a 2DEG is qualitatively different from the corresponding experiment in metals. At higher magnetic fields the resemblance to the classical focusing spectrum is lost; see Fig.\ \ref{fig54}. A Fourier transform of the spectrum for $B \geq 0.8\,{\rm T}$ (inset in Fig.\ \ref{fig54}) shows that the large-amplitude high­field oscillations have a dominant periodicity of 0.1 T, which is approximately the same as the periodicity $B_{\rm focus}$ of the much smaller focusing peaks at low magnetic fields ($B_{\rm focus}$ in Fig.\ \ref{fig54} differs from Fig.\ \ref{fig53} because of a smaller
$L=1.5\,\mu \mathrm{m})$. This dominant periodicity can be explained in terms of quantum
interference between the different skipping orbits from injector to collector or
in terms of interference of coherently excited edge channels, as we discuss in
the following subsection. The experimental implication is that the injector
acts as a {\it coherent\/} point source with the coherence maintained over a distance
of several microns to the collector.

\subsubsection{\label{sec14b} Theory}

To explain the characteristic features of the coherent electron focusing
experiments we have described, we must go beyond the classical description.\cite{ref80,ref347} As discussed in Section \ref{sec12}, quantum ballistic transport along
the 2DEG boundary in a magnetic field takes place via magnetic edge states,
which form the propagating modes at the Fermi level. Since the injector has a
width below $\lambda_{\mathrm{F}}$, it excites these modes coherently. For $k_{\mathrm{F}}L\gg  1$ the interference of modes at the collector is dominated by their rapidly varying phase
factors $ \exp(ik_{n}L)$. The wave number $k_{n}$ corresponds classically to the
separation of the center of the cyclotron orbit from the 2DEG boundary [Eq.\
(\ref{eq12.5})]. In the Landau gauge ${\bf A} =(0, Bx, 0)$ (with the axis chosen as in Fig.\ \ref{fig52})
one has $k_{n}=k_{\mathrm{F}}\sin\alpha_{n}$, where $\alpha$ is the angle with the $x$-axis under which the
cyclotron orbit is reflected from the boundary. The quantized values $\alpha_{n}$ follow
in this semiclassical description from the Bohr-Sommerfeld quantization
rule (\ref{eq12.6}) that the flux enclosed by the cyclotron orbit and the boundary
equals $(n- \frac{1}{4})h/e$ [the phase shift $\gamma$ in Eq.\ (\ref{eq12.6}) equals $\pi/2$ for an edge state at
an infinite barrier potential]. Simple geometry shows that this requires that
\be
\frac{\pi}{2}-\alpha_{n}-\frac{1}{2}\sin 2\alpha_{n}=\frac{2\pi}{k_{\mathrm{F}}l_{\mathrm{cycl}}}\left(n-\frac{1}{4}\right),\;\; n=1,2, \ldots, N.   \label{eq14.3}
\ee

\begin{figure}
\centerline{\includegraphics[width=8cm]{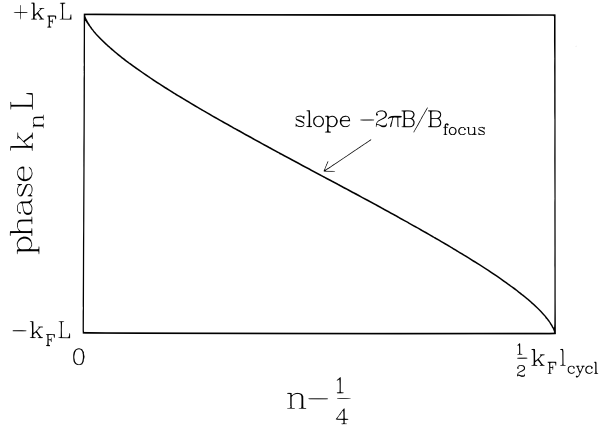}}
\caption{
Phase $k_{n}L$ of the edge channels at the collector, calculated from Eq.\ (\ref{eq14.3}). Note the domain of approximately linear $n$-dependence of the phase, responsible for the oscillations with $B_{\rm focus}$-periodicity. Taken from H. van Houten et al., Phys.\ Rev.\ B {\bf 39}, 8556 (1989).
\label{fig55}
}
\end{figure}

\begin{figure}
\centerline{\includegraphics[width=8cm]{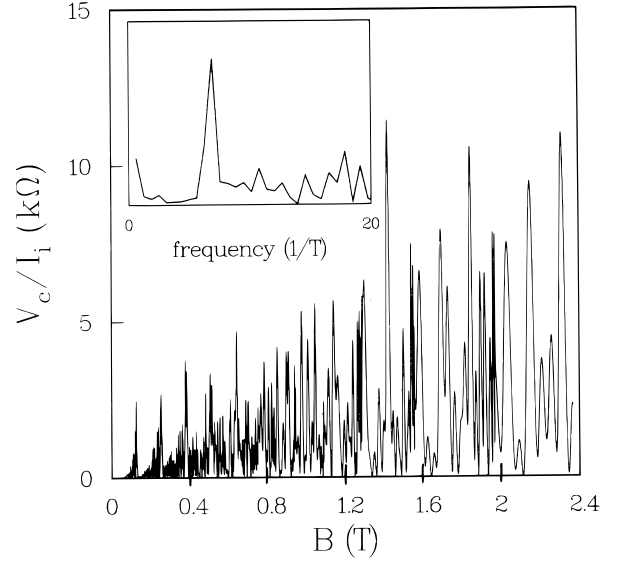}}
\caption{
Focusing spectrum calculated from Eq.\ (\ref{eq14.5}), for parameters corresponding to the experimental Fig.\ \ref{fig54}. The inset shows the Fourier transform for $B\geq 0.8\,{\rm T}$. Infinitesimally small point contact widths are assumed in the calculation. Taken from C. W. J. Beenakker et al., Festk\"{o}rperprobleme {\bf 9}, 299 (1989).
\label{fig56}
}
\end{figure}

As plotted in Fig.\ \ref{fig55}, the dependence on $n$ of the phase $k_{n}L$ is close to
linear in a broad interval. This also follows from expansion of Eq.\ (\ref{eq14.3})
around $\alpha_{n}=0$, which gives
\be
k_{n}L= \mathrm{constant}-2\pi n\frac{B}{B_{\mathrm{focus}}}+k_{\mathrm{F}}L\times {\rm order} \left( \frac{N-2n}{N}\right)^{3}. \label{eq14.4}
\ee
If $B/B_{\mathrm{focus}}$ is an integer, a fraction of order $(1/k_{\mathrm{F}}L)^{1/3}$ of the $N$ edge states
interfere constructively at the collector. Because of the $1/3$ power, this is a
substantial fraction even for the large $k_{\mathrm{F}}L\sim 10^{2}$ of the experiment. The
resulting mode interference oscillations with $B_{\mathrm{focus}}$-periodicity can become
much larger than the classical focusing peaks. This has been shown in Refs.\
\onlinecite{ref347} and \onlinecite{ref80}, where the transmission probability $T_{\mathrm{i}\rightarrow \mathrm{c}}$ was calculated in the
WKB approximation with neglect of the finite width of the injector and
detector. From Eq.\ (\ref{eq14.2}) the focusing spectrum is then obtained in the form
\be
\frac{V_{\mathrm{c}}}{I_{\mathrm{i}}}=\frac{h}{2e^{2}}\left|\frac{1}{N}\sum_{n=1}^{N}\mathrm{e}^{ik_{n}L}\right|^{2}, \label{eq14.5}
\ee 
which is plotted in Fig.\ \ref{fig56} for parameter values corresponding to the
experimental Fig.\ \ref{fig54}. The inset shows the Fourier transform for $B\geq 0.8\,{\rm T}$.

There is no detailed one-to-one correspondence between the experimental
and theoretical spectra. No such correspondence was to be expected in view
of the sensitivity of the experimental spectrum to small variations in the
voltage on the gate defining the point contacts and the 2DEG boundary.
Those features of the experimental spectrum that are insensitive to the precise
measurement conditions are, however, well reproduced by the calculation:
We recognize in Fig.\ \ref{fig56} the low-field focusing peaks and the large-amplitude
high-field oscillations with the same $B_{\mathrm{focus}}$-periodicity. The high-field oscillations range from about 0 to $10\, \mathrm{k}\Omega$ in both theory and experiment. The
maximum amplitude is not far below the theoretical upper bound of
$h/2e^{2}\approx 13\,\mathrm{k}\Omega$, which follows from Eq.\ (\ref{eq14.5}) if we assume that {\it all\/} the modes
interfere constructively. This indicates that a {\it maximal phase coherence\/} is
realized in the experiment and implies that the experimental injector and
collector point contacts resemble the idealized point source detector in the
calculation.

\subsubsection{\label{sec14c} Scattering and electron focusing}

\begin{figure}
\centerline{\includegraphics[width=8cm]{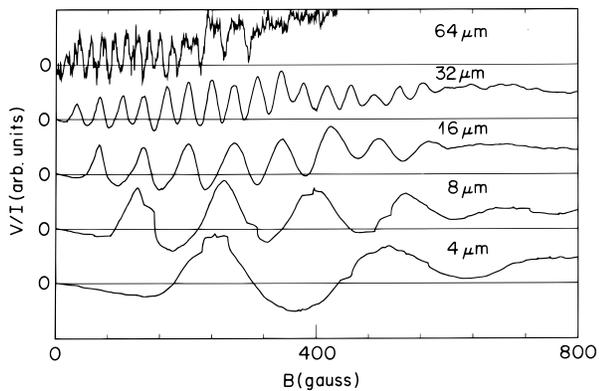}}
\caption{
Experimental electron focusing spectra (in the generalized longitudinal resistance configuration) at 0.3 K for five different injector-collector separations in a very high mobility material. The vertical scale varies among the curves. Taken from J. Spector et al., Surf.\ Sci.\ {\bf 228}, 283 (1990).
\label{fig57}
}
\end{figure}

\begin{figure}
\centerline{\includegraphics[width=8cm]{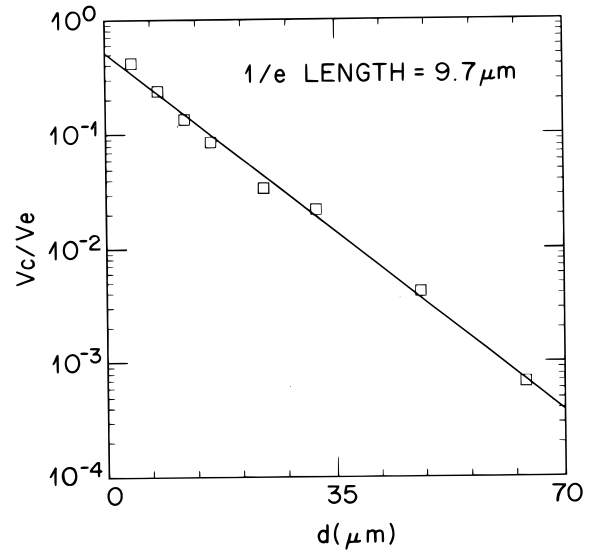}}
\caption{
Exponential decay of the oscillation amplitude of the collector voltage (normalized by the injector voltage) as a function of injector-collector separation $d$ (denoted by $L$ in the text). Taken from J. Spector et al., Surf.\ Sci.\ {\bf 228}, 283 (1990).
\label{fig58}
}
\end{figure}

Scattering events other than specular boundary scattering can be largely
ignored for the relatively small point contact separations $L\leq 3\,\mu \mathrm{m}$ in the
experiments discussed earlier.\cite{ref59,ref80} (any other inelastic or elastic scattering
events would have been detected as a reduction of the oscillations with $B_{\mathrm{focus}}$-periodicity below the theoretical estimate). Spector et al.\cite{ref349}  have repeated the
experiments for larger $L$ to study scattering processes in an ultrahigh
mobility\cite{ref353,ref354} 2DEG ($\mu_{\mathrm{e}}=5.5\times 10^{6}\,\mathrm{cm}^{2}/\mathrm{V}\mathrm{s}$). They used relatively wide
point contacts (about 1 $\mu \mathrm{m}$) so that electron focusing was in the classical
regime. In Fig.\ \ref{fig57} we reproduce their experimental results for point contact
separations up to 64 $\mu \mathrm{m}$. The peaks in the focusing spectrum for a given $L$
have a roughly constant amplitude, indicating that scattering at the boundary is mostly specular rather than diffusive --- in agreement with the experiments of Ref.\ \onlinecite{ref59}. Spector et al.\cite{ref349} find that the amplitude of the focusing
peaks decreases exponentially with increasing $L$, due to scattering in the
electron gas (see Fig.\ \ref{fig58}). The decay $\exp(- L/L_{0})$ with $L_{0}\approx 10\,\mu \mathrm{m}$ implies an
effective mean free path (measured along the arc of the skipping orbits) of
$L_{0}\pi/2\approx 15\,\mu \mathrm{m}$. This is smaller than the transport mean free path derived
from the conductivity by about a factor of 2, which may point to a greater
sensitivity of electron focusing to forward scattering.

\begin{figure}
\centerline{\includegraphics[width=8cm]{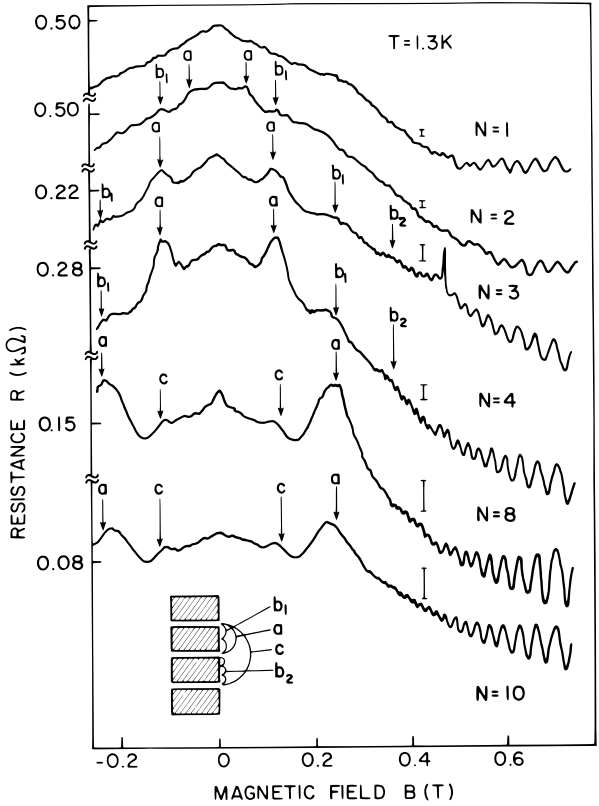}}
\caption{
Magnetoresistance of $N$ constrictions in parallel at 1.3 K. The arrows indicate the oscillations due to electron focusing, according to the mechanisms illustrated in the inset. The resistance scale is indicated by $10\,\Omega$ bars. Taken from K. Nakamura et al., Appl.\ Phys.\ Lett.\ {\bf 56}, 385 (1990).
\label{fig59}
}
\end{figure}

Electron focusing by a magnetic field may also play a role in geometries
other than the double-point contact geometry of Fig.\ \ref{fig52}. One example is
mentioned in the context of junction scattering in a cross geometry in Section
\ref{sec16}. Another example is the experiment by Nakamura et al.\cite{ref108} on the
magnetoresistance of equally spaced narrow channels in parallel (see Fig.\ \ref{fig59}).
Resistance peaks occur in this experiment when electrons that are transmitted through one of the channels are focused back through another
channel. The resistance peaks occur at $B=(n/m)B_{\mathrm{focus}}$, where $B_{\mathrm{focus}}$ is given
by Eq.\ (\ref{eq14.1}) with $L$ the spacing of adjacent channels. The identification of the
various peaks in Fig.\ \ref{fig59} is given in the inset. Nakamura et al.\cite{ref108} conclude
from the rapidly diminishing height of consecutive focusing peaks (which
require an increasing number of specular reflections) that there is a large
probability of diffuse boundary scattering. The reason for the difference with
the experiments discussed previously is that the boundary in the experiment
of Fig.\ \ref{fig59} is defined by focused ion beam lithography, rather than electrostatically by means of a gate. As discussed in Section \ref{sec5}, the former technique may
introduce a considerable boundary roughness.

Electron focusing has been used by Williamson et al.\cite{ref355}  to study scattering
processes for ``hot'' electrons, with an energy in excess of the Fermi energy,
and for ``cool'' holes, or empty states in the conduction band below the Fermi
level (see Ref.\ \onlinecite{ref307} for a review). An interesting aspect of hot-electron focusing
is that it allows a measurement of the local electrostatic potential drop across
a current-carrying quantum point contact,\cite{ref355} something that is not possible
using conventional resistance measurements, where the sum of electrostatic
and chemical potentials is measured. The importance of such alternative
techniques to study electrical conduction has been stressed by Landauer.\cite{ref356}

\subsection{\label{sec15} Collimation}

The subject of this section is the collimation of electrons injected by a
point contact\cite{ref329} and its effect on transport measurements in geometries
involving two opposite point contacts.\cite{ref327,ref357}  Collimation (i.e., the narrowing
of the angular injection distributions) follows from the constraints on the
electron momentum imposed by the potential barrier in the point contact
({\it barrier collimation}), and by the gradual widening of the point contact at its
entrance and exit ({\it horn collimation}). We summarize the theory in Section \ref{sec15a}.
The effect was originally proposed\cite{ref329} to explain the remarkable observation
of Wharam et al.\cite{ref357} that the series resistance of two opposite point contacts is
considerably less than the sum of the two individual resistances (Section \ref{sec15c}).
A direct experimental proof of collimation was provided by Molenkamp et
al.,\cite{ref327} who measured the deflection of the injected beam of electrons in a
magnetic field (Section \ref{sec15b}). A related experiment by Sivan et al.,\cite{ref350} aimed at
the demonstration of the focusing action of an electrostatic lens, is also
discussed in this subsection. The collimation effect has an importance in
ballistic transport that goes beyond the point contact geometry. It will be
shown in Section \ref{sec16} that the phenomenon is at the origin of a variety of
magnetoresistance anomalies in narrow multiprobe conductors.\cite{ref358,ref359,ref360}

\subsubsection{\label{sec15a} Theory}

\begin{figure}
\centerline{\includegraphics[width=8cm]{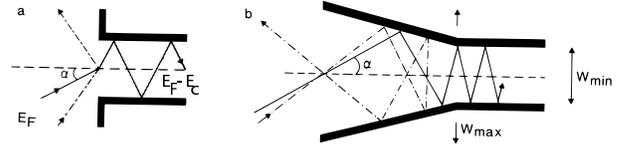}}
\caption{
Illustration of the collimation effect for an abrupt constriction (a) containing a potential barrier of height $E_{\rm c}$ and for a horn-shaped constriction (b) that is flared from a width $W_{\rm min}$ to $W_{\rm max}$. The dash-dotted trajectories approaching at an angle $\alpha$ outside the injection/acceptance cone are reflected. Taken from H. van Houten and C. W. J. Beenakker, in ``Nanostructure Physics and Fabrication'' (M. Reed and W. P. Kirk, eds.). Academic, New York, 1989.
\label{fig60}
}
\end{figure}

Since collimation follows from classical mechanics, a semiclassical theory
is sufficient to describe the essential phenomena, as we now discuss (following
Refs.\ \onlinecite{ref329} and \onlinecite{ref311}). Semiclassically, collimation results from the adiabatic
invariance of the product of channel width $W$ and absolute value of the
transverse momentum $\hbar k_{y}$ (this product is proportional to the action for
motion transverse to the channel\cite{ref361}). Therefore, if the electrostatic potential
in the point contact region is sufficiently smooth, the quantity $S=|k_{y}|W$ is
approximately constant from point contact entrance to exit. Note that $S/\pi$
corresponds to the quantum mechanical 1D subband index $n$. The quantum
mechanical criterion for adiabatic transport was derived by Yacoby and
Imry\cite{ref326} (see Section \ref{sec13}). As was discussed there, adiabatic transport breaks
down at the exit of the point contact, where it widens abruptly into a 2DEG
of essentially infinite width. Collimation reduces the {\em injection/acceptance cone\/}
of the point contact from its original value of $\pi$ to a value of $2\alpha_{\max}$. This effect
is illustrated in Fig.\ \ref{fig60}. Electrons incident at an angle $|\alpha|>\alpha_{\max}$ from normal
incidence are reflected. (The geometry of Fig.\ \ref{fig60}b is known in optics as a
{\it conical\/} reflector.\cite{ref362}.) Vice versa, all electrons leave the constriction at an angle
$|\alpha|<\alpha_{\max}$ (i.e., the injected electrons form a collimated beam of angular
opening $2\alpha_{\max}$).

To obtain an analytic expression for the collimation effect, we describe the
shape of the potential in the point contact region by three parameters: $W_{\min}$,
$W_{\max}$, and $E_{\mathrm{c}}$ (see Fig.\ \ref{fig60}). We consider the case that the point contact has its
minimal width $W_{\min}$ at the point where the barrier has its maximal height $E_{\mathrm{c}}$
above the bottom of the conduction band in the broad regions. At that point
the largest possible value of $S$ is
\[
S_{1}\equiv(2m/\hbar^{2})^{1/2}(E_{\mathrm{F}}-E_{\mathrm{c}})^{1/2}W_{\min}.
\]
We assume that adiabatic transport (i.e., $S=$ constant) holds up to a point of
zero barrier height and maximal width $W_{\max}$. The abrupt separation of
adiabatic and nonadiabatic regions is a simplification that can be, and has
been, tested by numerical calculations (see below). At the point contact exit,
the largest possible value of $S$ is
\[
S_{2}\equiv(2m/\hbar^{2})^{1/2}(E_{\mathrm{F}})^{1/2}\sin\alpha_{\max}\,W_{\max}.
\]
The invariance of $S$ implies that $S_{1}=S_{2}$; hence,
\be
\alpha_{\max}=\arcsin\left(\frac{1}{f}\right);\;\; f \equiv\left(\frac{E_{\mathrm{F}}}{E_{\mathrm{F}}-E_{\mathrm{c}}}\right)^{1/2}\frac{W_{\max}}{W_{\min}}. \label{eq15.1}
\ee
The {\it collimation factor\/} $f\geq 1$ is the product of a term describing the
collimating effect of a barrier of height $E_{\mathrm{c}}$ (barrier collimation) and a term
describing collimation due to a gradual widening of the point contact width
from $W_{\min}$ to $W_{\max}$ (horn collimation). In the adiabatic approximation, the
angular injection distribution $P(\alpha)$ is proportional to $\cos\alpha$ with an abrupt
truncation at $\pm\alpha_{\max}$. The cosine angular dependence follows from the cosine
distribution of the incident flux in combination with time-reversal symmetry
and is thus not affected by the reduction of the injection/acceptance cone.
We therefore conclude that in the adiabatic approximation $P(\alpha)$ (normalized
to unity) is given by
\be
P( \alpha)=\left\{\begin{array}{ll}
\frac{1}{2}f\cos\alpha&{\rm if}\;\;|\alpha|<\arcsin(1/f),\\
=0,&{\rm otherwise}.
\end{array}\right. \label{eq15.2}
\ee
We defer to Section \ref{sec15b} a comparison of the analytical result (\ref{eq15.2}) with a
numerical calculation.

Barrier collimation does not require adiabaticity. For an abrupt barrier,
collimation simply results from transverse momentum conservation, as in
Fig.\ \ref{fig60}a, leading directly to Eq.\ (\ref{eq15.2}). (The total external reflection at an
abrupt barrier for trajectories outside the collimation cone is similar to the
optical effect of total internal reflection at a boundary separating a region of
high refractive index from a region of small refractive index; see the end of
Section \ref{sec15b}.) A related collimation effect resulting from transverse momentum conservation occurs if electrons tunnel through a potential barrier.
Since the tunneling probability through a high potential barrier is only
weakly dependent on energy, it follows that the strongest collimation is to be
expected if the barrier height equals the Fermi energy. On lowering the
barrier below $E_{\mathrm{F}}$ ballistic transport over the barrier dominates, and the
collimation cone widens according to Eq.\ (\ref{eq15.2}). A quantum mechanical
calculation of barrier collimation may be found in Ref.\ \onlinecite{ref363}.

The injection distribution (\ref{eq15.2}) can be used to obtain (in the semiclassical
limit) the direct transmission probability $T_{\mathrm{d}}$ between two opposite identical
point contacts separated by a large distance $L$. To this end, first note that
$T_{\mathrm{d}}/N$ is the fraction of the injected current that reaches the opposite point
contact (since the transmission probability through the first point contact is
$N$, for $N$ occupied subbands in the point contact). Electrons injected within a
cone of opening angle $W_{\max}/L$ centered at $\alpha=0$ reach the opposite point
contact and are transmitted. If this opening angle is much smaller than the
total opening angle $2\alpha_{\max}$ of the beam, then the distribution function $P(\alpha)$ can
be approximated by $P(0)$ within this cone. This approximation requires
$W_{\max}/L\ll 1/f$, which is satisfied experimentally in devices with a sufficiently
large point contact separation. We thus obtain $T_{\mathrm{d}}/N=P(0)W_{\max}/L$, which,
using Eq.\ (\ref{eq15.2}), can be written as\cite{ref329}
\be
T_{\mathrm{d}}=f(W_{\max}/2L)N. \label{eq15.3}
\ee 
This simple analytical formula can be used to describe the experiments on
transport through identical opposite point contacts in terms of one empirical
parameter $f$, as discussed in the following subsections.

\subsubsection{\label{sec15b} Magnetic deflection of a collimated electron beam}

\begin{figure}
\centerline{\includegraphics[width=8cm]{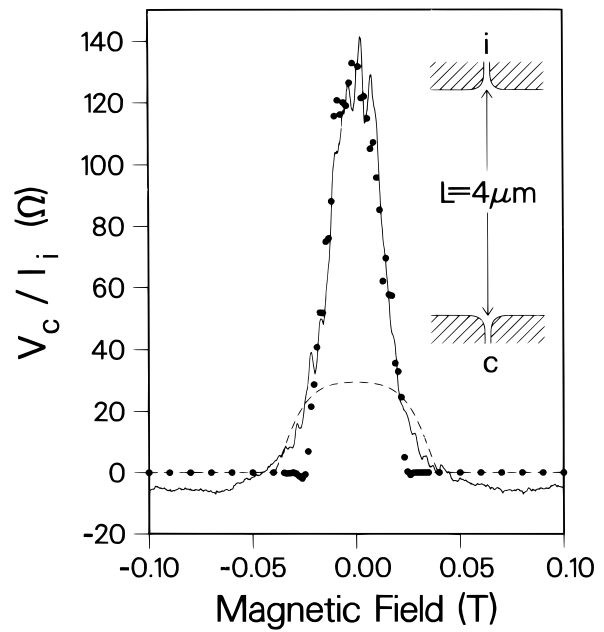}}
\caption{
Detection of a collimated electron beam over a distance of $4\,\mu{\rm m}$. In this four-terminal measurement, two ohmic contacts to the 2DEG region between the point contacts are used: One of these acts as a drain for the current $I_{\rm i}$ through the injector, and the other is used as a zero­reference for the voltage $V_{\rm c}$ on the collector. The drawn curve is the experimental data at $T = 1.8\,{\rm K}$. The black dots are the result of a semiclassical simulation, using a hard-wall potential with contours as shown in the inset. The dashed curve results from a simulation without collimation (corresponding to rectangular corners in the potential contour). Taken from L. W. Molenkamp et al., Phys.\ Rev.\ B {\bf 41}, 1274 (1990).
\label{fig61}
}
\end{figure}

A method\cite{ref311,ref329} to sensitively detect the collimated electron beam
injected by a point contact is to sweep the beam past a second opposite point
contact by means of a magnetic field. The geometry is shown in Fig.\ \ref{fig61} (inset).
The current $I_{\mathrm{i}}$ through the injecting point contact is drained to ground at one
or two (the difference is not essential) ends of the 2DEG channel separating
the point contacts. The opposite point contact, the collector, serves as a
voltage probe (with the voltage $V_{\mathrm{c}}$ being measured relative to ground). In the
case that both ends of the 2DEG channel are grounded, the collector voltage
divided by the injected current is given by
\be
\frac{V_{\mathrm{c}}}{I_{\mathrm{i}}}=\frac{1}{G}\frac{T_{\mathrm{d}}}{N},\;\; T_{\mathrm{d}}\ll N, \label{eq15.4}
\ee 
with $G=(2e^{2}/h)N$ the two-terminal conductance of the individual point
contact (both point contacts are assumed to be identical) and $T_{\mathrm{d}}$ the direct
transmission probability between the two point contacts calculated in
Section \ref{sec15a}. Equation (\ref{eq15.4}) can be obtained from the Landauer-B\"{u}ttiker
formalism (as done in Ref.\ \onlinecite{ref311}) or simply by noting that the current $I_{\mathrm{i}}T_{\mathrm{d}}/N$
incident on the collector has to be counterbalanced by an equal outgoing
current $GV_{\mathrm{c}}$. In the absence of a magnetic field, we obtain [using Equation
(\ref{eq15.3}) for the direct transmission probability]
\be
\frac{V_{\mathrm{c}}}{I_{\mathrm{i}}}=\frac{h}{2e^{2}}f^{2}\frac{\pi}{2k_{\mathrm{F}}L}, \label{eq15.5}
\ee 
where $k_{\mathrm{F}}$ is the Fermi wave vector in the region between the point contacts. In
an experimental situation $L$ and $k_{\mathrm{F}}$ are known, so the collimation factor $f$ can
be directly determined from the collector voltage by means of Eq.\ (\ref{eq15.5}).

The result (\ref{eq15.5}) holds in the absence of a magnetic field. A small magnetic
field $B$ will deflect the collimated electron beam past the collector. Simple
geometry leads to the criterion $L/2l_{\mathrm{cycl}}=\alpha_{\max}$ for the cyclotron radius at
which $T_{\mathrm{d}}$ is reduced to zero by the Lorentz force (assuming that $L\gg  W_{\max}$).
One would thus expect to see in $V_{\mathrm{c}}/I_{\mathrm{i}}$ a peak around zero field, of height given
by Eq.\ (\ref{eq15.5}) and of width
\be
\Delta B=(4\hbar k_{\rm F}/eL)\arcsin(1/f), \label{eq15.6}
\ee 
according to Eq.\ (\ref{eq15.1}).

In Fig.\ \ref{fig61} this collimation peak is shown (solid curve), as measured by
Molenkamp et al.\cite{ref327} at $T=1.2\,\mathrm{K}$ in a device with a $L=4.0$-$\mu \mathrm{m}$ separation
between injector and collector. In this measurement only one end of the
region between the point contacts was grounded --- a measurement configuration referred to in narrow Hall bar geometries as a {\it bend resistance\/}
measurement\cite{ref289,ref364} (cf.\ Section \ref{sec16}). One can show, using the Landauer-B\"{u}ttiker formalism,\cite{ref5} that the height of the collimation peak is still given by
Eq.\ (\ref{eq15.5}) if one replaces\cite{ref327} $f^{2}$ by $f^{2}- \frac{1}{2}$. The expression (\ref{eq15.6}) for the width
is not modified. The experimental result in Fig.\ \ref{fig61} shows a peak height of
$\approx 150\,\Omega$ (measured relative to the background resistance at large magnetic
fields). Using $L=4.0\,\mu \mathrm{m}$ and the value $k_{\mathrm{F}}=1.1\times 10^{8}\,\mathrm{m}^{-1}$ obtained from
Hall resistance measurements in the channel between the point contacts, one
deduces a collimation factor $f\approx 1.85$. The corresponding opening angle of the
injection/acceptance cone is $2\alpha_{\max}\approx 65^{\circ}$. The calculated value of $f$ would
imply a width $\Delta B\approx 0.04\,\mathrm{T}$, which is not far from the measured full width at
half maximum of $\approx 0.03\,{\rm T}$.

\begin{figure}
\centerline{\includegraphics[width=8cm]{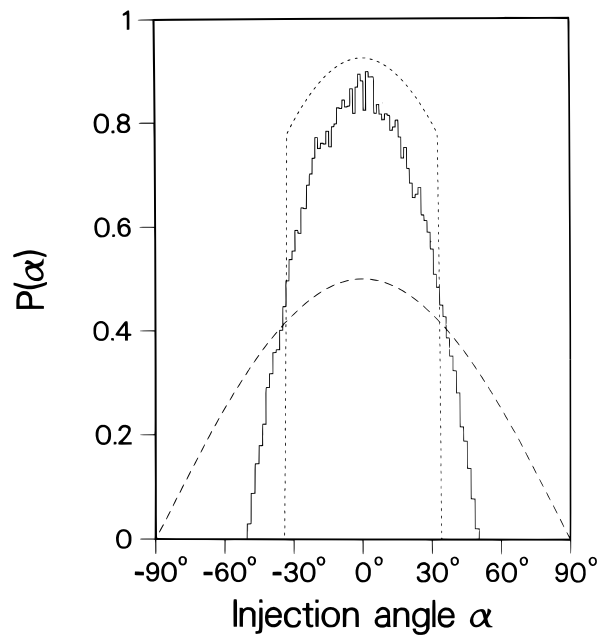}}
\caption{
Calculated angular injection distributions in zero magnetic field. The solid histogram is the result of a simulation of the classical trajectories at the Fermi energy in the geometry shown in the inset of Fig.\ \ref{fig61}. The dotted curve follows from the adiabatic approximation (\ref{eq15.2}), with the experimental collimation factor $f = 1.85$. The dashed curve is the cosine distribution in the absence of any collimation. Taken from L. W. Molenkamp et al., Phys.\ Rev.\ B. {\bf 41}, 1274 (1990).
\label{fig62}
}
\end{figure}

The experimental data in Fig.\ \ref{fig61} are compared with the result\cite{ref327} from a
numerical simulation of classical trajectories of the electrons at the Fermi
level (following the method of Ref.\ \onlinecite{ref329}). This semiclassical calculation was
performed in order to relax the assumption of adiabatic transport in the point
contact region, and of small $T_{\mathrm{d}}/N$, on which  Eqs.\ (\ref{eq15.3}) and (\ref{eq15.5}) are based.
The dashed curve is for point contacts defined by hardwall contours with
straight corners (no collimation); the dots are for the smooth hardwall
contours shown in the inset, which lead to collimation via the horn effect (cf.\
Fig.\ \ref{fig60}b; the barrier collimation of Fig.\ \ref{fig60}a is presumably unimportant at the
small gate voltage used in the experiment and is not taken into account in the
numerical simulation). The angular injection distributions $P(\alpha)$ that follow
from these numerical simulations are compared in Fig.\ \ref{fig62} (solid histogram)
with the result (\ref{eq15.2}) from the adiabatic approximation for $f=1.85$ (dotted
curve). The uncollimated distribution $P(\alpha)=(\cos\alpha)/2$ is also shown for
comparison (dashed curve). Taken together, Figs.\ \ref{fig61} and \ref{fig62} unequivocally
demonstrate the importance of collimation for the transport properties, as
well as the adequateness of the adiabatic approximation as an estimator of
the collimation cone.

Once the point contact width becomes less than a wavelength, diffraction
inhibits collimation of the electron beam. In the limit $k_{\mathrm{F}}W\ll 1$, the injection
distribution becomes proportional to $\cos^{2}\alpha$ for all $\alpha$, independent of the
shape of the potential in the point contact region.\cite{ref80,ref313}  The coherent electron
focusing experiments\cite{ref59,ref80} discussed in Sections \ref{sec14a} and \ref{sec14b} were performed
in this limit.

\begin{figure}
\centerline{\includegraphics[width=6cm]{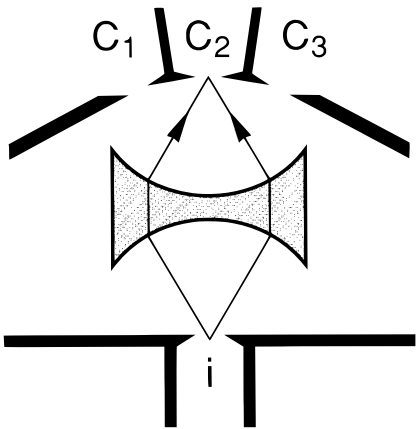}}
\caption{
Electrostatic focusing onto a collector ($C_{2}$) of an injected electron beam (at i) by means of a concave lens corresponding to a region of reduced electron density. Focusing in such an arrangement was detected experimentally.\cite{ref350}
\label{fig63}
}
\end{figure}

We conclude this subsection by briefly discussing an alternative way to
increase the transmission probability between two opposite point contacts,
which is {\it focusing\/} of the injected electron beam onto the collector. Magnetic
focusing, discussed in Section \ref{sec14} for adjacent point contacts, cannot be used
for opposite point contacts in two dimensions (unlike in three dimensions,
where a magnetic field along the line connecting the point contacts will focus
the beam\cite{ref296}). A succesful demonstration of {\it electrostatic\/} focusing was recently
reported by Sivan et al.\ and by Spector et al.\cite{ref350} The focusing is achieved by
means of a potential barrier of a concave shape, created as a region of reduced
density in the 2DEG by means of a gate between the injector and the collector
(see Fig.\ \ref{fig63}). A focusing lens for electrons is {\it concave\/} because electrons
approaching a potential barrier are deflected in a direction perpendicular to
the normal. This is an amusing difference with light, which is deflected toward
the normal on entering a more dense medium, so an optical focusing lens is
{\it convex}. The different dispersion laws are the origin of this different behavior
of light and electrons.\cite{ref350}

\subsubsection{\label{sec15c} Series resistance}

The first experimental study of ballistic transport through two opposite
point contacts was carried out by Wharam et al.,\cite{ref357} who discovered that the
series resistance is considerably less than the sum of the two individual
resistances. Sugsequent experiments confirmed this result.\cite{ref365,ref366} The
theoretical explanation\cite{ref329} of this observation is that collimation of the
electrons injected by a point contact enhances the direct transmission
probability through the opposite point contact, thereby significantly reducing the series resistance below its ohmic value. We will discuss the transport
and magnetotransport in this geometry. We will not consider the alternative
geometry of two adjacent point contacts in parallel (studied in Refs.\ \onlinecite{ref367,ref368,ref369}).
In that geometry the collimation effect cannot enhance the coupling of
the two point contacts, so only small deviations from Ohm's law are to be
expected.

\begin{figure}
\centerline{\includegraphics[width=8cm]{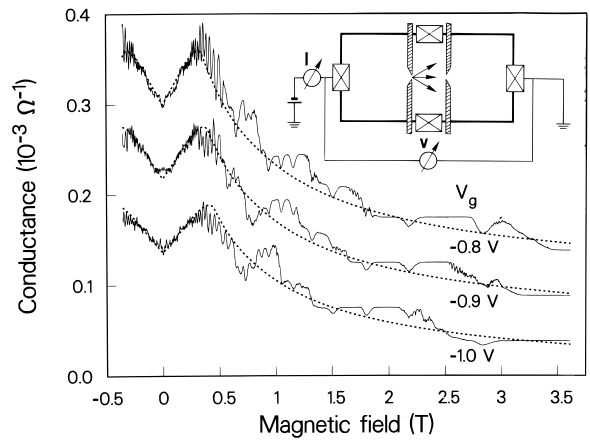}}
\caption{
Magnetic field dependence of the series conductance of two opposite point contacts (measured as shown in the inset; the point contact separation is $L = 1.0\,\mu{\rm m}$) for three different values of the gate voltage (solid curves) at $T = 100\,{\rm mK}$. For clarity, subsequent curves from bottom to top are offset by $0.5 \times 10^{-4}\,\Omega^{-1}$, with the lowest curve shown at its actual value. The dotted curves are calculated from Eqs.\ (\ref{eq15.10}) and (\ref{eq10.8}), with the point contact width as adjustable parameter. Taken from A. A. M. Staring et al., Phys.\ Rev.\ B. {\bf 41}, 8461 (1990).
\label{fig64}
}
\end{figure}

The expression for the two-terminal series resistance of two identical
opposite point contacts in terms of the direct transmission probability can be
obtained from the Landauer-B\"{u}ttiker formalism,\cite{ref5} as was done in Ref.\ \onlinecite{ref329}.
We give here an equivalent, somewhat more intuitive derivation. Consider
the geometry shown in Fig.\ \ref{fig64} (inset). A fraction $T_{\mathrm{d}}/N$ of the current $GV$
injected through the first point contact by the current source is directly
transmitted through the second point contact (and then drained to ground).
Here $G=(2e^{2}/h)N$ is the conductance of the individual point contact, and $V$
is the source-drain voltage. The remaining fraction $1 - T_{\mathrm{d}}/N$ equilibrates in
the region between the point contacts, as a result of inelastic scattering (elastic
scattering is sufficient if phase coherence does not play a role). Since that
region cannot drain charge (the attached contacts are not connected to
ground), these electrons will eventually leave via one of the two point
contacts. For a symmetric structure we may assume that the fraction
$\frac{1}{2}(1-T_{\mathrm{d}}/N)$ of the injected current $GV$ is transmitted through the second
point contact after equilibration. The total source-drain current $I$ is the sum
of the direct and indirect contributions:
\be
I={\textstyle\frac{1}{2}}(1+T_{\mathrm{d}}/N)GV. \label{eq15.7}
\ee
The series conductance $G_{\mathrm{series}}=I/V$ becomes
\be
G_{\mathrm{series}}= {\textstyle\frac{1}{2}}G(1+T_{\mathrm{d}}/N). \label{eq15.8}
\ee 
In the absence of direct transmission $(T_{\mathrm{d}}=0)$, one recovers the ohmic
addition law for the resistance, as expected for the case of complete
intervening equilibration (cf.\ the related analysis by B\"{u}ttiker of tunneling in
series barriers\cite{ref370,ref371}). At the opposite extreme, if all transmission is
direct ($T_{\mathrm{d}}=N$), the series conductance is identical to that of the single point
contact. Substituting (\ref{eq15.3}) into Eq.\ (\ref{eq15.8}), we obtain the result\cite{ref329} for small
but nonzero direct transmission:
\be
G_{\mathrm{series}}= {\textstyle\frac{1}{2}}G(1+f(W_{\max}/2L)). \label{eq15.9}
\ee 

The quantized plateaus in the series resistance, observed experimentally,\cite{ref357} are of course not obtained in the semiclassical calculation leading to
Eq.\ (\ref{eq15.9}). However, since the nonadditivity is essentially a semiclassical
collimation effect, the present analysis should give a reasonably reliable
estimate of deviations from additivity for not too narrow point contacts. For
a comparison with experiments we refer to Refs.\ \onlinecite{ref307} and \onlinecite{ref329}. A fully
quantum mechanical calculation of the series resistance has been carried out
numerically by Baranger (reported in Ref.\ \onlinecite{ref306}) for two closely spaced
constrictions.

So far we have only considered the case of a zero magnetic field. In a weak
magnetic field $(2l_{\mathrm{cycl}}>L)$ the situation is rather complicated. As discussed in
detail in Ref.\ \onlinecite{ref329}, there are two competing effects in weak fields: On the one
hand, the deflection of the electron beam by the Lorentz force reduces the
direct transmission probability, with the effect of decreasing the series
conductance. On the other hand, the magnetic field enhances the indirect
transmission, with the opposite effect. The result is an initial {\it decrease\/} in the
series conductance for small magnetic fields in the case of strong collimation
and an {\it increase\/} in the case of weak collimation. This is expected to be a
relatively small effect compared with the effects at stronger fields that are
discussed below.

In stronger fields $(2l_{\mathrm{cycl}}<L)$, the direct transmission probability vanishes,
which greatly simplifies the situation. If we assume that all transmission
between the opposite point contacts is with intervening equilibration, then
the result is\cite{ref329}
\be
G_{\mathrm{series}}= \frac{2e^{2}}{h}\left(\frac{2}{N}-\frac{1}{N_{\mathrm{wide}}}\right)^{-1}.   \label{eq15.10}
\ee
Here $N$ is the ($B$-dependent) number of occupied subbands in the point
contacts, and $N_{\mathrm{wide}}$ is the number of occupied Landau levels in the 2DEG
between the point contacts. The physical origin of the simple addition rule
(\ref{eq15.10}) is additivity of the four-terminal longitudinal resistance (\ref{eq13.7}). From
this additivity it follows that for $n$ different point contacts in series, Eq.\ (\ref{eq15.10})
generalizes to
\be
\frac{1}{G_{\mathrm{series}}}-\frac{h}{2e^{2}}\frac{1}{N_{\mathrm{wide}}}=\sum_{i=1}^{n}R_{\mathrm{L}}(i), \label{eq15.11}
\ee 
where
\be
R_{\mathrm{L}}(i)=\left( \frac{h}{2e^{2}}\right)\left(\frac{1}{N_{i}}-\frac{1}{N_{\mathrm{wide}}}\right)   \label{eq15.12}
\ee
is the four-terminal longitudinal resistance of point contact $i$. Equation
(\ref{eq15.10}) predicts a nonmonotonic $B$-dependence for $G_{\mathrm{series}}$. This can most
easily be seen by disregarding the discreteness of $N$ and $N_{\mathrm{wide}}$. We then have
$N_{\mathrm{L}}\approx E_{\mathrm{F}}/\hbar\omega_{\mathrm{c}}$, while the magnetic field dependence of $N$ (for a square-well
confining potential in the point contacts) is given by Eq.\ (\ref{eq10.8}). The resulting
$B$-dependence of $G_{\mathrm{series}}$ is shown in Fig.\ \ref{fig64} (dotted curves). The nonmonotonic behavior is due to the delayed depopulation of subbands in the point
contacts compared with the broad 2DEG. While the number of occupied
Landau levels $N_{\mathrm{wide}}$ in the region between the point contacts decreases
steadily with $B$ for $2l_{\mathrm{cycl}}<L$, the number $N$ of occupied subbands in the
point contacts remains approximately constant until $2l_{\mathrm{c},\min}\approx W_{\min}$, with
$l_{\mathrm{c},\min}\equiv l_{\mathrm{cycl}}(1-E_{\mathrm{c}}/E_{\mathrm{F}})^{1/2}$ denoting the cyclotron radius in the point contact
region. In this field interval $G_{\mathrm{series}}$ increases with $B$, according to Eq.\ (\ref{eq15.10}).
For stronger fields, depopulation in the point contacts begins to dominate
$G_{\mathrm{series}}$, leading finally to a decreasing conductance (as is the rule for single
point contacts; see Section \ref{sec13b}). The peak in $G_{\mathrm{series}}$ thus occurs at
$2l_{\mathrm{c},\min}\approx W_{\min}$.

The remarkable camelback shape of $G_{\mathrm{series}}$ versus $B$ predicted by Eq.\
(\ref{eq15.10}) has been observed experimentally by Staring et al.\cite{ref372} The data are
shown in Fig.\ \ref{fig64} (solid curves) for three values of the gate voltage $V_{\mathrm{g}}$ at
$T=100\,\mathrm{mK}$. The measurement configuration is as shown in the inset, with a
point contact separation $L=1.0\,\mu \mathrm{m}$. The dotted curves in Fig.\ \ref{fig64} are the
result of a one-parameter fit to the theoretical expression. It is seen that Eq.\
(\ref{eq15.10}) provides a good description of the overall magnetoresistance behavior
from low magnetic fields up to the quantum Hall effect regime. The
additional structure in the experimental curves has several different origins,
for which we refer to the paper by Staring et al.\cite{ref372} Similar structure in the
two-terminal resistance of a single point contact will be discussed in detail in
Section \ref{sec21}.

We emphasize that Eq.\ (\ref{eq15.10}) is based on the assumption of complete
equilibration of the current-carrying edge states in the region between the
point contacts. In a quantizing magnetic field, local equilibrium is reached by
inter-Landau level scattering. If the potential landscape (both in the point
contacts themselves and in the 2DEG region in between) varies by less than
the Landau level separation $\hbar\omega_{\mathrm{c}}$ on the length scale of the magnetic length
$(\hbar/eB)^{1/2}$, then inter-Landau level scattering is suppressed in the absence of
other scattering mechanisms (see Section \ref{sec18}). This means that the transport
from one point contact to the other is adiabatic. The series conductance is
then simply $G_{\mathrm{series}}=(2e^{2}/h)N$ for two identical point contacts
[$N \equiv\min(N_{1}, N_{2})$ for two different point contacts in series]. This expression
differs from Eq.\ (\ref{eq15.10}) if a barrier is present in the point contacts, since that
causes the number $N$ of occupied Landau levels in the point contact to be less
than the number $N_{\mathrm{wide}}$ of occupied levels in the wide 2DEG. [In a strong
magnetic field, $N\approx(E_{\mathrm{F}}-E_{\mathrm{c}})/\hbar\omega_{\mathrm{c}}$, while $N_{\mathrm{wide}}\approx E_{\mathrm{F}}/\hbar\omega_{\mathrm{c}}.$] Adiabatic transport in a magnetic field through two point contacts in series has been studied
experimentally by Kouwenhoven et al.\cite{ref373} and by Main et al.\cite{ref374}

\subsection{\label{sec16} Junction scattering}

In the regime of diffusive transport, the Hall bar geometry (a straight
current-carrying channel with small side contacts for voltage drop measurements) is very convenient, since it allows an independent determination of the
various components of the resistivity tensor. A downscaled Hall bar was
therefore a natural first choice as a geometry to study ballistic transport in a
2DEG.\cite{ref67,ref68,ref74,ref139,ref178,ref364} The resistances measured in narrow-channel
geometries are mainly determined by scattering at the junctions with the side
probes.\cite{ref289} These scattering processes depend strongly on the junction shape.
This is in contrast to the point contact geometry; compare the very similar
results of van Wees et al.\cite{ref6} and Wharam et al.\cite{ref7}  on the quantized conductance
of point contacts of a rather different design. The strong dependence of the
low-field Hall resistance on the junction shape was demonstrated theoretically by Baranger and Stone\cite{ref358} and experimentally by Ford et al.\cite{ref77} and Chang
et al.\cite{ref375} These results superseded many earlier attempts (listed in Ref.\ \onlinecite{ref360}) to
explain the discovery by Roukes et al.\cite{ref67} of the {\it quenching of the Hall effect\/}
without modeling the shape of the junction realistically. Baranger and
Stone\cite{ref358} argued that the rounded corners (present in a realistic situation) at
the junction between the main channel and the side branches lead to a
suppression (quenching) of the Hall resistance at low magnetic fields as a
consequence of the horn collimation effect discussed in Section \ref{sec15a}. A Hall
bar with straight corners, in contrast, does not show a generic suppression of
the Hall resistance,\cite{ref376,ref377,ref378} although quenching can occur for special parameter values if only a few subbands are occupied in the channel.

The quenched Hall effect\cite{ref67,ref77,ref375,ref379} is just one of a whole variety of
magnetoresistance anomalies observed in narrow Hall bars. Other anomalies
are the {\it last Hall plateau},\cite{ref67,ref68,ref77,ref139,ref178,ref379} reminiscent of quantum Hall
plateaus, but occurring at much lower fields; the {\it negative Hall resistance},\cite{ref77} as
if the carriers were holes rather than electrons; the {\it bend resistance},\cite{ref289,ref306,ref364,ref380} a longitudinal resistance associated with a current bend,
which is negative at small magnetic fields and zero at large fields, with an
overshoot to a positive value at intermediate fields; and more.

In Refs.\ \onlinecite{ref359} and \onlinecite{ref360} we have shown that all these phenomena can be
qualitatively explained in terms of a few simple semiclassical mechanisms
(reviewed in Section \ref{sec16a}). The effects of quantum interference and of
quantization of the lateral motion in the narrow conductor are not essential.
These magnetoresistance anomalies can thus be characterized as classical
magneto size effects in the ballistic regime. In Section \ref{sec5}, we have discussed
classical size effects in the quasi-ballistic regime, where the mean free path is
larger than the channel width but smaller than the separation between the
voltage probes. In that regime, the size effects found in a 2DEG were known
from work on metal films and wires. These earlier investigations had not
anticipated the diversity of magnetoresistance anomalies due to junction
scattering in the ballistic regime. That is not surprising, considering that the
theoretical formalism to describe a resistance measurement within a mean
free path had not been developed in that context. Indeed, this Landauer-B\"{u}ttiker formalism (described in Section \ref{sec12}) found one of its earliest
applications\cite{ref268} in the context of the quenching of the Hall effect, and the
success with which the experimental magnetoresistance anomalies can be
described by means of this formalism forms strong evidence for its validity.

\subsubsection{\label{sec16a} Mechanisms}

\begin{figure}
\centerline{\includegraphics[width=8cm]{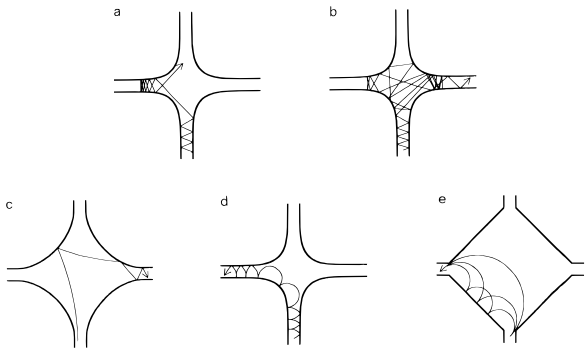}}
\caption{
Classical trajectories in an electron billiard, illustrating the collimation (a), scrambling (b), rebound (c), magnetic guiding (d), and electron focusing (e) effects. Taken from C. W. J. Beenakker and H. van Houten, in ``Electronic Properties of Multilayers and Low-Dimensional Semiconductor Structures'' (J. M. Chamberlain, L. Eaves, and J. C. Portal, eds.). Plenum, London, 1990.
\label{fig65}
}
\end{figure}

The variety of magnetoresistance anomalies mentioned can be understood
in terms of a few simple characteristics of the curved trajectories of electrons
in a classical billiard in the presence of a perpendicular magnetic field.\cite{ref359,ref360}
At very small magnetic fields, {\it collimation\/} and {\it scrambling\/} are the key concepts.
The gradual widening of the channel on approaching thejunction reduces the
injection/acceptance cone, which is the cone of angles with the channel axis
within which an electron is injected into the junction or within which an
electron can enter the channel coming from the junction. This is the horn
collimation effect\cite{ref329} discussed in Section \ref{sec15a} (see Fig.\ \ref{fig65}a). If the injection/acceptance cone is smaller than $90^{\circ}$, then the cones of two channels at right
angles do not overlap. That means that an electron approaching the side
probe coming from the main channel will be reflected (Fig.\ \ref{fig65}a) and will then
typically undergo multiple reflections in the junction region (Fig.\ \ref{fig65}b). The
trajectory is thus scrambled, whereby the probability for the electron to enter
the left or right side probe in a weak magnetic field is equalized. This
suppresses the Hall voltage. This ``scrambling'' mechanism for the quenching
of the Hall effect requires a weaker collimation than the ``nozzle'' mechanism
put forward by Baranger and Stone\cite{ref358} (we return to both these mechanisms
in Section \ref{sec16c}). Scrambling is not effective in the geometry shown in Fig.\ \ref{fig65}c,
in which a large portion of the boundary in the junction is oriented at
approximately $45^{\circ}$ with the channel axis. An electron reflected from a side
probe at this boundary has a large probability of entering the opposite side
probe. This is the ``rebound'' mechanism for a negative Hall resistance
proposed by Ford et al.\cite{ref77}

At somewhat larger magnetic fields, {\it guiding\/} takes over. As illustrated in
Fig.\ \ref{fig65}d, the electron is guided by the magnetic field along equipotentials
around the corner. Guiding is fully effective when the cyclotron radius $l_{\mathrm{cycl}}$
becomes smaller than the minimal radius of curvature $r_{\min}$ of the corner ---
that is, for magnetic fields greater than the guiding field $B_{\mathrm{g}}\equiv \hbar k_{\rm F}/er_{\min}$. In the
regime $B\gtrsim B_{\mathrm{g}}$ the junction cannot scatter the electron back into the channel
from which it came. The absence of backscattering in this case is an entirely
classical, weak-field phenomenon (cf.\ Section \ref{sec13b}). Because of the absence of
backscattering, the longitudinal resistance vanishes, and the Hall resistance
$R_{\mathrm{H}}$ becomes equal to the contact resistance of the channel, just as in the
quantum Hall effect, but without quantization of $R_{\mathrm{H}}$. The contact resistance
$R_{\mathrm{contact}}\approx(h/2e^{2})(\pi/k_{\mathrm{F}}W)$ is approximately independent of the magnetic field
for fields such that the cyclotron diameter $2l_{\mathrm{cycl}}$ is greater than the channel
width $W$, that is, for fields below $B_{\mathrm{crit}}\equiv 2\hbar k_{\rm F}/eW$ (see Sections \ref{sec12} and \ref{sec13}).
This explains the occurrence of the ``last plateau'' in $R_{\mathrm{H}}$ for $B_{\mathrm{g}}\sim<B\lesssim B_{\mathrm{crit}}$ as a
classical effect. At the low-field end of the plateau, the Hall resistance is
sensitive to {\it geometrical resonances\/} that increase the fraction of electrons
guided around the corner into the side probe. Figure \ref{fig65}e illustrates the
occurrence of one such a geometrical resonance as a result of the magnetic
focusing of electrons into the side probe, at magnetic fields such that the
separation of the two perpendicular channels is an integer multiple of the
cyclotron diameter. This is in direct analogy with electron focusing in a
double-point contact geometry (see Section \ref{sec14}) and leads to periodic
oscillations superimposed on the Hall plateau. Another geometrical resonance with similar effect is discussed in Ref.\ \onlinecite{ref360}.

These mechanisms for oscillations in the resistance depend on a commensurability between the cyclotron radius and a characteristic dimension of
the junction, but do not involve the wavelength of the electrons as an
independent length scale. This distinguishes these geometrical resonances
conceptually from the quantum resonances due to bound states in the
junction considered in Refs.\ \onlinecite{ref376,ref377}, and \onlinecite{ref380,ref381,ref382}.

\subsubsection{\label{sec16b} Magnetoresistance anomalies}

In this subsection we compare, following Ref.\ \onlinecite{ref360}, the semiclassical theory
with representative experiments on laterally confined two-dimensional
electron gases in high-mobility GaAs-AlGaAs heterostructures. The calculations are based on a simulation of the classical trajectories of a large
number (typically $10^{4}$) of electrons with the Fermi energy, to determine the
classical transmission probabilities. The resistances then follow from the
B\"{u}ttiker formula (\ref{eq12.12}). We refer to Refs.\ \onlinecite{ref359} and \onlinecite{ref360} for details on the
method of calculation. We first discuss the Hall resistance $R_{\mathrm{H}}$.

\begin{figure}
\centerline{\includegraphics[width=8cm]{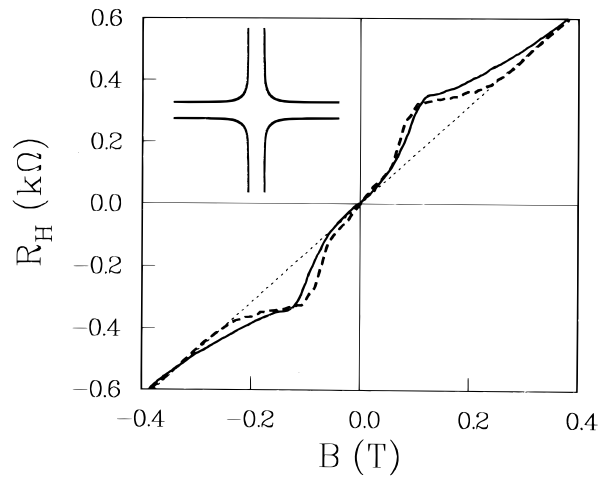}}
\caption{
Hall resistance as measured (solid curve) by Simmons et al.,\cite{ref178} and as calculated (dashed curve) for the hard-wall geometry in the inset ($W= 0.8\,\mu{\rm m}$ and $E_{\rm F} = 14\,{\rm meV}$). The dotted line is $R_{\rm H}$ in a bulk 2DEG. Taken from C. W. J. Beenakker and H. van Houten, in ``Electronic Properties of Multilayers and Low-Dimensional Semiconductor Structures'' (J. M. Chamberlain, L. Eaves, and J. C. Portal, eds.). Plenum, London, 1990.
\label{fig66}
}
\end{figure}

Figure \ref{fig66} shows the precursor of the classical Hall plateau (the ``last
plateau'') in a relatively wide Hall cross. The experimental data (solid curve) is
from a paper by Simmons et al.\cite{ref178} The semiclassical calculation (dashed
curve) is for a square-well confining potential of channel width $W=0.8\,\mu \mathrm{m}$
(as estimated in the experimental paper) and with the relatively sharp corners
shown in the inset. The Fermi energy used in the calculation is $E_{\mathrm{F}}=14\,\mathrm{meV}$,
which corresponds (via $n_{\mathrm{s}}=E_{\mathrm{F}}m/\pi \hbar^{2}$) to a sheet density in the channel of
$n_{\mathrm{s}}=3.9\times 10^{15}\,\mathrm{m}^{-2}$, somewhat below the value of $4.9\times 10^{15}\,\mathrm{m}^{-2}$ of the bulk
material in the experiment. Good agreement between theory and experiment
is seen in Fig.\ \ref{fig66}. Near zero magnetic field, the Hall resistance in this
geometry is close to the linear result $R_{\mathrm{H}}=B/en_{\mathrm{s}}$ for a bulk 2DEG (dotted
line). The corners are sufficiently smooth to generate a Hall plateau via the
guiding mechanism discussed in Section \ref{sec16a}. The horn collimation effect,
however, is not sufficiently large to suppress $R_{\mathrm{H}}$ at small $B$. Indeed, the
injection/acceptance cone for this junction is considerably wider (about
$115^{\circ})$ than the maximal angular opening of $90^{\circ}$ required for quenching of the
Hall effect via the scrambling mechanism described in Section \ref{sec16a}.

\begin{figure}
\centerline{\includegraphics[width=6cm]{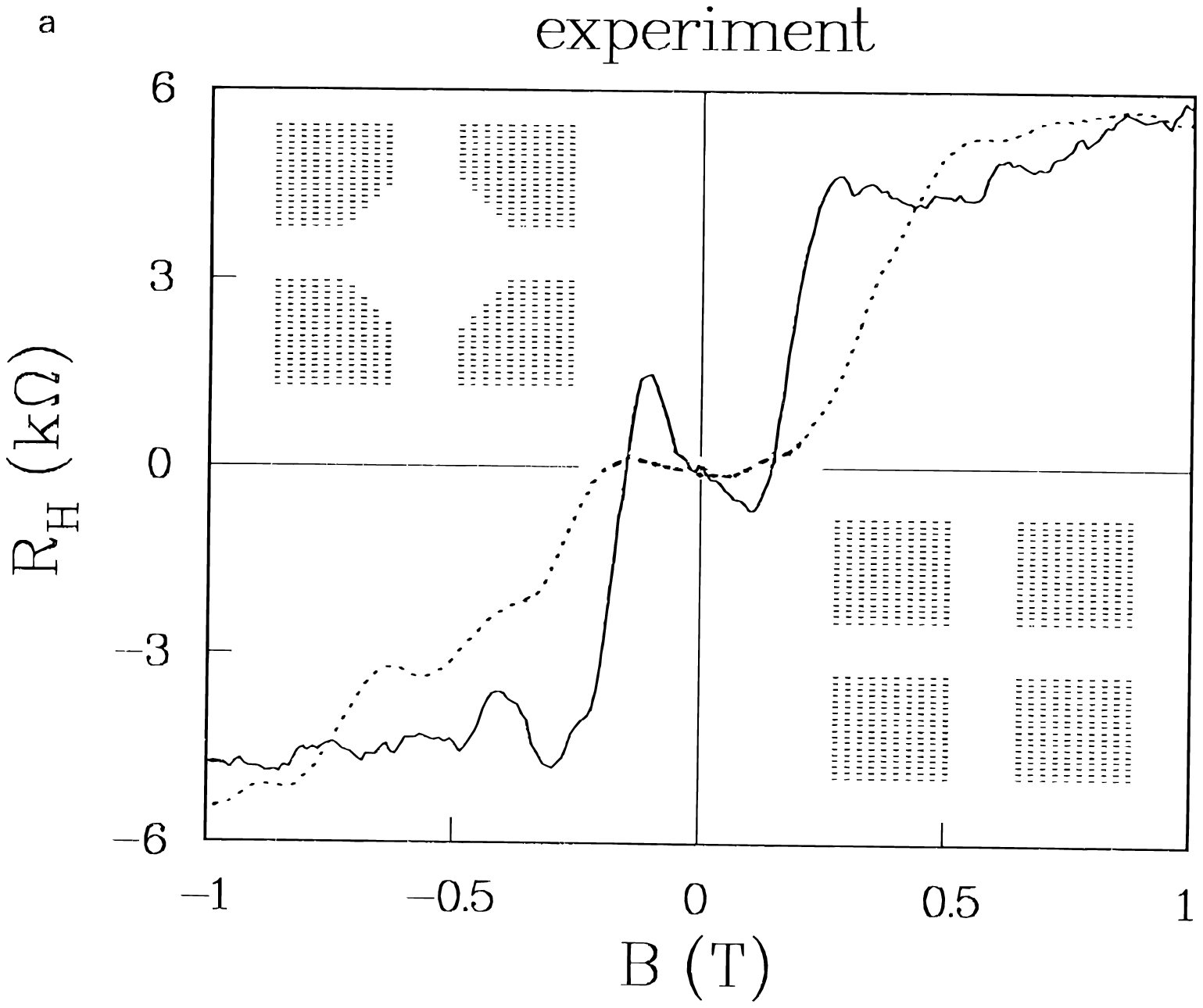}}

\centerline{\includegraphics[width=6cm]{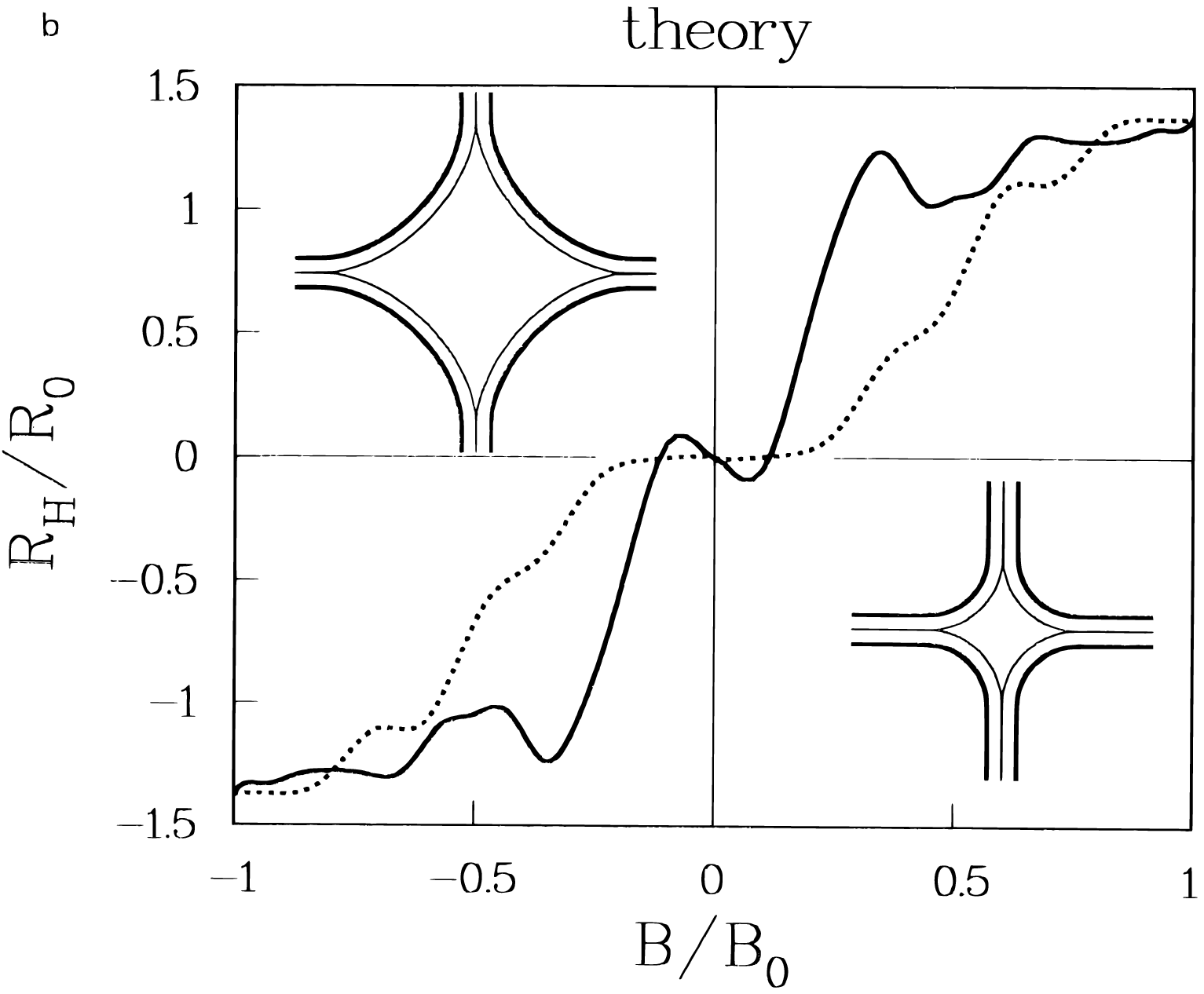}}
\caption{
Hall resistance as measured (a) by Ford et al.\cite{ref77} and as calculated (b). In (a) as well as in (b), the solid curve corresponds to the geometry in the upper left inset, the dotted curve to the geometry in the lower right inset. The insets in (a) indicate the shape of the gates, not the actual confining potential. The insets in (b) show equipotentials of the confining potential at $E_{\rm F}$ (thick contour) and 0 (thin contour). The potential rises parabolically from 0 to $E_{\rm F}$ and vanishes in the diamond-shaped region at the center of the junction. Taken from C. W. J. Beenakker and H. van Houten, in ``Electronic Properties of Multilayers and Low-Dimensional Semiconductor Structures'' (J. M. Chamberlain, L. Eaves, and J. C. Portal, eds.). Plenum, London, 1990.
\label{fig67}
}
\end{figure}

The low-field Hall resistance changes drastically if the channel width
becomes smaller, relative to the radius of curvature of the corners. Figure \ref{fig67}a
shows experimental data by Ford et al.\cite{ref77} The solid and dotted curves are for
the geometries shown respectively in the upper left and lower right insets of
Fig.\ \ref{fig67}a. Note that these insets indicate the gates with which the Hall crosses
are defined electrostatically. The equipotentials in the 2DEG will be
smoother than the contours of the gates. The experiment shows a well-developed Hall plateau with superimposed fine structure. At small positive
fields $R_{\mathrm{H}}$ is either quenched or negative, depending on the geometry. The
geometry is seen to affect also the width of the Hall plateau but not the height.
In Fig.\ \ref{fig67}b we give the results of the semiclassical theory for the two
geometries in the insets, which should be reasonable representations of the
confining potential induced by the gates in the experiment. In the theoretical
plot the resistance and the magnetic field are given in units of
\be
R_{0} \equiv\frac{h}{2e^{2}}\frac{\pi}{k_{\mathrm{F}}W},\;\;B_{0} \equiv\frac{\hbar k_{\rm F}}{eW}, \label{eq16.1}
\ee 
where the channel width $W$ for the parabolic confinement used is defined as
the separation of the equipotentials at the Fermi energy ($W_{\mathrm{par}}$ in Section \ref{sec10}).
The experimental estimates $W\approx 90\,\mathrm{nm}$, $n_{\mathrm{s}}\approx 1.2\times 10^{15}\,\mathrm{m}^{-2}$ imply
$R_{0}=5.2\,\mathrm{k}\Omega$, $B_{0}=0.64\,{\rm T}$. With these parameters the calculated resistance
and field scales do not agree well with the experiment, which may be due in
part to the uncertainties in the modeling of the shape of the experimental
confining potential. The $\pm B$ asymmetry in the experimental plot is undoubtedly due to asymmetries in the cross geometry [in the calculation the
geometry has fourfold symmetry, which leads automatically to
$R_{\mathrm{H}}(B)=-R_{\mathrm{H}}(-B)]$. Apart from these differences, there is agreement in all
the important features: the appearance of quenched and negative Hall
resistances, the independence of the height of the last Hall plateau on the
smoothness of the corners, and the shift of the onset of the last plateau to
lower fields for smoother corners. The oscillations on the last plateau in the
calculation (which, as we discussed in Section \ref{sec16a}, are due to geometrical
resonances) are also quite similar to those in the experiment, indicating that
these are classical rather than quantum resonances.

\begin{figure}
\centerline{\includegraphics[width=6cm]{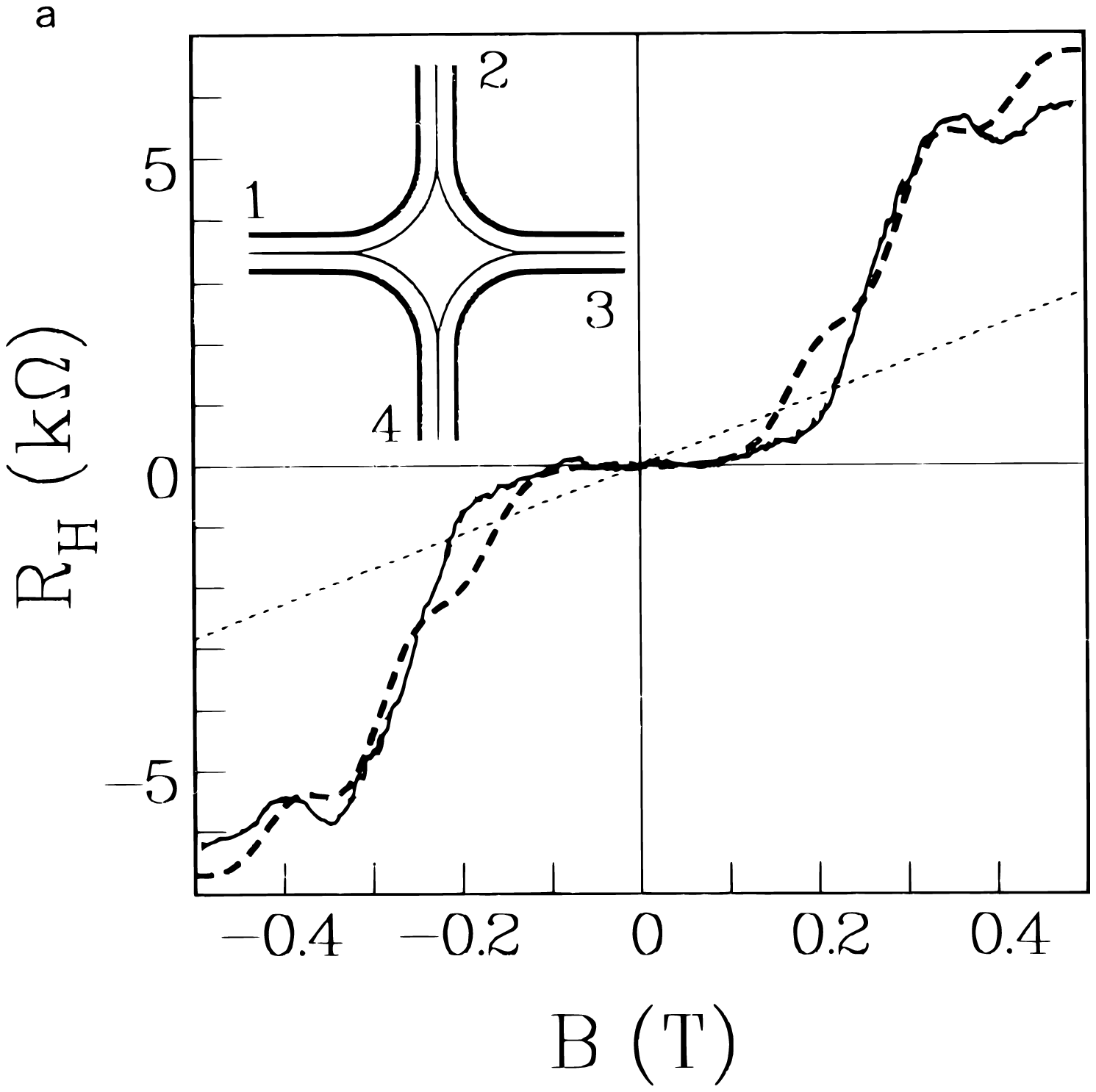}}

\centerline{\includegraphics[width=6cm]{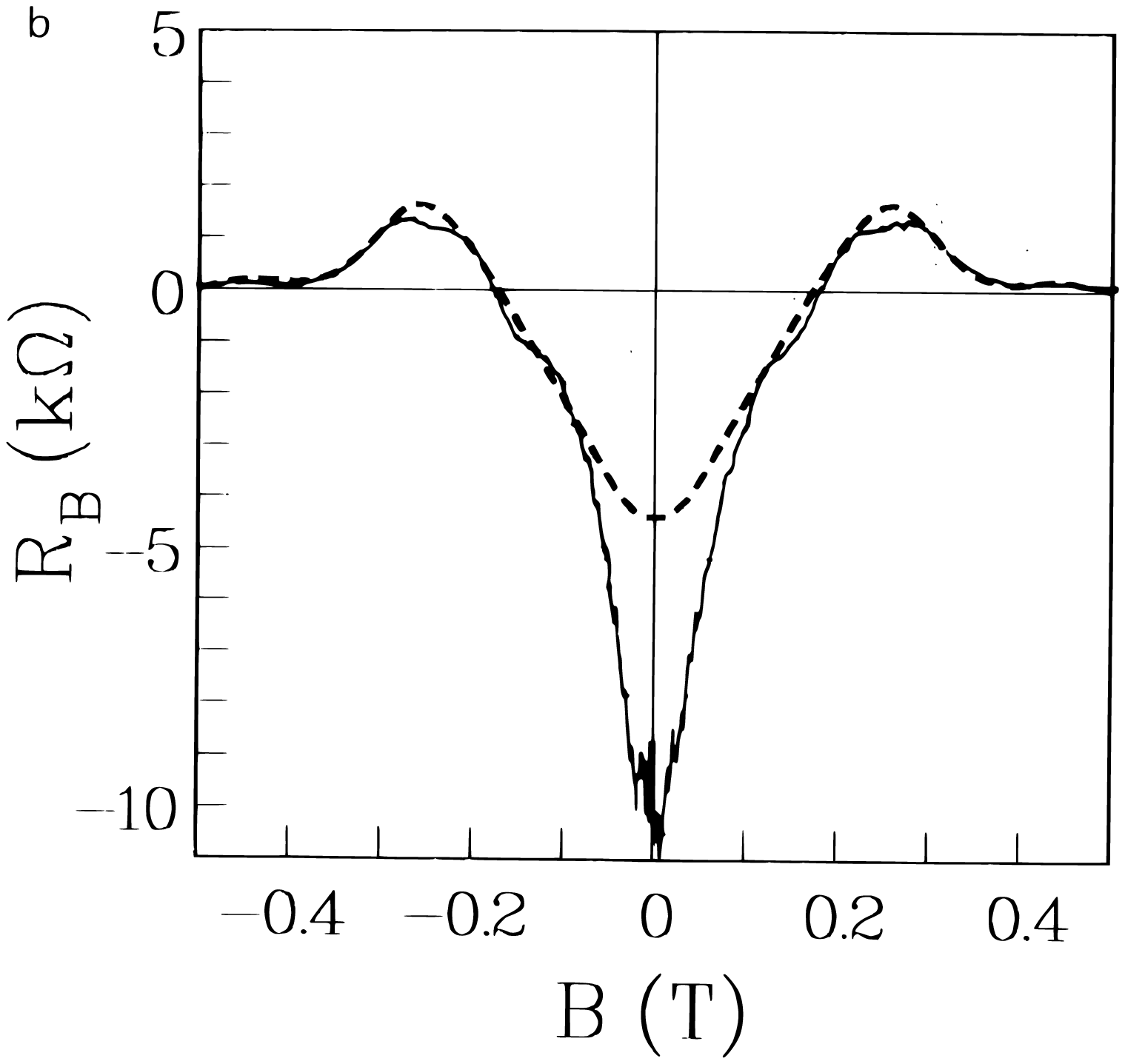}}
\caption{
Hall resistance $R_{\rm H} \equiv R_{13,24}$ (a) and bend resistance $R_{\rm B} \equiv R_{12,43}$ (b), as measured (solid curves) by Timp et al.\cite{ref306} and as calculated (dashed curves) for the geometry in the inset (consisting of a parabolic confining potential with the equipotentials at $E_{\rm F}$ and 0 shown respectively as thick and thin contours; the parameters are $W = 100\,{\rm nm}$ and $E_{\rm F} = 3.9\,{\rm meV}$). The dotted line in (a) is $R_{\rm H}$ in a bulk 2DEG. Taken from C. W. J. Beenakker and H. van Houten, in ``Electronic Properties of Multilayers and Low-Dimensional Semiconductor Structures'' (J. M. Chamberlain, L. Eaves, and J. C. Portal, eds.). Plenum, London, 1990.
\label{fig68}
}
\end{figure}

We now turn to the bend resistance $R_{\mathbf{B}}$. In Fig.\ \ref{fig68} we show experimental
data by Timp et al.\cite{ref306} (solid curves) on $R_{\mathbf{B}}\equiv R_{12,43}$ and $R_{\mathrm{H}}\equiv R_{13,24}$
measured in the same Hall cross (defined by gates of a shape similar to that in
the lower right inset of Fig.\ \ref{fig67}a; see the inset of Fig.\ \ref{fig68}a for the numbering of
the channels). The dashed curves are calculated for a parabolic confining
potential in the channels (with the experimental values $W=100\,\mathrm{nm}$,
$E_{\mathrm{F}}=3.9\,\mathrm{meV})$ and with corners as shown in the inset of Fig.\ \ref{fig68}a. The
calculated quenching of the Hall resistance and the onset of the last plateau
are in good agreement with the experiment, and also the observed overshoot
of the bend resistance around $0.2\, \mathrm{T}$ as well as the width of the negative peak in
$R_{\mathbf{B}}$ around zero field are well described by the calculation. The calculated
height of the negative peak, however, is too small by more than a factor of 2.
We consider this disagreement to be significant in view of the quantitative
agreement with the other features in both $R_{\mathbf{B}}$ and $R_{\mathrm{H}}$. The negative peak in $R_{\mathbf{B}}$
is due to the fact that the collimation effect couples the current source 1 more
strongly to voltage probe 3 than to voltage probe 4, so $R_{\mathbf{B}}\propto V_{4}-V_{3}$ is
negative for small magnetic fields (at larger fields the Lorentz force destroys
collimation by bending the trajectories, so $R_{\mathbf{B}}$ shoots up to a positive value
until guiding takes over and brings $R_{\mathbf{B}}$ down to zero by eliminating
backscattering at the junction). The discrepancy in Fig.\ \ref{fig68}b thus seems to
indicate that the semiclassical calculation underestimates the collimation
effect in this geometry. The positive overshoot of $R_{\mathbf{B}}$ seen in Fig.\ \ref{fig68}b is found
only for rounded corners. This explains the near absence of the effect in the
calculation of Kirczenow\cite{ref381} for a junction with straight corners.

For a discussion of the temperature dependence of the magnetoresistance
anomalies, we refer to Ref.\ \onlinecite{ref360}. Here it suffices to note that the experiments
discussed were carried out at temperatures around $1\, \mathrm{K}$, for which we expect
the zero-temperature semiclassical calculation to be appropriate. At lower
temperatures the effects of quantum mechanical phase coherence that have
been neglected will become more important. At higher temperatures the
thermal average smears out the magnetoresistance anomalies and eventually
inelastic scattering causes a transition to the diffusive transport regime in
which the resistances have their normal $B$-dependence.

\subsubsection{\label{sec16c} Electron waveguide versus electron billiard}

The overall agreement between the experiments and the semiclassical
calculations is remarkable in view of the fact that the channel width in the
narrowest structures considered is comparable to the Fermi wavelength.
When the first experiments on these ``electron waveguides'' appeared, it was
expected that the presence of only a small number of occupied transverse
waveguide modes would fundamentally alter the nature of electron transport.\cite{ref68} The results of Refs.\ \onlinecite{ref359} and \onlinecite{ref356} show instead that the modal structure
plays only a minor role and that the magnetoresistance anomalies observed
are characteristic for the {\it classical\/} ballistic transport regime. The reason that a
phenomenon such as the quenching of the Hall effect has been observed only
in Hall crosses with narrow channels is simply that the radius of curvature of
the corners at the junction is too small compared with the channel width in
wider structures. This is not an essential limitation, and the various
magnetoresistance anomalies discussed here should be observable in macrocopic Hall bars with artificially smoothed corners, provided of course that
the dimensions of the junction remain well below the mean free path. Ballistic
transport is essential, but a small number of occupied modes is not.

Although we believe that the characteristic features of the magnetoresistance anomalies are now understood, several interesting points of disagreement between theory and experiment remain that merit further investigation.
One of these is the discrepancy in the magnitude of the negative bend
resistance at zero magnetic field noted before. The disappearance of a region
of quenched Hall resistance at low electron density is another unexpected
observation by Chang et al.\cite{ref375} and Roukes et al.\cite{ref383}  The semiclassical theory
predicts a universal behavior (for a given geometry) if the resistance and
magnetic field are scaled by $R_{0}$ and $B_{0}$ defined in Eq.\ (\ref{eq16.1}). For a square-well
confining potential the channel width $W$ is the same at each energy, and since
$B_{0}\propto k_{\mathrm{F}}$ one would expect the field region of quenched Hall resistance to vary
with the electron density as $\sqrt{n_{\mathrm{s}}}$. For a more realistic smooth confining
potential, $W$ depends on $E_{\mathrm{F}}$ and thus on $n_{\mathrm{s}}$ as well, in a way that is difficult to
estimate reliably. In any case, the experiments point to a systematic
disappearance of the quench at the lowest densities, which is not accounted
for by the present theory (and has been attributed by Chang et al.\cite{ref375} to
enhanced diffraction at low electron density as a result of the increase in the
Fermi wavelength). For a detailed investigation of departures from classical
scaling, we refer to a paper by Roukes et al.\cite{ref384} As a third point, we mention
the curious density dependence of the quenching observed in approximately
straight junctions by Roukes et al.,\cite{ref383} who find a low-field suppression of $R_{\mathrm{H}}$
that occurs only at or near certain specific values of the electron density. The
semiclassical model applied to a straight Hall cross (either defined by a
square well or by a parabolic confining potential) gives a low-field slope of $R_{\mathrm{H}}$
close to its bulk 2D value. The fully quantum mechanical calculations for a
straight junction\cite{ref376,ref381} do give quenching at special parameter values, but
not for the many-mode channels in this experiment (in which quenching
occurs with as many as 10 modes occupied, whereas in the calculations a
straight cross with more than 3 occupied modes in the channel does not show
a quench).

In addition to the points of disagreement discussed, there are fine details in
the measured magnetoresistances, expecially at the lowest temperatures
(below $100\, \mathrm{mK}$), which are not obtained in the semiclassical approximation.
The quantum mechanical calculations\cite{ref358,ref376,ref377,ref381} show a great deal of fine
structure due to interference of the waves scattered by the junction. The fine
structure in most experiments is not quite as pronounced as in the
calculations presumably partly as a result of a loss of phase coherence after
many multiple scatterings in the junction. The limited degree of phase
coherence in the experiments and the smoothing effect of a finite temperature
help to make the semiclassical model work so well even for the narrowest
channels. We draw attention to the fact that classical {\it chaotic\/} scattering can
also be a source of irregular resistance fluctuations (see Ref.\ \onlinecite{ref360}).

Some of the most pronounced features in the quantum mechanical
calculations are due to transmission resonances that result from the presence
of bound states in the junction.\cite{ref376,ref377,ref380,ref381,ref382} In Section \ref{sec16a} we have
discussed a different mechanism for transmission resonances that has a
classical, rather than a quantum mechanical, origin. As mentioned in Section
\ref{sec16b}, the oscillations on the last Hall plateau observed experimentally are
quite well accounted for by these geometrical resonances. One way to
distinguish experimentally between these resonance mechanisms is by means
of the temperature dependence, which should be much weaker for the
classical than for the quantum effect. One would thus conclude that the
fluctuations in Fig.\ \ref{fig67}a, measured by Ford et al.\cite{ref77} at $4.2\, \mathrm{K}$, have a classical
origin, while the fine structure that Ford et al.\cite{ref385} observe only at $\mathrm{mK}$
temperatures (see below) is intrinsically quantum mechanical.

The differences between the semiclassical and the quantum mechanical
models may best be illustrated by considering once again the quenching of
the Hall effect, which has the most subtle explanation and is the most
sensitive to the geometry among the magnetoresistance anomalies observed
in the ballistic regime. The classical scrambling of the trajectories after
multiple reflections suppresses the asymmetry between the transmission
probabilities $t_{\rm l}$ and $t_{\rm r}$ to enter the left or right voltage probe, and without this
transmission asymmetry there can be no Hall voltage. We emphasize that this
{\it scrambling mechanism\/} is consistent with the original findings of Baranger and
Stone\cite{ref358} that quenching requires collimation. The point is that the collimation effect leads to nonoverlapping injection/acceptance cones of two perpendicular channels, which ensures that electrons cannot enter the voltage
probe from the current source directly, but rather only after multiple
reflections (cf.\ Section \ref{sec16a}). In this way a rather weak collimation to within
an injection/acceptance cone of about $90^{\circ}$ angular opening is sufficient to
induce a suppression of the Hall resistance via the scrambling mechanism.

Collimation can also suppress $R_{\mathrm{H}}$ directly by strongly reducing $t_{\rm l}$ and $t_{\rm r}$
relative to $t_{\mathrm{s}}$ (the probability for transmission straight through the junction).
This {\it nozzle mechanism}, introduced by Baranger and Stone,\cite{ref358} requires a
strong collimation of the injected beam in order to affect $R_{\mathrm{H}}$ appreciably. In
the geometries considered here, we find that quenching of $R_{\mathrm{H}}$ is due
predominantly to scrambling and not to the nozzle mechanism ($t_{\rm l}$ and $t_{\rm r}$ each
remain more than 30\% of $t_{\mathrm{s}}$), but data by Baranger and Stone\cite{ref358} show that
both mechanisms can play an important role.

\begin{figure}
\centerline{\includegraphics[width=8cm]{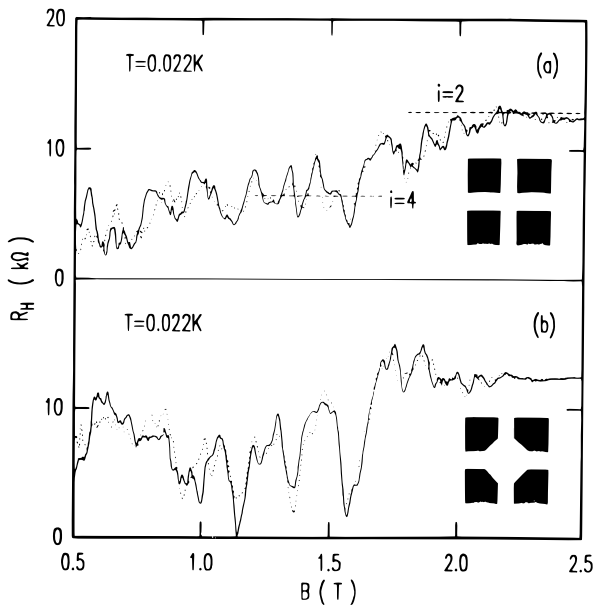}}
\caption{
Measured Hall resistance in an abrupt (a) and in a widened (b) cross as a function of $B$ in the strong field regime. Large fluctuations are resolved at the low temperature of $22\,{\rm mK}$. The dotted curves indicate the reproducibility of the measurement. Taken from C. J. B. Ford et al., Surf.\ Sci.\ {\bf 229}, 298 (1990).
\label{fig69}
}
\end{figure}

There is a third proposed mechanism for the quenching of the Hall
effect,\cite{ref376,ref377} which is the reduction of the transmission asymmetry due to a
bound state in the junction. The {\it bound state mechanism\/} is purely quantum
mechanical and does not require collimation (in contrast to the classical
scrambling and nozzle mechanisms). Numerical calculations have shown that
it is only effective in straight Hall crosses with very narrow channels (not
more than three modes occupied), and even then for special values of the
Fermi energy only. Although this mechanism cannot account for the
experiments performed thus far, it may become of importance in future work.
A resonant suppression of the Hall resistance may also occur in strong
magnetic fields, in the regime where the Hall resistance in wide Hall crosses
would be quantized. Such an effect is intimately related to the high-field
Aharonov-Bohm magnetoresistance oscillations in a singly connected
geometry (see Section \ref{sec21}). Ford et al.\cite{ref385} have observed oscillations superimposed on quantized Hall plateaux at low temperatures in very narrow crosses
of two different shapes (see Fig.\ \ref{fig69}). The strong temperature dependence
indicates that these oscillations are resonances due to the formation of bound
states in the cross.\cite{ref306,ref385,ref386}

\subsection{\label{sec17} Tunneling}

In this section we review recent experiments on tunneling through
potential barriers in a two-dimensional electron gas. Subsection \ref{sec17a} deals
with {\it resonant\/} tunneling through a bound state in the region between two
barriers. Resonant tunneling has previously been studied extensively in
layered semiconductor heterostructures for transport perpendicular to the
layers.\cite{ref387,ref388,ref389} For example, a thin AlGaAs layer embedded between two
GaAs layers forms a potential barrier, whose height and width can be tailored
with great precision by means of advanced growth techniques (such as
molecular beam epitaxy). Because of the free motion in the plane of the layers,
one can only realize bound states with respect to one direction. Tunneling
resonances are consequently smeared out over a broad energy range. A
2DEG offers the possibility of confinement in all directions and thus of a
sharp resonance. A gate allows one to define potential barriers of adjustable
height in the 2DEG. In contrast, the heterostructure layers form fixed
potential barriers, so one needs to study a current-voltage characteristic to
tune the system through a resonance (observable as a peak in the $I-V$ curve).
The gate-induced barriers in a 2DEG offer a useful additional degree of
freedom, allowing a study of resonant tunneling in the linear response regime
of small applied voltages (to which we limit the discussion in this review). A
drawback of these barriers is that their shape cannot be precisely controlled,
or modeled, so that a description of the tunneling process will of necessity be
qualitative.

Subsection \ref{sec17b} deals with the effects of Coulomb repulsion on tunneling
in a 2DEG. The electrostatic effects of charge buildup in the 1D potential well
formed by heterostructure layers have received considerable attention in
recent years.\cite{ref389,ref390} Because of the large capacitance of the potential well in
this case (resulting from the large surface area of the layers) these are
macroscopic effects, involving a large number of electrons. The 3D potential
well in a 2DEG nanostructure, in contrast, can have a very small capacitance
and may contain a few electrons only. The tunneling of a single electron into
the well will then have a considerable effect on the electrostatic potential
difference with the surrounding 2DEG. For a small applied voltage this effect
of the Coulomb repulsion can completely suppress the tunneling current. In
metals this ``Coulomb blockade'' of tunneling has been studied extensively.\cite{ref391} In those systems a semiclassical description suffices. The large Fermi
wavelength in a 2DEG should allow the study of quantum mechanical effects
on the Coulomb blockade or, more generally, of the interplay between
electron-electron interactions and resonant tunneling.\cite{ref318,ref392,ref393}

\subsubsection{\label{sec17a} Resonant tunneling}

The simplest geometry in which one might expect to observe transmission
resonances is formed by a single potential barrier across a 2DEG channel.
Such a geometry was studied by Washburn et al.\cite{ref394} in a GaAs-AlGaAs
heterostructure containing a 2-$\mu \mathrm{m}$-wide channel with a 45-nm-long gate on
top of the heterostructure. At low temperatures (around $20\, \mathrm{mK}$) an irregular
set of peaks was found in the conductance as a function of gate voltage in the
region close to the depletion threshold. The amplitude of the peaks was on
the order of $e^{2}/h$. The origin of the effect could not be pinned down. The
authors examine the possibility that transmission resonances associated with
a square potential barrier are responsible for the oscillations in the conductance, but also note that the actual barrier is more likely to be smooth on
the scale of the wavelength. For such a smooth barrier the transmission
probability as a function of energy does not show oscillations. It seems most
likely that the effect is disorder-related. Davies and Nixon\cite{ref395} have suggested
that some of the structure observed in this experiment could be due to
potential fluctuations in the region under the gate. These fluctuations can be
rather pronounced close to the depletion threshold, due to the lack of
screening in the low-density electron gas. A quantum mechanical calculation
of transmission through such a fluctuating barrier has not been performed.
As discussed below, conductance peaks of order $e^{2}/h$ occur in the case of
resonant tunneling via localized states in the barrier (associated with
impurities), a mechanism that might well play a role in the experiment of
Washburn et al.\cite{ref394}

\begin{figure}
\centerline{\includegraphics[width=8cm]{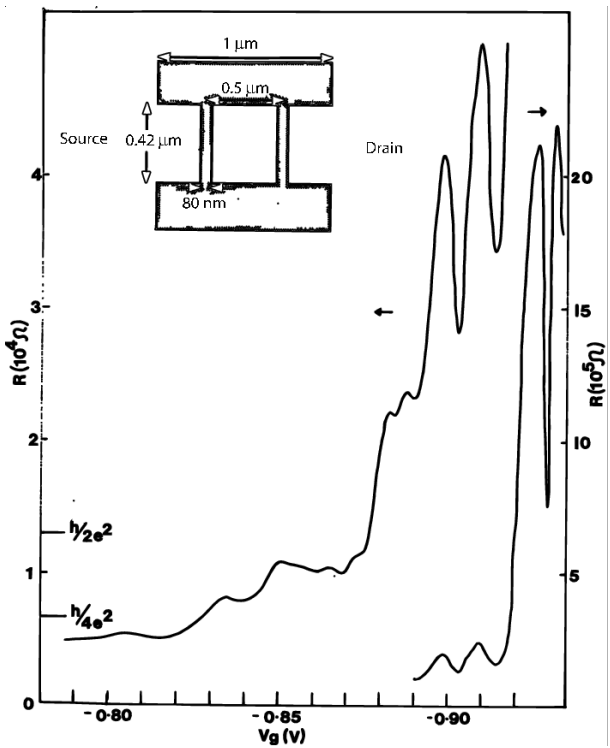}}
\caption{
Resistance versus gate voltage of a cavity (defined by gates on top of a GaAs-AlGaAs heterostructure; see inset), showing plateau like features (for $R\lesssim h/2e^{2}$) and tunneling resonances (for $R\gtrsim h/2e^{2}$). The left- and right-hand curves refer to the adjacent resistance scales. Taken from C. G. Smith et al., Surf.\ Sci.\ {\bf 228}, 387 (1990).
\label{fig70}
}
\end{figure}

In pursuit of resonant tunneling in a 2DEG, Chou et al.\cite{ref396}  have fabricated
double-barrier devices involving two closely spaced short gates across a wide
GaAs-AlGaAs heterostructure. Both the spacing and the length of the gates
were $100\, \mathrm{nm}$. They observed a peak in the transconductance (the derivative of
the channel current with respect to the gate voltage), which was attributed to
resonant tunneling through a quasi-bound state in the 2D potential well
between the barriers. Palevski et al.\cite{ref397} have also investigated transport
through two closely spaced potential barriers in a double-gate structure, but
they did not find evidence for transmission resonances.
A 3D potential well has truly bound states and is expected to show
the strongest transmission resonances. Transport through such a cavity or
``quantum box'' has been studied theoretically by several
authors.\cite{ref318,ref333,ref382,ref398} Experiments have been performed by Smith et
al.\cite{ref399,ref400,ref401}  Their device is based on a quantum point contact, but contains two
potential barriers that separate the constriction from the wide 2DEG regions
(see the inset of Fig.\ \ref{fig70}). As the negative gate voltage is increased, a potential
well is formed between the two barriers, resulting in confinement in all
directions. The tunneling regime corresponds to a resistance $R$ that is greater
than $h/2e^{2}$. It is also possible to study the ballistic regime $R<h/2e^{2}$ when the
height of the potential barriers is less than the Fermi energy. In this regime
the transmission resonances are similar to the resonances in long quantum
point contacts (these are determined by an interplay of tunneling through
evanescent modes and reflection at the entrance and exit of the point contact;
cf.\ Section \ref{sec13}). Results of Smith et al.\cite{ref399,ref400,ref401} for the resistance as a function of
gate voltage at $330\, \mathrm{mK}$ are reproduced in Fig.\ \ref{fig70}. In the tunneling regime
($R>h/2e^{2}$) giant resistance oscillations are observed. A regular series of
smaller resistance peaks is found in the ballistic regime ($R<h/2e^{2}$). Martin-Moreno and Smith\cite{ref333} have modeled the electrostatic potential in the device
of Refs.\ \onlinecite{ref399,ref400,ref401} and have performed a quantum mechanical calculation of
the resistance. Very reasonable agreement with the experimental data in the
ballistic regime was obtained. The tunneling regime was not compared in
detail with the experimental data. The results were found to depend rather
critically on the assumed chape of the potential, in particular on the rounding
of the tops of the potential barriers. Martin-Moreno and Smith also
investigated the effects of asymmetries in the device structure on the tunneling
resonances and found in particular that small differences in the two barrier
heights (of order 10\%) lead to a sharp suppression of the resonances, a finding
that sheds light on the fact that they were observed in certain devices only.
Experimentally, the effect of a magnetic field on the oscillations in the
resistance versus gate voltage was also investigated.\cite{ref399,ref400,ref401} A strong suppression of the peaks was found in relatively weak magnetic fields (of about
$0.3\, \mathrm{T}$).

\begin{figure}
\centerline{\includegraphics[width=8cm]{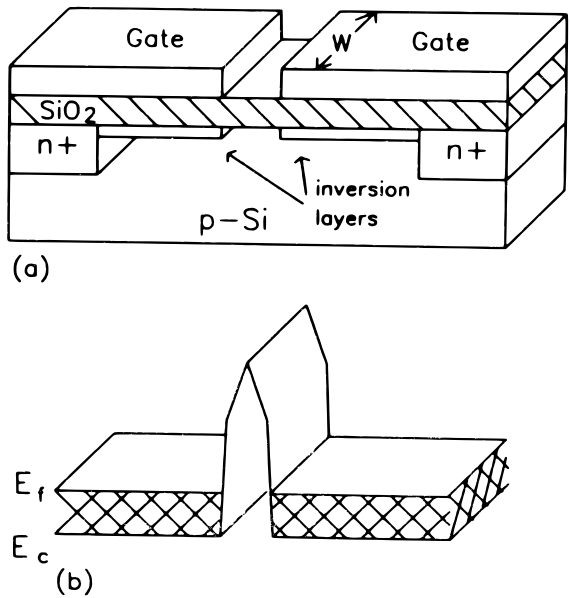}}
\caption{
Schematic diagram of a Si MOSFET with a split gate (a), which creates a potential barrier in the inversion layer (b). Taken from T. E. Kopley et al. Phys.\ Rev.\ Lett.\ {\bf 61}, 1654 (1988).
\label{fig71}
}
\end{figure}

\begin{figure}
\centerline{\includegraphics[width=8cm]{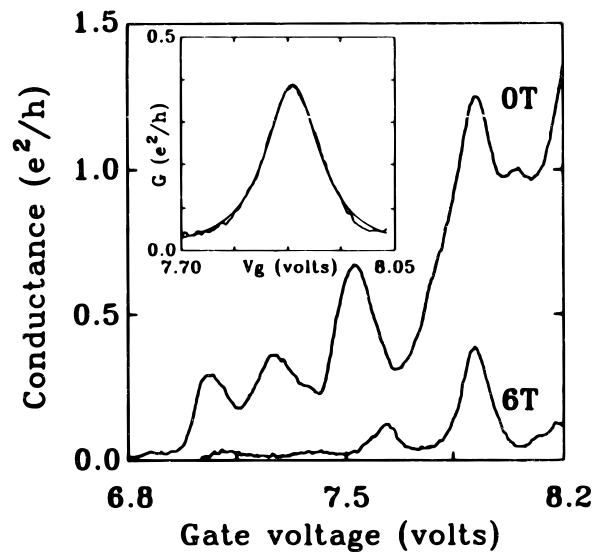}}
\caption{
Oscillations in the conductance as a function of gate voltage at 0.5 K are attributed to resonant tunneling through localized states in the potential barrier. A second trace is shown for a magnetic field of 6 T (with a horizontal offset of $- 0.04\,{\rm V}$). The inset is a close-up of the largest peak at 6 T, together with a Lorentzian fit. Taken from T. E. Kopley et al. Phys.\ Rev.\ Lett.\ {\bf 61}, 1654 (1988).
\label{fig72}
}
\end{figure}

Tunneling through a cavity, as in the experiment by Smith et al.,\cite{ref399,ref400,ref401} is
formally equivalent to tunneling through an impurity state (see, e.g., Refs.\ \onlinecite{ref402}
and \onlinecite{ref403}). The dramatic subthreshold structure found in the conductance of
quasi-one-dimensional MOSFETs has been interpreted in terms of resonant
tunneling through a series of localized states.\cite{ref32,ref35,ref36,ref37} Kopley et al.\cite{ref404} have
observed large conductance peaks in a MOSFET with a split gate (see Fig.\
\ref{fig71}). Below the 200-nm-wide slot in the gate, the inversion layer is interrupted
by a potential barrier. Pronounced conductance peaks were seen at $0.5\, \mathrm{K}$ as
the gate voltage was varied in the region close to threshold (see Fig.\ \ref{fig72}). No
clear correlation was found between the channel width and the peak spacing
or amplitude. The peaks were attributed to resonant transmission through
single localized states associated with bound states in the Si band gap in the
noninverted region under the gate.

The theory of resonant tunneling of noninteracting electrons through
localized states between two-dimensional reservoirs was developed by Xue
and Lee\cite{ref405} (see also Refs.\ \onlinecite{ref159} and \onlinecite{ref406}). If the resonances are well separated in
energy, a single localized state will give the dominant contribution to the
transmission probability. The maximum conductance on resonance is then
$e^{2}/h$ (for one spin direction), regardless of the number of channels $N$ in the
reservoirs.\cite{ref405,ref406} This maximum (which may be interpreted as a contact
resistance, similar to that of a quantum point contact) is attained if the
localized state has identical leak rates $\Gamma_{\mathrm{L}}/\hbar$ and $\Gamma_{\mathrm{R}}/\hbar$ to the left and right
reservoirs. Provided these leak rates are small (cf.\ Section \ref{sec21}) the conductance
$G$ as a function of Fermi energy $E_{\mathrm{F}}$ is a Lorentzian centered around the
resonance energy $E_{0}$:
\be
G(E_{\mathrm{F}})= \frac{e^{2}}{h}\frac{\Gamma_{\mathrm{L}}\Gamma_{\mathrm{R}}}{(E_{\mathrm{F}}-E_{0})^{2}+\frac{1}{4}(\Gamma_{\mathrm{L}}+\Gamma_{\mathrm{R}})^{2}}. \label{eq17.1}
\ee 
This is the Breit-Wigner formula of nuclear physics.\cite{ref93} For an asymmetrically placed impurity the peak height is reduced below $e^{2}/h$ (by up to a factor
$4\Gamma_{\mathrm{R}}/\Gamma_{\mathrm{L}}$  if $\Gamma_{\mathrm{L}}\gg  \Gamma_{\mathrm{R}}$).

The amplitudes of the peaks observed by Kopley et al.\cite{ref404} were found to be
in agreement with this prediction, while the line shape of an isolated peak
could be well described by a Lorentzian (see inset of Fig.\ \ref{fig72}). (Most of the
peaks overlapped, hampering a line-shape analysis). In addition, they studied
the effect of a strong magnetic field on the conductance peaks and found that
the amplitudes of most peaks were substantially suppressed. This was
interpreted as a reduction of the leak rates because of a reduced overlap
between the wave functions on the impurity and the reservoirs. The
amplitude of one particular peak was found to be unaffected by the field,
indicative of a symmetrically placed impurity in the barrier $(\Gamma_{\mathrm{R}}=\Gamma_{\mathrm{L}})$, while
the width of that peak was reduced, in agreement with Eq.\ (\ref{eq17.1}). This study
therefore exhibits many characteristic features of resonant tunneling through
a single localized state.

\begin{figure}
\centerline{\includegraphics[width=8cm]{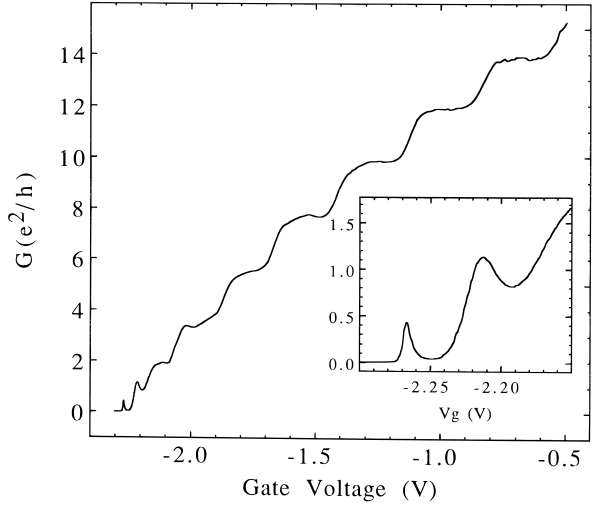}}
\caption{
Conductance as a function of gate voltage for a quantum point contact at 0.55 K. The inset is a close-up of the low-conductance regime, showing peaks attributed to transmission resonances associated with impurity states in the constriction. Taken from P. L. McEuen et al., Surf.\ Sci. {\bf 229}, 312 (1990).
\label{fig73}
}
\end{figure}

Transmission resonances due to an impurity in a quantum point contact
or narrow channel have been studied theoretically in Refs.\ \onlinecite{ref241,ref407}, and \onlinecite{ref408}.
In an experiment it may be difficult to distinguish these resonances from
those associated with reflection at the entrance and exit of the quantum point
contact (discussed in Section \ref{sec13}). A conductance peak associated with
resonant tunneling through an impurity state in a quantum point contact was
reported by McEuen et al.\cite{ref409} The experimental results are shown in Fig.\ \ref{fig73}.
The resonant tunneling peak is observed near the onset of the first
conductance plateau, where $G<2e^{2}/h$. A second peak seen in Fig.\ \ref{fig73} was
conjectured to be a signature of resonant scattering, in analog with similar
processes known in atomic physics.\cite{ref410}

\begin{figure}
\centerline{\includegraphics[width=8cm]{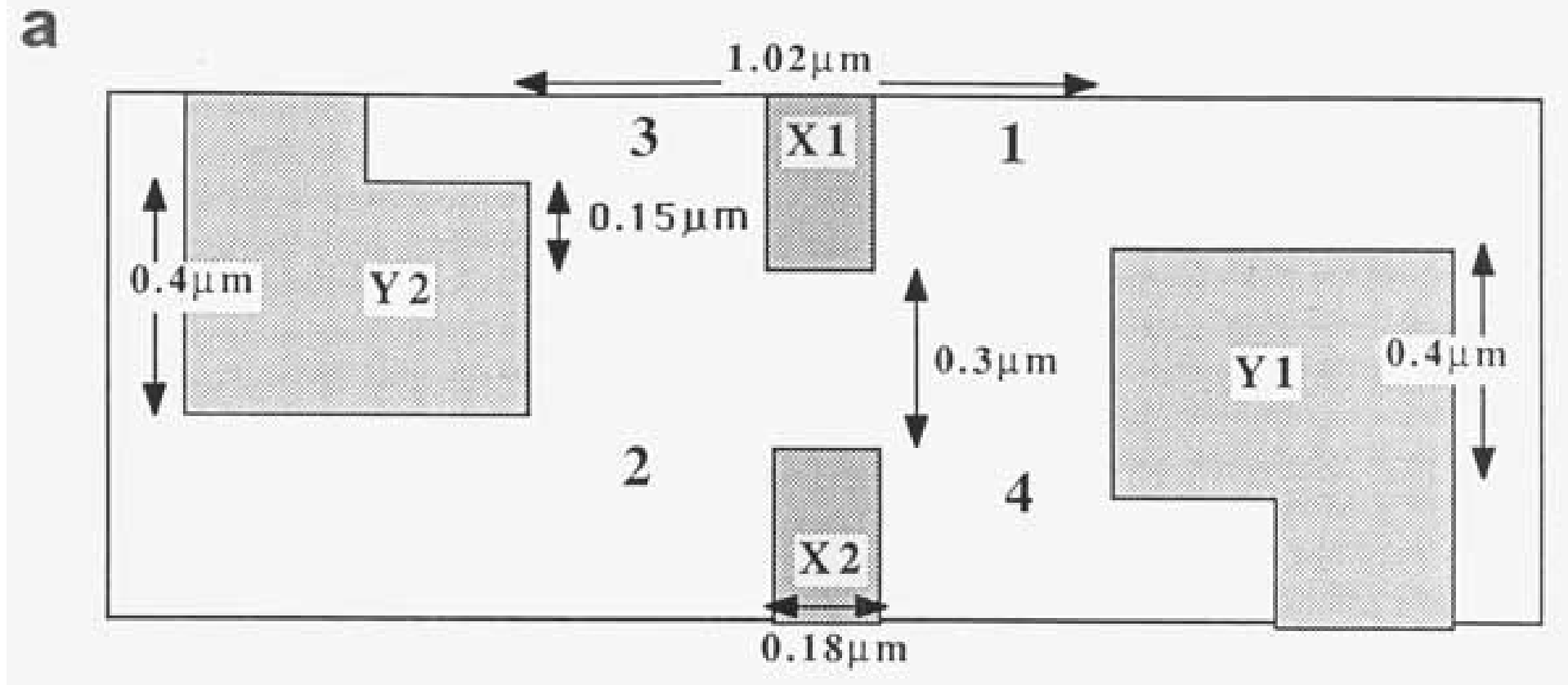}}

\centerline{\includegraphics[width=8cm]{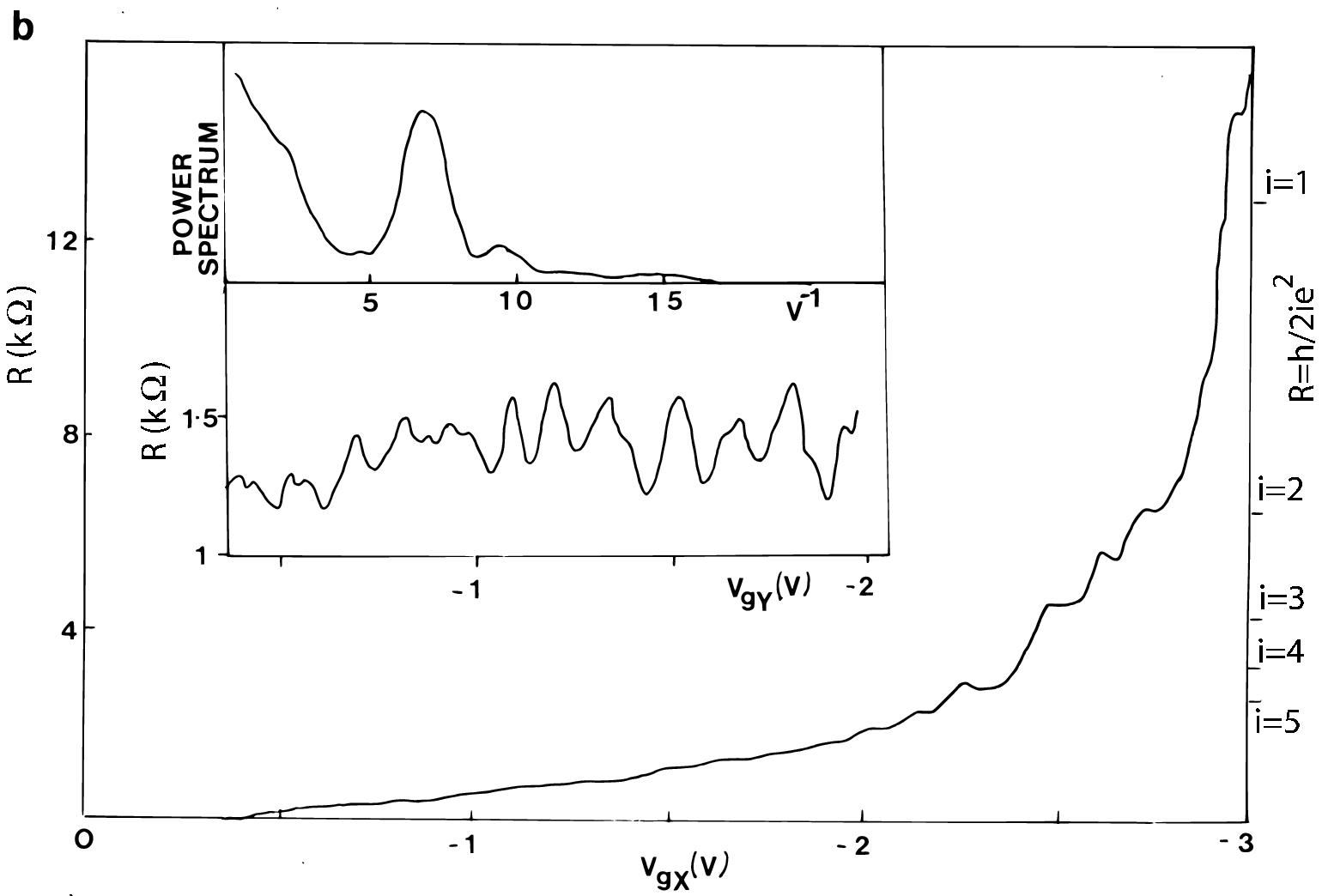}}
\caption{
(a) Schematic diagram of a constriction with two adjustable external reflectors defined by gates on top of a GaAs-AlGaAs heterostructure. (b) Plot of the constriction resistance as a function of gate voltage with the external reflector gates (Y1, Y2) grounded. Inset: Fabry-Perot-type transmission resonances due to a variation of the gate voltage on the reflectors (Y1, Y2) (bottom panel), and Fourier power spectrum (top panel). Taken from C. G. Smith et al., Surf.\ Sci. {\bf 228}, 387 (1990).
\label{fig74}
}
\end{figure}

We want to conclude this subsection on transmission resonances by
discussing an experiment by Smith et al.\cite{ref401,ref411} on what is essentially a
Fabry-Perot interferometer. The device consists of a point contact with
external reflectors in front of its entrance and exit. The reflectors are potential
barriers erected by means of two additional gate electrodes (see Fig.\ \ref{fig74}a). By
varying the gate voltage on the external reflectors of this device, Smith et al.\
could tune the effective cavity length without changing the width of the
narrow section. This experiment is therefore more controlled than the
quantum dot experiment\cite{ref399,ref400,ref401} discussed earlier. The resulting periodic
transmission resonances are reproduced in Fig.\ \ref{fig74}b. A new oscillation
appears each time the separation between the reflectors increases by $\lambda_{\mathrm{F}}/2$. A
numerical calculation for a similar geometry was performed by Avishai et
al.\cite{ref412} The significance of this experiment is that it is the first clear realization
of an electrostatically tuned electron interferometer. Such a device has
potential transistor applications. Other attempts to fabricate an electrostatic
interferometer have been less succesful. The electrostatic Aharonov-Bohm
effect in a ring was discussed in Section \ref{sec8}. The solid-state analogue of the
microwave stub tuner (proposed by Sols et al.\cite{ref413} and by Datta\cite{ref414}) was
studied experimentally by Miller et al.\cite{ref415} The idea is to modify the
transmission through a narrow channel by changing the length of a side
branch (by means of a gate across the side branch). Miller et al.\ have
fabricated such a $\mathrm{T}$-shaped conductor and found some evidence for the
desired effect. Much of the structure was due, however, to disorder-related
conductance fluctuations. The electrostatic Aharonov-Bohm effect had
similar problems. Transport in a long and narrow channel is simply not fully
ballistic, because of partially diffuse boundary scattering and impurity
scattering. The device studied by Smith et al.\ worked because it made use of a
very short constriction (a quantum point contact), while the modulation of
the interferometer length was done externally in the wide 2DEG, where the
effects of disorder are much less severe (in high-mobility material).

\subsubsection{\label{sec17b} Coulomb blockade}

\begin{figure}
\centerline{\includegraphics[width=6cm]{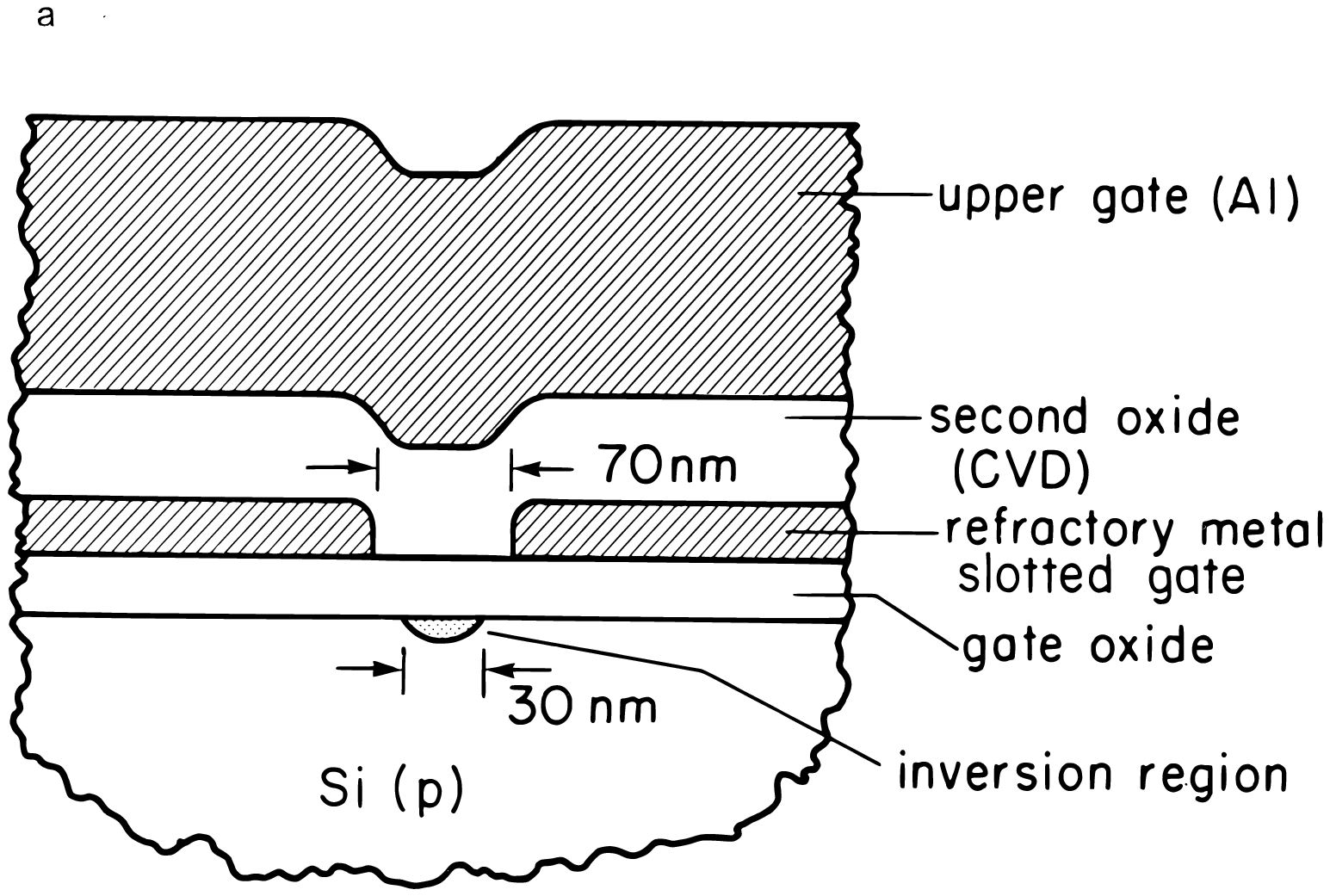}}

\centerline{\includegraphics[width=6cm]{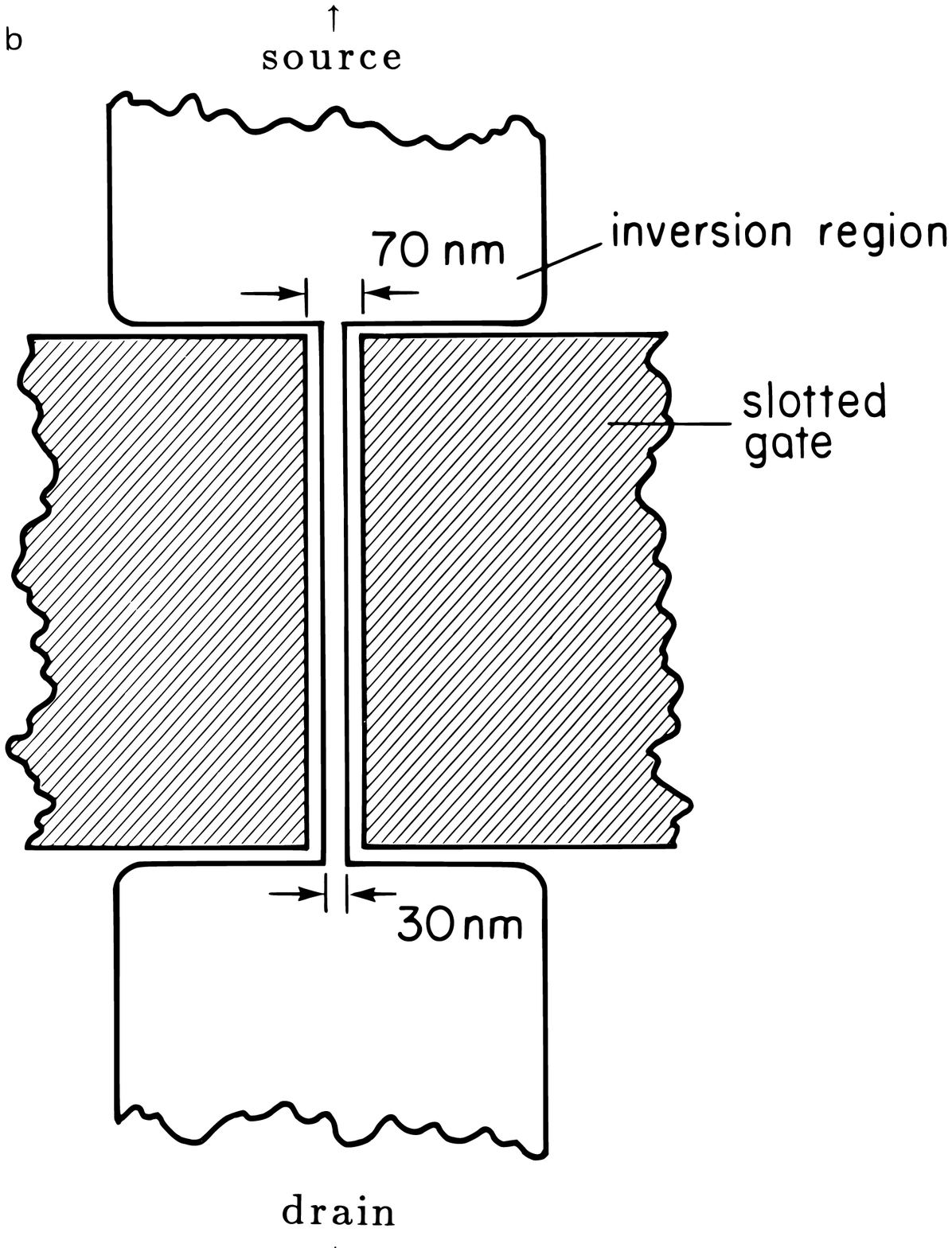}}
\caption{
Schematic cross sectional (a) and top (b) view of a double-gate Si MOSFET device. The lower split gate is at a negative voltage, confining the inversion layer (due to the positive voltage on the upper gate) to a narrow channel. Taken from J. H. F. Scott-Thomas et al., Phys.\ Rev.\ Lett.\ {\bf 62}, 583 (1989).
\label{fig75}
}
\end{figure}

\begin{figure}
\centerline{\includegraphics[width=8cm]{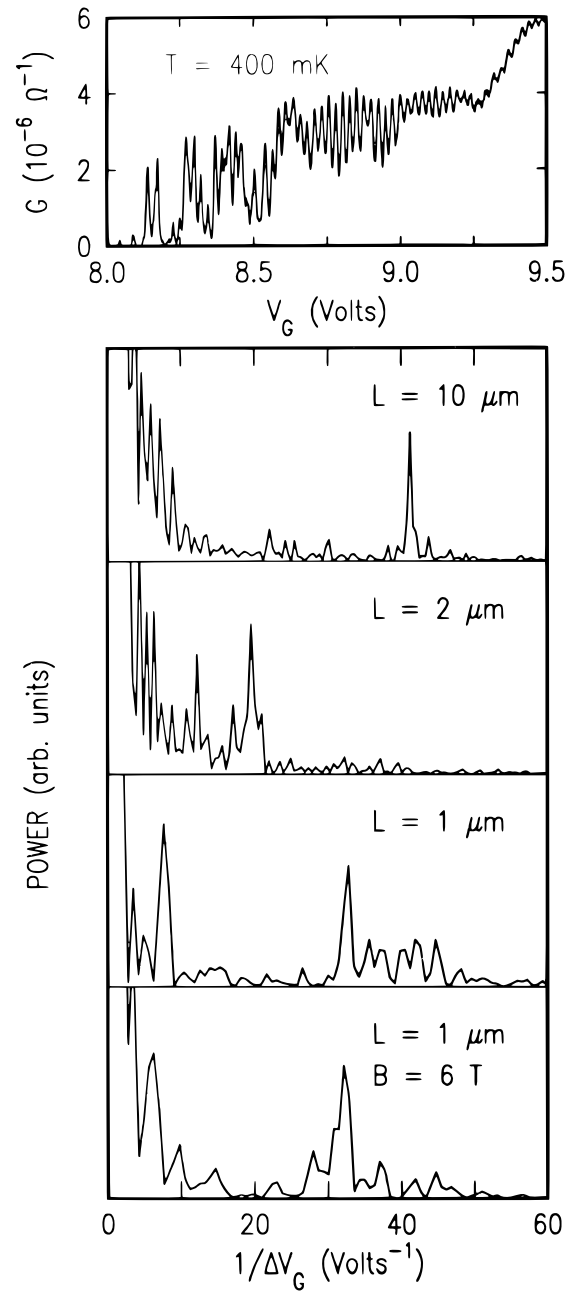}}
\caption{
Top panel: Periodic oscillations in the conductance versus gate voltage at 0.4 K for a 10-$\mu{\rm m}$-long inversion channel. Next three panels: Fourier power spectra of this curve and of data obtained for 2- and 1-$\mu{\rm m}$-long channels. Bottom panel: Fourier spectrum for the 1-$\mu{\rm m}$-long device in a magnetic field of 6 T. Taken from J. H. F. Scott-Thomas et al., Phys.\ Rev.\ Lett.\ {\bf 62}, 583 (1989).
\label{fig76}
}
\end{figure}

In this subsection we would like to speculate on the effects of electron-electron interactions on tunneling through impurities in narrow semiconductor channels, in relation to a recent paper in which Scott-Thomas et al.\cite{ref416}
announced the discovery of conductance oscillations periodic in the density
of a narrow Si inversion layer. The device features a continuous gate on top of
a split gate, as illustrated schematically in Fig.\ \ref{fig75}. In the experiment, the
voltage on the upper gate is varied while the split-gate voltage is kept
constant. Figure \ref{fig76} shows the conductance as a function of gate voltage at
$0.4\, \mathrm{K}$, as well as a set of Fourier power spectra obtained for devices of
different length. A striking pattern of rapid periodic oscillations is seen. No
correlation is found between the periodicity of the oscillations and the
channel length, in contrast to the transmission resonances in ballistic
constrictions discussed in Sections \ref{sec13} and \ref{sec17a}. The oscillations die out as the
channel conductance increases toward $e^{2}/h\approx 4\times 10^{-5}\,\Omega^{-1}$. The conductance peaks are relatively insensitive to a change in temperature, while the
minima depend exponentially on temperature as $\exp(- E_{\mathrm{a}}/k_{\mathbf{B}}T)$, with an
activation energy $E_{\mathrm{a}}\approx 50\,\mu \mathrm{eV}$. Pronounced nonlinearities occur in the
current as a function of source-drain voltage. An interpretation in terms of
pinned charge density waves was suggested,\cite{ref416} based on a model due to
Larkin and Lee\cite{ref417} and Lee and Rice.\cite{ref418} In such a model, one expects the
conductance to be thermally activated, because of the pinning of the charge
density wave by impurities in the one-dimensional channel. The activation
energy is determined by the most strongly pinned segment in the channel, and
periodic oscillations in the conductance as a function of gate voltage
correspond to the condition that an integer number of electrons is contained
between the two impurities delimiting that specific segment. The same
interpretation has been given to a similar effect observed in a narrow channel
in a GaAs-AlGaAs heterostructure by Meirav et al.\cite{ref85}

\begin{figure}
\centerline{\includegraphics[width=8cm]{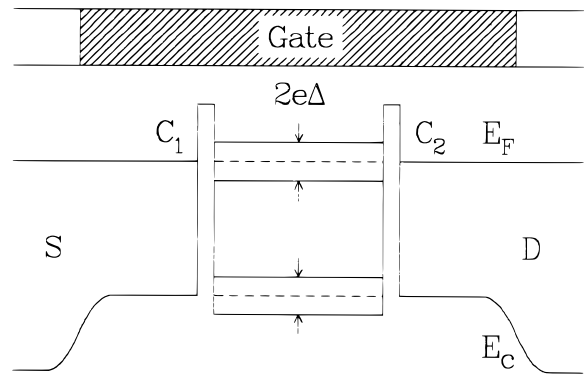}}
\caption{
Schematic representation of the bottom of the conduction band $E_{\rm c}$, and Fermi energy $E_{\rm F}$ in the device of Fig.\ \ref{fig76} along the channel. The band bending at the connections of the narrow channel to the wide source S and drain D regions arises from the higher threshold for the electrostatic creation of a narrow inversion layer by a gate (shaded part). Tunnel barriers associated with two scattering centers are shown. The maximum Fermi energy difference sustainable by the Coulomb blockade, $\Delta E_{\rm F}=\pm e\Delta$ (where $\Delta=e/2C$, with $C = C_{1} + C_{2}$), is indicated. Taken from H. van Houten and C. W. J. Beenakker, Phys.\ Rev.\ Lett.\ {\bf 63}, 1893 (1989).
\label{fig77}
}
\end{figure}

We have proposed\cite{ref419} an alternative {\it single-electron\/} explanation of the
remarkable effect discovered by Scott-Thomas et al.,\cite{ref416} based upon the
concept of the Coulomb blockade of tunneling mentioned at the beginning of
this section. Likharev\cite{ref391} and Mullen et al.\cite{ref420} have studied theoretically the
possibility of removing the Coulomb blockade by capacitive charging (by
means of a gate electrode) of the region between two tunnel barriers. They
found that the conductance of this system exhibits periodic peaks as a
function of gate voltage, due to the modulation of the net charge (mod $e$) on
the interbarrier region. Following the theoretical papers,\cite{ref391,ref420} the authors
in Ref.\ \onlinecite{ref419} proposed that the current through the channel in the experiment
of Scott-Thomas et al.\cite{ref416} is limited by tunneling through potential barriers
constituted by two dominant scattering centers that delimit a segment of the
channel (see Fig.\ \ref{fig77}). Because the number of electrons localized in the region
between the two barriers is necessarily an integer, a charge imbalance, and
hence an electrostatic potential difference, arises between this region and the
adjacent regions connected to wide electron gas reservoirs. As the gate
voltage is varied, the resulting Fermi level difference $\Delta E_{\mathrm{F}}$ oscillates in a
sawtooth pattern between $\pm e\Delta$, where $\Delta=e/2C$ and $C=C_{1}+C_{2}$ is the
effective capacitance of the region between the two barriers. The single-electron charging energy $e^{2}/2C$ maintains the Fermi level difference until
$\Delta E_{\mathrm{F}}=\pm e\Delta$ (this is the Coulomb blockade). When $\Delta E_{\mathrm{F}}=\pm e\Delta$, the energy
required for the transfer of a single electron to (or from) the region between
the two barriers vanishes so that the Coulomb blockade is removed. The
conductance then shows a maximum at low temperatures $T$ and source-drain voltages $V$ ($k_{B}T/e, V\lesssim\Delta$). We note that in the case of very different
tunneling rates through the two barriers, one would expect steps in the
current as a function of source-drain voltage, which are not observed in the
experiments.\cite{ref85,ref416} For two similar barriers this ``Coulomb staircase'' is
suppressed.\cite{ref420} The oscillation of the Fermi energy as the gate voltage is
varied thus leads to a sequence of conductance peaks. The periodicity of the
oscillations corresponds to the addition of a single electron to the region
between the two scattering centers forming the tunnel barriers, so the
oscillations are periodic in the density, as in the experiment. This single-electron tunneling mechanism also explains the observed activation of the
conductance minima and the insensitivity to a magnetic field.\cite{ref85,ref416} The
capacitance associated with the region between the scattering centers is hard
to ascertain. The experimental value of the activation energy $E_{\mathrm{a}}\approx 50\,\mu \mathrm{eV}$
would imply $C\approx e^{2}/2E_{\mathrm{a}}\approx 10^{-15}\,{\rm F}$. Kastner et al.\cite{ref421} argue that the
capacitance in the device is smaller than this amount by an order of
magnitude (the increase in the effective capacitance due to the presence of the
gate electrodes is taken into account in their estimate). In addition, they point
to a discrepancy between the value for the Coulomb blockade inferred from
the nonlinear conductance and that from the thermal activation energy. The
temperature dependence of the oscillatory conductance was found to be
qualitatively different in the experiment by Meirav et al.\cite{ref85} At elevated
temperatures an exponential $T$-dependence was found, but at low temperatures the data suggest a much weaker $T$-dependence. It is clear that more
experimental and theoretical work is needed to arrive at a definitive
interpretation of this intriguing phenomenon.

It would be of interest to study the effects of the Coulomb blockade of
tunneling in a more controlled fashion in a structure with two adjustable
potential barriers. Such an experiment was proposed by Glazman and
Shekter,\cite{ref422} who studied theoretically a system similar to the cavity of the
experiments by Smith et al.\cite{ref399,ref400,ref401} (discussed in Section \ref{sec17a}). A difficulty
with this type of device is, as pointed out in Ref.\ \onlinecite{ref422}, that a variation in gate
voltage affects the barrier height (and thus their transparency) as well as the
charge in the cavity. This is expected to lead to an exponential damping of the
oscillations due to the Coulomb blockade.\cite{ref391,ref420} A characteristic feature of
these oscillations is their insensitivity to an applied magnetic field, which can
serve to distinguish the effect from oscillations due to resonant tunneling
(Section \ref{sec17a}). The field dependence of the peaks observed by Smith et
al.\cite{ref399,ref400,ref401} in the tunneling regime was not reported, so the question of
whether or not the Coulomb oscillations are observed in their experiment
remains unanswered. In our opinion, substantial progress could be made
with the development of thin tunnel barriers of larger height, which would be
less sensitive to the application of an external gate voltage. If our interpretation of the experiments by Scott-Thomas et al.\cite{ref416} and Meirav et al.\cite{ref85} is
correct, such tunneling barriers might be formed by the incorporation of
negatively charged impurities (e.g., ionized acceptors) in a narrow electron
gas channel. This speculation is based on the fact that such acceptor
impurities are present in the Si inversion layers of the experiment of Scott-Thomas et al.,\cite{ref416} as well as in the {\it p}-{\it n\/} junctions employed for lateral
confinement by Meirav et al.\cite{ref85}

As we were completing this review, we learned of several experiments that
demonstrate the Coulomb blockade in split-gate confined GaAs-AlGaAs
heterostructures.\cite{ref423,ref424,ref425} These experiments should open the way for the
controlled study of the effects of Coulomb interactions on tunneling in
semiconductor nanostructures.

\section{\label{secIV} Adiabatic transport}
\subsection{\label{sec18} Edge channels and the quantum Hall effect}

In this section we give an overview of the characteristics of adiabatic
transport via edge channels in the regime of the quantum Hall effect as a
background to the following sections. We restrict ourselves here to the {\it integer}
quantum Hall effect, where the edge channels can be described by single-electron states. Recent developments on adiabatic transport in the regime of
the {\it fractional\/} quantum Hall effect (which is fundamentally a many-body
effect) will be considered in Section \ref{sec20}.

\subsubsection{\label{sec18a} Introduction}

Both the quantum Hall effect (QHE) and the quantized conductance of a
ballistic point contact are described by the same relation, $G=Ne^{2}/h$,
between the conductance $G$ and the number $N$ of propagating modes at the
Fermi level (counting both spin directions separately). The smooth transition
from zero-field quantization to QHE that follows from this relation is evident
from Fig.\ \ref{fig48}. The nature of the modes is very different, however, in weak and
strong magnetic fields. As we discussed in Section \ref{sec12a}, the propagating
modes in a strong magnetic field consist of edge states, which interact with
one of the sample edges only. Edge states with the same mode index are
referred to collectively as an {\it edge channel}. Edge channels at opposite edges
propagate in opposite directions. In a weak magnetic field, in contrast, the
modes consist of magnetoelectric subbands that interact with both edges. In
that case there is no spatial separation of modes propagating in opposite
directions.

The different spatial extension of edge channels and magnetoelectric
subbands leads to an entirely different sensitivity to scattering processes in
weak and strong magnetic fields. Firstly, the zero-field conductance
quantization is destroyed by a small amount of elastic scattering (due to
impurities or roughness of the channel boundaries; cf.\ Refs.\ \onlinecite{ref313,ref316,ref317,ref407},
and \onlinecite{ref408}), while the QHE is robust to scattering.\cite{ref97} This difference is a
consequence of the {\it suppression of backscattering\/} by a magnetic field discussed
in Section \ref{sec13b}, which itself follows from the spatial separation at opposite
edges of edge channels moving in opposite directions. Second, the spatial
separation of edge channels at the {\it same\/} edge in the case of a smooth confining
potential opens up the possibility of {\it adiabatic transport\/} (i.e., the full
suppression of interedge channel scattering). In weak magnetic fields,
adiabaticity is of importance within a point contact, but not on longer length
scales (cf.\ Sections \ref{sec13a} and \ref{sec15a}). In a wide 2DEG region, scattering among
the modes in weak fields establishes local equilibrium on a length scale given
by the inelastic scattering length (which in a high-mobility GaAs-AlGaAs
heterostructure is presumably not much longer than the elastic scattering
length $l\sim 10\,\mu \mathrm{m}$). The situation is strikingly different in a strong magnetic
field, where the {\it selective\/} population and detection of edge channels observed
by van Wees et al.\cite{ref426} has demonstrated the persistence of adiabaticity outside
the point contact.

In the absence of interedge channel scattering the various edge channels at
the same boundary can be occupied up to different energies and consequently
carry different amounts of current. The electron gas at the edge of the sample
is then not in {\it local\/} equilibrium. Over some long distance (which is not yet
known precisely) adiabaticity breaks down, leading to a partial equilibration
of the edge channels. However, as demonstrated by Komiyama et al.\cite{ref427} and
by others,\cite{ref307,ref428,ref429,ref430} local equilibrium is not fully established even on
macroscopic length scales exceeding $0.25\, \mathrm{mm}$. Since local equilibrium is a
prerequisite for the use of a local resistivity tensor, these findings imply a
nonlocality of the transport that had not been anticipated in theories of the
QHE (which are commonly expressed in terms of a local resistivity).\cite{ref97}

A theory of the QHE that is able to explain anomalies resulting from the
absence of local equilibrium has to take into account the properties of the
current and voltage contacts used to measure the Hall resistance. That is not
necessary if local equilibrium is established at the voltage contacts, for the
fundamental reason that two systems in equilibrium that are in contact have
identical electrochemical potentials. In the Landauer-B\"{u}ttiker formalism
described in Section \ref{sec12b}, the contacts are modeled by electron gas reservoirs
and the resistances are expressed in terms of transmission probabilities of
propagating modes at the Fermi level from one reservoir to the other. This
formalism is not restricted to zero or weak magnetic fields, but can equally
well be applied to the QHE, where edge channels form the modes. In this way
B\"{u}ttiker could show\cite{ref112} that the nonideality of the coupling of the reservoirs
to the conductor affects the accuracy of the QHE in the absence of local
equilibrium. An {\it ideal\/} contact in the QHE is one that establishes an
equilibrium population among the outgoing edge channels by distributing
the injected current equally among these propagating modes (this is the
equipartitioning of current discussed for an ideal electron waveguide in
Section \ref{sec12b}). A quantum point contact that selectively populates certain
edge channels\cite{ref426} can thus be seen as an extreme example of a nonideal, or
{\it disordered}, contact.

\subsubsection{\label{sec18b} Edge channels in a disordered conductor}

After this general introduction, let us now discuss in some detail how edge
channels are formed at the boundary of a 2DEG in a strong magnetic field. In
Section \ref{sec12a} we discussed the edge states in the case of a narrow channel
without disorder, relevant for the point contact geometry. Edge states were
seen to originate from Landau levels, which in the bulk lie below the Fermi
level but rise in energy on approaching the sample boundary (cf.\ Fig.\ \ref{fig40}b).
The point of intersection of the $n$th Landau level $(n=1,2, \ldots)$ with the Fermi
level forms the site of edge states belonging to the $n$th edge channel. The
number $N$ of edge channels at $E_{\mathrm{F}}$ is equal to the number of bulk Landau
levels below $E_{\mathrm{F}}$. This description can easily be generalized to the case of a
slowly varying potential energy landscape $V(x, y)$ in the 2DEG, in which case
a semiclassical analysis can be applied.\cite{ref431} The energy $E_{\mathrm{F}}$ of an electron at the
Fermi level in a strong magnetic field contains a part $(n-\frac{1}{2})\hbar\omega_{\mathrm{c}}$ due to the
quantized cyclotron motion and a part $\pm\frac{1}{2} g\mu_{\mathbf{B}}B$ (depending on the spin
direction) from spin splitting. The remainder is the energy $E_{\mathrm{G}}$ due to the
electrostatic potential
\be
E_{\mathrm{G}}=E_{\mathrm{F}}-(n- {\textstyle\frac{1}{2}})\hbar\omega_{\mathrm{c}}\pm{\textstyle\frac{1}{2}}g\mu_{\mathbf{B}}B. \label{eq18.1}
\ee
The cyclotron orbit center $\mathbf{R}$ is guided along equipotentials of $V$ at the
guiding center energy $E_{\mathrm{G}}$. As derived in Section \ref{sec11b}, the drift velocity $\mathbf{v}_{\mathrm{drift}}$ of
the orbit center (known as the guiding center drift or $\mathbf{E}\times \mathbf{B}$ drift) is given by
\be
 \mathbf{v}_{\mathrm{drift}}(\mathbf{R})=\frac{1}{eB^{2}}\nabla V(\mathbf{R})\times \mathbf{B}, \label{eq18.2}
\ee 
which indeed is parallel to the equipotentials. An important distinction with
the weak-field case of Section \ref{sec11b} is that the spatial extension of the
cyclotron orbit can now be neglected, so $V$ is evaluated at the position of the
orbit center in Eq.\ (\ref{eq18.2}) [compared with Eq.\ (\ref{eq11.1})]. The guiding center drift
contributes a kinetic energy 
$\frac{1}{2}mv_{\mathrm{drift}}^{2}$ to the energy of the electron, which is
small for large $B$ and smooth $V$. (More precisely,
$\frac{1}{2}mv_{\mathrm{drift}}^{2}\ll \hbar\omega_{\mathrm{c}}$ if
$|\nabla V|\ll \hbar\omega_{\mathrm{c}}/l_{\mathrm{m}}$, with $l_{\mathrm{m}}$ the magnetic length defined as $l_{\mathrm{m}}\equiv(h/eB)^{1/2}.)$ This
kinetic energy term has therefore not been included in Eq.\ (\ref{eq18.1}).

The simplicity of the guiding center drift along equipotentials has been
originally used in the percolation  theory\cite{ref432,ref433,ref434} of the QHE, soon after its
experimental discovery.\cite{ref8} In this theory the existence of edge states is ignored,
so the Hall resistance is not expressed in terms of equilibrium properties of
the 2DEG (in contrast to the edge channel formulation that will be discussed).
The physical requirements on the smoothness of the disorder potential have
received considerable attention\cite{ref435,ref436} in the context of the percolation
theory and, more recently,\cite{ref437,ref438,ref439} in the context of adiabatic transport in
edge channels. Strong potential variations should occur on a spatial scale
that is large compared with the magnetic length $l_{\mathrm{m}}$ ($l_{\mathrm{m}}$ corresponds to the
cyclotron radius in the QHE, $l_{\mathrm{cycl}}\equiv l_{\mathrm{m}}(2n-1)^{1/2}\approx l_{\mathrm{m}}$ if the Landau level
index $n\approx 1$). More rapid potential fluctuations may be present provided their
amplitude is much less than $\hbar\omega_{\mathrm{c}}$ (the energy separation of Landau levels).

\begin{figure}
\centerline{\includegraphics[width=8cm]{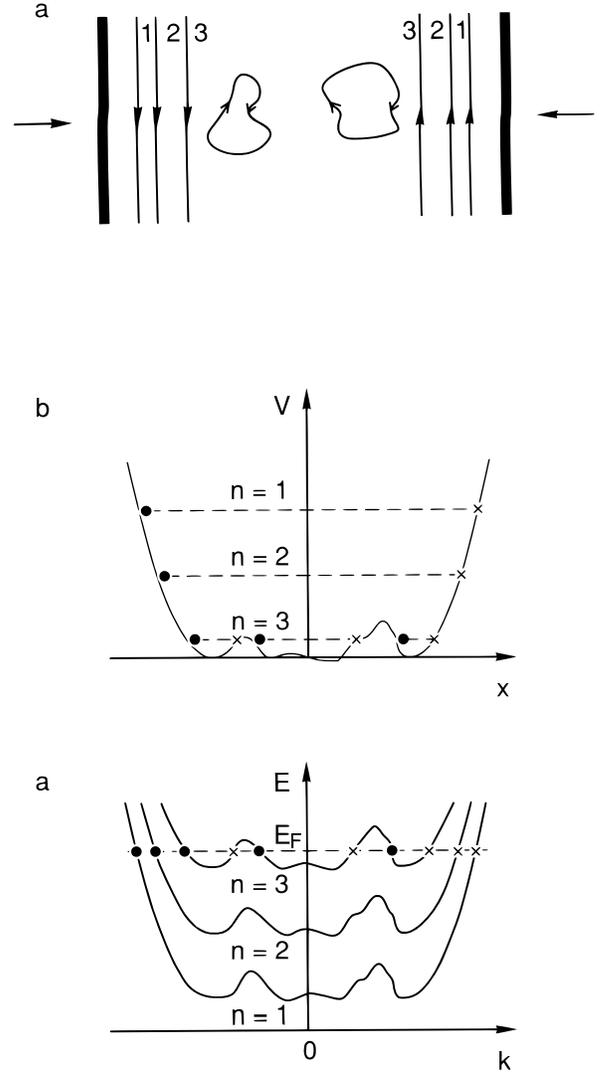}}
\caption{
Formation of edge channels in a disordered potential, from various viewpoints discussed in the text.
\label{fig78}
}
\end{figure}

In Fig.\ \ref{fig78} we have illustrated the formation of edge channels in a smooth
potential energy landscape from various viewpoints. The wave functions of
states at the Fermi level are extended along equipotentials at the guiding
center energy (\ref{eq18.1}), as shown in Fig.\ \ref{fig78}a (for Landau level index $n=1,2,3$
and a single spin direction). One can distinguish between {\it extended\/} states near
the sample boundaries and {\it localized\/} states encircling potential maxima and
minima in the bulk. The extended states at the Fermi level form the edge
channels. The edge channel with the smallest index $n$ is closest to the sample
boundary, because it has the largest $E_{\mathrm{G}}$ [Eq.\ (\ref{eq18.1})]. This is seen more clearly
in the cross-sectional plot of $V(x, y)$ in Fig.\ \ref{fig78}b (along the line connecting the
two arrows in Fig.\ \ref{fig78}a). The location of the states at the Fermi level is
indicated by dots and crosses (depending on the direction of motion). The
value of $E_{\mathrm{G}}$ for each $n$ is indicated by the dashed line. If the peaks and dips of
the potential in the bulk have amplitudes below $\hbar\omega_{\mathrm{c}}/2$, then only states with
highest Landau level index can exist in the bulk at the Fermi level. This is
obvious from Fig.\ \ref{fig78}c, which shows the total energy of a state
$E_{\mathrm{G}}+(n- \frac{1}{2})\hbar\omega_{\mathrm{c}}$ along the same cross section as Fig.\ \ref{fig78}b. If one identifies
$k=-xeB/\hbar$, this plot can be compared with Fig.\ \ref{fig40}b of the dispersion
relation $E_{n}(k)$ for a disorder-free electron waveguide in strong magnetic field.

\begin{figure}
\centerline{\includegraphics[width=8cm]{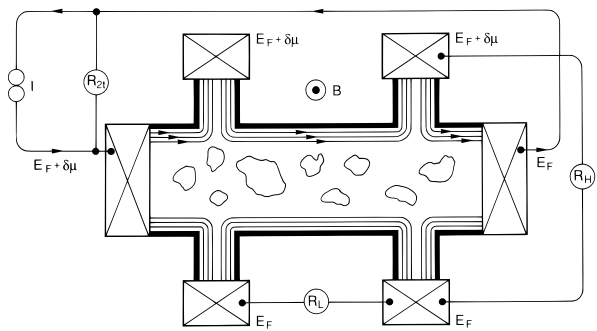}}
\caption{
Measurement configuration for the two-terminal resistance $R_{2{\rm t}}$, the four-terminal Hall resistance $R_{\rm H}$, and the longitudinal resistance $R_{\rm L}$. The edge channels at the Fermi level are indicated; arrows point in the direction of motion of edge channels filled by the source contact at chemical potential $E_{\rm F} + \delta\mu$. The current is equipartitioned among the edge channels at the upper edge, corresponding to the case of local equilibrium.
\label{fig79}
}
\end{figure}

A description of the QHE based on extended edge states and localized
bulk states, as in Fig.\ \ref{fig78}, was first put forward by Halperin\cite{ref440} and further
developed by several authors.\cite{ref441,ref442,ref443,ref444} In these papers a local equilibrium is
assumed at each edge. In the presence of a chemical potential difference $\delta\mu$
between the edges, each edge channel carries a current $(e/h)\delta\mu$ and thus
contributes $e^{2}/h$ to the Hall conductance (cf.\ the derivation of Landauer's
formula in Section \ref{sec12b}). In this case of local equilibrium the two-terminal
resistance $R_{2\mathrm{t}}$ of the Hall bar is the same as the four-terminal Hall resistance
$R_{\mathrm{H}}=R_{2\mathrm{t}}=h/e^{2}N$ (see Fig.\ \ref{fig79}). The longitudinal resistance vanishes, $R_{\mathrm{L}}=0$.
The distinction between a longitudinal and Hall resistance is topological: A
four-terminal resistance measurement gives $R_{\mathrm{H}}$ if current and voltage
contacts alternate along the boundary of the conductor, and $R_{\mathrm{L}}$ if that is not
the case. There is no need to further characterize the contacts in the case of
local equilibrium at the edge.

If the edges are not in local equilibrium, the measured resistance depends
on the properties of the contacts. Consider, for example, a situation in which
the edge channels at the lower edge are in equilibrium at chemical potential
$E_{\mathrm{F}}$, while the edge channels at the upper edge are not in local equilibrium.
The current at the upper edge is then not equipartitioned among the $N$
modes. Let $f_{n}$ be the fraction of the total current $I$ that is carried by states
above $E_{\mathrm{F}}$ in the $n$th edge channel at the upper edge, $I_{n}=f_{n}I$. The voltage
contact at the lower edge measures a chemical potential $E_{\mathrm{F}}$ regardless of its
properties. The voltage contact at the upper edge, however, will measure a
chemical potential that depends on how it couples to each of the edge
channels. The transmission probability $T_{n}$ is the fraction of $I_{n}$ that is
transmitted through the voltage probe to a reservoir at chemical potential
$E_{\mathrm{F}}+\delta\mu$. The incoming current
\be
I_{\mathrm{in}}= \sum_{n=1}^{N}T_{n}f_{n}I,\;\;{\rm with}\;\;
\sum_{n=1}^{N}f_{n}=1, \label{eq18.3}
\ee
has to be balanced by an outgoing current
\be
I_{\mathrm{out}}= \frac{e}{h}\delta\mu(N-R)=\frac{e}{h}\delta\mu\sum_{n=1}^{N}T_{n}  \label{eq18.4}
\ee
of equal magnitude, so that the voltage probe draws no net current. (In Eq.\
(\ref{eq18.4}) we have applied Eq.\ (\ref{eq12.14}) to identify the total transmission probability $N-R$ of outgoing edge channels with the sum of transmission
probabilities $T_{n}$ of incoming edge channels.) The requirement $I_{\mathrm{in}}=I_{\mathrm{out}}$
determines $\delta\mu$ and hence the Hall resistance $R_{\mathrm{H}}=\delta\mu/eI$:
\be
R_{\mathrm{H}}= \frac{h}{e^{2}}\left(\sum_{n=1}^{N}T_{n}f_{n}\right)\left(\sum_{n=1}^{N}T_{n}\right)^{-1}.   \label{eq18.5}
\ee
The Hall resistance has its regular quantized value $R_{\mathrm{H}}=h/e^{2}N$ only if {\it either}
$f_{n}=1/N$ {\it or\/} $T_{n}=1$, for $n=1,2, \ldots, N$. The first case corresponds to local
equilibrium (the current is equipartitioned among the modes), the second case
to an ideal contact (all edge channels are fully transmitted). The Landauer-B\"{u}ttiker formalism discussed in Section \ref{sec12b} forms the basis on which
anomalies in the QHE due to the absence of local equilibrium in combination
with nonideal contacts can be treated theoretically.\cite{ref112}

A nonequilibrium population of the edge channels is generally the result of
{\it selective backscattering}. Because edge channels at opposite edges of the
sample move in opposite directions, backscattering requires scattering from
one edge to the other. Selective backscattering of edge channels with $n\geq n_{0}$ is
induced by a potential barrier across the sample,\cite{ref113,ref339,ref340,ref427}  if its height is
between the guiding center energies of edge channel $n_{0}$ and $n_{0}-1$ (note that
the edge channel with a larger index $n$ has a smaller value of $E_{\mathrm{G}}$). The
anomalous Shubnikov-De Haas effect,\cite{ref428} to be discussed in Section \ref{sec19}, has
demonstrated that selective backscattering can also occur {\it naturally\/} in the
absence of an imposed potential barrier. The edge channel with the highest
index $n=N$ is selectively backscattered when the Fermi level approaches the
energy $(N- \frac{1}{2})\hbar\omega_{\mathrm{c}}$ of the $N$th bulk Landau level. The guiding center energy of
the $N$th edge channel then approaches zero, and backscattering either by
tunneling or by thermally activated processes becomes effective, but for that
edge channel only, which remains almost completely decoupled from the
other $N-1$ edge channels over distances as large as 250 $\mu \mathrm{m}$ (although on
that length scale the edge channels with $n\leq N-1$ have equilibrated to a
large extent).\cite{ref429}

\subsubsection{\label{sec18c} Current distribution}

The edge channel theory has been criticized on the grounds that experiments measure a nonzero current in the bulk of a Hall bar.\cite{ref445} In this
subsection we want to point out that a measurement of the current
distribution cannot be used to prove or disprove the edge channel formulation of the QHE.

The fact that the Hall resistance can be expressed in terms of the
transmission probabilities of edge states at the Fermi level does {\it not\/} imply that
these few states carry a macroscopic current, {\it nor\/} does it imply that the
current flows at the edges. A determination of the spatial current distribution
$i(\mathbf{r})$, rather than just the total current $I$, requires consideration of all the states
below the Fermi level, which acquire a net drift velocity because of the Hall
field. As we discussed in Section \ref{sec12b}, knowledge of $i(\mathbf{r})$ is not necessary to
know the resistances in the regime of linear response, because the Einstein
relation allows one to obtain the resistance from the diffusion constant. Edge
channels tell you where the current flows if the electrochemical potential
difference $\delta\mu$ is entirely due to a density difference, relevant for the diffusion
problem. Edge channels have nothing to say about where the current flows if
$\delta\mu$ is mainly of electrostatic origin, relevant for the problem of electrical
conduction. The ratio $\delta\mu/I$ is the same for both problems, but $i(\mathbf{r})$ is not.

With this in mind, it remains an interesting problem to find out just how
the current is distributed in a Hall bar, or, alternatively, what is the
electrostatic potential profile. This problem has been treated theoretically in
many papers.\cite{ref446,ref447,ref448,ref449,ref450,ref451,ref452,ref453,ref454,ref455} In the case of a 3D conductor, a linearly varying
potential and uniform current density are produced by a surface charge. As
noted by MacDonald et al.,\cite{ref446} the electrostatics is qualitatively different in
the 2D case because an edge charge $\delta(x-W/2)$ produces a potential
proportional to $\ln|x-W/2|$, which is weighted toward the edge, and hence a
concentration of current at the edge.

Experiments aimed at measuring the electrostatic potential distribution
were originally carried out by attaching contacts to the interior of the Hall
bar and measuring the voltage differences between adjacent contacts.\cite{ref456,ref457,ref458,ref459,ref460}
It was learned from these studies that relatively small inhomogeneities in the
density of the 2DEG have a large effect on these voltage differences in the
QHE regime. The main difficulty in the interpretation of such experiments is
that the voltage difference measured between two contacts is the difference in
electrochemical potential, not the line integral of the electric field. B\"{u}ttiker\cite{ref461}
has argued that the voltage measured at an interior contact can exhibit large
variations for a small increase in magnetic field without an appreciable
change in the current distribution. Contactless measurements of the QHE
from the absorption of microwave radiation\cite{ref462} are one alternative to interior
contacts, which might be used to determine the potential (or current)
distribution.

\begin{figure}
\centerline{\includegraphics[width=8cm]{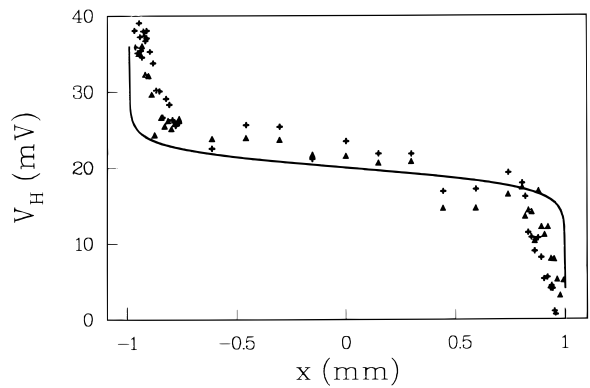}}
\caption{
Electrostatic potential $V_{\rm H}$ induced by passing a current through a Hall bar. The sample edges are at $x = \pm 1\,{\rm mm}$. The data points are from the experiment of Fontein et a1.,\cite{ref463} at two magnetic field values on the $R_{\rm H} = h/4e^{2}$ quantized Hall plateau (triangles: $B = 5\,{\rm T}$; crosses: $B = 5.25\,{\rm T}$). The solid curve is calculated from Eq.\ (\ref{eq18.9}), assuming an impurity-free Hall bar with four filled Landau levels. The theory contains no adjustable parameters.
\label{fig80}
}
\end{figure}

Fontein et al.\cite{ref463} have used the birefringence of GaAs induced by an
electric field to perform a contactless measurement of the electrostatic
potential distribution in a Hall bar. They measure the Hall potential profile
$V_{\mathrm{H}}(x)$ as a change in the local electrostatic potential if a current is passed
through the Hall bar. The data points shown in Fig.\ \ref{fig80} were taken at $1.5\, \mathrm{K}$
for two magnetic field values on the plateau of quantized Hall resistance at
$\frac{1}{4}h/e^{2}$. The potential varies steeply at the edges (at $x=\pm 1\,\mathrm{mm}$ in Fig.\ \ref{fig80}) and
is approximately linear in the bulk. The spatial resolution of the experiment
was 70 $\mu \mathrm{m}$, limited by the laser beam used to measure the birefringence. The
current distribution is not directly measured, but can be estimated from the
guiding center drift (\ref{eq18.2}) (this assumes a slowly varying potential). The
nonequilibrium current density $i(x)$ along the Hall bar is then given by
\be
i( x)=\frac{en_{\mathrm{s}}}{B}\frac{dV_{\rm H}(x)}{dx}. \label{eq18.6}
\ee 
Fontein et al.\ thus estimate that under the conditions of their experiment two
thirds of the total imposed current $I=5\,\mu{\rm A}$ flows within 70 $\mu \mathrm{m}$ from the
edges while the remainder is uniformly distributed in the bulk.

This experimental data can be modeled\cite{ref464} by means of an integral
equation derived by MacDonald et al.\cite{ref446} for the self-consistent potential
profile in an ideal impurity-free sample with $N$ completely filled (spin-split)
Landau levels. The electron charge density $\rho_{\mathrm{e}}(x)$ in the 2DEG is given by
\be
\rho_{\mathrm{e}}(x)=-en_{\mathrm{s}}\left[1-\frac{el_{\mathrm{m}}^{2}}{\hbar\omega_{\mathrm{c}}}V_{\mathrm{H}}^{\prime\prime}(x)\right]. \label{eq18.7}
\ee 
This equation follows from the Schr\"{o}dinger equation in a smoothly varying
electrostatic potential, so the factor between brackets is close to unity.
Substitution of the net charge density $en_{\mathrm{s}}+\rho_{\mathrm{e}}(x)$ into the Poisson equation
gives\cite{ref446}
\be
V_{\mathrm{H}}( x)=-\xi\int_{-W/2}^{+W/2}dx^{\prime}\ln\left(\frac{2}{W}|x-x^{\prime}|\right)V_{\mathrm{H}}^{\prime\prime}(x^{\prime}). \label{eq18.8}
\ee 
The characteristic length $\xi\equiv Nl_{\mathrm{m}}^{2}/\pi a^{*}$ is defined in terms of the magnetic
length $l_{\mathrm{m}}$ and the effective Bohr radius $a^{*}\equiv\epsilon \hbar^{2}/me^{2}$ (with $\epsilon$ the dielectric
constant).

The integral equation (\ref{eq18.8}) was solved numerically by MacDonald et
al.\cite{ref446} and analytically by means of the Wiener-Hopf technique by Thouless.\cite{ref448} Here we describe a somewhat simpler approach,\cite{ref464} which is sufficiently accurate for the present purpose. For magnetic field strengths in the
QHE regime the length $\xi$ is very small. For example, if $N=4$, $l_{\mathrm{m}}=11.5\,\mathrm{nm}$
(for $B=5\,\mathrm{T}$), $a^{*}=10\,\mathrm{nm}$ (for GaAs with $\epsilon=13\,\epsilon_{0}$ and $m=0.067\,m_{\mathrm{e}}$), then
$\xi=17\,\mathrm{nm}$. It is therefore meaningful to look for a solution of Eq.\ (\ref{eq18.8}) in the
limit $\xi\ll W$. The result is that $V_{\mathrm{H}}(x)=\mathrm{constant}\times\ln|(x-W/2)/(x+W/2)|$ if
$|x|\leq W/2-\xi$, with a linear extrapolation from $|x|=W/2-\xi$ to $|x|=W/2$.
One may verify that this is indeed the answer, by substituting the preceding
expression into Eq.\ (\ref{eq18.8}) and performing one partial integration. The
arbitrary constant in the expression for $V_{\mathrm{H}}$ may be eliminated in favor of the
total current $I$ flowing through the Hall bar, by applying Eq.\ (\ref{eq18.6}) to the case
of $N$ filled spin-split Landau levels. This gives the final answer
\begin{eqnarray}
V_{\mathrm{H}}( x)&=&\frac{1}{2}IR_{\mathrm{H}}\left(1+\ln\frac{W}{\xi}\right)^{-1}\ln\left|\frac{x-W/2}{x+W/2}\right|\nonumber\\
&&{\rm if}\;\;|x| \leq\frac{W}{2}-\xi, \label{eq18.9}
\end{eqnarray}
with a linear extrapolation of $V_{\mathrm{H}}$ to $\pm\frac{1}{2}IR_{\mathrm{H}}$ in the interval within $\xi$ from the
edge. The Hall resistance is $R_{\mathrm{H}}=h/Ne^{2}$. The approximation (\ref{eq18.9}) is
equivalent for small $\xi$ to the analytical solution of Thouless, and is close to
the numerical solutions given by MacDonald et al., even for a relatively large
value $\xi/W=0.1$.

In Fig.\ \ref{fig80} the result (\ref{eq18.9}) has been plotted (solid curve) for the parameters
of the experiment by Fontein et al.\ ($\xi/W=0.85\times 10^{-5}$ for $N=4$, $B=5\,\mathrm{T}$,
and $W=2\,\mathrm{mm}$). The agreement with experiment is quite satisfactory in view
of the fact that the theory contains {\it no\/} adjustable parameters. The theoretical
profile is steeper at the edges than in the experiment, which may be due to the
limited experimental resolution of 70 $\mu \mathrm{m}$. The total voltage drop between the
two edges in the calculation ($hI/Ne^{2}\approx 32\,\mathrm{mV}$ for $I=5\,\mu \mathrm{A}$ and $N=4$) agrees
with the measured Hall voltage of $\approx 30\,\mathrm{mV}$, but the optically determined
value of $40\, \mathrm{mV}$ is somewhat larger for a reason that we do not understand.

We have discussed this topic of the current distribution in the QHE in
some detail to convince the reader that the concentration of the potential
drop (and hence of the current) near the edges can be understood from the
electrostatics of edge {\it charges}, but cannot be used to test the validity of a
linear response formulation of the QHE in terms of edge {\it states}. Indeed, edge
states were completely neglected in the foregoing theoretical analysis, which
nonetheless captures the essential features of the experiment.

\subsection{\label{sec19} Selective population and detection of edge channels}

The absence of local equilibrium at the current or voltage contacts leads to
anomalies in the quantum Hall effect, unless the contacts are ideal (in the
sense that each edge channel at the Fermi level is transmitted through the
contact with probability 1). Ideal versus disordered contacts are dealt with in
Sections \ref{sec19a} and \ref{sec19b}. A quantum point contact can be seen as an extreme
example of a disordered contact, as discussed in Section \ref{sec19c}. Anomalies in
the Shubnikov-De Haas effect due to the absence of local equilibrium are the
subject of Section \ref{sec19d}.

\subsubsection{\label{sec19a} Ideal contacts}

In a two-terminal measurement of the quantum Hall effect the contact
resistances of the current source and drain are measured in series with the
Hall resistance. For this reason precision measurements of the QHE are
usually performed in a four-terminal measurement configuration, in which
the voltage contacts do not carry a current.\cite{ref445} Contact resistances then do
not play a role, provided that local equilibrium is established near the voltage
contacts [or, by virtue of the reciprocity relation (\ref{eq12.16}), near the current
contacts]. As we mentioned in Section \ref{sec18}, local equilibrium can be grossly
violated in the QHE. Accurate quantization then requires that either the
current or the voltage contacts are {\it ideal}, in the sense that the edge states at
the Fermi level have unit transmission probability through the contacts.
In this subsection we return to the four-terminal measurements on a
quantum point contact considered in Section \ref{sec13b}, but now in the QHE
regime where the earlier assumption of local equilibrium near the voltage
contacts is no longer applicable in general. We assume strong magnetic fields
so that the four-terminal longitudinal resistance $R_{\mathrm{L}}$ of the quantum point
contact is determined by the potential barrier in the constriction (rather than
by its width).

\begin{figure}
\centerline{\includegraphics[width=8cm]{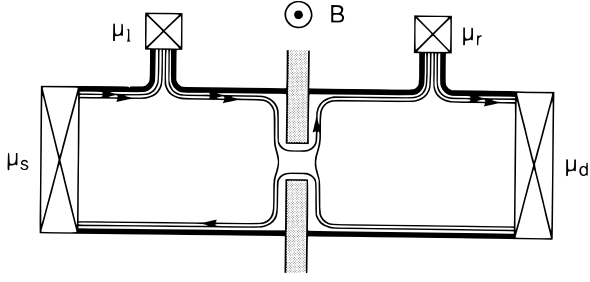}}
\caption{
Motion along equipotentials in the QHE regime, in a four-terminal geometry with a saddle-shaped potential formed by a split gate (shaded). Ideal contacts are assumed. The thin lines indicate the location of the edge channels at the Fermi level, with the arrows pointing in the direction of motion of edge channels that are populated by the contacts (crossed squares). Taken from H. van Houten et al., in Ref.\ \onlinecite{ref9}.
\label{fig81}
}
\end{figure}

Let us apply the Landauer-B\"{u}ttiker formalism to the geometry of Fig.\ \ref{fig81}.
As in Section \ref{sec13b}, the number of spin-degenerate edge channels in the wide
2DEG and in the constriction are denoted by $N_{\mathrm{wide}}$ and $N_{\min}$, respectively.
An ideal contact to the wide 2DEG perfectly transmits $N_{\mathrm{wide}}$ channels,
whereas the constriction transmits only $N_{\min}$ channels. The remaining
$N_{\mathrm{wide}}-N_{\min}$ channels are reflected back along the opposite 2DEG boundary
(cf.\ Fig.\ \ref{fig81}). We denote by $\mu_{\mathrm{l}}$ and $\mu_{\mathrm{r}}$ the chemical potentials of adjacent
voltage probes to the left and to the right of the constriction. The current
source is at $\mu_{\mathrm{s}}$, and the drain at $\mu_{\mathrm{d}}$. Applying Eq.\ (\ref{eq12.12}) to this case, using
$I_{\mathrm{s}}=-I_{\mathrm{d}}\equiv I$, $I_{\mathrm{r}}=I_{\mathrm{l}}=0$, one finds for the magnetic field direction indicated in Fig.\ \ref{fig81},
\begin{subequations}
\label{eq19.1}
\begin{eqnarray}
(h/2e)I&=&N_{\mathrm{wide}}\mu_{\mathrm{s}}-(N_{\mathrm{wide}}-N_{\min})\mu_{\mathrm{l}}, \label{eq19.1a}\\
0&=&N_{\mathrm{wide}}\mu_{\mathrm{l}}-N_{\mathrm{wide}}\mu_{\mathrm{s}}, \label{eq19.1b}\\
0&=&N_{\mathrm{wide}}\mu_{\mathrm{r}}-N_{\min}\mu_{\mathrm{l}}. \label{eq19.1c}
\end{eqnarray}
\end{subequations}
We have used the freedom to choose the zero level of chemical potential by
fixing $\mu_{\mathrm{d}}=0$, so we have three independent (rather than four dependent)
equations. The two-terminal resistance $R_{2\mathrm{t}}\equiv\mu_{\mathrm{s}}/eI$ following from Eq.\ (\ref{eq19.1})
is
\be
R_{2\mathrm{t}}= \frac{h}{2e^{2}}\frac{1}{N_{\min}}, \label{eq19.2}
\ee 
unaffected by the presence of the additional voltage probes in Fig.\ \ref{fig81}. The
four-terminal longitudinal resistance $R_{\mathrm{L}}\equiv(\mu_{\mathrm{l}}-\mu_{\mathrm{r}})/eI$ is
\be
R_{\mathrm{L}}= \frac{h}{2e^{2}}\left(\frac{1}{N_{\min}}-\frac{1}{N_{\mathrm{wide}}}\right). \label{eq19.3}
\ee 
In the reversed field direction the same result is obtained. Equation (\ref{eq19.3}),
derived for ideal contacts without assuming local equilibrium near the
contacts, is identical to Eq.\ (\ref{eq13.7}), derived for the case of local equilibrium.

\begin{figure}
\centerline{\includegraphics[width=8cm]{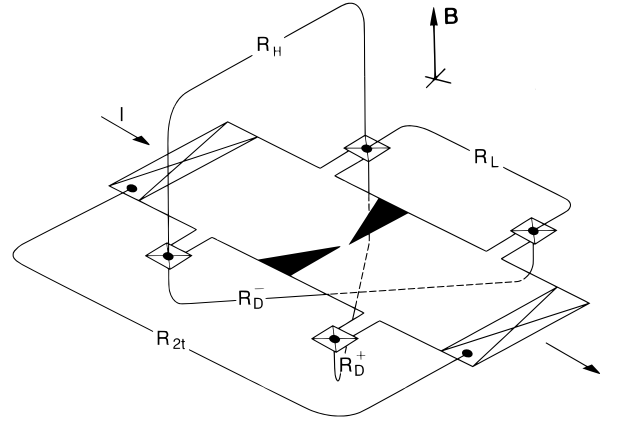}}
\caption{
Perspective view of a six-terminal Hall bar containing a point contact, showing the various two- and four-terminal resistances mentioned in the text. Taken from H. van Houten et al., in Ref.\ \onlinecite{ref9}.
\label{fig82}
}
\end{figure}

In a six-terminal measurement geometry (see Fig.\ \ref{fig82}), one can also
measure the Hall resistance in the wide regions, which is simply
$R_{\mathrm{H}}=R_{2\mathrm{t}}-R_{\mathrm{L}}$ or
\be
R_{\mathrm{H}}= \frac{h}{2e^{2}}\frac{1}{N_{\mathrm{wide}}}, \label{eq19.4}
\ee 
which is unaffected by the presence of the constriction. This is a consequence
of our assumption of ideal voltage probes. One can also measure the two
four-terminal {\it diagonal\/} resistances $R_{\mathrm{D}}^{+}$ and $R_{\mathrm{D}}^{-}$ across the constriction in such
a way that the two voltage probes are on opposite edges of the 2DEG, on
either side of the constriction (see Fig.\ \ref{fig82}). Additivity of voltages on contacts
tells us that $R_{\mathrm{D}}^{\pm}=R_{\mathrm{H}}\pm R_{\mathrm{L}}$ (for the magnetic field direction of Fig.\ \ref{fig82}); thus,
\be
R_{\mathrm{D}}^{+}= \frac{h}{2e^{2}}\frac{1}{N_{\min}};\;\;R_{\mathrm{D}}^{-}= \frac{h}{2e^{2}}\left(\frac{2}{N_{\mathrm{wide}}}-\frac{1}{N_{\min}}\right). \label{eq19.5}
\ee 
On field reversal, $R_{\mathrm{D}}^{+}$ and $R_{\mathrm{D}}^{-}$ are interchanged. Thus, a four-terminal
resistance [$R_{\mathrm{D}}^{+}$ in Eq.\ (\ref{eq19.5})] can in principle be equal to the two-terminal
resistance [$R_{2\mathrm{t}}$ in Eq.\ (\ref{eq19.2})]. The main difference between these two
quantities is that an additive contribution of the ohmic contact resistance
(and of a part of the diffusive background resistance in weak magnetic fields)
is eliminated in the four-terminal resistance measurement.

The fundamental reason that the assumption of local equilibrium made in
Section \ref{sec13b} (appropriate for weak magnetic fields) and that of ideal contacts
made in this section (for strong fields) yield identical answers is that an ideal
contact attached to the wide 2DEG regions {\it induces\/} a local equilibrium by
equipartitioning the outgoing current among the edge channels. (This is
illustrated in Fig.\ \ref{fig81}, where the current entering the voltage probe to the right
of the constriction is carried by a singe edge channel, while the equally large
current flowing out of that probe is equipartitioned over the two edge
channels available for transport in the wide region.) In weaker magnetic
fields, when the cyclotron radius exceeds the width of the narrow 2DEG
region connecting the voltage probe to the Hall bar, not all edge channels in
the wide 2DEG region are transmitted into the voltage probe (even if it does
not contain a potential barrier). This probe is then not effective in equipartitioning the current. That is the reason that the weak-field analysis in Section
\ref{sec13b} required the assumption of a local equilibrium in the wide 2DEG near
the contacts.

\begin{figure}
\centerline{\includegraphics[width=8cm]{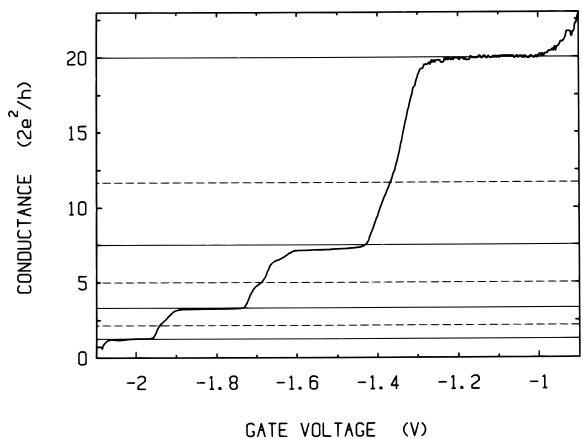}}
\caption{
``Fractional'' quantization in the integer QHE of the four-terminal longitudinal conductance $R_{\rm L}^{-1}$ of a point contact in a magnetic field of 1.4 T at $T = 0.6\,{\rm K}$. The solid horizontal lines indicate the quantized plateaus predicted by Eq.\ (\ref{eq19.3}), with $N_{\rm wide} = 5$ and $N_{\rm min} = 1,2,3,4$. The dashed lines give the location of the spin-split plateaux, which are not well resolved at this magnetic field value. Taken from L. P. Kouwenhoven, Master's thesis, Delft University of Technology, 1988.
\label{fig83}
}
\end{figure}

We now discuss some experimental results, which confirm the behavior
predicted by Eq.\ (\ref{eq19.3}) in the QHE regime, to complement the weak-field
experiments discussed in Section \ref{sec13b}. Measurements on a quantum point
contact by Kouwenhoven et al.\cite{ref307,ref465} in Fig.\ \ref{fig83} show the quantization of the
longitudinal conductance $R_{\mathrm{L}}^{-1}$ in {\it fractions\/} of $2e^{2}/h$ (for unresolved spin
degeneracy). The magnetic field is kept fixed at $1.4\, \mathrm{T}$ (such that $N_{\mathrm{wide}}=5$) and
the gate voltage is varied (such that $N_{\min}$ ranges from 1 to 4). Conductance
plateaux close to 5/4, 10/3, 15/2, and $20\times(2e^{2}/h)$ (solid horizontal lines) are
observed, in accord with Eq.\ (\ref{eq19.3}). Spin-split plateaux (dashed lines) are
barely resolved at this rather low magnetic field. Similar data were reported
by Snell et al.\cite{ref342} Observations of such a ``fractional'' quantization due to the
integer QHE were made before on wide Hall bars with regions of different
electron density in series,\cite{ref466,ref467} but the theoretical explanation\cite{ref468} given at
that time was less straightforward than Eq.\ (\ref{eq19.3}).

\begin{figure}
\centerline{\includegraphics[width=6cm]{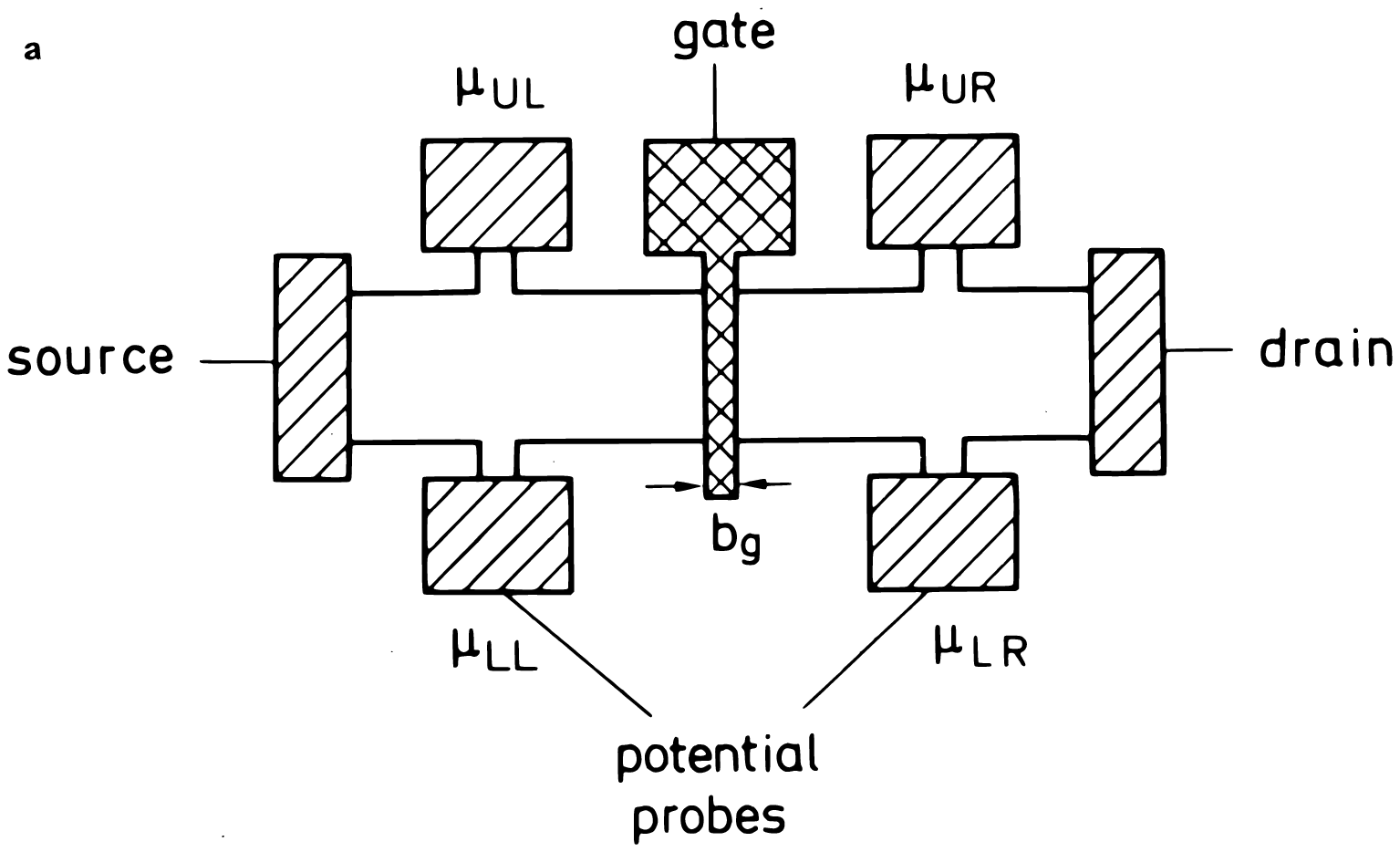}}

\centerline{\includegraphics[width=6cm]{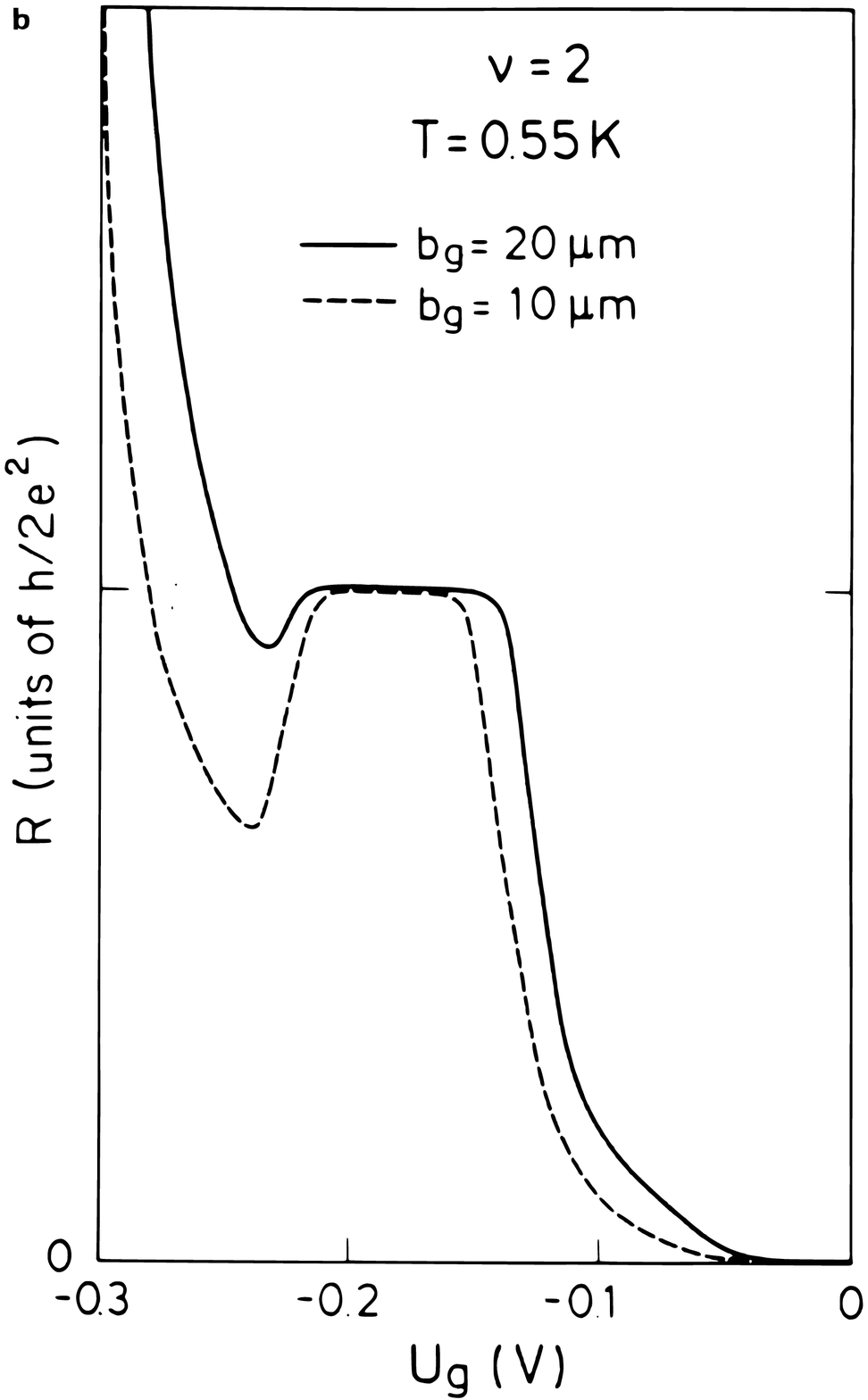}}
\caption{
(a) Schematic view of a wide Hall bar containing a potential barrier imposed by a gate electrode of length $b_{\rm g}$. (b) Longitudinal resistance as a function of gate voltage in the QHE regime (two spin-split Landau levels are occupied in the unperturbed electron gas regions). The plateau shown is at $R_{\rm L} = h/2e^{2}$, in agreement with Eq.\ (\ref{eq19.3}). Results for $b_{\rm g} = 10\,\mu{\rm m}$ and $20\,\mu{\rm m}$ are compared. A pronounced dip develops in the device with the shortest gate length. Taken from R. J. Haug et al., Phys.\ Rev.\ B {\bf 39}, 10892 (1989).
\label{fig84}
}
\end{figure}

\begin{figure}
\centerline{\includegraphics[width=8cm]{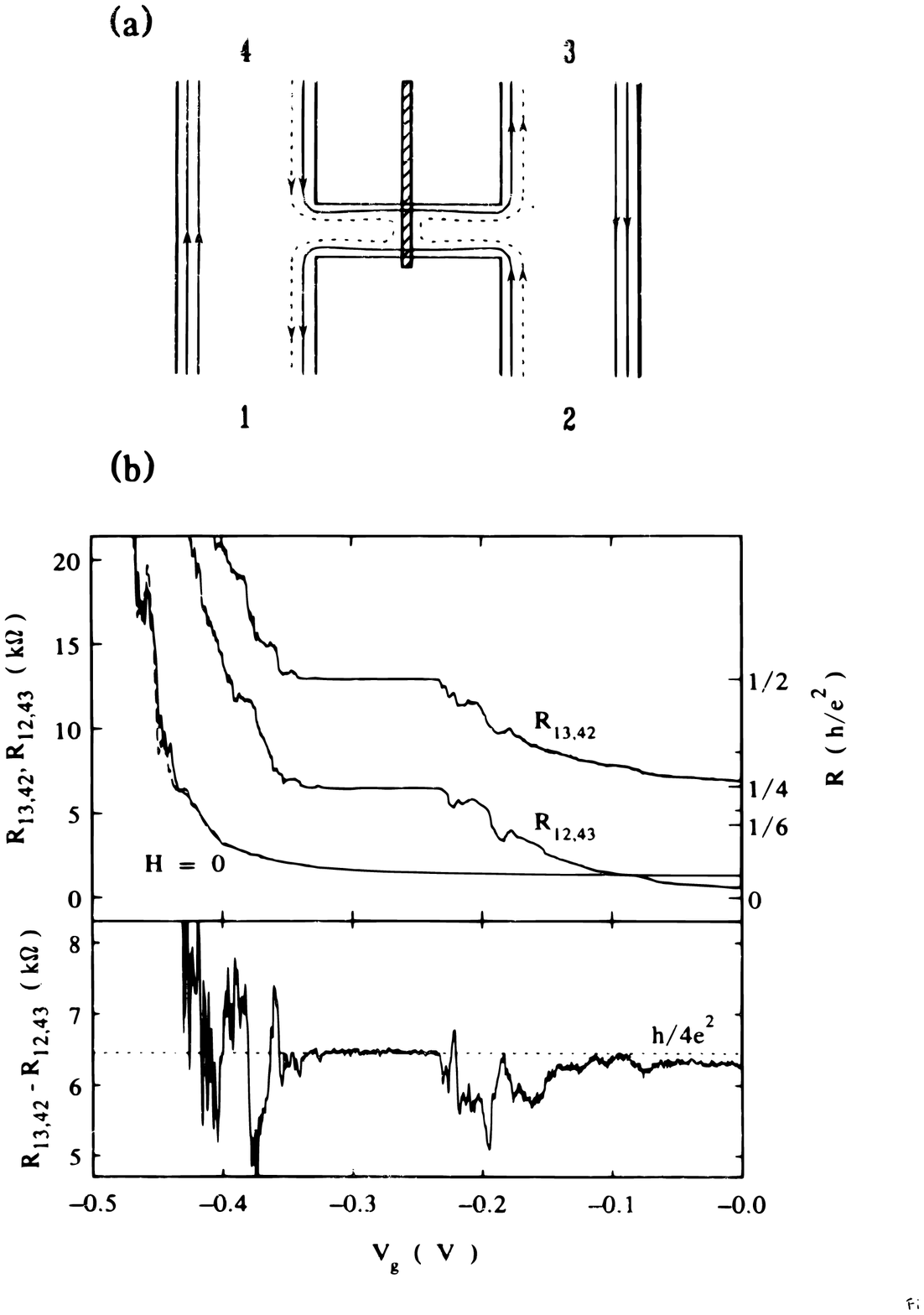}}
\caption{
(a) Schematic view of a $2$-$\mu{\rm m}$-wide channel containing a potential barrier imposed by a $0.1$-$\mu{\rm m}$-long gate. (b) Top: diagonal resistance $R_{13,42} \equiv R_{\rm D}^{+}$ and longitudinal resistance $R_{12,43} \equiv R_{\rm L}$ as a function of gate voltage in a strong magnetic field ($B = 5.2\,{\rm T}$), showing a quantized plateau in agreement with Eqs.\ (\ref{eq19.5}) and (\ref{eq19.3}), respectively. For comparison also the two zero-field traces are shown, which are almost identical. Bottom: Difference $R_{\rm D}^{+}- R_{\rm L} = R_{\rm H}$ at 5.2 T. A normal quantum Hall plateau is found, with oscillatory structure superimposed in gate voltage regions where $R_{\rm D}^{+}$ and $R_{\rm L}$ are not quantized. Taken from S. Washburn et al., Phys.\ Rev.\ Lett.\ {\bf 61}, 2801 (1988).
\label{fig85}
}
\end{figure}

In the high-field regime the point contact geometry of Fig.\ \ref{fig81} is essentially
equivalent to a geometry in which a potential barrier is present across the
entire width of the Hall bar (created by means of a narrow continuous gate).
The latter geometry was studied by Haug et al.\cite{ref340} and by Washburn et al.\cite{ref339}
The geometries of both experiments\cite{ref339,ref340} are the same (see Figs.\ \ref{fig84} and \ref{fig85}),
but the results exhibit some interesting differences because of the different
dimensions of gate and channel. Hauge et al.\cite{ref340} used a sample of macroscopic dimensions, the channel width being 100 $\mu \mathrm{m}$ and the gate length 10
and 20 $\mu \mathrm{m}$. Results are shown in Fig.\ \ref{fig84}. As the gate voltage is varied, a
quantized plateau at $h/2e^{2}$ is seen in the longitudinal resistance at fixed
magnetic field, in agreement with Eq.\ (\ref{eq19.3}) (the plateau occurs for two spin-split Landau levels in the wide region and one spin-split level under the gate).
A qualitatively different aspect of the data in Fig.\ \ref{fig84}, compared with Fig.\ \ref{fig83},
is the presence of a resistance minimum. Equation (\ref{eq19.3}), in contrast, predicts
that $R_{\mathrm{L}}$ varies monotonically with barrier height, and thus with gate voltage.
A model for the effect has been proposed in a different paper by Haug et
al.,\cite{ref341} based on a competition between backscattering and tunneling through
localized states in the barrier region. They find that edge states that are
totally reflected at a given barrier height may be partially transmitted if the
barrier height is further increased. The importance of tunneling is consistent
with the increase of the amplitude of the dip as the gate length is reduced from
20 to 10 $\mu \mathrm{m}$. A related theoretical study was performed by Zhu et al.\cite{ref469}

Washburn et al.\cite{ref339} studied the longitudinal resistance of a barrier defined
by a $0.1$-$\mu \mathrm{m}$-long gate across a $2$-$\mu \mathrm{m}$-wide channel. The relevant dimensions
are thus nearly two orders of magnitude smaller than in the experiment of
Haug et al. Again, the resistance is studied as a function of gate voltage at
fixed magnetic field. The longitudinal $(R_{\mathrm{L}}\equiv R_{12,43})$ and diagonal
$(R_{\mathrm{D}}^{+}\equiv R_{13,42})$ resistances are shown in Fig.\ \ref{fig85}, as well as their difference
[which according to  Eqs.\ (\ref{eq19.3}) and (\ref{eq19.5}) would equal the Hall resistance
$R_{\mathrm{H}}$]. In this small sample the quantized plateaux predicted by Eq.\ (\ref{eq19.3}) are
clearly seen, but the resistance dips of the large sample of Haug et al.\ are not.
We recall that resistance dips were not observed in the quantum point
contact experiment of Fig.\ \ref{fig83} either. The model of Haug et al.\cite{ref341}  would imply
that localized states do not form in barriers of small area. Washburn et al.\
find weak resistance fluctuations in the gate voltage intervals between
quantized plateaux. These fluctuations are presumably due to some form of
quantum interference, but have not been further identified.
Related experiments on the quantum Hall effect in a 2DEG with a
potential barrier have been performed by Hirai et al.\ and by Komiyama et
al.\cite{ref427,ref470,ref471,ref472} These studies have focused on the role of nonideal contacts in
the QHE, which is the subject of the next subsection.

\subsubsection{\label{sec19b} Disordered contacts}

The validity of Eqs.\ (\ref{eq19.2}--\ref{eq19.5}) in the QHE regime breaks down for
nonideal contacts if local equilibrium near the contacts is not established. The
treatment of Section \ref{sec19a} for ideal contacts implies that the Hall voltage over
the wide 2DEG regions adjacent to the constriction is {\it unaffected\/} by the
presence of the constriction or potential barrier. Experiments by Komiyama
et al.\cite{ref427,ref472} have demonstrated that this is no longer true if one or more
contacts are disordered. The analysis of their experiments is rather involved,\cite{ref472} which is why we do not give a detailed discussion here. Instead we
review a different experiment,\cite{ref113}  which shows a deviating Hall resistance in a
sample with a constriction and a singe disordered contact. This experiment
can be analyzed in a relatively simple way,\cite{ref307} following the work of
B\"{u}ttiker\cite{ref112} and Komiyama et al.\cite{ref427,ref470,ref471,ref472}

\begin{figure}
\centerline{\includegraphics[width=8cm]{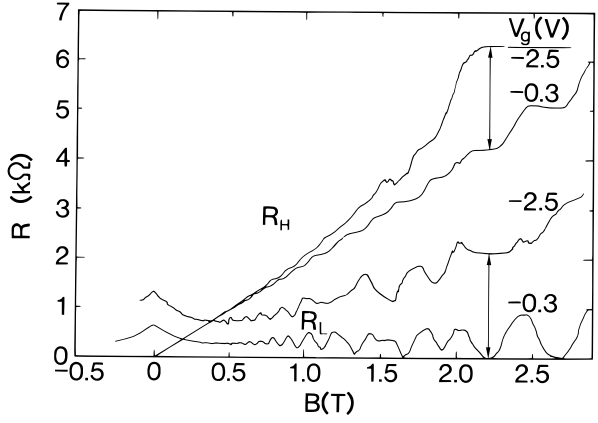}}
\caption{
Nonvanishing Shubnikov-De Haas minima in the longitudinal resistance $R_{\rm L}$ and anomalous quantum Hall resistance $R_{\rm H}$, measured in the point contact geometry of Fig.\ \ref{fig82} at 50 mK. These experimental results are extensions to higher fields of the weak-field traces shown in Fig.\ \ref{fig50}. The Hall resistance has been measured across the wide region, more than $100\,\mu{\rm m}$ away from the constriction, yet $R_{\rm H}$ is seen to increase if the gate voltage is raised from $- 0.3\,{\rm V}$ to $-2.5\,{\rm V}$. The magnitude at $B = 2.2\,{\rm T}$ of the deviation in $R_{\rm H}$ and of the Shubnikov-De Haas minimum in $R_{\rm L}$ are indicated by arrows, which both for $R_{\rm H}$ and $R_{\rm L}$ have a length of $(h/2e^{2})(\frac{1}{2}-\frac{1}{3})$, in agreement with the analysis given in the text. Taken from H. van Houten et al., in Ref.\ \onlinecite{ref9}.
\label{fig86}
}
\end{figure}

The sample geometry is that of Fig.\ \ref{fig82}. In Fig.\ \ref{fig86} the four-terminal
longitudinal resistance $R_{\mathrm{L}}$ and Hall resistance $R_{\mathrm{H}}$ are shown for both a small
voltage ($-0.3$ V) and a large voltage ($-2.5$ V) on the gate defining the
constriction. The longitudinal resistance decreases in weak fields because of
reduction of backscattering, as discussed in Section \ref{sec13b}. At larger fields
Shubnikov-De Haas oscillations develop. The data for $V_{\mathrm{g}}=-0.3\,\mathrm{V}$ exhibit
zero minima in the Shubnikov-De Haas oscillations in $R_{\mathrm{L}}$ and the normal
quantum Hall resistance $R_{\mathrm{H}}=(h/2e^{2})N_{\mathrm{wide}}^{-1}$, determined by the number of
Landau levels occupied in the wide regions ($N_{\mathrm{wide}}$ can be obtained from the
quantum Hall effect measured in the absence of the constriction or from the
periodicity of the Shubnikov-De Haas oscillations).

At the higher gate voltage $V_{\mathrm{g}}=-2.5\,\mathrm{V}$, nonvanishing minima in $R_{\mathrm{L}}$ are
seen in Fig.\ \ref{fig86} as a result of the formation of a potential barrier in the
constriction. At the minima, $R_{\mathrm{L}}$ has the fractional quantization predicted by
Eq.\ (\ref{eq19.3}). For example, the plateau in $R_{\mathrm{L}}$ around $2.2\, \mathrm{T}$ for $V_{\mathrm{g}}=-2.5\,\mathrm{V}$ is
observed to be at $R_{\mathrm{L}}=2.1\,\mathrm{k}\Omega\approx(h/2e^{2})\times(\frac{1}{2}-\frac{1}{3})$, in agreement with the fact
that the two-terminal resistance yields $N_{\min}=2$ and the number of Landau
levels in the wide regions $N_{\mathrm{wide}}=3$. In spite of this agreement, and in
apparent conflict with Eq.\ (\ref{eq19.4}), the Hall resistance $R_{\mathrm{H}}$ has {\it increased\/} over its
value for small gate voltages. Indeed, around $2.2\, \mathrm{T}$ a Hall plateau at
$R_{\mathrm{H}}=6.3\, \mathrm{k}\Omega\approx(h/2e^{2})\times\frac{1}{2}$ is found for $V_{\mathrm{g}}=-2.5\,\mathrm{V}$, as if the number of
occupied Landau levels was given by $N_{\min}=2$ rather than by $N_{\mathrm{wide}}=3$. This
unexpected deviation was noted in Ref.\ \onlinecite{ref113}, but was not understood at the
time. At higher magnetic fields (not shown in Fig.\ \ref{fig86}) the $N=1$ plateau is
reached, and the deviation in the Hall resistance vanishes.

\begin{figure}
\centerline{\includegraphics[width=8cm]{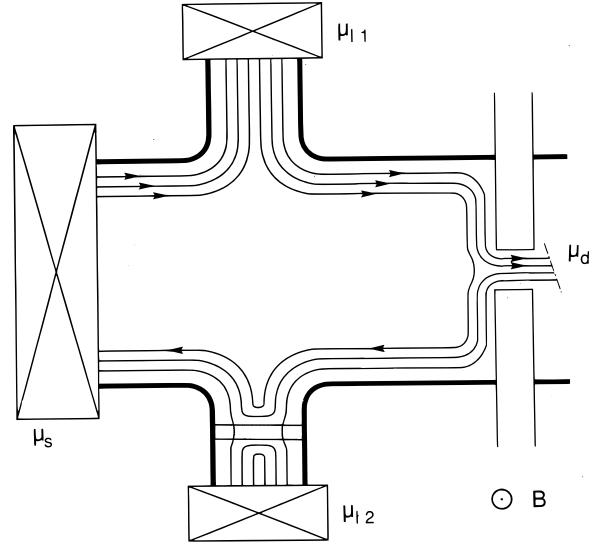}}
\caption{
Illustration of the flow of edge channels along equipotentials in a sample with a constriction (defined by the shaded gates) and a disordered voltage probe (a potential barrier in the probe is indicated by the shaded bar). Taken from H. van Houten et al., in Ref.\ \onlinecite{ref9}.
\label{fig87}
}
\end{figure}

As pointed out in Ref.\ \onlinecite{ref307}, the likely explanation of the data of Fig.\ \ref{fig86} is
that one of the ohmic contacts used to measure the Hall voltage is {\it disordered}
in the sense of B\"{u}ttiker\cite{ref112} that not all edge channels have unit transmission probability into the voltage probe. The disordered contact can be
modeled by a potential barrier in the lead with a height not below that of the
barrier in the constriction, as illustrated in Fig.\ \ref{fig87}. A net current $I$ flows
through the constriction, determined by its two-terminal resistance according
to $I=(2e/h)N_{\min}\mu_{\mathrm{s}}$, with $\mu_{\mathrm{s}}$ the chemical potential of the source reservoir (the
chemical potential of the drain reservoir $\mu_{\mathrm{d}}$ is taken as a zero reference).
Equation (\ref{eq12.12}) applied to the two opposite Hall probes $l_{1}$ and $l_{2}$ in Fig.\ \ref{fig87}
takes the form (using $I_{l_{1}}=I_{l_{2}}=0$, $\mu_{\mathrm{s}}=(h/2e)I/N_{\min}$, and $\mu_{\mathrm{d}}=0$)
\begin{subequations}
\label{eq19.6}
\begin{eqnarray}
0&=&N_{\mathrm{wide}} \mu_{l_{1}}-T_{\mathrm{s}\rightarrow l_{1}}\frac{h}{2e}\frac{I}{N_{\min}}-T_{l_{2}\rightarrow l_{1}}\mu_{l_{2}}, \label{eq19.6a}\\
0&=&N_{l_{2}} \mu_{l_{2}}-T_{\mathrm{s}\rightarrow l_{2}}\frac{h}{2e}\frac{I}{N_{\min}}-T_{l_{1}\rightarrow l_{2}}\mu_{l_{1}}, \label{eq19.6b}
\end{eqnarray}
\end{subequations}
where we have assumed that the disordered Hall probe $l_{2}$ transmits only
$N_{l_{2}}<N_{\mathrm{wide}}$ edge channels because of the barrier in the lead. For the field
direction shown in Fig.\ \ref{fig87} one has, under the assumption of no inter-edge-channel scattering from constriction to probe $l_{2}$, $T_{\mathrm{s}\rightarrow l_{1}}=N_{\mathrm{wide}}$,
$T_{\mathrm{s}\rightarrow l_{2}}=T_{l_{2}\rightarrow l_{1}}=0$, and $T_{l_{1}\rightarrow l_{2}}= \max(0, N_{l_{2}}-N_{\min})$. Equation (\ref{eq19.6}) then
leads to a Hall resistance $R_{\mathrm{H}}\equiv(\mu_{l_{1}}-\mu_{l_{2}})/eI$ given by
\be
R_{\mathrm{H}}=\frac{h}{2e^{2}}\frac{1}{\max(N_{l_{2}}, N_{\min})}.\label{eq19.7}
\ee
In the opposite field direction the normal Hall resistance $R_{\mathrm{H}}=(h/2e^{2})N_{\mathrm{wide}}^{-1}$ is
recovered.

The assumption of a single disordered probe, plus absence of interedge
channel scattering from constriction to probe, thus explains the observation
in Fig.\ \ref{fig86} of an anomalously high quantum Hall resistance for large gate
voltages, such that $N_{\min}<N_{\mathrm{wide}}$. Indeed, the experimental Hall resistance for
$V_{\mathrm{g}}=-2.5\,\mathrm{V}$ has a plateau around $2.2\, \mathrm{T}$ close to the value $R_{\mathrm{H}}=(h/2e^{2})N_{\min}^{-1}$
(with $N_{\min}=2$), in agreement with Eq.\ (\ref{eq19.7}) if $N_{l_{2}}\leq N_{\min}$ at this gate voltage.
This observation demonstrates the absence of interedge channel scattering
over 100 $\mu \mathrm{m}$ (the separation of constriction and probe), but only between the
highest-index channel (with index $n=N_{\mathrm{wide}}=3$) and the two lower-index
channels. Since the $n=1$ and $n=2$ edge channels are either both empty or
both filled (cf.\ Fig.\ \ref{fig87}, where these two edge channels lie closest to the sample
boundary), any scattering between $n=1$ and 2 would have no measurable
effect on the resistances. As discussed in Section \ref{sec19c}, we know from the work
of Alphenaar et al.\cite{ref429} that (at least in the present samples) the edge channels
with $n\leq N_{\mathrm{wide}}-1$ do in fact equilibrate to a large extent on a length scale of
$100\,\mu \mathrm{m}$.

In the absence of a constriction, or at small gate voltages (where the
constriction is just defined), one has $N_{\min}=N_{\mathrm{wide}}$ so that the normal Hall
effect is observed in both field directions. This is the situation realized in the
experimental trace for $V_{\mathrm{g}}=-0.3\,\mathrm{V}$ in Fig.\ \ref{fig86}. In very strong fields such that
$N_{\min}=N_{l_{2}}=N_{\mathrm{wide}}=1$ (still assuming nonresolved spin splitting), the
normal result $R_{\mathrm{H}}=h/2e^{2}$ would follow even if the contacts contain a
potential barrier, in agreement with the experiment (not shown in Fig.\ \ref{fig86}).
This is a more general result, which holds also for a barrier that only partially
transmits the $n=1$ edge channel.\cite{ref112,ref308,ref472,ref473,ref474,ref475}

A similar analysis as the foregoing predicts that the longitudinal resistance
measured on the edge of the sample that contains ideal contacts retains its
regular value (\ref{eq19.3}). On the opposite sample edge the measurement would
involve the disordered contact, and one finds instead
\be
R_{\mathrm{L}}= \frac{h}{2e^{2}}\left(\frac{1}{N_{\min}}-\frac{1}{\max(N_{l_{2}},N_{\min})}\right)   \label{eq19.8}
\ee
for the field direction shown in Fig.\ \ref{fig87}, while Eq.\ (\ref{eq19.3}) is recovered for the
other field direction. The observation in the experiment of Fig.\ \ref{fig86} for
$V_{\mathrm{g}}=-2.5\,\mathrm{V}$ of a regular longitudinal resistance [in agreement with Eq.\
(\ref{eq19.3})], along with an anomalous quantum Hall resistance is thus consistent
with this analysis.

The experiments\cite{ref426,ref429} discussed in the following subsection are topologically equivalent to the geometry of Fig.\ \ref{fig87}, but involve quantum point
contacts rather than ohmic contacts. This gives the possibility of populating
and detecting edge channels selectively, thereby enabling a study of the effects
of a nonequilibrium population of edge channels in a controlled manner.

\subsubsection{\label{sec19c} Quantum point contacts}

\begin{figure}
\centerline{\includegraphics[width=6cm]{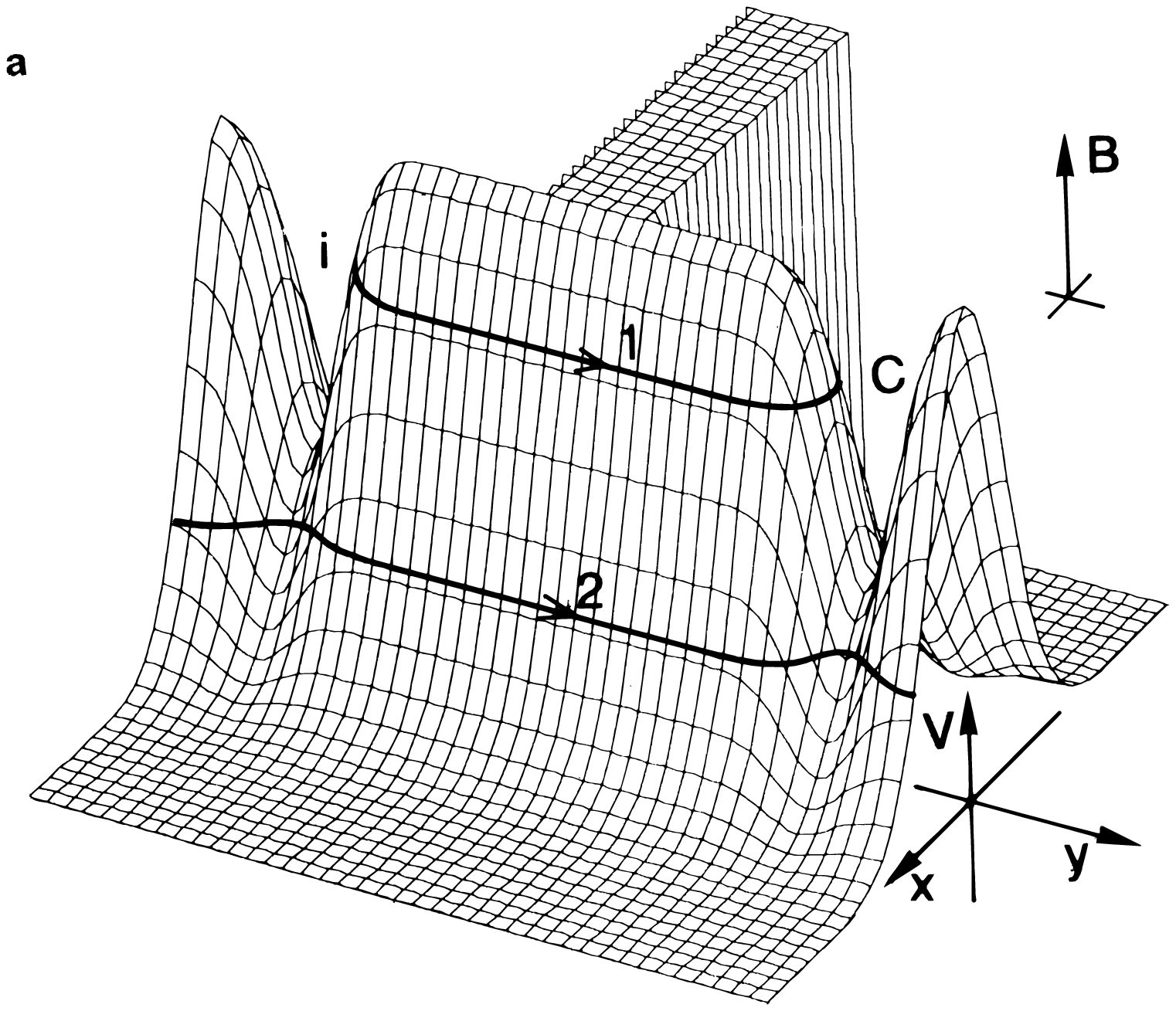}}

\centerline{\includegraphics[width=6cm]{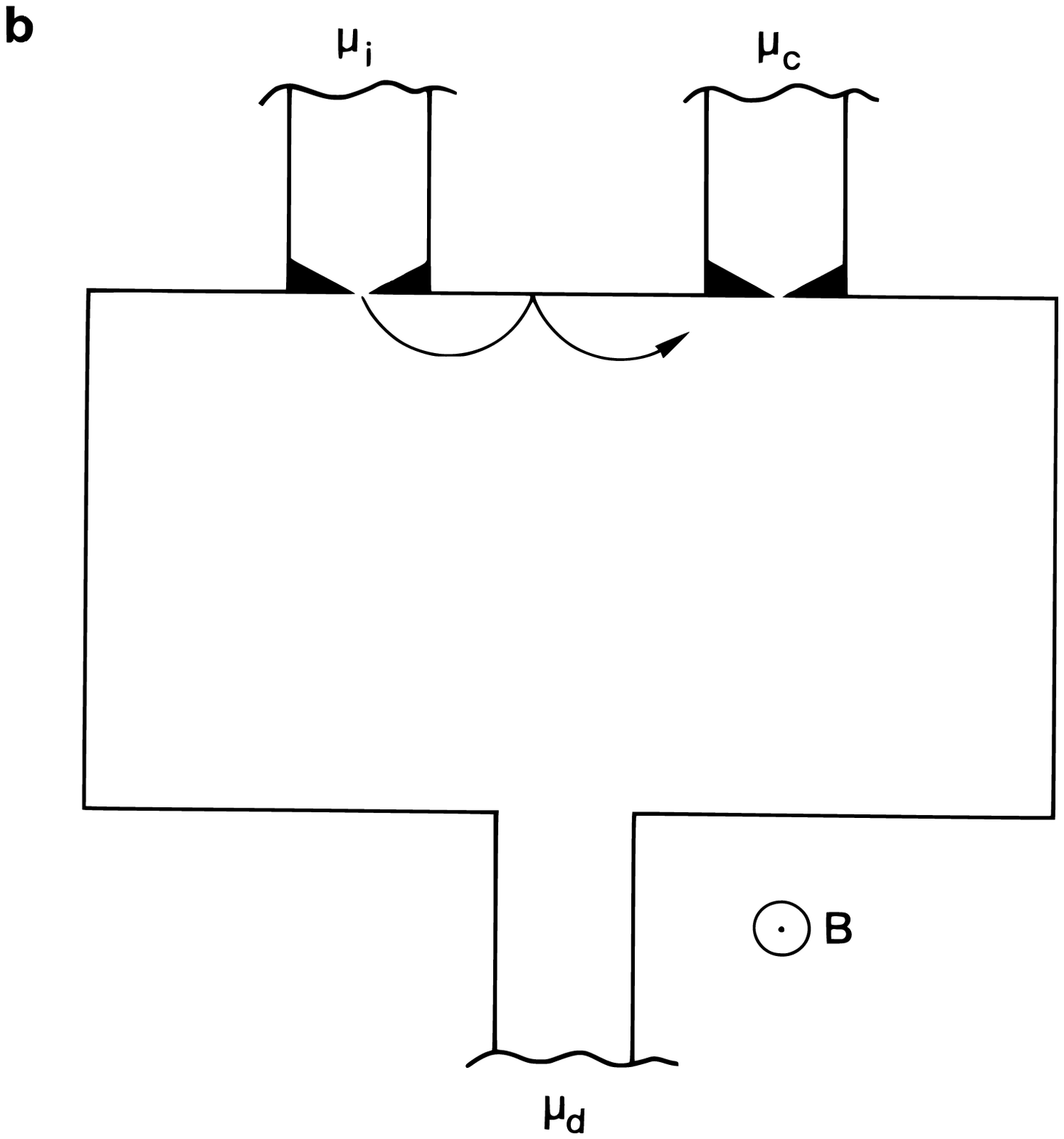}}
\caption{
(a) Schematic potential landscape, showing the 2DEG boundary and the saddle­shaped injector and collector point contacts. In a strong magnetic field the edge channels are extended along equipotentials at the guiding center energy, as indicated here for edge channels with index $n = 1,2$ (the arrows point in the direction of motion). In this case a Hall conductance of ($2e^{2}/h)N$ with $N = 1$ would be measured by the point contacts, in spite of the presence of two occupied spin-degenerate Landau levels in the bulk 2DEG. Taken from C. W. J. Beenakker et al., Festk\"{o}rperprobleme {\bf 29}, 299 (1989). (b) Three-terminal conductor in the electron focusing geometry. Taken from H. van Houten et al., Phys.\ Rev.\ B {\bf 39}, 8556 (1989).
\label{fig88}
}
\end{figure}

In Section \ref{sec14} we have seen how a quantum point contact can inject a
{\it coherent\/} superposition of edge channels at the 2DEG boundary, in the
coherent electron focusing experiment.\cite{ref59} In that section we restricted
ourselves to weak magnetic fields. Here we discuss the experiment by van
Wees et al.,\cite{ref426} which shows how in the QHE regime the point contacts can be
operated in a different way as {\it selective\/} injectors (and detectors) of edge
channels. We recall that electron focusing can be measured as a generalized
Hall resistance, in which case the pronounced peaked structure due to mode
interference is superimposed on the weak-field Hall resistance (cf.\ Fig.\ \ref{fig53}). If
the weak-field electron-focusing experiments are extended to stronger magnetic fields, a transition is observed to the quantum Hall effect, provided the
injecting and detecting point contacts are not too strongy pinched off.\cite{ref59}  The
oscillations characteristic of mode interference disappear in this field regime,
suggesting that the coupling of the edge channels (which form the propagating modes from injector to collector) is suppressed, and adiabatic transport is
realized. It is now no longer sufficient to model the point contacts by a point
source-detector of infinitesimal width (as was done in Section \ref{sec14}), but a
somewhat more detailed description of the electrostatic potential $V(x, y)$
defining the point contacts and the 2DEG boundary between them is
required. Schematically, $V(x, y)$ is represented in Fig.\ \ref{fig88}a. Fringing fields from
the split gate create a potential barrier in the point contacts, so $V$ has a saddle
form as shown. The heights of the barriers $E_{\mathrm{i}}, E_{\mathrm{c}}$ in the injector and collector
are separately adjustable by means of the voltages on the split gates and can
be determined from the two-terminal conductances of the individual point
contacts. The point contact separation in the experiment of Ref.\ \onlinecite{ref426} is small
(1.5 $\mu \mathrm{m}$), so one can assume fully adiabatic transport from injector to
collector in strong magnetic fields. As discussed in Section \ref{sec18}, adiabatic
transport is along equipotentials at the guiding center energy $E_{\mathrm{G}}$. Note that
the edge channel with the smallest index $n$ has the largest guiding center
energy [according to Eq.\ (\ref{eq18.1})]. In the absence of inter-edge-channel
scattering, edge channels can only be transmitted through a point contact if
$E_{\mathrm{G}}$ exceeds the potential barrier height (disregarding tunneling through the
barrier). The injector thus injects $N_{\mathrm{i}}\approx(E_{\mathrm{F}}-E_{\mathrm{i}})/\hbar\omega_{\mathrm{c}}$ edge channels into the
2DEG, while the collector is capable of detecting $N_{\mathrm{c}}\approx(E_{\mathrm{F}}-E_{\mathrm{c}})/\hbar\omega_{\mathrm{c}}$
channels. Along the boundary of the 2DEG, however, a larger number of
$N_{\mathrm{wide}}\approx E_{\mathrm{F}}/\hbar\omega_{\mathrm{c}}$ edge channels, equal to the number of occupied bulk Landau
levels in the 2DEG, are available for transport at the Fermi level. The
selective population and detection of Landau levels leads to deviations from
the normal Hall resistance.

These considerations can be put on a theoretical basis by applying the
Landauer-B\"{u}ttiker formalism discussed in Section \ref{sec12} to the electron-focusing geometry.\cite{ref80} We consider a three-terminal conductor as shown in
Fig.\ \ref{fig88}b, with point contacts in two of the probes (injector $\mathrm{i}$ and collector $\mathrm{c}$),
and a wide ideal drain contact $\mathrm{d}$. The collector acts as a voltage probe,
drawing no net current, so that $I_{\mathrm{c}}=0$ and $I_{\mathrm{d}}=-I_{\mathrm{i}}$. The zero of energy is
chosen such that $\mu_{\mathrm{d}}=0$. One then finds from Eq.\ (\ref{eq12.12}) the two equations
\begin{subequations}
\label{eq19.9}
\begin{eqnarray}
0&=&(N_{\mathrm{c}}-R_{\mathrm{c}})\mu_{\mathrm{c}}-T_{\mathrm{i}\rightarrow \mathrm{c}}\mu_{\mathrm{i}}, \label{19.9a}\\
(h/2e)I_{\mathrm{i}}&=&(N_{\mathrm{i}}-R_{\mathrm{i}})\mu_{\mathrm{i}}-T_{\mathrm{c}\rightarrow \mathrm{i}}\mu_{\mathrm{c}}, \label{eq19.9b}
\end{eqnarray}
\end{subequations}
and obtains for the ratio of collector voltage $V_{\mathrm{c}}=\mu_{\mathrm{c}}/e$ (measured relative to
the voltage of the current drain) to injected current $I_{\mathrm{i}}$ the result
\be
\frac{V_{\mathrm{c}}}{I_{\mathrm{i}}}=\frac{2e^{2}}{h}\frac{T_{\mathrm{i}\rightarrow \mathrm{c}}}{G_{\mathrm{i}}G_{\mathrm{c}}-\delta}. \label{eq19.10}
\ee 
Here $\delta\equiv(2e^{2}/h)^{2}T_{\mathrm{i}\rightarrow \mathrm{c}}T_{\mathrm{c}\rightarrow \mathrm{i}}$, and $G_{\mathrm{i}}\equiv(2e^{2}/h)(N_{\mathrm{i}}-R_{\mathrm{i}})$, $G_{\mathrm{c}}\equiv(2e^{2}/h)(N_{\mathrm{c}}-R_{\mathrm{c}})$
denote the conductances of injector and collector point contact.

\begin{figure}
\centerline{\includegraphics[width=8cm]{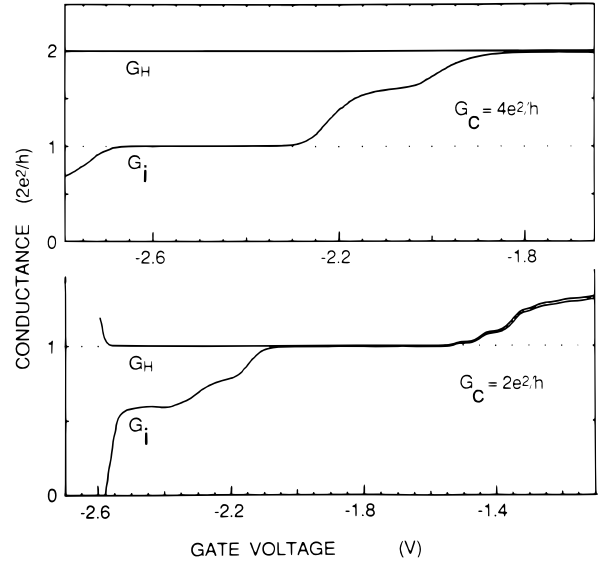}}
\caption{
Experimental correlation between the conductances $G_{\rm i}$, $G_{\rm c}$ of injector and collector, and the Hall conductance $G_{\rm H}\equiv I_{\rm i}/V_{\rm c}$, shown to demonstrate the validity of Eq.\ (\ref{eq19.11}) ($T = 1.3\,{\rm K}$, point contact separation is $1.5\,\mu{\rm m}$). The magnetic field was kept fixed (top: $B = 2.5\,{\rm T}$, bottom: $B = 3.8\,{\rm T}$, corresponding to a number of occupied bulk Landau levels $N = 3$ and $2$, respectively). By increasing the gate voltage on one half of the split-gate defining the injector, $G_{\rm i}$ was varied at constant $G_{\rm c}$. Taken from B. J. van Wees et al., Phys.\ Rev.\ Lett.\ {\bf 62}, 1181 (1989).
\label{fig89}
}
\end{figure}

For the magnetic field direction indicated in Fig.\ \ref{fig88}, the term $\delta$ in Eq.\
(\ref{eq19.10}) can be neglected since $T_{\mathrm{c}\rightarrow \mathrm{i}}\approx 0$ [the resulting Eq.\ (\ref{eq14.2}) was used in
Section \ref{sec14}]. An additional simplification is possible in the adiabatic transport
regime. We consider the case that the barrier in one of the two point contacts
is sufficiently higher than in the other, to ensure that electrons that are
transmitted over the highest barrier will have a negligible probability of being
reflected at the lowest barrier. Then $T_{\mathrm{i}\rightarrow \mathrm{c}}$ is dominated by the transmission
probability over the highest barrier, $T_{\mathrm{i}\rightarrow \mathrm{c}} \approx\min(N_{\mathrm{i}}-R_{\mathrm{i}}, N_{\mathrm{c}}-R_{\mathrm{c}})$. Substitution in Eq.\ (\ref{eq19.10}) gives the remarkable result\cite{ref426} that the {\it Hall conductance\/}
$G_{\mathrm{H}}\equiv I_{\mathrm{i}}/V_{\mathrm{c}}$ measured in the electron focusing geometry can be expressed
entirely in terms of the {\it contact conductances\/} $G_{\mathrm{i}}$ and $G_{\mathrm{c}}$:
\be
G_{\mathrm{H}} \approx\max(G_{\mathrm{i}}, G_{\mathrm{c}}). \label{eq19.11}
\ee 
Equation (\ref{eq19.11}) tells us that quantized values of $G_{\mathrm{H}}$ occur not at
$(2e^{2}/h)N_{\mathrm{wide}}$, as one would expect from the $N_{\mathrm{wide}}$ populated Landau levels in
the 2DEG but at the smaller value of $(2e^{2}/h) \max(N_{\mathrm{i}}, N_{\mathrm{c}})$. As shown in Fig.\ \ref{fig89}
this is indeed observed experimentally.\cite{ref426} Notice in particular how any
deviation from quantization in
$\max(G_{\mathrm{i}}, G_{\mathrm{c}})$ is faithfully reproduced in $G_{\mathrm{H}}$, in
complete agreement with Eq.\ (\ref{eq19.11}).

\begin{figure}
\centerline{\includegraphics[width=8cm]{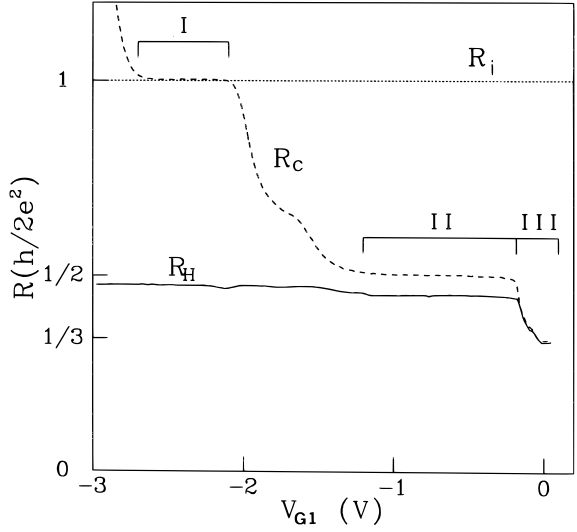}}
\caption{
Results of an experiment similar to that of Fig.\ \ref{fig89}, but with a much larger separation of $80\,\mu{\rm m}$ between injector and collector. Shown are $R_{\rm i} = G_{\rm i}^{-1}$, $R_{\rm c} = G_{\rm c}^{-1}$, and $R_{\rm H} = G_{\rm H}^{-1}$, as a function of the gate voltage on the collector. ($T = 0.45\,{\rm K}$, $B = 2.8\,{\rm T}$; the normal quantized Hall resistance is $\frac{1}{3}(h/2e^{2})$.) Regimes I, II, and III are discussed in the text. Taken from B. W. Alphenaar et al., Phys.\ Rev.\ Lett.\ {\bf 64}, 677 (1990).
\label{fig90}
}
\end{figure}

The experiment of Ref.\ \onlinecite{ref426} was repeated by Alphenaar et al.\cite{ref429} for much
larger point contact separations $(\approx 100\,\mu \mathrm{m})$, allowing a study of the length
scale for equilibration of edge channels at the 2DEG boundary. Even after
such a long distance, no complete equilibration of the edge channels was
found, as manifested by a dependence of the Hall resistance on the gate
voltage used to vary the number of edge channels transmitted through the
point contact voltage probe (see Fig.\ \ref{fig90}). As discussed in Section \ref{sec18b}, a
dependence of the resistance on the properties of the contacts is only possible
in the absence of local equilibrium. In contrast to the experiment by van Wees
et al.,\cite{ref426} and in disagreement with Eq.\ (\ref{eq19.11}), the Hall resistance in Fig.\ \ref{fig90}
does not simply follow the smallest of the contact resistances of current and
voltage probe. This implies that the assumption of fully adiabatic transport
has broken down on a length scale of $100\,\mu \mathrm{m}$.

\begin{figure}
\centerline{\includegraphics[width=8cm]{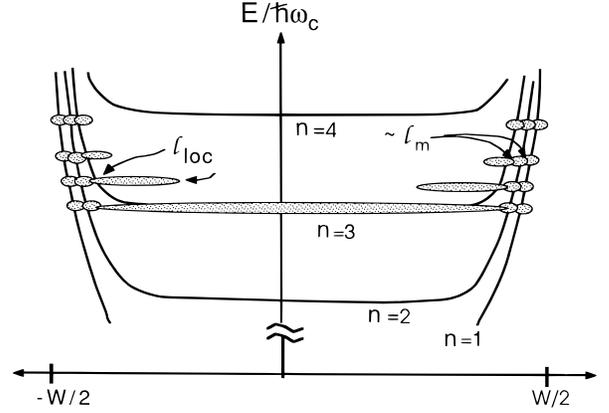}}
\caption{
Illustration of the spatial extension (shaded ellipsoids) of edge channels for four different values of the Fermi energy. The $n = 3$ edge channel can penetrate into the bulk by hybridizing with the $n = 3$ bulk Landau level, coexisting at the Fermi level. This would explain the absence of equilibration between the $n = 3$ and $n = 1,2$ edge channels. The penetration depth $l_{\rm loc}$ and the magnetic length are indicated. Taken from B. W. Alphenaar et al., Phys.\ Rev.\ Lett.\ {\bf 64}, 677 (1990).
\label{fig91}
}
\end{figure}

In the experiment a magnetic field was applied such that three edge
channels were available at the Fermi level. The contact resistance of the
injector was adjusted to $R_{i}=h/2e^{2}$, so current was injected in a single edge
channel $(n=1)$ only. The gate voltage defining the collector point contact
was varied. In Fig.\ \ref{fig90} the contact resistances of injector $(R_{\mathrm{i}})$ and collector $(R_{\mathrm{c}})$
are plotted as a function of this gate voltage, together with the Hall resistance
$R_{\mathrm{H}}$. At zero gate voltage the Hall resistance takes on its normal quantized
value [$R_{\mathrm{H}}= \frac{1}{3}(h/2e^{2})$]. On increasing the negative gate voltage three regions
of interest are traversed (labeled III to I in Fig.\ \ref{fig90}). In region III edge
channels 1 and 2 are completely transmitted through the collector, but the
$n=3$ channel is partially reflected. In agreement with Eq.\ (\ref{eq19.11}), $R_{\mathrm{H}}$
increases following $R_{\mathrm{c}}$. As region II is entered, $R_{\mathrm{H}}$ levels off while $R_{\mathrm{c}}$ continues
to increase up to the $\frac{1}{2}(h/2e^{2})$ quantized value. The fact that $R_{\mathrm{H}}$ stops slightly
short of this value proves that some scattering between the $n=3$ and $n=1,2$
channels has occurred. On increasing the gate voltage further, $R_{\mathrm{c}}$ rises to
$h/2e^{2}$ in region I. However, $R_{\mathrm{H}}$ shows hardly any increase with respect to its
value in region II. This demonstrates that the $n=2$ and $n=1$ edge channels
have almost fully equilibrated. A quantitative analysis\cite{ref429} shows that, in fact,
$92\%$ of the current originally injected into the $n=1$ edge channel is
redistributed equally over the $n=1$ and $n=2$ channels, whereas only $8\%$ is
transferred to the $n=3$ edge channel. The suppression of scattering between
the highest-index $n=N$ edge channel and the group of edge channels with
$n\leq N-1$ was found to exist only if the Fermi level lies in (or near) the $N\mathrm{th}$
bulk Landau level. As a qualitative explanation it was suggested\cite{ref429,ref476} that
the $N\mathrm{th}$ edge channel hybridizes with the $N\mathrm{th}$ bulk Landau level when both
types of states coexist at the Fermi level. Such a coexistence does not occur
for $n\leq N-1$ if the potential fluctuations are small compared with $\hbar\omega_{\mathrm{c}}$ (cf.\
Fig.\ \ref{fig78}). The spatial extension of the wave functions of the edge channels is
illustrated in Fig.\ \ref{fig91} (shaded ellipsoids) for various values of the Fermi level
between the $n=3$ and $n=4$ bulk Landau levels. As the Fermi level
approaches the $n=3$ bulk Landau level, the corresponding edge channel
penetrates into the bulk, so the overlap with the wave functions of lower-index edge channels decreases. This would explain the decoupling of the
$n=3$ and $n=1,2$ edge channels.

These experiments thus point the way in which the transition from
microscopic to macroscopic behavior takes place in the QHE, while they also
demonstrate that quite large samples will be required before truly macroscopic behavior sets in.

\subsubsection{\label{sec19d} Suppression of the Shubnikov-De Haas oscillations}

Shubnikov-De Haas magnetoresistance oscillations were discussed in
Sections \ref{sec4c} and \ref{sec10}. In weak magnetic fields, where a theoretical description
in terms of a local resistivity tensor applies, a satisfactory agreement between
theory and experiment is obtained.\cite{ref20} As we now know, in strong magnetic
fields the concept of a local resistivity tensor may break down entirely
because of the absence of local equilibrium. A theory of the Shubnikov-De
Haas effect then has to take into account explicitly the properties of the
contacts used for the measurement. The resulting anomalies are considered in
this subsection.

\begin{figure}
\centerline{\includegraphics[width=8cm]{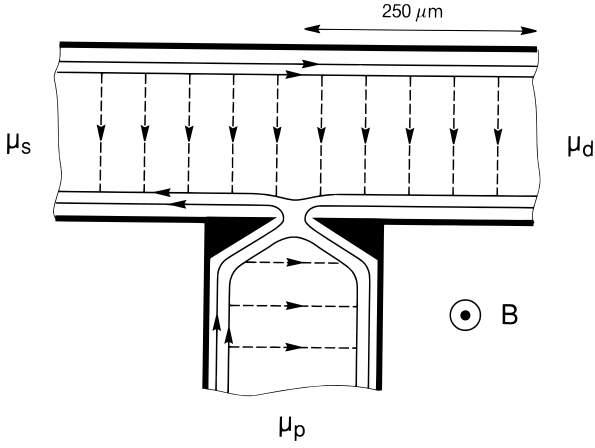}}
\caption{
Illustration of the mechanism for the suppression of Shubnikov-De Haas oscillations due to selective detection of edge channels. The black area denotes the split-gate point contact in the voltage probe, which is at a distance of $250\,\mu{\rm m}$ from the drain reservoir. Dashed arrows indicate symbolically the selective back scattering in the highest-index edge channel, via states in the highest bulk Landau level that coexist at the Fermi level. Taken from H. van Houten et al., in Ref.\ \onlinecite{ref9}.
\label{fig92}
}
\end{figure}

Van Wees et al.\cite{ref428} found that the amplitude of the high-field Shubnikov-De Haas oscillations was suppressed if a quantum point contact was used as
a voltage probe. To discuss this anomalous Shubnikov-De Haas effect, we
consider the three-terminal geometry of Fig.\ \ref{fig92}, where a single voltage
contact is present on the boundary between source and drain contacts. (An
alternative two-terminal measurement configuration is also possible; see Ref.\
\onlinecite{ref428}.) The voltage probe $\mathrm{p}$ is formed by a quantum point contact, while source
$\mathrm{s}$ and drain $\mathrm{d}$ are normal ohmic contacts. (Note that {\it two\/} special contacts were
required for the anomalous quantum Hall effect of Section \ref{sec19c}.) One
straightforwardly finds from Eq.\ (\ref{eq12.12}) that the three-terminal resistance
$R_{3\mathrm{t}}\equiv(\mu_{\mathrm{p}}-\mu_{\mathrm{d}})/eI$ measured between point contact probe and drain is given
by
\be
R_{3\mathrm{t}}= \frac{h}{2e^{2}}\frac{T_{\mathrm{s}\rightarrow \mathrm{p}}}{(N_{\mathrm{s}}-R_{\mathrm{s}})(N_{\mathrm{p}}-R_{\mathrm{p}})-T_{\mathrm{p}\rightarrow \mathrm{s}}T_{\mathrm{s}\rightarrow \mathrm{p}}}. \label{eq19.12}
\ee 
This three-terminal resistance corresponds to a generalized {\it longitudinal\/}
resistance if the magnetic field has the direction of Fig.\ \ref{fig92}. In the absence of
backscattering in the 2DEG, one has $T_{\mathrm{s}\rightarrow \mathrm{p}}=0$, so $R_{3\mathrm{t}}$ vanishes, as it should
for a longitudinal resistance in a strong magnetic field.

\begin{figure}
\centerline{\includegraphics[width=8cm]{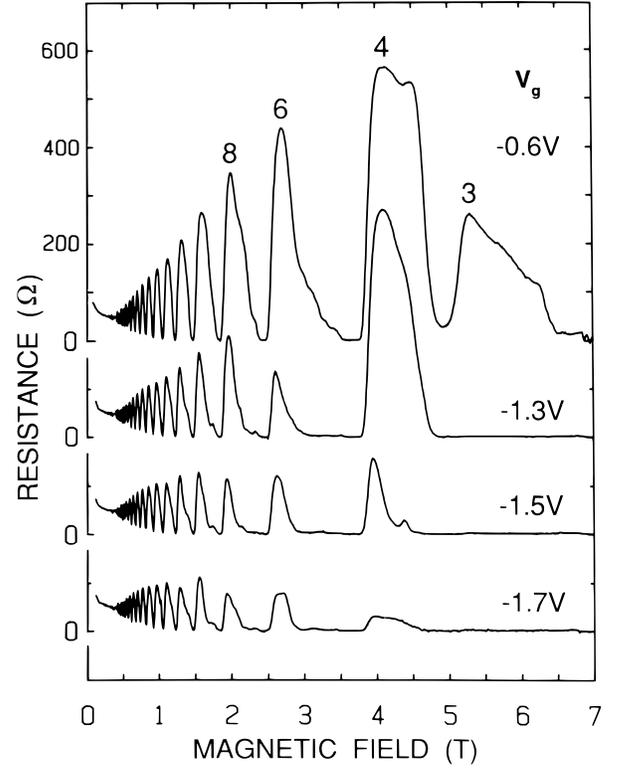}}
\caption{
Measurement of the anomalous Shubnikov-De Haas oscillations in the geometry of Fig.\ \ref{fig92}. The plotted longitudinal resistance is the voltage drop between contacts p and d divided by the current from s to d. At high magnetic fields the oscillations are increasingly suppressed as the point contact in the voltage probe is pinched off by increasing the negative gate voltage. The number of occupied spin-split Landau levels in the bulk is indicated at several of the Shubnikov-De Haas maxima. Taken from B. J. van Wees et al., Phys.\ Rev.\ B {\bf 39}, 8066 (1989).
\label{fig93}
}
\end{figure}

Shubnikov-De Haas oscillations in the longitudinal resistance arise when
backscattering leads to $T_{\mathrm{s}\rightarrow \mathrm{p}}\neq 0$. The resistance reaches a maximum when
the Fermi level lies in a bulk Landau level, corresponding to a maximum
probability for backscattering (which requires scattering from one edge to the
other across the bulk of the sample, as indicated by the dashed lines in Fig.\
\ref{fig92}). From the preceding discussion of the anomalous quantum Hall effect, we
know that the point contact voltage probe in a high magnetic field functions
as a selective detector of edge channels with index $n$ less than some value
determined by the barrier height in the point contact. If backscattering itself
occurs selectively for the channel with the highest index $n=N$, and if the edge
channels with $n\leq N-1$ do not scatter to that edge channel, then a
suppression of the Shubnikov-De Haas oscillations is to be expected when
$R_{3\mathrm{t}}$ is measured with a point contact containing a sufficiently high potential
barrier. This was indeed observed experimentally,\cite{ref428} as shown in Fig.\ \ref{fig93}. The
Shubnikov-De Haas maximum at $5.2\, \mathrm{T}$, for example, is found to disappear at
gate voltages such that the point contact conductance is equal to, or smaller
than $2e^{2}/h$, which means that the point contact only transmits two spin-split
edge channels. The number of occupied spin-split Landau levels in the bulk at
this magnetic field value is 3. This experiment thus demonstrates that the
Shubnikov-De Haas oscillations result from the highest-index edge channel
only, presumably because that edge channel can penetrate into the bulk via
states in the bulk Landau level with the same index that coexist at the Fermi
level (cf.\ Section \ref{sec19c}). Moreover, it is found that this edge channel does not
scatter to the lower-index edge channels over the distance of 250 $\mu \mathrm{m}$ from
probe $\mathrm{p}$ to drain $\mathrm{d}$, consistent with the experiment of Alphenaar et al.\cite{ref429}

In Section \ref{sec19a} we discussed how an ``ideal'' contact at the 2DEG
boundary {\it induces\/} a local equilibrium by equipartitioning the outgoing
current equally among the edge channels. The anomalous Shubnikov-De
Haas effect provides a direct way to study this contact-induced equilibration
by means of a second point contact between the point contact voltage probe
$\mathrm{p}$ and the current drain $\mathrm{d}$ in Fig.\ \ref{fig92}. This experiment was also carried out by
van Wees et al., as described in Ref.\ \onlinecite{ref308}. Once again, use was made of the
double-split-gate point contact device (Fig.\ \ref{fig5}b), in this case with a $1.5$-$\mu \mathrm{m}$
separation between point contact $\mathrm{p}$ and the second point contact. It is found
that the Shubnikov-De Haas oscillations in $R_{3\mathrm{t}}$ are suppressed only if the
second point contact has a conductance of $(2e^{2}/h)(N_{\mathrm{wide}}-1)$ or smaller. At
larger conductances the oscillations in $R_{3\mathrm{t}}$ return, because this point contact
can now couple to the highest-index edge channel and distribute the
backscattered electrons over the lower-index edge channels. The point
contact positioned between contacts $\mathrm{p}$ and $\mathrm{d}$ thus functions as a controllable
``edge channel mixer.''

The conclusions of the previous paragraph have interesting implications
for the Shubnikov-De Haas oscillations in the strong-field regime even if
measured with contacts that do {\it not\/} selectively detect certain edge channels
only.\cite{ref307} Consider again the geometry of Fig.\ \ref{fig92}, in the low-gate voltage limit
where the point contact voltage probe transmits all edge channels with unit
probability. (This is the case of an ``ideal'' contact; cf.\ Section \ref{sec18b}.) To simplify
expression (\ref{eq19.12}) for the three-terminal longitudinal resistance $R_{3\mathrm{t}}$, we use
the fact that the transmission and reflection probabilities $T_{\mathrm{s}\rightarrow \mathrm{p}}$, $R_{\mathrm{s}}$, and $R_{\mathrm{p}}$
refer to the highest-index edge channel only (with index $n=N$), under the
assumptions of selective backscattering and absence of scattering to lower-index edge channels discussed earlier. As a consequence, $T_{\mathrm{s}\rightarrow \mathrm{p}}, R_{\mathrm{s}}$, and $R_{\mathrm{p}}$ are
each at most equal to 1; thus, up to corrections smaller by a factor $N^{-1}$, we
may put these terms equal to zero in the denominator on the right-hand side
of Eq.\ (\ref{eq19.12}). In the numerator, the transmission probability $T_{\mathrm{s}\rightarrow \mathrm{p}}$ may be
replaced by the backscattering probability $t_{\mathrm{bs}}\leq 1$, which is the probability
that the highest-index edge channel injected by the source contact reaches the
point contact probe following scattering across the wide 2DEG (dashed lines
in Fig.\ \ref{fig92}). With these simplifications Eq.\ (\ref{eq19.12}) takes the form (assuming
spin degeneracy)
\be
R_{3\mathrm{t}}= \frac{h}{2e^{2}}\frac{t_{\mathrm{bs}}}{N^{2}}\times(1+ {\rm order}\,N^{-1}). \label{eq19.13}
\ee
Only if $t_{\mathrm{bs}}\ll 1$ may the backscattering probability be expected to scale
linearly with the separation of the two contacts $\mathrm{p}$ and $\mathrm{d}$ (between which the
voltage drop is measured). If $t_{\mathrm{bs}}$ is not small, then the upper limit $t_{\mathrm{bs}^{<}}1$ leads
to the prediction of a {\it maximum\/} possible amplitude\cite{ref307}
\be
R_{\max}=\frac{h}{2e^{2}}\frac{1}{N^{2}}\times(1+{\rm order}\,N^{-1}) \label{eq19.14}
\ee
of the Shubnikov-De Haas resistance oscillations in a given large magnetic
field, independently of the length of the segment over which the voltage drop
is measured, provided equilibration does not occur on this segment. Equilibration might result, for example, from the presence of additional contacts
between the voltage probes, as discussed before. One easily verifies that the
high-field Shubnikov-De Haas oscillations in Fig.\ \ref{fig93} at $V_{\mathrm{g}}=-0.6\,\mathrm{V}$ (when
the point contact is just defined, so that the potential barrier is small) lie well
below the upper limit (\ref{eq19.14}). For example, the peak around $2\, \mathrm{T}$ corresponds
to the case of four occupied spin-degenerate Landau levels, so the theoretical
upper limit is $(h/2e^{2}) \times\frac{1}{16}\approx 800\,\Omega$, well above the observed peak value of
about $350\,\Omega$. The prediction of a maximum longitudinal resistance implies
that the linear scaling of the amplitude of the Shubnikov-De Haas oscillations with the distance between voltage probes found in the weak-field
regime, and expected on the basis of a description in terms of a local
resistivity tensor,\cite{ref20} breaks down in strong magnetic fields. Anomalous
scaling of the Shubnikov-De Haas effect has been observed experimentally\cite{ref457,ref460,ref466} and has recently also been interpreted\cite{ref430} in terms of a
nonequilibrium between the edge channels. A quantitative experimental and
theoretical investigation of these issues has now been carried out by McEuen
et al.\cite{ref477}

Selective backscattering and the absence of local equilibrium have
consequences as well for the two-terminal resistance in strong magnetic
fields.\cite{ref307}  In weak fields one usually observes in two-terminal measurements a
superposition of the Shubnikov-De Haas longitudinal resistance oscillations
and the quantized Hall resistance. This superposition shows up as a
characteristic ``overshoot'' of the two-terminal resistance as a function of the
magnetic field as it increases from one quantized Hall plateau to the next (the
plateaux coincide with minima of the Shubnikov-De Haas oscillations). In
the strong-field regime (in the absence of equilibration between source and
drain contacts), no such superposition is to be expected. Instead, the two-terminal resistance would increase monotonically from $(h/2e^{2})N^{-1}$ to
$(h/2e^{2})(N-1)^{-1}$ as the transmission probability from source to drain
decreases from $N$ to $N-1$. We are not aware of an experimental test of this
prediction.

The foregoing analysis assumes that the length $L$ of the conductor is much
greater than its width $W$, so edge channels are the only states at the Fermi
level that extend from source to drain. If $L\ll W$, additional extended states
may appear in the bulk of the 2DEG, whenever the Fermi level lies in a bulk
Landau level. An experiment by Fang et al.\ in this short-channel regime, to
which our analysis does not apply, is discussed by B\"{u}ttiker.\cite{ref386}

\subsection{\label{sec20} Fractional quantum Hall effect}

Microscopically, quantization of the Hall conductance $G_{\mathrm{H}}$ in fractional
multiples of $e^{2}/h$ is entirely different from quantization in integer multiples.
While the {\it integer\/} quantum Hall effect\cite{ref8} can be explained satisfactorily in
terms of the states of noninteracting electrons in a magnetic field (see Section
\ref{sec18}), the {\it fractional\/} quantum Hall effect\cite{ref478} exists only because of electron-electron interactions.\cite{ref479} Phenomenologically, however, the two effects are
quite similar. Several experiments on edge channel transport in the integer
QHE,\cite{ref339,ref340,ref426} reviewed in Section \ref{sec19} have been repeated\cite{ref480,ref481} for the
fractional QHE with a similar outcome. The interpretation of Section \ref{sec19} in
terms of selective population and detection of edge channels cannot be
applied in that form to the fractional QHE. Edge channels in the integer
QHE are defined in one-to-one correspondence to bulk Landau levels
(Section \ref{sec18b}). The fractional QHE requires a generalization of the concept of
edge channels that allows for independent current channels within the same
Landau level. Two recent papers have addressed this problem\cite{ref482,ref483} and
have obtained different answers. The present status of theory and experiment
on transport in ``fractional'' edge channels is reviewed in Section \ref{sec20b},
preceded by a brief introduction to the fractional QHE.

\subsubsection{\label{sec20a} Introduction}

Excellent high-level introductions to the fractional QHE in an unbounded
2DEG can be found in Refs.\ \onlinecite{ref97} and \onlinecite{ref484}. The following is an oversimplification of Laughlin's theory\cite{ref479} of the effect and is only intended to introduce the
reader to some of the concepts that play a role in edge channel transport in
the fractional QHE.

It is instructive to first consider the motion of two interacting electrons in a
strong magnetic field.\cite{ref485} The dynamics of the relative coordinate $\mathbf{r}$ decouples
from that of the center of mass. Semiclassically, $\mathbf{r}$ moves along equipotentials
of the Coulomb potential $e^{2}/\epsilon r$ (this is the guiding center drift discussed in
Section \ref{sec18b}). The relative coordinate thus executes a circular motion around
the origin, corresponding to the two electrons orbiting around their center of
mass. The phase shift acquired on one complete revolution,
\be
\Delta\phi=\frac{e}{\hbar}\oint d{\bf l}\cdot {\bf A} = \frac{e}{\hbar}B\pi r^{2}, \label{eq20.1}
\ee
should be an integer multiple of $2\pi$ so that
\be
r=l_{\mathrm{m}}\sqrt{2q},\;\;q=1,2, \ldots. \label{eq20.2}
\ee 
The interparticle separation in units of the magnetic length $l_{\mathrm{m}}\equiv(\hbar/eB)^{1/2}$ is
quantized. In the field regime where the fractional QHE is observed, only one
spin-split Landau level is occupied in general. If the electrons have the same
spin, the wave function should change sign when two coordinates are
interchanged. In the case considered here of two electrons, an interchange of
the coordinates is equivalent to $\mathbf{r}\rightarrow-\mathbf{r}$. A change of sign is then obtained if
the phase shift for one half revolution is an odd multiple of $\pi$ (i.e., for $\Delta\phi$ an
odd multiple of $2\pi$). The Pauli principle thus restricts the integer $q$ in Eq.\
(\ref{eq20.2}) to {\it odd\/} values.

The interparticle separation of a system of more than two electrons is not
quantized. Still, one might surmise that the energy at densities $n_{\mathrm{s}}\approx 1/\pi \bar{r}^{2}$
corresponding to an average separation $\bar{r}$ in accord with Eq.\ (\ref{eq20.2}) would be
particularly low. This occurs when the Landau level filling factor $\nu\equiv hn_{\mathrm{s}}/eB$
equals $\nu\approx 1/q$. Theoretical work by Laughlin, Haldane, and Halperin\cite{ref479,ref486,ref487} shows that the energy density $u(\nu)$ of a uniform 2DEG in a
strong magnetic field has downward {\it cusps\/} at these values of $\nu$ as well as at
other fractions, given generally by
\be
\nu=p/q, \label{eq20.3}
\ee 
with $p$ and $q$ mutually prime integers and $q$ odd. The cusp in $u$ at {\it integer\/} $\nu$ is a
consequence solely of Landau level quantization, according to
\be
du/dn_{\mathrm{s}}=( \mathrm{Int}[\nu]+{\textstyle\frac{1}{2}})\hbar\omega_{\mathrm{c}}. \label{eq20.4}
\ee 
Because of the cusp in $u$, the chemical potential $du/dn_{\mathrm{s}}$ has a discontinuity
$\Delta\mu=\hbar\omega_{\mathrm{c}}$ at integer $\nu$. At these values of the filling factor an infinitesimal
increase in electron density costs a finite amount of energy, so the electron gas
can be said to be {\it incompressible}. The cusp in $u$ at {\it fractional\/} $\nu$ exists because of
the Coulomb interaction. The discontinuity $\Delta\mu$ is now approximately
$\Delta\mu\approx e^{2}/\epsilon l_{\mathrm{m}}\propto \sqrt{B}$, which at a typical field of $6\, \mathrm{T}$ in GaAs is $10\, \mathrm{meV}$, of the
same magnitude as the Landau level separation $\hbar\omega_{\mathrm{c}}\propto B$.

The incompressibility of the 2DEG at $\nu=p/q$ implies that a nonzero
minimal energy is required to add charge to the system. An important
consequence of Laughlin's theory is that charge can be added only in the
form of quasiparticle excitations of {\it fractional\/} charge $e^{*}=e/q$. The discontinuity $\Delta\mu$ in the chemical potential equals the energy that it costs to
create $p$ pairs of oppositely charged quasiparticles (widely separated from
each other), $\Delta\mu=p\times 2\Delta$ with $\Delta$ the quasiparticle creation energy.

The fractional QHE in a disordered macroscopic sample occurs because
the quasiparticles are localized by potential fluctuations in the bulk of the
2DEG. A variation of the filling factor $\nu=p/q+\delta \nu$ in an interval around the
fractional value changes the density of localized quasiparticles without
changing the Hall conductance, which retains the value $G_{\mathrm{H}}=(p/q)e^{2}/h$. The
precision of the QHE has been explained by Laughlin\cite{ref488} in terms of the
quantization of the quasiparticle charge $e^{*}$, which is argued to imply
quantization of $G_{\mathrm{H}}$ at integer multiples of $ee^{*}/h$.

\subsubsection{\label{sec20b} Fractional edge channels}

In a small sample the fractional QHE can occur in the absence of disorder
and can show deviations from precise quantization. Moreover, in special
geometries\cite{ref481} $G_{\mathrm{H}}$ can take on quantized values that are not simply related to
$e^{*}$. These observations cannot be easily understood within the conventional
description of the fractional QHE, as outlined in the previous subsection. An
approach along the lines of the edge channel formulation of the integer QHE
(Sections \ref{sec18} and \ref{sec19}) seems more promising. In Ref.\ \onlinecite{ref482} the concept of an edge
channel was generalized to the fractional QHE, and a generalized Landauer
formula relating the conductance to the transmission probabilities of the edge
channels was derived. We review this theory and the application to experiments. A different edge channel theory by MacDonald\cite{ref483} is discussed toward
the end of this subsection.

The edge channels for the conductance in the linear transport regime are
defined in terms of properties of the equilibrium state of the system. If the
electrostatic potential energy $V(x, y)$ varies slowly in the 2DEG, then the
equilibrium density distribution $n(x, y)$ follows by requiring that the local
electrochemical potential $V(\mathbf{r})+du/dn$ has the same value $\mu$ at each point $\mathbf{r}$ in
the 2DEG. Here $du/dn$ is the chemical potential of the {\it uniform\/} 2DEG with
density $n(\mathbf{r})$. As discussed in Section \ref{sec20a}, the internal energy density $u(n)$ of a
uniform interacting 2DEG in a strong magnetic field has downward cusps at
densities $n=\nu_{p}Be/h$ corresponding to certain fractional filling factors $\nu_{p}$. As a
result, the chemical potential $du/dn$ has a discontinuity (an energy gap) at
$\nu=\nu_{p}$, with $du_{p}^{+}/dn$ and $du_{p}^{-}/dn$ the two limiting values as $\nu\rightarrow \nu_{p}$. As noted
by Halperin,\cite{ref489} when $\mu-V$ lies in the energy gap the filling factor is pinned
at the value $\nu_{p}$. The equilibrium electron density is thus given by\cite{ref489}
\be
n=\left\{\begin{array}{l}
\nu_{p}Be/h,\;\;{\rm if}\;\;du_{p}^{-}/dn<\mu-V<du_{p}^{+}/dn,\\
du/dn+V(\mathbf{r})=\mu,\;\;{\rm otherwise}.
\end{array}\right.\label{eq20.5}
\ee
Note that $V(\mathbf{r})$ itself depends on $n(\mathbf{r})$ and thus has to be determined self-consistently from Eq.\ (\ref{eq20.5}), taking the electrostatic screening in the 2DEG
into account. We do not need to solve explicitly for $n(\mathbf{r})$, but we can identify
the edge channels from the following general considerations.\cite{ref482}

\begin{figure}
\centerline{\includegraphics[width=8cm]{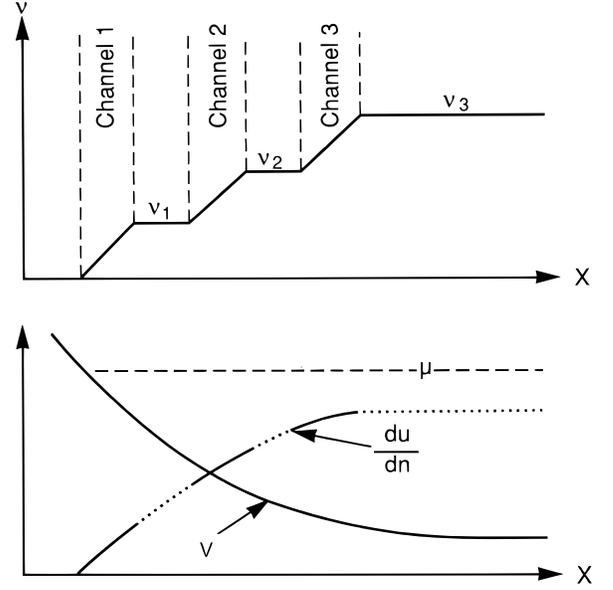}}
\caption{
Schematic drawing of the variation in filling factor $\nu$, electrostatic potential $V$, and chemical potential $du/dn$, at a smooth boundary in a 2DEG. The dashed line in the bottom panel denotes the constant electrochemical potential $\mu=V+du/dn$. The dotted intervals indicate a discontinuity (energy gap) in $du/dn$ and correspond in the top panel to regions of constant fractional filling factor $\nu_{p}$ that spatially separate the edge channels. The width of the edge channel regions shrinks to zero in the integer QHE, since the compressibility $\chi$ of these regions is infinitely large in that case. Taken from C. W. J. Beenakker, Phys.\ Rev.\ Lett.\ {\bf 64}, 216 (1990).
\label{fig94}
}
\end{figure}

At the edge of the 2DEG, the electron density decreases from its bulk value
to zero. Eq.\ (\ref{eq20.5}) implies that this decrease is stepwise, as illustrated in Fig.\ \ref{fig94}. The requirement on the smoothness of $V$ for the appearance of a well-defined region at the edge in which $\nu$ is pinned at the fractional value $\nu_{p}$ is that
the change in $V$ within the magnetic length $l_{\mathrm{m}}$ is small compared with the
energy gap $du_{p}^{+}/dn-du_{p}^{-}/dn$. This ensures that the width of this region is
large compared with $l_{\mathrm{m}}$, which is a necessary (and presumably sufficient)
condition for the formation of the incompressible state. Depending on the
smoothness of $V$, one thus obtains a series of steps at $\nu=\nu_{p}$ ($p=1,2, \ldots, P$) as
one moves from the edge toward the bulk. The series terminates in the filling
factor $\nu_{P}=\nu_{\mathrm{bulk}}$ of the bulk, assuming that in the bulk the chemical potential
$\mu-V$ lies in an energy gap. The regions of constant $\nu$ at the edge form bands
extending along the wire. These {\it incompressible bands\/} [in which the compressibility $\chi\equiv(n^{2}d^{2}u/dn^{2})^{-1}=0]$ alternate with bands in which $\mu-V$ does not lie
in an energy gap. The latter compressible bands (in which $\chi>0$) may be
identified as the {\it edge channels\/} of the transport problem, as will be discussed
later. To resolve a misunderstanding,\cite{ref490} we note that the particular potential
and density profile illustrated in Fig.\ \ref{fig94} (in which the edge channels have a
nonzero width) assumes that the compressibility of the edge channels is not
infinitely large, but the subsequent analysis is independent of this assumption
(requiring only that the edge channels are flanked by bands of zero
compressibility). Indeed, the analysis is applicable also to the integer QHE,
where the edge channels have an infinitely large compressibility and hence an
infinitesimally small width (limited only by the magnetic length).

\begin{figure}
\centerline{\includegraphics[width=8cm]{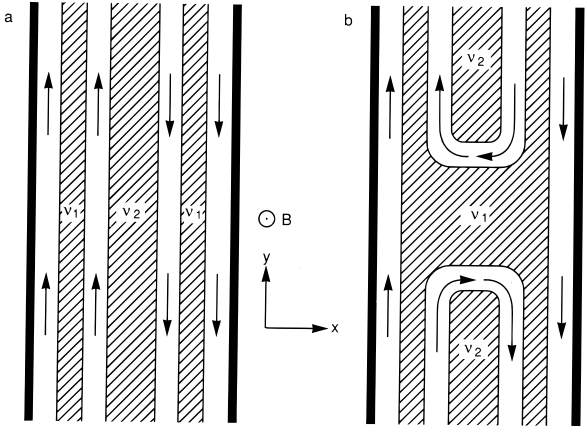}}
\caption{
Schematic drawing of the incompressible bands (hatched) of fractional filling factor $\nu_{p}$, alternating with the edge channels (arrows indicate the direction of electron motion in each channel). (a) A uniform conductor. (b) A conductor containing a barrier of reduced filling factor. Taken from C. W. J. Beenakker, Phys.\ Rev.\ Lett.\ {\bf 64}, 216 (1990).
\label{fig95}
}
\end{figure}

The conductance is calculated by bringing one end of the conductor in
contact with a reservoir at a slightly higher electrochemical potential $\mu+\Delta\mu$
without changing $V$ (as in the derivation of the usual Landauer formula; cf.\
Section \ref{sec12b}). The resulting change $\Delta n$ in electron density is
\be
\Delta n=\left(\frac{\delta n}{\delta\mu}\right)_{V}\Delta\mu=-\left(\frac{\delta n}{\delta V}\right)_{\mu}\Delta\mu, \label{eq20.6}
\ee 
where $\delta$ denotes a functional derivative. In the second equality in Eq.\ (\ref{eq20.6}),
we used the fact that $n$ is a functional of $\mu-V$, by virtue of Eq.\ (\ref{eq20.5}). In a
strong magnetic field, this excess density moves along equipotentials with the
guiding-center-drift velocity $E/B$ ($\mathbf{E}\equiv\partial V/e\partial \mathbf{r}$ being the electric field). The
component $v_{\mathrm{drift}}$ of the drift velocity in the $y$-direction (along the conductor) is
\be
v_{\mathrm{drift}}= \hat{\mathbf{y}}\cdot\left(\mathbf{E}\times\frac{\mathbf{B}}{B^{2}}\right)=-\frac{1}{eB}\frac{\partial V}{\partial x}. \label{eq20.7}
\ee 
The current density $j=-e\Delta nv_{\mathrm{drift}}$ becomes simply
\be
j=- \frac{e}{h}\Delta\mu\frac{\partial v}{\partial x}. \label{eq20.8}
\ee 
It follows from Eq.\ (\ref{eq20.8}) that the incompressible bands of constant $\nu=\nu_{p}$
do not contribute to $j$. The reservoir injects the current into the compressible
bands at one edge of the conductor only (for which the sign of $\partial \nu/\partial x$ is such
that $j$ moves away from the reservoir). The edge channel with index $p=1,2,
\ldots, P$ is defined as that compressible band that is flanked by incompressible
bands at filling factors $\nu_{p}$ and $\nu_{p-1}$. The outermost band from the center of the
conductor, which is the $p=1$ edge channel, is included by defining formally
$\nu_{0}\equiv 0$. The arrangement of alternating edge channels and compressible
bands is illustrated in Fig.\ \ref{fig95}a. Note that different edges may have a different
series of edge channels at the same magnetic field value, depending on the
smoothness of the potential $V$ at the edge (which, as discussed before,
determines the incompressible bands that exist at the edge). This is in contrast
to the situation in the integer QHE, where a one-to-one correspondence
exists between edge channels and bulk Landau levels (Section \ref{sec18b}). In the
fractional QHE an infinite hierarchy of energy gaps exists, in principle,
corresponding to an infinite number of possible edge channels, of which only
a small number (corresponding to the largest energy gaps) will be realized in
practice.

The current $I_{p}=(e/h)\Delta\mu(\nu_{p}-\nu_{p-1})$ injected into edge channel $p$ by the
reservoir follows directly from Eq.\ (\ref{eq20.8}) on integration over $x$. The total
current $I$ through the wire is $I= \sum_{p=1}^{P}I_{p}T_{p}$, if a fraction $T_{p}$ of the injected
current $I_{p}$ is transmitted to the reservoir at the other end of the wire (the
remainder returning via the opposite edge). For the conductance $G\equiv eI/\Delta\mu$,
one thus obtains the generalized Landuer formula for a two-terminal
conductor,\cite{ref482}
\be
G= \frac{e^{2}}{h}\sum_{p=1}^{P}T_{p}\Delta \nu_{p}, \label{eq20.9}
\ee 
which differs from the usual Landauer formula by the presence of the
fractional weight factors $\Delta \nu_{p}\equiv \nu_{p}-\nu_{p-1}$. In the integer QHE, $\Delta \nu_{p}=1$ for all
$p$ so that the usual Landauer formula with unit weight factor is recovered.

A multiterminal generalization of Eq.\ (\ref{eq20.9}) for a two-terminal conductor
is easily constructed, following B\"{u}ttiker\cite{ref5} (cf.\ Section \ref{sec12b}):
\begin{subequations}
\label{eq20.10}
\begin{eqnarray}
I_{\alpha}&=& \frac{e}{h}\nu_{\alpha}\mu_{\alpha}-\frac{e}{h}\sum_{\beta}T_{\alpha\beta}\mu_{\beta}, \label{eq20.10a}\\
T_{\alpha\beta}&=&\sum_{p=1}^{P_{\beta}}T_{p,\alpha\beta}\Delta \nu_{p}. \label{eq20.10b}
\end{eqnarray}
\end{subequations}
Here $I_{\alpha}$ is the current in lead $\alpha$ connected to a reservoir at electrochemical
potential $\mu_{\alpha}$ and fractional filling factor $\nu_{\alpha}$. Equation (\ref{eq20.10b}) defines the
transmission probability $T_{\alpha\beta}$ from reservoir $\beta$ to reservoir $\alpha$ (or the reflection
probability for $\alpha=\beta$) in terms of a sum over the generalized edge channels in
lead $\beta$. The contribution from each edge channel $p=1,2, \ldots ,P_{\beta}$ contains the
weight factor $\Delta \nu_{p}\equiv \nu_{p}-\nu_{p-1}$ and the fraction $T_{p,\alpha\beta}$ of the current injected by
reservoir $\beta$ into the $p\mathrm{th}$ edge channel of lead $\beta$ that reaches reservoir $\alpha$. Apart
from the fractional weight factors, the structure of Eq.\ (\ref{eq20.10}) is the same as
that of the usual B\"{u}ttiker formula (\ref{eq12.12}).

Applying the generalized Landauer formula (\ref{eq20.9}) to the ideal conductor
in Fig.\ \ref{fig95}a, where $T_{p}=1$ for all $p$, one finds the quantized two-terminal
conductance
\be
G= \frac{e^{2}}{h}\sum_{p=1}^{P}\Delta \nu_{p}=\frac{e^{2}}{h}\nu_{P}. \label{eq20.11}
\ee 
The four-terminal Hall conductance $G_{\mathrm{H}}$ has the same value, because each
edge is in local equilibrium. In the presence of disorder this edge channel
formulation of the fractional QHE is generalized in an analogous way as in
the integer QHE by including localized states in the bulk. In a smoothly
varying disorder potential, these localized states take the form of circulating
edge channels, as in Figs.\ \ref{fig78} and \ref{fig79}. In this way the filling factor of the bulk
can locally deviate from $\nu_{P}$ without a change in the Hall conductance, leading
to the formation of a plateau in the magnetic field dependence of $G_{\mathrm{H}}$. In a
narrow channel, localized states are not required for a finite plateau width
because the edge channels make it possible for the chemical potential to lie in
an energy gap for a finite-magnetic-field interval. The Hall conductance then
remains quantized at $\nu_{P}(e^{2}/h)$ as long as $\mu-V$ in the bulk lies between
$du_{P}^{+}/dn$ and $du_{P}^{-}/dn$.

We now turn to a discussion of experiments on the fractional QHE in
semiconductor nanostructures. Timp et al.\cite{ref491} have measured the fractionally
quantized four-terminal Hall conductance $G_{\mathrm{H}}$ in a narrow cross geometry
(defined by two sets of split gates). The channel width $W\approx 90\,\mathrm{nm}$ is greater
than, but comparable to, the correlation length $l_{\mathrm{m}}$ of the incompressible state
in this experiment ($l_{\mathrm{m}}\approx 9\,\mathrm{nm}$ at $B=8\,\mathrm{T}$), so one may expect the fractional
QHE to be modified by the lateral confinement.\cite{ref492} Timp et al.\ find, in
addition to quantized plateaux near $\frac{1}{3}$, $\frac{2}{5}$, and $\frac{2}{3}\times e^{2}/h$, a plateau-like feature
around $\frac{1}{2}\times e^{2}/h$. This even-denominator fraction is not observed as a Hall
plateau in a bulk 2DEG.\cite{ref493} The plateaux in $G_{\mathrm{H}}$ correlate with dips in a four-terminal longitudinal resistance (the bend resistance defined in Section \ref{sec16}).

\begin{figure}
\centerline{\includegraphics[width=8cm]{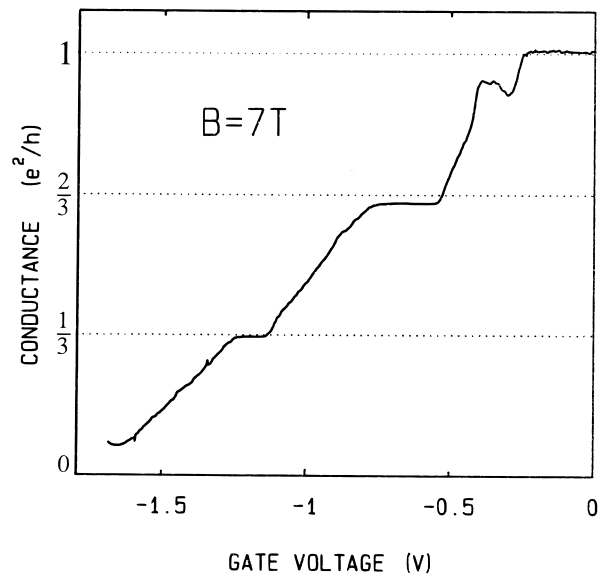}}
\caption{
Two-terminal conductance of a constriction containing a potential barrier, as a function of the voltage on the split gate defining the constriction, at a fixed magnetic field of 7 T. The conductance is quantized according to Eq.\ (\ref{eq20.12}). Taken from L. P. Kouwenhoven et al., unpublished.
\label{fig96}
}
\end{figure}

Consider now a conductor containing a potential barrier. The potential
barrier corresponds to a region of reduced filling factor $\nu_{P_{\min}}\equiv \nu_{\min}$ separating
two regions of filling factor $\nu_{P_{\mathrm{m}}}\equiv \nu_{\max}$. The arrangement of edge channels
and incompressible bands is illustrated in Fig.\ \ref{fig95}b. We assume that the
potential barrier is sufficiently smooth that scattering between the edge
channels at opposite edges can be neglected. All transmission probabilities
are then either 0 or 1: $T_{p}=1$ for $1\leq p\leq P_{\min}$, and $T_{p}=0$ for
$P_{\min}<p\leq P_{\max}$. Equation (\ref{eq20.9}) then tells us that the two-terminal conductance is
\be
G=(e^{2}/h)\nu_{\min}. \label{eq20.12}
\ee 
In Fig.\ \ref{fig96} we show experimental data by Kouwenhoven et al.\cite{ref481} of the
fractionally quantized two-terminal conductance of a constriction containing
a potential barrier. The constriction (or point contact) is defined by a split
gate on top of a GaAs-AlGaAs heterostructure. The conductance in Fig.\ \ref{fig96}
is shown for a fixed magnetic field of $7\, \mathrm{T}$ as a function of the gate voltage.
Increasing the negative gate voltage increases the barrier height, thereby
reducing $G$ below the Hall conductance corresponding to $\nu_{\max}=1$ in the wide
2DEG. The curve in Fig.\ \ref{fig96} shows plateaux corresponding to $\nu_{\min}=1$, $\frac{2}{3}$, and
$\frac{1}{3}$ in Eq.\ (\ref{eq20.12}). The $\frac{2}{3}$ plateau is not exactly quantized, but is too low by a few
percent. The constriction width on this plateau is estimated\cite{ref481} at $500\, \mathrm{nm}$,
which is a factor of 50 larger than the magnetic length at $B=7$ T. It would
seem that scattering between fractional edge channels at opposite edges
(necessary to reduce the conductance below its quantized value) can only
occur via states in the bulk for this large ratio of $W/l_{\mathrm{m}}$.

\begin{figure}
\centerline{\includegraphics[width=8cm]{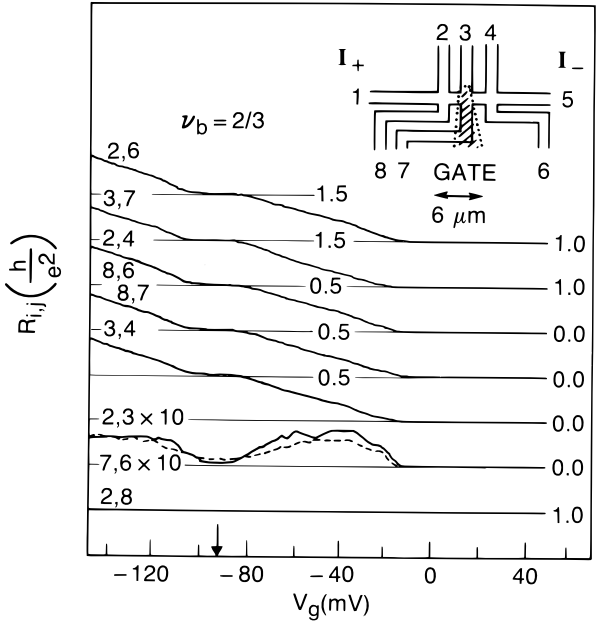}}
\caption{
Four-terminal resistances of a 2DEG channel containing a potential barrier, as a function of the gate voltage ($B = 0.114\,{\rm T}$, $T= 70\,{\rm mK}$). The current flows from contact 1 to contact 5 (see inset), the resistance curves are labeled by the contacts $i$ and $j$ between which the voltage is measured. (The curves for $i,j = 2,4$ and $8,6$ are identical.) The magnetic field points outward. This measurement corresponds to the case $\nu_{\rm max} = 1$ and $\nu_{\rm min}=\nu_{\rm b}$ varying from 1 at $V_{\rm g}\geq-10\,{\rm mV}$ to $2/3$ at $V_{\rm g}\approx-90\,{\rm mV}$ (arrow). The resistances $R_{\rm L} \equiv R_{2,4} = R_{8,6}$ and $R_{\rm D}^{+} \equiv R_{2,6}$ are quantized according to Eqs.\ (\ref{eq20.13}) and (\ref{eq20.14}), respectively. The resistances $R_{3,7}$ and $R_{2,8}$ are the Hall resistances in the gated and ungated regions, respectively. From Eq.\ (\ref{eq20.10}) one can also derive that $R_{8,7} = R_{3,4} = R_{\rm L}$ and $R_{2,3} = R_{7,6} = 0$ on the quantized plateaux, as observed experimentally. Taken from A. M. Chang and J. E. Cunningham, Surf.\ Sci.\ {\bf 229}, 216 (1990).
\label{fig97}
}
\end{figure}

A four-terminal measurement of the fractional QHE in a conductor
containing a potential barrier can be analyzed by means of Eq.\ (\ref{eq20.10}),
analogously to the case of the integer QHE discussed in Section \ref{sec19}. The four-terminal longitudinal resistance $R_{\mathrm{L}}$ (in the geometry of Fig.\ \ref{fig82}) is given by the
analog of Eq.\ (\ref{eq19.3}),
\be
R_{\mathrm{L}}= \frac{h}{e^{2}}\left(\frac{1}{\nu_{\min}}-\frac{1}{\nu_{\max}}\right), \label{eq20.13}
\ee 
provided that {\it either\/} the edge channels transmitted across the barrier have
equilibrated with the extra edge channels available outside the barrier region
{\it or\/} the voltage contacts are ideal; that is, they have unit transmission
probability for all fractional edge channels. Similarly, the four-terminal
diagonal resistances $R_{\mathrm{D}}^{\pm}$ defined in Fig.\ \ref{fig82} are given by [cf.\ Eq.\ (\ref{eq19.5})]
\be
R_{\mathrm{D}}^{+}= \frac{h}{e^{2}}\frac{1}{v_{\min}};\;\;R_{\mathrm{D}}^{-}= \frac{h}{e^{2}}\left(\frac{2}{v_{\max}}-\frac{1}{v_{\min}}\right). \label{eq20.14}
\ee 
Chang and Cunningham\cite{ref480} have measured $R_{\mathrm{L}}$ and $R_{\mathrm{D}}$ in the fractional QHE,
using a $1.5$-$\mu \mathrm{m}$-wide 2DEG channel with a gate across a segment of the
channel (the gate length is also approximately 1.5 $\mu \mathrm{m}$). Ohmic contacts to the
gated and ungated regions allowed $\nu_{\min}$ and $\nu_{\max}$ to be determined independently. Equations (\ref{eq20.13}) and (\ref{eq20.14}) were found to hold to within $0.5\%$
accuracy. This is illustrated in Fig.\ \ref{fig97} for the case that $\nu_{\max}=1$ and $\nu_{\min}$
varying from 1 to 2/3 on increasing the negative gate voltage (at a fixed
magnetic field of $0.114\,\mathrm{T}$). Similar results were obtained\cite{ref480} for the case that
$\nu_{\max}= \frac{2}{3}$ and $\nu_{\min}$ varies from $\frac{2}{3}$ to $\frac{1}{3}$.

\begin{figure}
\centerline{\includegraphics[width=8cm]{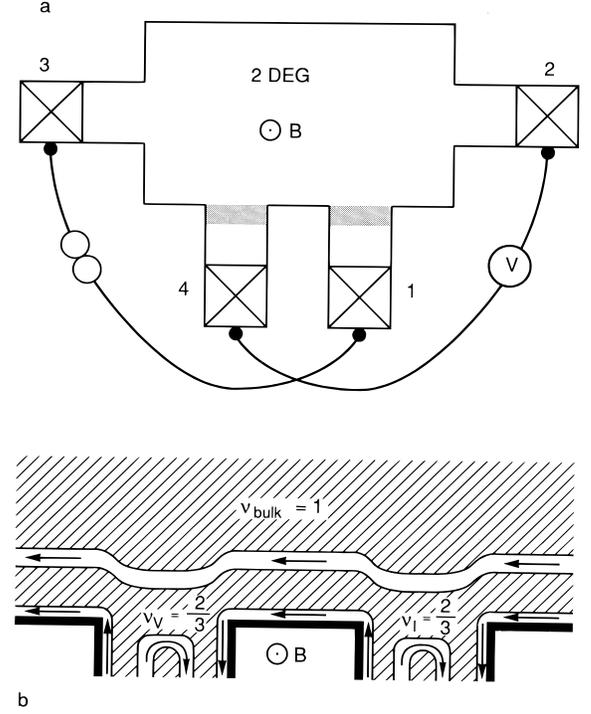}}
\caption{
(a) Schematic of the experimental geometry of Kouwenhoven et al.\cite{ref481} The crossed squares are contacts to the 2DEG. One current lead and one voltage lead contain a barrier (shaded), of which the height can be adjusted by means of a gate (not drawn). The current $I$ flows between contacts 1 and 3; the voltage $V$ is measured between contacts 2 and 4. (b) Arrangement of incompressible bands (hatched) and edge channels near the two barriers. In the absence of scattering between the two fractional edge channels, one would measure a Hall conductance $G_{\rm H} \equiv I/V$ that is fractionally quantized at $\frac{2}{3}\times e^{2}/h$, although the bulk has unit filling factor. Taken from C. W. J. Beenakker, Phys.\ Rev.\ Lett.\ {\bf 64}, 216 (1990).
\label{fig98}
}
\end{figure}

\begin{figure}
\centerline{\includegraphics[width=8cm]{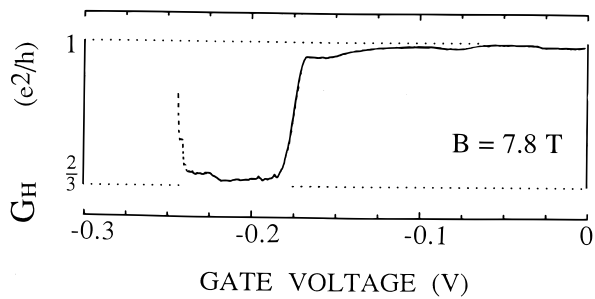}}
\caption{
Anomalously quantized Hall conductance in the geometry of Fig.\ \ref{fig98}, in accord with Eq.\ (\ref{eq20.16}) ($\nu_{\rm bulk} = 1$, $\nu_{\rm I} = \nu_{\rm V}$ decreases from 1 to $2/3$ as the negative gate voltage is increased). The temperature is 20 mK. The rapidly rising part (dotted) is an artifact due to barrier pinch-off. Taken from L. P. Kouwenhoven et al., Phys.\ Rev.\ Lett.\ {\bf 64}, 685 (1990).
\label{fig99}
}
\end{figure}

Adiabatic transport in the fractional QHE can be studied by the selective
population and detection of fractional edge channels, achieved by means of
barriers in two closely separated current and voltage contacts (Fig.\ \ref{fig98}a). The
analysis using Eq.\ (\ref{eq20.10}) is completely analogous to the analysis of the
experiment in the integer QHE,\cite{ref426} discussed in Section \ref{sec19}. Figure \ref{fig98}b
illustrates the arrangement of edge channels and incompressible bands for the
case that the chemical potential lies in an energy gap for the bulk 2DEG (at
$\nu=\nu_{\mathrm{bulk}})$, as well as for the two barriers (at $\nu_{\mathrm{I}}$ and $\nu_{\mathrm{V}}$ for the barrier in the
current and voltage lead, respectively). Adiabatic transport is assumed over
the barrier, as well as from barrier I to barrier V (for the magnetic field
direction indicated in Fig.\ \ref{fig98}). Equation (\ref{eq20.10}) for this case reduces to
\be
I= \frac{e}{h}\nu_{\mathrm{I}}\mu_{\mathrm{I}},\;\;
0= \frac{e}{h}\nu_{\mathrm{V}}\mu_{\mathrm{V}}-\frac{e}{h}\min(\nu_{\mathrm{I}}, \nu_{\mathrm{V}})\mu_{\mathrm{I}}, \label{eq20.15}
\ee 
so the Hall conductance $G_{\mathrm{H}}=eI/\mu_{\mathrm{V}}$ becomes
\be
G_{\mathrm{H}}= \frac{e^{2}}{h}\max(\nu_{\mathrm{I}}, \nu_{\mathrm{V}}) \leq\frac{e^{2}}{h}\nu_{\mathrm{bulk}}.
\label{eq20.16}
\ee
The quantized Hall plateaux are determined by the fractional filling factors of
the current and voltage leads, not of the bulk 2DEG. Kouwenhoven et al.\cite{ref481}
have demonstrated the selective population and detection of fractional edge
channels in a device with a $2$-$\mu \mathrm{m}$ separation of the gates in the current and
voltage leads. The gates extended over a length of 40 $\mu \mathrm{m}$ along the 2DEG
boundary. In Fig.\ \ref{fig99} we reproduce one of the experimental traces of
Kouwenhoven et al.\ The Hall conductance is shown for a fixed magnetic field
of $7.8\, \mathrm{T}$ as a function of the gate voltage (all gates being at the same voltage).
As the barrier heights in the two leads are increased, the Hall conductance
decreases from the bulk value $1\times e^{2}/h$ to the value $\frac{2}{3}\times e^{2}/h$ determined by the
leads, in accord with Eq.\ (\ref{eq20.16}). A more general formula for $G_{\mathrm{H}}$ valid also in
between the quantized plateaux is shown in Ref.\ \onlinecite{ref481} to be in quantitative
agreement with the experiment.

MacDonald has, independent of Ref.\ \onlinecite{ref482}, proposed a different generalized
Landauer formula for the fractional QHE.\cite{ref483} The difference with Eq.\ (\ref{eq20.9}) is
that the weight factors in MacDonald's formula can take on both positive {\it and negative\/} values (corresponding to electron and hole channels). In the case of
local equilibrium at the edge, the sum of weight factors is such that the two
formulations give identical results. The results differ in the absence of local
equilibrium if fractional edge channels are selectively populated and detected.
For example, MacDonald predicts a {\it negative\/} longitudinal resistance in a
conductor at filling factor $\nu= \frac{2}{3}$ containing a segment at $\nu=1$. Another
implication of Ref.\ \onlinecite{ref483} is that the two-terminal conductance $G$ of a conductor
at $\nu_{\max}=1$ containing a potential barrier at filling factor $\nu_{\min}$ is reduced to $\frac{1}{3}\times e^{2}/h$ if $\nu_{\min}=\frac{1}{3}$ [in accord with Eq.\ (\ref{eq20.12})], but remains at $1\times e^{2}/h$ if
$\nu_{\min}=2/3$. That this is not observed experimentally (cf.\ Fig.\ \ref{fig96}) could be due
to interedge channel scattering, as argued by MacDonald. The experiment by
Kouwenhoven et al.\cite{ref481} (Fig.\ \ref{fig99}), however, is apparently in the adiabatic
regime, and was interpreted in Fig.\ \ref{fig98} in terms of an edge channel of weight $\frac{1}{3}$
at the edge of a conductor at $\nu=1$. In MacDonald's formulation, the
conductor at $\nu=1$ has only a singe edge channel of weight 1. This would
need to be reconciled with the experimental observation of quantization of
the Hall conductance at $\frac{2}{3}\times e^{2}/h$.

We conclude this section by briefly addressing the question: What charge
does the resistance measure? The fractional quantization of the conductance
in the experiments discussed is understood as a consequence of the fractional
weight factors in the generalized Landauer formula (\ref{eq20.9}). These weight
factors $\Delta \nu_{p}=\nu_{p}-\nu_{p-1}$ are {\it not\/} in general equal to $e^{*}/e$, with $e^{*}$ the fractional
charge of the quasiparticle excitations of Laughlin's incompressible state (cf.\
Section \ref{sec20a}). The reason for the absence of a one-to-one correspondence
between $\Delta \nu_{p}$ and $e^{*}$ is that the edge channels themselves are not incompressible.\cite{ref482} The transmission probabilities in Eq.\ (\ref{eq20.9}) refer to charged ``gapless''
excitations of the edge channels, which are not identical to the charge $e^{*}$
excitations above the energy gap in the incompressible bands (the latter
charge might be obtained from thermal activation measurements; cf.\ Ref.\
\onlinecite{ref494}). It is an interesting and (to date) unsolved problem to determine the
charge of the edge channel excitations. Kivelson and Pokrovsky\cite{ref495} have
suggested performing tunneling experiments in the fractional QHE regime
for such a purpose, by using the charge dependence of the magnetic length
$(\hbar/eB)^{1/2}$ (which determines the penetration of the wave function in a tunnel
barrier and, hence, the transmission probability through the barrier). Alternatively, one could use the $h/e$ periodicity of the Aharanov-Bohm magnetoresistance oscillations as a measure of the edge channel charge. Simmons
et al.\cite{ref496} find that the characteristic field scale of quasiperiodic resistance
fluctuations in a $2$-$\mu \mathrm{m}$-wide Hall bar increases from $0.016\,{\rm T}\pm 30\%$ near
$\nu=1,2,3,4$ to $0.05\, \mathrm{T}\pm 30\%$ near $\nu= \frac{1}{3}$. This is suggestive of a reduction in
charge from $e$ to $e/3$, but not conclusive since the area for the Aharonov-
Bohm effect is not well defined in a Hall bar (cf.\ Section \ref{sec21}).

\subsection{\label{sec21} Aharonov-Bohm effect in strong magnetic fields}

As mentioned briefly in Section \ref{sec8}, the Aharonov-Bohm oscillations in the
magnetoresistance of a ring are gradually suppressed in strong magnetic
fields. This suppression provides additional support for edge channel trans-
port in the quantum Hall effect regime (Section \ref{sec21a}). Entirely new mechanisms for the Aharonov-Bohm effect become operative in strong magnetic
fields. These mechanisms, resonant tunneling and resonant reflection of edge
channels, do not require a ring geometry. Theory and experiments on
Aharonov-Bohm oscillations in singly connected geometries are the subject
of Section \ref{sec21b}.

\subsubsection{\label{sec21a} Suppression of the Aharonov-Bohm effect in a ring}

\begin{figure}
\centerline{\includegraphics[width=6cm]{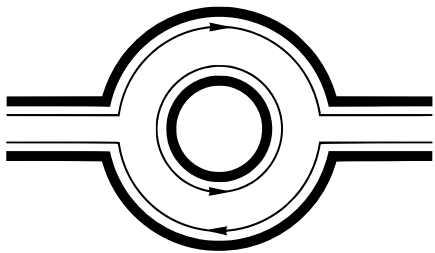}}
\caption{
Illustration of a localized edge channel circulating along the inner perimeter of a ring, and of extended edge channels on the
leads and on the outer perimeter. No Aharonov-Bohm magnetoresistance oscillations can occur in the absence of scattering between these two types of edge channels.
\label{fig100}
}
\end{figure}

In Section \ref{sec8} we have seen how the quantum interference of clockwise and
counterclockwise trajectories in a ring in the diffusive transport regime leads
to magnetoresistance oscillations with two different periodicities: the fundamental Aharonov-Bohm effect with $\Delta B=(h/e)S^{-1}$ periodicity, and the
harmonic with $\Delta B=(h/2e)S^{-1}$ periodicity, where $S$ is the area of the ring. In
arrays of rings only the $h/2e$ effect is observable, since the $h/e$ effect has a
sample specific phase and is averaged to zero. In experiments by Timp et al.\cite{ref69}
and by Ford et al.\cite{ref74} on single rings in the 2DEG of high-mobility GaAs-AlGaAs heterostructures, the $h/e$ effect was found predominantly. The
amplitude of these oscillations is strongly reduced\cite{ref69,ref74,ref195,ref497} by a large
magnetic field (cf.\ the magnetoresistance traces shown in Fig.\ \ref{fig26}). This
suppression was found to occur for fields such that $2l_{\mathrm{cycl}}<W$, where $W$ is the
width of the arms of the ring. The reason is that in strong magnetic fields the
states at the Fermi level that can propagate through the ring are edge states at
the outer perimeter. These states do not complete a revolution around the
ring (see Fig.\ \ref{fig100}). Scattering between opposite edges is required to complete
a revolution, but such backscattering would also lead to a nonzero longitudinal resistance. This argument\cite{ref112,ref498} explains the absence of Aharonov-Bohm oscillations on the quantized Hall plateaux, where the longitudinal
resistance is zero. Magnetoresistance oscillations return between the plateaux
in the Hall resistance, but at a larger value of $\Delta B$ than in weak fields. Timp et
al.\cite{ref497} have argued that the Aharonov-Bohm oscillations in a ring in strong
magnetic fields are associated with scattering from the outer edge to edge
states circulating along the inner perimeter of the ring. The smaller area
enclosed by the inner perimeter explains the increase in $\Delta B$. This interpretation is supported by numerical calculations.\cite{ref497}

\subsubsection{\label{sec21b} Aharonov-Bohm effect in singly connected geometries}

\begin{figure}
\centerline{\includegraphics[width=8cm]{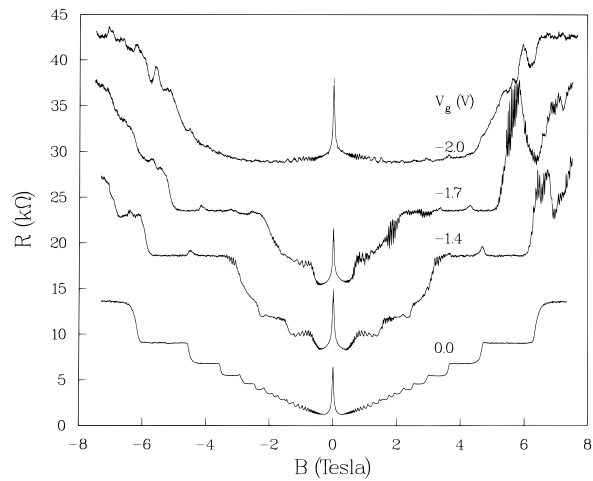}}
\caption{
Two-terminal magnetoresistance of a point contact for a series of gate voltages at $T= 50\,{\rm mK}$, showing oscillations that are periodic in $B$ between the quantum Hall plateaux. The second, third, and fourth curves from the bottom have offsets of, respectively, 5, 10, and 15 ${\rm k}\Omega$. The rapid oscillations below 1 T are Shubnikov-De Haas oscillations periodic in $1/B$, originating from the wide 2DEG regions. The sharp peak around $B = 0\,{\rm T}$ originates from the ohmic contacts. Taken from P. H. M. van Loosdrecht et al., Phys.\ Rev.\ B {\bf 38}, 10162 (1988).
\label{fig101}
}
\end{figure}

\begin{figure}
\centerline{\includegraphics[width=8cm]{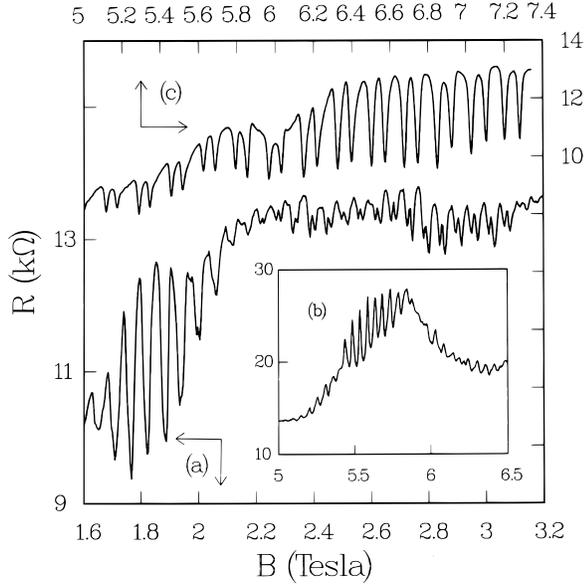}}
\caption{
Curves a and b are close-ups of the curve for $V_{\rm g} = -1.7\,{\rm V}$ in Fig.\ \ref{fig101}. Curve c is a separate measurement on the same device (note the different field scale due to a change in electron density in the constriction). Taken from P. H. M. van Loosdrecht et al., Phys.\ Rev.\ B {\bf 38}, 10162 (1988).
\label{fig102}
}
\end{figure}

{\bf (a) Point contact.} Aharonov-Bohm oscillations in the magnetoresistance of
a quantum point contact were discovered by van Loosdrecht et al.\cite{ref292} The
magnetic field dependence of the two-terminal resistance is shown in Fig.\ \ref{fig101},
for various gate voltages. The periodic oscillations occur predominantly
between quantum Hall plateaux, in a limited range of gate voltages, and only
at low temperatures (in Fig.\ \ref{fig101}, $T=50\,\mathrm{mK}$; the effect has disappeared at
$1\, \mathrm{K}$). The fine structure is very well reproducible if the sample is kept in the
cold, but changes after cycling to room temperature. As one can see from the
enlargements in Fig.\ \ref{fig102}, a splitting of the peaks occurs in a range of magnetic
fields, presumably as spin splitting becomes resolved. A curious aspect of the
effect (which has remained unexplained) is that the oscillations have a much
larger amplitude in one field direction than in the other (see Fig.\ \ref{fig101}), in
apparent conflict with the $\pm B$ symmetry of the two-terminal resistance
required by the reciprocity relation (\ref{eq12.16}) in the absence of magnetic
impurities. Other devices of the same design did not show oscillations of well-defined periodicity and had a two-terminal resistance that was approximately
$\pm B$ symmetric.

\begin{figure}
\centerline{\includegraphics[width=8cm]{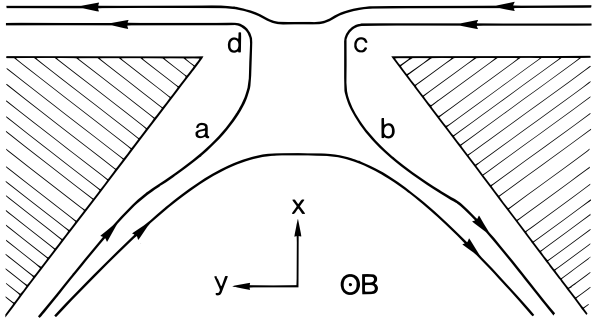}}
\caption{
Equipotentials at the guiding center energy in the saddle-shaped potential created by a split gate (shaded). Aharonov-Bohm oscillations in the point contact magnetoresistance result from the interference of tunneling paths ab and adcb. Tunneling from $a$ to $b$ may be assisted by an impurity at the entrance of the constriction. Taken from P. H. M. van Loosdrecht et al., Phys.\ Rev.\ B {\bf 38}, 10162 (1988).
\label{fig103}
}
\end{figure}

Figure \ref{fig103} illustrates the tunneling mechanism for the periodic magnetoresistance oscillations as it was originally proposed\cite{ref292} to explain the
observations. Because of the presence of a barrier in the point contact, the
electrostatic potential has a saddle form. Equipotentials at the guiding center
energy (\ref{eq18.1}) are drawn schematically in Fig.\ \ref{fig103} (arrows indicate the
direction of motion along the equipotential). An electron that enters the
constriction at $a$ can be reflected back into the broad region by tunneling to
the opposite edge, either at the potential step at the entrance of the
constriction (from $a$ to $b$) or at its exit (from $d$ to $c$). These two tunneling paths
acquire an Aharonov-Bohm phase difference\cite{ref499} of $eBS/\hbar$ (were $S$ is the
enclosed area {\it abcd}), leading to periodic magnetoresistance oscillations. (Note
that the periodicity $\Delta B$ may differ\cite{ref438,ref500} somewhat from the usual expression
$\Delta B=h/eS$, since $S$ itself is $B$-dependent due to the $B$-dependence of the
guiding center energy.) This mechanism shows how an Aharonov-Bohm
effect is possible in principle in a singly connected geometry: The point
contact behaves as if it were multiply connected, by virtue of the spatial
separation of edge channels moving in opposite directions. (Related mechanisms, based on circulating edge currents, have been considered for
Aharonov-Bohm effects in small conductors.\cite{ref473,ref474,ref501,ref502,ref503}) The oscillations
periodic in $B$ are only observed at large magnetic fields (above about $1\, \mathrm{T}$; the
oscillations at lower fields are Shubnikov-De Haas oscillations periodic in
$1/B$, due to the series resistance of the wide 2DEG regions). At low magnetic
fields the spatial separation of edge channels responsible for the Aharanov-Bohm effect is not yet effective. The spatial separation can also be destroyed
by a large negative gate voltage (top curve in Fig.\ \ref{fig101}), when the width of the
point contact becomes so small that the wave functions of edge states at
opposite edges overlap.

Although the mechanism illustrated in Fig.\ \ref{fig103} is attractive because it is
an intrinsic consequence of the point contact geometry, the observed well-defined periodicity of the magnetoresistance oscillations requires that the
potential induced by the split gate varies rapidly over a short distance (in
order to have a well-defined area $S$). A smooth saddle potential seems more
realistic. Moreover, one would expect the periodicity to vary more strongly
with gate voltage than the small $10\%$ variation observed experimentally as $V_{\mathrm{g}}$
is changed from $-1.4$ to $-1.7$ V. Glazman and Jonson\cite{ref438} have proposed
that one of the two tunneling processes (from $a$ to $b$ in Fig.\ \ref{fig103}) is mediated
by an impurity outside but close to the constriction. The combination of
impurity and point contact introduces a well-defined area even for a smooth
saddle potential, which moreover will not be strongly gate-voltage-dependent. Such an impurity-assisted Aharonov-Bohm effect in a quantum
point contact has been reported by Wharam et al.\cite{ref504} In order to study the
Aharonov-Bohm effect due to interedge channel tunneling under more
controlled conditions, a double-point contact device is required, as discussed
below.

\begin{figure}
\centerline{\includegraphics[width=8cm]{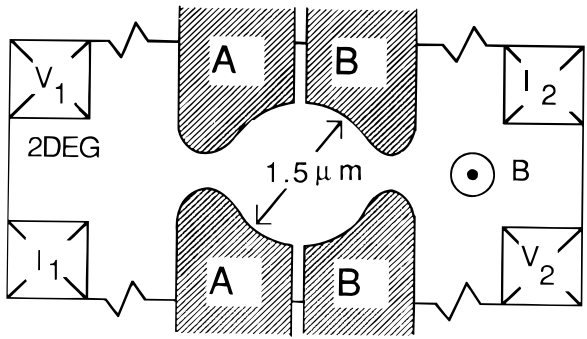}}
\caption{
Cavity (of $1.5\,\mu{\rm m}$ diameter) defined by a double set of split gates A and B. For large negative gate voltages the 2DEG region under the narrow gap between gates A and B is fully depleted, while transmission remains possible over the potential barrier in the wider openings at the left and right of the cavity. Taken from B. J. van Wees et al., Phys.\ Rev.\ Lett.\ {\bf 62}, 2523 (1989).
\label{fig104}
}
\end{figure}

\begin{figure}
\centerline{\includegraphics[width=8cm]{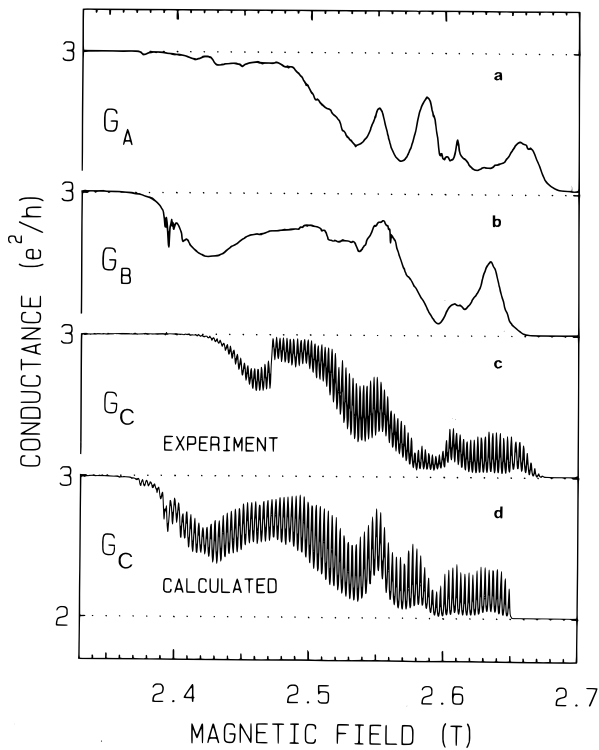}}
\caption{
Magnetoconductance experiments on the device of Fig.\ \ref{fig104} at 6 mK, for a fixed gate voltage of $-0.35\,{\rm V}$. (a) Conductance of point contact A, measured with gate B grounded. (b) Conductance of point contact B (gate A grounded). (c) Measured conductance of the entire cavity. (d) Calculated conductance of the cavity, obtained from Eqs.\ (\ref{eq21.1}) and (\ref{eq21.2}) with the measured $G_{\rm A}$ and $G_{\rm B}$ as input. Taken from B. J. van Wees et al., Phys.\ Rev.\ Lett.\ {\bf 62}, 2523 (1989).
\label{fig105}
}
\end{figure}

\begin{figure}
\centerline{\includegraphics[width=8cm]{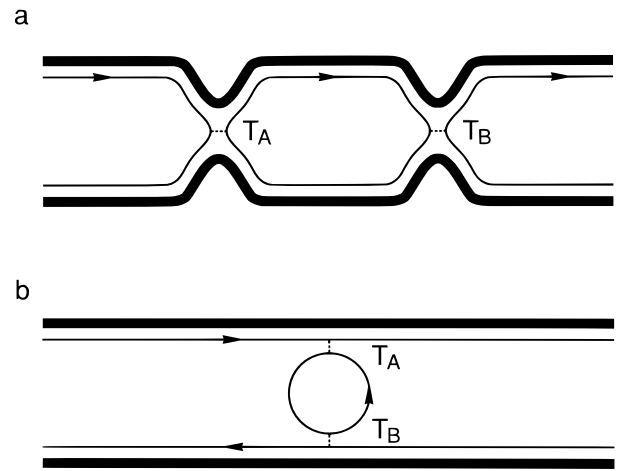}}
\caption{
Illustration of mechanisms leading to Aharonov-Bohm oscillations in singly connected geometries. (a) Cavity containing a circulating edge state. Tunneling through the left and right barriers (as indicated by dashed lines) occurs with transmission probabilities $T_{\rm A}$ and $T_{\rm B}$. On increasing the magnetic field, resonant tunneling through the cavity occurs periodically each time the flux $\Phi$ enclosed by the circulating edge state increases by one flux quantum $h/e$. (b) A circulating edge state bound on a local potential maximum causes resonant backscattering, rather than resonant transmission.
\label{fig106}
}
\end{figure}

{\bf (b) Cavity.} Van Wees et al.\cite{ref500} performed magnetoresistance experiments in
a geometry shown schematically in Fig.\ \ref{fig104}. A cavity with two opposite point
contact openings is defined in the 2DEG by split gates. The diameter of the
cavity is approximately $1.5\,\mu \mathrm{m}$. The conductances $G_{\mathrm{A}}$ and $G_{\mathrm{B}}$ of the two point
contacts A and B can be measured independently (by grounding one set of
gates), with the results plotted in Fig.\ \ref{fig105}a,b (for $V_{\mathrm{g}}=-0.35\,\mathrm{V}$ on either gate
A or B). The conductance $G_{\mathrm{C}}$ of the cavity (for $V_{\mathrm{g}}=-0.35\,\mathrm{V}$ on both the split
gates) is plotted in Fig.\ \ref{fig105}c. A long series of periodic oscillations is observed
between two quantum Hall plateaux. Similar series of oscillations (but with a
different periodicity) have been observed between other quantum Hall
plateaux. The oscillations are suppressed on the plateaux themselves. The
amplitude of the oscilIations is comparable to that observed in the experiment on a single point contact\cite{ref292} (discussed before), but the period is much
smaller (consistent with a larger effective area in the double-point contact
device), and no splitting of the peaks is observed (presumably due to a fully
resolved spin degeneracy). No gross $\pm B$ asymmetries were found in the
present experiment, although an accurate test of the symmetry on field
reversal was not possible because of difficulties with the reproducibility. The
oscillations are quite fragile, disappearing when the temperature is raised
above $200\, \mathrm{mK}$ or when the voltage across the device exceeds $40\,\mu \mathrm{V}$ (the data
in Fig.\ \ref{fig105} were taken at $6\, \mathrm{mK}$ and 6 $\mu \mathrm{V}$). The experimental data are well
described by resonant transmission through a circulating edge state in the
cavity,\cite{ref500} as illustrated in Fig.\ \ref{fig106}a and described in detail later. Aharonov-Bohm oscillations due to resonant transmission through a similar structure
have been reported by Brown et al.\cite{ref505} and analyzed theoretically by
Yosephin and Kaveh.\cite{ref506}

{\bf (c) Resonant transmission and reflection of edge channels.} The electrostatic
 potential in a point contact has a saddle shape (cf.\ Fig.\ \ref{fig103}), due to the
combination of the lateral confinement and the potential barrier. The height
of the barrier can be adjusted by means of the gate voltage. An edge state with
a guiding center energy below the barrier height is a bound state in the cavity
formed by two opposite point contacts, as is illustrated in Fig.\ \ref{fig106}a.
Tunneling of edge channels through the cavity via this bound state occurs
with transmission probability $T_{\mathrm{AB}}$, which for a singe edge channel is given
by\cite{ref474,ref498}
\begin{eqnarray}
T_{\mathrm{AB}}&=&\left| \frac{t_{\mathrm{A}}t_{\mathrm{B}}}{1-r_{\mathrm{A}}r_{\mathrm{B}} \exp(\mathrm{i}\Phi e/h)}\right|^{2}\nonumber\\
&=&\frac{T_{\mathrm{A}}T_{\mathrm{B}}}{1+R_{\mathrm{A}}R_{\mathrm{B}}-2(R_{\mathrm{A}}R_{\mathrm{B}})^{1/2}\cos(\phi_{0}+\Phi e/\hbar)}.\nonumber\\
&&
\label{eq21.1}
\end{eqnarray}
Here $t_{\mathrm{A}}$ and $r_{\mathrm{A}}$ are the transmission and reflection probability amplitudes
through point contact $\mathrm{A}$, $T_{\mathrm{A}}\equiv|t_{\mathrm{A}}|^{2}$, and $R_{\mathrm{A}}\equiv|r_{\mathrm{A}}|^{2}=1-T_{\mathrm{A}}$ are the transmission and reflection probabilities, and $t_{\mathrm{B}}, r_{\mathrm{B}}, T_{\mathrm{B}}, R_{\mathrm{B}}$ denote the corresponding quantities for point contact B. In Eq.\ (\ref{eq21.1}) the phase acquired by the
electron on one revolution around the cavity is the sum of the phase $\phi_{0}$ from
the reflection probability amplitudes (which can be assumed to be only
weakly $B$-dependent) and of the Aharonov-Bohm phase $\Phi\equiv BS$, which
varies rapidly with $B$ ($\Phi$ is the flux through the area $S$ enclosed by the
equipotential along which the circulating edge state is extended). Resonant
transmission occurs periodically with $B$, whenever $\phi_{0}+\Phi e/\hbar$ is a multiple of
$2\pi$. In the weak coupling limit ($T_{\mathrm{A}}, T_{\mathrm{B}}\ll 1$), Eq.\ (\ref{eq21.1}) is equivalent to the
Breit-Wigner resonant tunneling formula (\ref{eq17.1}). This equivalence has been
discussed by B\"{u}ttiker,\cite{ref386} who has also pointed out that the Breit-Wigner
formula is more generally applicable to the case that several edge channels
tunnel through the cavity via the same bound state.

In the case that only a single (spin-split) edge channel is occupied in the
2DEG, the conductance $G_{\mathrm{C}}=(e^{2}/h)T_{\mathrm{AB}}$ of the cavity follows directly from Eq.\
(\ref{eq21.1}). The transmission and reflection probabilities can be determined
independently from the individual point contact conductances $G_{\mathrm{A}}=(e^{2}/h)T_{\mathrm{A}}$
(and similarly for $G_{\mathrm{B}}$), at least if one may assume that the presence of the
cavity has no effect on $T_{\mathrm{A}}$ and $T_{\mathrm{B}}$ itself (but only on the total transmission
probability $T_{\mathrm{AB}}$). If $N>1$ spin-split edge channels are occupied and the
$N-1$ lowest-index edge channels are fully transmitted, one can write
\begin{eqnarray}
&&G_{\mathrm{C}}= \frac{e^{2}}{h}(N-1+T_{\mathrm{AB}}),\;\;G_{\mathrm{A}}= \frac{e^{2}}{h}(N-1+T_{\mathrm{A}}),\nonumber\\
&& G_{\mathrm{B}}= \frac{e^{2}}{h}(N-1+T_{\mathrm{B}}).
\label{eq21.2}
\end{eqnarray}
Van Wees et al.\cite{ref500}  have compared this simple model with their experimental
data, as shown in Fig.\ \ref{fig105}. The trace in Fig.\ \ref{fig105}d has been calculated from
Eqs.\ (\ref{eq21.1}) and (\ref{eq21.2}) by using the individual point contact conductances in
Fig.\ \ref{fig105}a,b as input for $T_{\mathrm{A}}$ and $T_{\mathrm{B}}$. The flux $\Phi$ has been adjusted to the
experimental periodicity of $3\, \mathrm{mT}$, and the phase $\phi_{0}$ in Eq.\ (\ref{eq21.1}) has been
ignored (since that would only amount to a phase shift of the oscillations).
Energy averaging due to the finite temperature and voltage has been taken
into account in the calculation. The agreement with experimental trace (Fig.\
\ref{fig105}c) is quite satisfactory.

Resonant reflection of an edge channel can occur in addition to the
resonant transmission already considered. Aharonov-Bohm oscillations due
to interference of the reflections at the entrance and exit of a point contact,
illustrated in Fig.\ \ref{fig103}, are one example of resonant reflection.\cite{ref292}
Jain\cite{ref498} has
considered resonant reflection via a localized state circulating around a
potential maximum, as in Fig.\ \ref{fig106}b. Such a maximum may result naturally
from a repulsive scatterer or artificially in a ring geometry (cf.\ Fig.\ \ref{fig100}).
Tunneling of an edge state at each of the channel boundaries through the
localized state occurs with probabilities $T_{\mathrm{A}}$ and $T_{\mathrm{B}}$. The reflection probability
of the edge channel is still given by $T_{\mathrm{AB}}$ in Eq.\ (\ref{eq20.1}), but the channel
conductance $G_{\mathrm{C}}$ is now a decreasing function of $T_{\mathrm{AB}}$, according to
\be
G_{\mathrm{C}}= \frac{e^{2}}{h}(N-T_{\mathrm{AB}}). \label{eq21.3}
\ee 
Quasi-periodic magnetoresistance oscillations have been observed in narrow
channels by several groups.\cite{ref70,ref496,ref507} These may occur by resonant reflection
via one or more localized states in the channel, as in Fig.\ \ref{fig106}b.

\subsection{\label{sec22} Magnetically induced band structure}

The one-dimensional nature of edge channel transport has recently been
exploited in an innovative way by Kouwenhoven et al.\cite{ref250} to realize a one-dimensional superlattice exhibiting band structure in strong magnetic fields.
The one-dimensionality results because only the highest-index edge channel
(with the smallest guiding center energy) has an appreciable backscattering
probability. The $N-1$ lower-index edge channels propagate adiabatically,
with approximately unit transmission probability. One-dimensionality in
zero magnetic fields cannot be achieved with present techniques. That is one
important reason why the zero-field superlattice experiments described in
Section \ref{sec11} could not provide conclusive evidence for a bandstructure effect.
The work by Kouwenhoven et al.\cite{ref250} is reviewed in Section \ref{sec22a}. The
magnetically induced band structure differs in an interesting way from the
zero-field band structure familiar from solid-state textbooks, as we show in
Section \ref{sec22b}.

\subsubsection{\label{sec22a} Magnetotransport through a one-dimensional superlattice}

\begin{figure}
\centerline{\includegraphics[width=8cm]{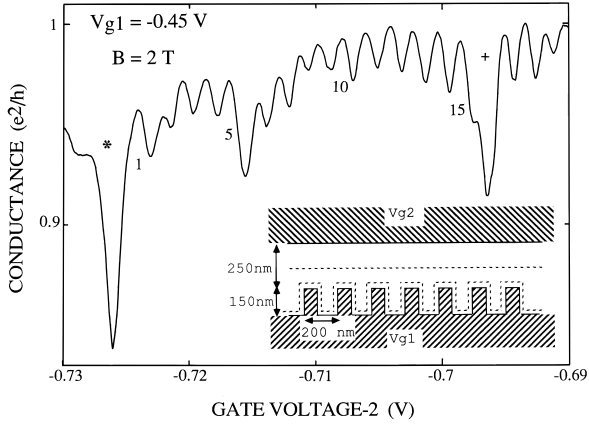}}
\caption{
Inset: Corrugated gate used to define a narrow channel with a one-dimensional periodic potential (the total number of barriers is 16, corresponding to 15 unit cells). Plotted is the conductance in a magnetic field of 2 T as a function of the voltage on the smooth gate at 10 mK. The deep conductance minima (marked by $+$ and $*$) are attributed to minigaps, and the 15 enclosed maxima to discrete states in the miniband. Taken from L. P. Kouwenhoven et al., Phys.\ Rev.\ Lett.\ {\bf 65}, 361 (1990).
\label{fig107}
}
\end{figure}

The device studied by Kouwenhoven et al.\cite{ref250} is shown in the inset of Fig.\ \ref{fig107}. A narrow channel is defined in the 2DEG of a GaAs-AlGaAs
heterostructure by two opposite gates. One of the gates is corrugated with
period $a=200\,\mathrm{nm}$, to introduce a periodic modulation of the confining
potential. At large negative gate voltages the channel consists of 15 cavities
[as in Section \ref{sec21b}(b)] coupled in series. The conductance of the channel was
measured at $10\, \mathrm{mK}$ in a fixed magnetic field of $2\, \mathrm{T}$, as a function of the voltage
on the gate that defines the smooth channel boundary. The results, reproduced in Fig.\ \ref{fig107}, show two pronounced conductance dips (of magnitude
$0.1\,e^{2}/h)$, with 15 oscillations in between of considerably smaller amplitude.
The two deep and widely spaced dips are attributed to minigaps, the more
rapid oscillations to discrete states in the miniband.

\begin{figure}
\centerline{\includegraphics[width=8cm]{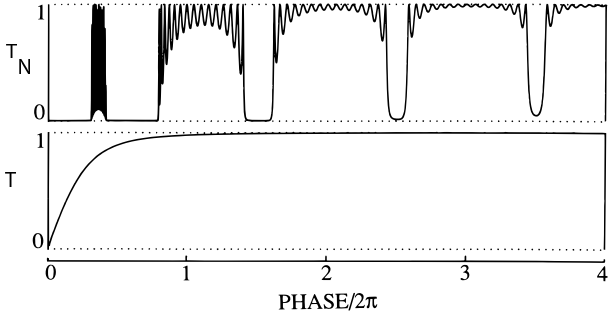}}
\caption{
Top: Calculated transmission probability $T_{N}$ of an edge channel through a periodic potential of $N = 15$ periods as a function of the Aharonov-Bohm phase $eBS/\hbar$ (with $S$ the area of one unit cell). The transmission probability through a single barrier is varied as shown in the bottom panel. Taken from L. P. Kouwenhoven et al., Phys.\ Rev.\ Lett.\ {\bf 65}, 361 (1990).
\label{fig108}
}
\end{figure}

This interpretation is supported in Ref.\ \onlinecite{ref250} by a calculation of the
transmission probability amplitude $t_{n}$ through $n$ cavities in series, given by
the recursion formula
\be
t_{n}= \frac{tt_{n-1}}{1-rr_{n-1} \exp(\mathrm{i}\phi)}. \label{eq22.1}
\ee 
Here $t$ and $r$ are transmission and reflection probability amplitudes of the
barrier separating two cavities (all cavitities are assumed to be identical), and
$\phi=eBS/h$ is the Aharonov-Bohm phase for a circulating edge state
enclosing area $S$. Equation (\ref{eq22.1}) is a generalization of Eq.\ (\ref{eq21.1}) for a single
cavity. The dependence on $\phi$ of $T_{n}=|t_{n}|^{2}$ shown in Fig.\ \ref{fig108} is indeed
qualitatively similar to the experiment. Deep minima in the transmission
probability occur with periodicity $\Delta\phi=2\pi$. Experimentally (where $S$ is
varied via the gate voltage at constant $B$) this would correspond to
oscillations with periodicity $\Delta S=h/eB$ of Aharonov-Bohm oscillations in a
single cavity. The 15 smaller oscillations between two deep minima have the
periodicity of Aharonov-Bohm oscillations in the entire area covered by the
15 cavities. The observation of such faster oscillations shows that phase
coherence is maintained in the experiment throughout the channel and
thereby provides conclusive evidence for band structure in a lateral
superlattice.

\subsubsection{\label{sec22b} Magnetically induced band structure}

{\bf (a) Skew minibands.} The band structure in the experiment of Kouwenhoven
et al.\cite{ref250} is present only in the quantum Hall effect regime and can thus be said
to be {\it magnetically induced}. The magnetic field breaks time-reversal symmetry.
Let us see what consequences that has for the band structure.

The hamiltonian in the Landau gauge ${\bf A}=(0, Bx, 0)$ is
\be
{\cal H}=\frac{p_{x}^{2}}{2m}+\frac{(p_{y}+eBx)^{2}}{2m}+V(x, y),\;\;V(x, y+a)=V(x, y), \label{eq22.2}
\ee 
where $V$ is the periodically modulated confining potential. Bloch's theorem is
not affected by the presence of the magnetic field, since ${\cal H}$ remains periodic in
$y$ (in the Landau gauge). The eigenstates $\Psi$ have the form
\be
\Psi_{nk}(x, y)=\mathrm{e}^{i\mathrm{k}y}f_{nk}(x, y),\;\;f_{nk}(x, y+a)=f_{nk}(x, y), \label{eq22.3}
\ee 
where the function $f$ is a solution periodic in $y$ of the eigenvalue problem
\begin{eqnarray}
\left( \frac{p_{x}^{2}}{2m}+\frac{(p_{y}+\hbar k+eBx)^{2}}{2m}+V(x, y)\right)f_{nk}(x, y)\nonumber\\
=E_{n}(k, B)f_{nk}(x, y). \label{eq22.4}
\end{eqnarray}
If the wave number $k$ is restricted to the first Brillouin zone $|k|<\pi/a$, the
index $n$ labels both the subbands from the lateral confinement and the
minibands from the periodic modulation. Since $E$ and $V$ are real, one finds by
taking the complex conjugate of Eq.\ (\ref{eq22.4}) that
\be
E_{n}(k, B)=E_{n}(-k, -B). \label{eq22.5}
\ee 
In zero magnetic fields the energy $E$ is an even function of $k$, regardless of the
symmetry of the potential $V$. This can be viewed as a consequence of time-reversal symmetry.\cite{ref508} In nonzero magnetic fields, however, $E$ is only even in
$k$ if the lateral confinement is symmetric:
\be
E_{n}(k, B)=E_{n}(-k, B)\;;\mbox{only if}\;\; V(x, y)=V(-x, y). \label{eq22.6}
\ee

\begin{figure}
\centerline{\includegraphics[width=8cm]{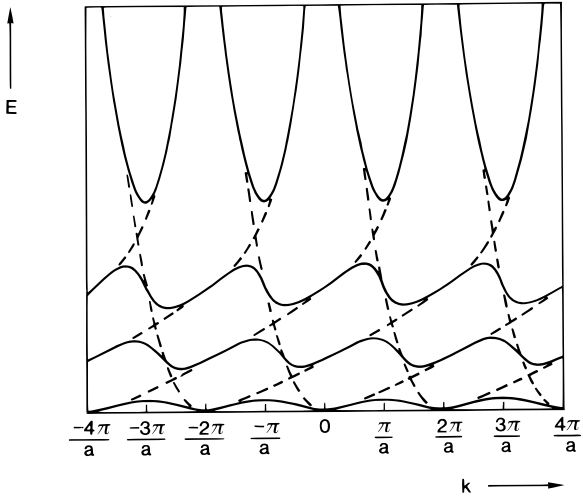}}
\caption{
Illustration of magnetically induced band structure in a narrow channel with a weak periodic modulation of the confining potential $V(x)$ (for the case $V(x)\neq V(-x)$). The dashed curves represent the unperturbed dispersion relation (\ref{eq22.7}) for a single Landau level. Skew minibands result from the broken time-reversal symmetry in a magnetic field.
\label{fig109}
}
\end{figure}

To illustrate the formation of {\it skew\/} minibands in a magnetically induced
band structure, we consider the case of a weak periodic modulation $V_{1}(y)$ of
the confining potential $V(x, y)=V_{0}(x)+V_{1}(x, y)$. The dispersion relation
$E_{n}^{0}(k)$ in the absence of the periodic modulation can be approximated by
\be
E_{n}^{0}(k)=(n- {\textstyle\frac{1}{2}})\hbar\omega_{\mathrm{c}}+V_{0}(x=-kl_{m}^{2}). \label{eq22.7}
\ee 
The index $n$ labels the Landau levels, and the wave number $k$ runs from $-\infty$
to $+\infty$. The semiclassical approximation (\ref{eq22.7}) is valid if the confining
potential $V_{0}$ is smooth on the scale of the magnetic length $l_{\mathrm{m}}\equiv(\hbar/eB)^{1/2}$.
[Equation (\ref{eq22.7}) follows from the guiding center energy (\ref{eq18.1}), using the
identity $x\equiv-k\hbar/eB$ between the guiding center coordinate and the wave
number; cf.\ Section \ref{sec12a}] For simplicity we restrict ourselves to the strictly
one-dimensional case of one Landau level and suppress the Landau level
index in what follows. To first order in the amplitude of the periodic
modulation $V_{1}$, the zeroth-order dispersion relation is modified only near the
points of degeneracy $K_{p}$ defined by
\be
E^{0}[K_{p}-p(2\pi/a)]=E^{0}(K_{p}),\;\;p=\pm 1, \pm 2, \ldots. \label{eq22.8}
\ee 
A gap opens near $K_{p}$, leading to the formation of a band structure as
illustrated in Fig.\ \ref{fig109}. The gaps do not occur at multiples of $\pi/a$, as in a
conventional 1D band structure. Moreover, the maxima and minima of two
subsequent bands occur at different $k$-values. This implies {\it indirect\/} optical
transitions between the bands if the Fermi level lies in the gap.

It is instructive to consider the special case of a parabolic confining
potential $V_{0}( x)=\frac{1}{2}m\omega_{0}^{2}x^{2}$ in more detail, for which the zeroth-order dispersion relation can be obtained exactly (Section \ref{sec10}). Since the confinement is
symmetric in $x$, the minigaps in this case occur at the Brillouin zone
boundaries $k=p\pi/a$. Other gaps at points where the periodic modulation
induces transitions between different 1D subbands are ignored for simplicity.
From Eq.\ (\ref{eq10.5}) one then finds that the Fermi energy lies in a minigap when
\be
E_{\mathrm{F}}=(n- {\textstyle\frac{1}{2}})\hbar\omega+\frac{\hbar^{2}}{2M}\left(\frac{p\pi}{a}\right)^{2}, \label{eq22.9}
\ee 
with the definitions $\omega\equiv(\omega_{\mathrm{c}}^{2}+\omega_{0}^{2})^{1/2}$, $M\equiv m\omega^{2}/\omega_{0}^{2}$. In the limiting case
$B=0,$ Eq.\ (\ref{eq22.9}) reduces to the usual condition\cite{ref249} that Bragg reflection
occurs when the longitudinal momentum $mv_{y}$ is a multiple of $\hbar\pi/a$. In the
opposite limit of strong magnetic fields $(\omega_{\mathrm{c}}\gg  \omega_{0}),$ Eq.\ (\ref{eq22.9}) becomes
\be
aW_{\mathrm{eff}}B=p \frac{h}{e},\;\;W_{\mathrm{eff}} \equiv 2\left(\frac{2E_{\mathrm{G}}}{m\omega_{0}^{2}}\right)^{1/2}. \label{eq22.10}
\ee
The effective width $W_{\mathrm{eff}}$ of the parabolic potential is the separation of the
equipotentials at the guiding center energy $E_{\mathrm{G}} \equiv E_{\mathrm{F}}-(n-\frac{1}{2})\hbar\omega_{\mathrm{c}}$.

The two-terminal conductance of the periodically modulated channel
drops by $e^{2}/h$ whenever $E_{\mathrm{F}}$ lies in a minigap. If the magnetic field dependence
of $W_{\mathrm{eff}}$ is small, then Eq.\ (\ref{eq22.10}) shows that the magnetoconductance
oscillations have approximately the periodicity $\Delta B\sim h/eaW_{\mathrm{eff}}$ of the
Aharonov-Bohm effect in a single unit cell, in agreement with the calculations of Kouwenhoven et al.\cite{ref250} (Note that in their experiment the Fermi
energy is tuned through the minigap by varying the gate voltage rather than
the magnetic field.) The foregoing analysis is for a channel of infinite length.
The interference of reflections at the entrance and exit of a finite superlattice
of length $L$ leads to transmission resonances\cite{ref249,ref387} whenever $k=p\pi/L$, as
described by Eqs.\ (\ref{eq22.9}) and (\ref{eq22.10}) after substituting $L$ for $a$. These
transmission resonances are observed by Kouwenhoven et al.\ as rapid
oscillations in the conductance. The number of conductance maxima between
two deep minima from the minigap equals approximately the number $L/a$ of
unit cells in the superlattice. The number of maxima may become somewhat
larger than $L/a$ if one takes into account reflections at the transition from a
narrow channel to a wide 2DEG. This might explain the observation in Ref.\
\onlinecite{ref250} of 16, rather than 15, conductance maxima between two minigaps in one
particular experiment on a 15-period superlattice.

{\bf (b) Bloch oscillations.} In zero magnetic fields, an oscillatory current has
been predicted to occur on application of a dc electric field to an electron gas
in a periodic potential.\cite{ref509} This {\it Bloch oscillation\/} would result from Bragg
reflection of electrons that, accelerated by the electric field, approach the
band gap. A necessary condition is that the field be sufficiently weak that
tunneling across the gap does not occur.\cite{ref510,ref511,ref512,ref513} The wave number increases
in time according to $\dot{k}=eE/\hbar$ in an electric field $E$. The time interval between
two Bragg reflections is $2\pi/a\dot{k}=h/eaE$. The oscillatory current thus would
have a frequency $\Delta Ve/h$, with $\Delta V=aE$ the electrostatic potential drop over
one unit cell. Bloch oscillations have so far eluded experimental observation.

The successful demonstration\cite{ref250} of miniband formation in strong magnetic fields naturally leads to the question of whether Bloch oscillations might
be observable in such a system. This question would appear to us to have a
negative answer. The reason is simple, and it illustrates another interesting
difference of magnetically induced band structure. In the quantum Hall effect
regime the electric field is perpendicular to the current, so no acceleration of
the electrons occurs. Since $\dot{k}=0$, no Bloch oscillations should be expected.

\end{document}